\documentclass[12pt,oneside,phd]{dms}

\usepackage[utf8]{inputenc}
\usepackage[T1]{fontenc}

\usepackage{lmodern}

\anglais

\def\sloppy{%
  \tolerance 500%
  \emergencystretch 3em%
  \hfuzz .5pt
  \vfuzz\hfuzz}
\usepackage{microtype}

\usepackage{graphicx}
\usepackage{subcaption}
\usepackage{float}
\usepackage{booktabs}
\usepackage{tabularx}
\usepackage{array}
\usepackage{multirow}
\usepackage{longtable}
\usepackage{ragged2e}
\usepackage{amssymb}

\usepackage{listings}
\usepackage{xcolor}

\usepackage{comment}

\DeclareRobustCommand{\added}[1]{#1}
\DeclareRobustCommand{\deleted}[1]{}

\excludecomment{deletedblock}

\usepackage{doi}

\usepackage{csquotes}
\usepackage{enumitem}
\setlist[itemize]{leftmargin=*,itemsep=0.2\baselineskip,topsep=0.2\baselineskip}
\setlist[enumerate]{leftmargin=*,itemsep=0.2\baselineskip,topsep=0.2\baselineskip}

\usepackage{tikz}
\usetikzlibrary{arrows.meta,positioning,shapes.misc,shapes,fit,calc}
\usepackage{adjustbox}
\usepackage{pdfpages}

\usepackage[authoryear,round]{natbib}

\newif\ifincludeappendices
\includeappendicestrue 

\usepackage[labelfont=bf, width=\linewidth]{caption}

\usepackage{placeins}
\counterwithin{figure}{chapter}
\AtBeginDocument{\renewcommand{\thefigure}{\thechapter.\arabic{figure}}}
\usepackage{etoolbox}
\pretocmd{\chapter}{\FloatBarrier}{}{}

\usepackage{hyperref}
\hypersetup{colorlinks=true,allcolors=black}
\usepackage{hypcap}
\usepackage{bookmark}

\usepackage[capitalize,noabbrev,nameinlink]{cleveref}

\usepackage[most]{tcolorbox}

\newcolumntype{L}{>{\raggedright\arraybackslash}X}
\newcolumntype{C}{>{\centering\arraybackslash}p{0.11\textwidth}}
\newcolumntype{P}[1]{>{\raggedright\arraybackslash}p{#1}}
\newcolumntype{Q}[1]{>{\centering\arraybackslash}p{#1}}
\newcolumntype{R}[1]{>{\raggedleft\arraybackslash}p{#1}}
\newcolumntype{Y}{>{\raggedright\arraybackslash}X}

\makeatletter
\let\Oldincludegraphics\includegraphics
\newcommand{\includegraphicsmaybe}[2][]{%
  \IfFileExists{\detokenize{#2}}{%
    \Oldincludegraphics[#1]{#2}%
  }{%
    \fbox{\parbox[c][0.22\textheight][c]{0.9\linewidth}{\centering Missing figure file: \texttt{\detokenize{#2}}}}%
  }%
}
\renewcommand{\includegraphics}[2][]{\includegraphicsmaybe[#1]{#2}}
\makeatother

\newcommand{\RightToAI}{Right to AI}
\newcommand{\DelibCell}{deliberative cell}
\newcommand{\DelibCells}{deliberative cells}
\newcommand{\ValueRegister}{value register}

\begin{document}

\version{1}
\pagenumbering{roman}

\title{The urban right to AI: Pluralistic co-design and governance of public space}
\author{Rashid Ahmad Mushkani}
\copyrightyear{2026}

\department{\'Ecole d'urbanisme et d'architecture de paysage\\Facult\'e de l'Am\'enagement\\Universit\'e de Montr\'eal}

\date{juillet 2026}

\sujet{interdisciplinaire en am\'enagement}

\president{Professeur Heather Braiden}
\directeur{Professeur Shin Koseki (remplacé par Èvelyne Brie)}

\membrejury{Professeur Michael Robert Doyle}
\examinateur{Professeur Tan Yigitcanlar}

\pdfbookmark[chapter]{Couverture}{PageUn}

\maketitle
\maketitle

\francais
\chapter*{R\'esum\'e}

Les syst\`emes d'intelligence artificielle m\'ediatisent de plus en plus la fa\c{c}on dont les villes per\c{c}oivent, \'evaluent et transforment l'espace public. Lorsque ces syst\`emes sont consid\'er\'es comme de simples outils (voire comme des outils neutres), la couche algorithmique qui rend certains ph\'enom\`enes visibles et gouvernables est laiss\'ee aux logiques de march\'e, aux choix des fournisseurs ou aux contraintes techniques. Cette th\`ese soutient que l'urbanisme contemporain doit \^etre gouvern\'e comme une double infrastructure~: la ville mat\'erielle et une couche \'epist\'emique et algorithmique qui produit des repr\'esentations, des scores et des alternatives g\'en\'eratives, et qui influence de plus en plus les d\'ecisions municipales.

La th\`ese d\'eveloppe un \emph{droit civique \`a l'IA} appliqu\'e \`a l'espace public et l'op\'erationnalise \`a travers l'\emph{alignement pluraliste}. Plut\^ot que d'imposer une norme universelle d'inclusion, l'alignement pluraliste exige que les villes fassent \'emerger des valeurs h\'et\'erog\`enes, les repr\'esentent sans les r\'eduire \`a une cible unique, puis n\'egocient des normes l\'egitimes pouvant \^etre traduites en code et inscrites dans les am\'enagements comme dans les syst\`emes d'IA.

La recherche r\'epond \`a cinq questions portant sur~: (1) la mani\`ere dont les municipalit\'es doivent conceptualiser la couche algorithmique fa\c{c}onnant l'espace public et les responsabilit\'es qui en d\'ecoulent~; (2) ce que le droit \`a l'IA exige et interdit dans l'espace public, notamment en mati\`ere de participation, de transparence et de recours~; (3) l'ampleur des divergences de perception des rues selon les positions sociales et les dimensions plus ou moins susceptibles de converger par la d\'elib\'eration~; (4) la construction d'une cha\^\i ne allant de la mesure \`a la gouvernance, capable de pr\'eserver l'h\'et\'erog\'en\'eit\'e sans la r\'eduire \`a un score unique, et l'effet de ces repr\'esentations sur ce qui devient gouvernable~; et (5) les pratiques de gouvernance du cycle de vie et d'approvisionnement permettant d'inscrire des normes n\'egoci\'ees dans les mod\`eles comme dans les rues au fil du temps, y compris des m\'ecanismes de recommissionnement, d'audit et de r\'eparation.

Empiriquement, la th\`ese s'appuie sur des \'etudes participatives \`a Montr\'eal combinant entretiens, groupes de discussion et notation et classement structur\'es d'images de rues. Une premi\`ere \'etude mobilise 12 participants qui \'evaluent 20 rues repr\'esent\'ees par 60 points de vue, en comparant l'inclusivit\'e, l'accessibilit\'e, l'esth\'etique et la praticit\'e entre des r\'esidents (post-occupation) et des nouveaux arrivants (avant occupation). Une seconde \'etude mobilise 28 r\'esidents \`a travers des entretiens puis des exercices de notation et de classement, incluant des discussions en petits groupes. Dans l'ensemble, l'accord est plus \'elev\'e pour certaines dimensions visuellement saillantes que pour des dimensions sociales telles que l'inclusivit\'e. La d\'elib\'eration structur\'ee augmente la convergence sur certains crit\`eres sans \'eliminer les d\'esaccords r\'esiduels, ce qui indique que ces d\'esaccords doivent \^etre document\'es et gouvern\'es.

Sur le plan technique, la th\`ese d\'eveloppe Street Review comme une infrastructure de mesure participative et montre que des \'etiquettes co-produites peuvent soutenir des mod\`eles pr\'edictifs sensibles aux sous-groupes et une cartographie \`a l'\'echelle de la ville. Le pipeline combine l'\'elaboration participative de descripteurs \'evaluatifs avec un mod\`ele capable de pr\'edire des \'evaluations multi-crit\`eres et de produire des cartes de chaleur \`a partir d'environ 45,000 images au niveau de la rue. Les r\'esultats indiquent que la mise \`a l'\'echelle est possible et peut faire appara\^\i tre des motifs spatiaux utiles pour l'audit et la priorisation, tout en montrant que la qualit\'e des images et les limites de repr\'esentation restreignent ce qui peut \^etre gouvern\'e \`a partir d'entr\'ees visuelles seules.

La th\`ese introduit aussi LIVS (Local Intersectional Visual Spaces), un jeu de donn\'ees d'alignement pluraliste pour l'espace public inclusif, d\'evelopp\'e sur deux ans avec 30 organismes communautaires. LIVS regroupe 37,710 comparaisons par paires portant sur 13,462 images, structur\'ees selon six crit\`eres d\'eriv\'es de 634 concepts d\'efinis par les communaut\'es. En utilisant l'Optimisation directe par pr\'ef\'erences (DPO), ce travail affine un mod\`ele Stable Diffusion~XL et l'\'evalue \`a travers des \'etudes de cas portant sur l'alignement des pr\'ef\'erences et la neutralit\'e. Lors d’un atelier d’évaluation mené avec des résidents, 2,100 annotations supplémentaires ont été recueillies : 700 jugements favorisent le modèle DPO, 300 le modèle de référence et 1,100 sont neutres. Ces r\'esultats indiquent que l'affinage par pr\'ef\'erences peut am\'eliorer l'alignement avec une partie des jugements tout en laissant une large zone d'ind\'etermination, qui doit \^etre interpr\'et\'ee comme un signal de gouvernance.

La th\`ese synth\'etise enfin ces r\'esultats dans une architecture de mise en \oe uvre~: un \emph{droit civique \`a l'IA} comme couche de gouvernance municipale, un cycle de vie co-produit avec des points de contr\^ole du cadrage \`a la maintenance, et des m\'ecanismes d'approvisionnement et de supervision assurant l'auditabilit\'e, le recommissionnement et les voies de recours.

\textbf{Mots-cl\'es~:} urbanisme~; espace public~; gouvernance de l'IA~; IA participative~; alignement pluraliste~; intersectionnalit\'e~; d\'elib\'eration~; apprentissage par pr\'ef\'erences.

\anglais
\chapter*{Abstract}

AI systems increasingly mediate how cities perceive, evaluate, and transform public space. When these systems are treated as neutral tools, the algorithmic layer that shapes what becomes legible and actionable in planning is left to market incentives, vendor defaults, or technical convenience. This dissertation argues that contemporary urbanism must be governed as a dual infrastructure: the material city and an epistemic, algorithmic layer that produces representations, scores, and generative alternatives that increasingly influence municipal decisions.

The dissertation develops a civic \emph{Right to AI} for public-space contexts and operationalizes it through \emph{pluralistic alignment}. Rather than imposing a universal standard for what counts as an inclusive place, pluralistic alignment requires cities to elicit heterogeneous values, represent them without collapsing them into a single target, and negotiate legitimate standards that can be encoded into both physical design and the AI systems that increasingly mediate urban decisions.

The research addresses five questions concerning: (1) how municipalities should conceptualize the algorithmic layer shaping public space and the responsibilities that follow from it; (2) what a civic Right to AI requires and prohibits in public space, notably with respect to participation, transparency, and avenues for recourse; (3) the extent to which perceptions of streets diverge across social positions and which dimensions are more or less likely to converge through deliberation; (4) how cities can elicit and represent heterogeneous values without collapsing them into a single score, and how such representations change what becomes governable; and (5) how co-produced lifecycle governance and procurement practices can encode negotiated standards into models and streets over time, including mechanisms for recommissioning, audit, and redress.

Empirically, the dissertation draws on participatory studies in Montr\'eal that combine interviews, focus groups, and structured ratings and rankings of streetscape imagery. In one study, 12 participants evaluated 20 streets represented through 60 vantage points, comparing perceptions of inclusivity, accessibility, aesthetics, and practicality between post-occupancy residents and pre-occupancy newcomers. In a second study, 28 residents participated in interviews and then in rating and ranking exercises, including small-group discussions. Across studies, agreement is higher for some visually legible dimensions than for social dimensions such as inclusivity. Structured deliberation increases convergence on selected criteria without eliminating residual contestation, indicating that disagreement should be documented and governed rather than treated as annotation noise.

Technically, the dissertation develops Street Review as a participatory measurement infrastructure and demonstrates how locally co-produced labels can support subgroup-aware predictive modeling and citywide mapping. The Street Review pipeline combines participatory elicitation of evaluative descriptors with a scalable model that predicts multi-criteria streetscape evaluations and produces citywide heatmaps from approximately 45,000 street-view images. The results show that scaling with co-produced labels can surface spatial patterns useful for auditing and prioritization, but also that image data quality and representational limits constrain what can be governed through vision-only inputs.

The dissertation also introduces LIVS (Local Intersectional Visual Spaces), a pluralistic alignment dataset for inclusive public spaces developed through a two-year participatory process with 30 community organizations. LIVS encodes 37,710 pairwise comparisons across 13,462 images, structured along six criteria derived from 634 community-defined concepts. Using Direct Preference Optimization (DPO), the work fine-tunes a Stable Diffusion XL model and evaluates the fine-tuned model through case studies that examine preference alignment and neutrality. As part of a resident evaluation workshop, 2,100 additional annotations were collected. Of these, 700 judgments favored the DPO model, 300 favored the baseline, and 1,100 were neutral. These results indicate that preference tuning can improve alignment on a subset of judgments while leaving a large indeterminate region that should be treated as a governance signal rather than forced into optimization.

The dissertation synthesizes these findings into an implementation architecture: a Right to AI as a municipal governance layer, an augmented co-produced AI lifecycle with checkpoints from co-framing to co-maintenance, and procurement and oversight mechanisms that enable auditability, recommissioning, and recourse. Together, these elements specify how cities can govern the algorithmic layer of public space while treating pluralism as an empirical condition and a requirement of legitimacy.

\textbf{Keywords:} urban planning; public space; AI governance; participatory AI; pluralistic alignment; intersectionality; deliberation; preference learning.


\chapter*{Lay summary}

Cities in the twenty-first century need two kinds of infrastructure. One is physical, like streets and buildings. The other is digital, including data, knowledge, and AI systems. These systems should be guided by a public right to AI that serves everyone. Instead of forcing one definition of what makes a street “good” or “inclusive,” cities should listen to different communities, compare their needs, and openly work through disagreements. The shared decisions that come out of this process should shape both how cities are built and how AI is used to make urban decisions.

This thesis argues that urban AI systems should be treated as civic infrastructure governed by a \emph{Right to AI}. This right requires transparency, participatory power, and contestability in how AI systems that affect public space are designed and used. The thesis operationalizes this claim through pluralistic alignment: a framework in which cities elicit, compare, and legitimately negotiate heterogeneous values and encode the negotiated outcomes into both physical designs and AI systems.

Empirically, the thesis develops \emph{Street Review}, a participatory method in which residents evaluate street images based on lived experience. These evaluations are used to train an AI model that predicts how different social groups perceive accessibility, comfort, and inclusivity across a city. The thesis also introduces \emph{LIVS}, a community-co-produced dataset for aligning generative models with local public-space values while explicitly preserving neutral and conflicting preferences.

The findings show that disagreement and neutrality are not failures of measurement, but signals of contested values that require deliberation rather than optimization. By treating both streets and AI systems as objects that must be commissioned, evaluated, and periodically re-commissioned, this thesis provides a concrete governance framework for cities seeking to use AI without collapsing pluralism into a single, misleading notion of what counts as a “good” or “inclusive” public space.

\cleardoublepage
\pdfbookmark[chapter]{\contentsname}{toc}
\tableofcontents

\cleardoublepage
\phantomsection
\listoftables

\cleardoublepage
\phantomsection
\listoffigures

\chapter*{List of Acronyms and Abbreviations}
\begin{twocolumnlist}{.2\textwidth}{.7\textwidth}
  AAAI & Association for the Advancement of Artificial Intelligence\\
  ACM & Association for Computing Machinery\\
  AGI & Artificial General Intelligence\\
  AI & Artificial Intelligence\\
  AIES & AAAI/ACM Conference on AI, Ethics, and Society\\
  CLIP & Contrastive Language--Image Pre-training\\
  DEI & Diversity, Equity, and Inclusion\\
  DPO & Direct Preference Optimization\\
  FRQ & Fonds de recherche du Qu\'ebec (Research Funds of Qu\'ebec)\\
  GeoAI & Geospatial Artificial Intelligence\\
  GPT-4 & Generative Pre-trained Transformer 4\\
  HCI & Human--Computer Interaction\\
  IA & Artificial Intelligence (French: \emph{intelligence artificielle})\\
  ICC & Intraclass Correlation Coefficient\\
  ICML & International Conference on Machine Learning\\
  IEEE & Institute of Electrical and Electronics Engineers\\
  INRS & Institut national de la recherche scientifique (National Institute of Scientific Research)\\
  IVADO & Institute for Data Valorization\\
  JUM & Journal of Urban Management\\
  KPI & Key Performance Indicator\\
  LGBTQ2+ & Lesbian, Gay, Bisexual, Transgender, Queer/Questioning, and Two-Spirit (plus)\\
  LIVS & Local Intersectional Visual Spaces\\
  ML & Machine Learning\\
  MLP & Multi-Layer Perceptron\\
  NIST & National Institute of Standards and Technology\\
  NLP & Natural Language Processing\\
  OBVIA & International Observatory on the Societal Impacts of AI and Digital Technology\\
  PAU & Pluralistic-Alignment Urbanism\\
  PLM & Product Lifecycle Management\\
  PMLR & Proceedings of Machine Learning Research\\
  Right to AI & Civic right governing the use of artificial intelligence in public space, ensuring participation, transparency, and recourse\\
  RLQ--QLN & R\'eseau LGBT du Qu\'ebec / Quebec LGBT Network\\
  RQ1 & Research Question 1: Urban theory and governance\\
  RQ2 & Research Question 2: Rights and legitimacy\\
  RQ3 & Research Question 3: Pluralism as an empirical condition\\
  RQ4 & Research Question 4: Measurement-to-governance pipeline\\
  RQ5 & Research Question 5: Implementation realism\\
  SDLC & Software Development Life Cycle\\
  SDXL & Stable Diffusion XL\\
  T2I & Text-to-Image\\
  UNESCO & United Nations Educational, Scientific and Cultural Organization\\
\end{twocolumnlist}

\cleardoublepage
\chapter*{Dedication}

\begin{flushright}
\emph{To Ayat,\\
who carries what comes next.}

\vspace{1em}

\emph{To Afghanistan,\\
which carries everything before.}
\end{flushright}

\chapter*{Acknowledgements}

Research on public space often treats participation as an object of study.
This dissertation treats it as a condition of production.
The work was made through sustained engagement, disagreement, and care, across streets, workshops, annotation sessions, and model evaluation.
Its arguments about inclusion are inseparable from the relations that made them possible.

I thank my supervisor, Professor Shin Koseki, for guiding this research thoughtfully and for encouraging me to pursue meaningful questions.
Shin supported a research program that insists on crossing boundaries, between urban planning and machine learning, between fieldwork and computation, and between normative commitments and empirical accountability.
He gave me a rare combination of freedom and structure: freedom to pursue questions that did not fit neatly into a single discipline, and structure to ensure that the work remained rigorous, legible, and grounded.

Several parts of this dissertation are inseparable from the collaborators who shaped them.
I thank Hugo Berard for being a foundational collaborator across multiple projects.
Hugo brings a steady kind of rigor: the ability to sharpen a concept without shrinking it, to improve a method without losing the story it is meant to tell, and to be both demanding and generous in the same sentence.

I am grateful to Allison Cohen for her intellectual clarity and for her feedback, which consistently improved not just the writing, but the moral precision of the claims.
I thank Shravan Nayak for contributions that bridged technical experimentation with the practical demands of participatory research.

I also thank Toumadher Ammar for their collaboration, support, and the many forms of labor that make participatory projects real: coordination, facilitation, synthesis, and the quiet work of keeping a complex process coherent.

A large part of this dissertation exists because residents of Montr\'eal chose to participate.
To everyone who joined an interview, a focus group, a workshop, or an annotation session, thank you for your time, your candor, and your willingness to make your experience legible to a research process.
I hope this dissertation reflects your contributions with the seriousness they deserve.

This research was also supported by community organizations that opened doors, offered guidance, challenged assumptions, and helped connect the work to local priorities.
I am grateful to the organizations that supported recruitment and contextualization, including
R\'EZO; Soci\'et\'e qu\'eb\'ecoise de la d\'eficience intellectuelle; Alpar; Agir Montr\'eal; Aide aux Trans du Qu\'ebec; RLQ--QLN; ALAC; Association musulmane de Montr\'eal-Nord; Maison d'Ha\"iti; and Afrique au F\'eminin.
For the LIVS project, I am especially grateful to the community organizations engaged through workshops and annotation activities, including
the Congolese Community Center of Montr\'eal, Altergo, La Maisonn\'ee, the Cummings Centre, Projet Changement, the Women's Center of Plateau Mont-Royal, the LGBTQ+ Community Center of Montr\'eal, the Marguerite-Bourgeoys Hub, L'Agence On est l\`a!, and the Montr\'eal Women's Groups Table.
What you offered was more than participation; it was co-production in the deepest sense.

I also thank the collaborators and partners who helped connect this work to practice, and who made it possible to build tools and processes rather than only describing them.
I appreciate the collaboration of partner organizations and architectural offices, including Enclume, Sid Lee Architecture, Dark Matter Labs, IVADO, OBVIA, and the Canadian Commission for UNESCO.
I thank Emmanuel Beaudry-Marchand for developing the foundational schema of the \textit{Aipithet} platform used during annotation.
I thank Jerome Solis for coordination support, and I am grateful to Sarah Tannir, Leandry Jieutsa, Ad\`ele Kremer, Fr\'ed\'erique Roy, and Roxane Kasprzyk for contributions across workshops, data preparation, and project operations.

This research was supported by funding and institutional programs that made long-horizon, community-engaged work possible.
I gratefully acknowledge support from the Quebec Research Fund (FRQ, \texttt{10.69777/347989}) and the Social Sciences and Humanities Research Council (NFRFR-2021-00397).
The LIVS project was funded by the Universit\'e de Montr\'eal program \textit{Soutien aux initiatives avec les collectivit\'es et les entreprises, Collaboration avec les organismes communautaires}, as well as the \textit{Bourses en intelligence artificielle des Études supérieures et postdoctorales (ESP)} (Universit\'e de Montr\'eal).
I also acknowledge Mitacs support.
I am grateful as well for the computing resources provided by Mila, and by the Digital Research Alliance of Canada and Calcul Qu\'ebec, which enabled experiments that would otherwise have remained purely conceptual.

Finally, I thank my family and friends for their ongoing support and encouragement throughout this journey.

\chapter*{Publications Details and Author Contributions}

This dissertation is based on peer-reviewed publications produced in collaboration with co-authors.
For each publication, the contributions of all authors are summarized below.
I was the primary author and took the lead in study design, data analysis, and writing.

\medskip
\noindent\textbf{Chapter 3 ``The Right to AI''}, by Mushkani, R., Berard, H., Cohen, A., and Koseki, S., is published in the \emph{Proceedings of the 42nd International Conference on Machine Learning (ICML)}, \emph{Proceedings of Machine Learning Research (PMLR)}, 267, 81876--81896.
I led the argument, framing, and writing.
Berard and Cohen contributed to the argumentation.
Koseki provided supervision.
(\url{https://proceedings.mlr.press/v267/mushkani25b.html}
)

\medskip
\noindent\textbf{Chapter 4 ``Co-Producing AI: Toward an Augmented, Participatory Lifecycle''}, by Mushkani, R., Berard, H., Ammar, T., Chatonnier, C., and Koseki, S., is published in \emph{Proceedings of the AAAI/ACM Conference on AI, Ethics, and Society (AIES)}, 8(2), 1785--1799.
I led the conceptual framing and writing.
Berard, Ammar, and Chatonnier contributed through data gathering and the design and facilitation of workshops that informed the manuscript.
Koseki provided supervision.
(\url{https://doi.org/10.1609/aies.v8i2.36674}
)

\medskip
\noindent\textbf{Chapter 5 ``Street review: A participatory AI-based framework for assessing streetscape inclusivity''}, by Mushkani, R. and Koseki, S., is published in \emph{Cities}, 170, 106602.
I led the study, developed the framework, conducted the analyses, and wrote the manuscript.
Koseki supervised the work.
(\url{https://doi.org/10.1016/j.cities.2025.106602}
)

\medskip
\noindent\textbf{Chapter 6 ``LIVS: A Pluralistic Alignment Dataset for Inclusive Public Spaces''}, by Mushkani, R., Nayak, S., Berard, H., Cohen, A., Koseki, S., and Bertrand, H., is published in the \emph{Proceedings of the 42nd International Conference on Machine Learning (ICML)}, \emph{Proceedings of Machine Learning Research (PMLR)}, 267, 45311--45341.
I led the project, dataset and study design, and writing.
Nayak contributed the image generation pipeline and DPO.
Berard contributed through data gathering and the design of workshops.
Cohen contributed to governance and management support.
Bertrand contributed technical input on modeling and evaluation.
Koseki provided supervision.
(\url{https://proceedings.mlr.press/v267/mushkani25a.html}
)

\medskip
\noindent In addition to the chapter-linked manuscripts above, this dissertation also includes (and draws upon) the following peer-reviewed articles, which report complementary empirical results and methodological developments used across the thesis:

\medskip
\noindent\textbf{``Intersecting perspectives: A participatory Street Review framework for urban inclusivity''}, by Mushkani, R. and Koseki, S., is published in \emph{Habitat International}, 164, 103536.
I led the study and wrote the paper.
Koseki provided supervision.
(\url{https://doi.org/10.1016/j.habitatint.2025.103536}
)

\medskip
\noindent\textbf{``Public perceptions of Montréal's streets: Implications for inclusive public space making and management''}, by Mushkani, R., Berard, H., Ammar, T., and Koseki, S., is published in the \emph{Journal of Urban Management}.
I led the study and wrote the paper.
Berard and Ammar contributed to data gathering and workshop activities.
Koseki provided supervision.
(\url{https://doi.org/10.1016/j.jum.2025.07.004}
)

\NoChapterPageNumber
\cleardoublepage
\pagenumbering{arabic}

\chapter{Introduction}
\label{ch:introduction}

Municipal decisions about streets and public space have traditionally been framed as decisions about material form and operation, including geometry, maintenance, enforcement, and capital investment. In many cities, these decisions are now mediated by a second layer of urban infrastructure: data sources, classification schemes, models, and interfaces that make public space legible to administrative processes and actionable through computation \citep{Kitchin2014a,Kitchin2014b,Batty2018}. This dissertation argues that this epistemic and algorithmic infrastructure is not ancillary to urbanism but is increasingly constitutive of what cities can justify, prioritize, and contest. Because public space is a contested civic good, governance of this algorithmic layer cannot rely on the assumption that a single universal metric can represent inclusion. The dissertation therefore develops a civic \RightToAI{} for public-space contexts and operationalizes it through pluralistic alignment, understood as a process of eliciting, representing, and legitimately negotiating heterogeneous values and encoding negotiated standards into both physical interventions and AI-mediated representations \citep{rawls1993political,Habermas1996,Mouffe2000}.

\section{Urbanism as dual infrastructure}
\label{sec:dual-infrastructure}

Urban planning has long treated streets, parks, and public facilities as primary mechanisms through which cities allocate mobility, safety, health, and access to public life \citep{Low2000}. Contemporary governance still depends on these material systems, but it increasingly relies on a second layer through which urban conditions are sensed, represented, scored, simulated, and communicated \citep{Batty2024}. This layer includes data capture practices, labeling and categorization conventions \citep{BowkerStar1999,Kitchin2014a}, modeling pipelines, and the interfaces through which outputs are interpreted and translated into administrative action \citep{Kitchin2014b,townsend2013smart,shelton2015actually,Mattern2017b,Mattern2017a}. In this dissertation, the term \emph{epistemic and algorithmic layer}, or \emph{algorithmic layer}, refers to this socio-technical infrastructure and to the institutional routines that stabilize it \citep{star1999ethnography,BowkerStar1999}. The emphasis is not on computation as a technical add-on but on the way it reorganizes what becomes visible to governance and what can be defended as evidence in planning decisions \citep{Kitchin2016,scott1998seeing,Porter1995,Jasanoff2004,Mattern2017a}.

To motivate why this layer must be treated as infrastructure, consider a municipal decision that is familiar in form but increasingly hybrid in its evidentiary basis. A borough is preparing a multi-year program to retrofit intersections and sidewalks in response to documented barriers for people with mobility impairments, caregivers with strollers, and older adults. Budget constraints make prioritization unavoidable, and prioritization becomes the political center of the program because it distributes safety, access, and investment across neighborhoods. To structure the prioritization, the project team assembles an intersection audit \citep{SahaEtAl2019}, integrates open street imagery \citep{Naik2014,Dubey2016}, and commissions a model to produce an \enquote{inclusivity heatmap} that scores streets by predicted ease of access and comfort \citep{Batty2024,shelton2015actually}. In parallel, a generative model is used to produce candidate redesign visualizations for public consultation because it can generate multiple scenarios with limited turnaround time \citep{Mattern2017a}. In practice, such scores and images can become the operational reference for meetings, documents, and timelines, not because they resolve value conflict, but because they reduce complex judgments to outputs that can be compared, ranked, and inserted into workflows \citep{Kitchin2014a,Batty2018}.

The resulting outputs can appear neutral because they are stable under repeated computation, presented as comparable across space, and formatted for incorporation into administrative documents and public-facing dashboards \citep{Batty2024,Merry2016}. Yet the production of these outputs requires choices about what to measure, which proxies represent the relevant civic values, which populations are treated as the reference for training or evaluation, and how disagreement and uncertainty are expressed \citep{Noble2018,Pasquale2015,selbst2019fairness}. These choices are not external to governance; they delimit the decision space by defining what the city can see and therefore what the city can plausibly claim to govern \citep{scott1998seeing}. This dissertation therefore treats the algorithmic layer as a governance object rather than as a tool, in the sense that its design and use shape what counts as evidence, how accountability is performed, and what forms of contestation are institutionally available \citep{Crawford2021,Eubanks2018,Benjamin2019,Noble2018,Pasquale2015,Reisman2018,selbst2021institutional}.

Two observations support the claim that the algorithmic layer behaves like infrastructure in municipal contexts. First, AI systems used in public-space decision-making are increasingly coupled to durable municipal processes \citep{Batty2024,yigitcanlar2025editorial}. They enter through procurement, consulting, platform partnerships, and interdepartmental modernization programs, and they persist through updates, maintenance arrangements, and the accumulation of institutional dependence \citep{Crawford2021,townsend2013smart,shelton2015actually}. Once embedded, they influence not only what a city decides, but also how it justifies decisions, what it audits, and what it is capable of disputing or revising when harms are alleged \citep{Power1997,Strathern2000}. Second, the algorithmic layer is normative because it converts contested public goals into measurement conventions and decision supports \citep{EspelandStevens1998,Mattern2017a}. It decides what counts as evidence by specifying which cues in an input modality are treated as relevant signals, it shapes organizational incentives by translating complex goals into indicators \citep{LarteyLaw2025,Merry2016}, and it generates representations that influence public deliberation by shaping what is shown, what is comparable, and what is framed as feasible \citep{Fishkin2009,DubeyEtAl2024,zhang2024t2i-survey}. These properties are not unique to AI, but AI systems can expand the scale of measurement and representation while obscuring how value judgments are embedded in data selection, labeling rules, and modeling choices \citep{Crawford2021,Noble2018,Pasquale2015,selbst2019fairness,Morozov2013}.

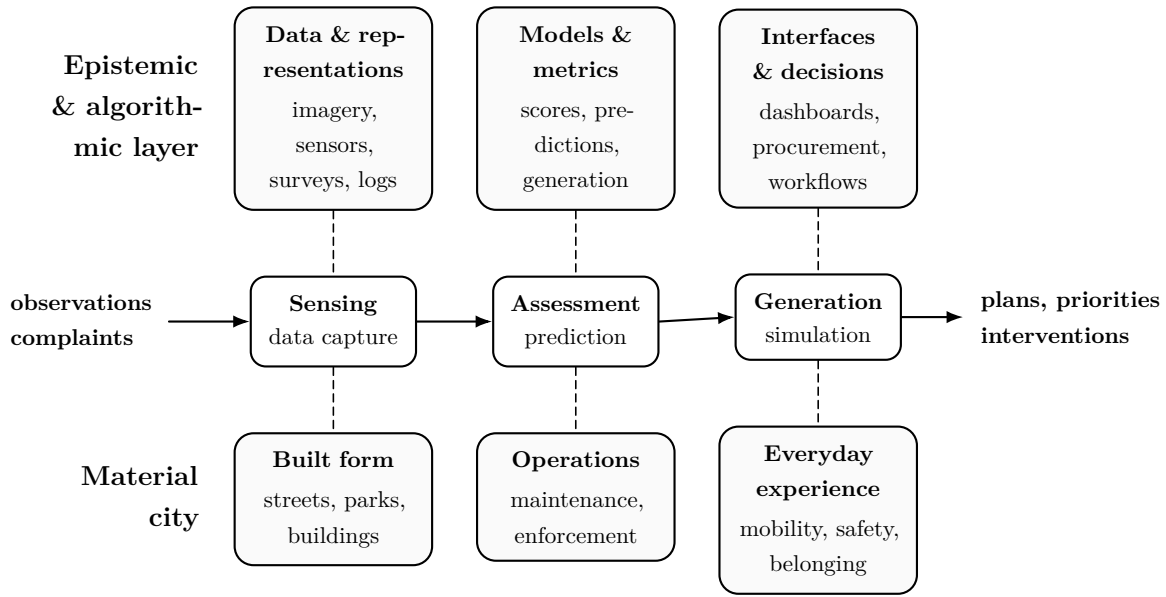
\begin{figure}[htbp]
\centering
\begin{tikzpicture}[
  scale=0.86,
  transform shape,
  font=\footnotesize,
  >=Latex,
  line cap=round,
  node distance=12mm
]

\tikzset{
  card/.style={
    draw, rounded corners=2.5mm, line width=0.8pt,
    fill=black!2, align=center,
    text width=0.15\linewidth,
    inner xsep=2.8mm, inner ysep=2.8mm
  },
  proc/.style={
    draw, rounded corners=2mm, line width=0.8pt,
    fill=white, align=center,
    text width=0.12\linewidth,
    inner xsep=2.8mm, inner ysep=2.8mm
  },
  flow/.style={-{Latex[length=2.2mm]}, line width=0.8pt},
  link/.style={densely dashed, line width=0.6pt},
  ext/.style={font=\footnotesize\itshape, align=left}
}

\coordinate (L) at (8mm,0);
\coordinate (R) at (\linewidth-8mm,0);

\path (L) -- (R)
  coordinate[pos=0.25] (c1)
  coordinate[pos=0.50] (c2)
  coordinate[pos=0.75] (c3);

\node[card] (t1) at (c1) {\textbf{Data \& representations}\\[2pt]imagery, sensors,\\ surveys, logs};
\node[card] (t2) at (c2) {\textbf{Models \& metrics}\\[2pt]scores, predictions,\\ generation};
\node[card] (t3) at (c3) {\textbf{Interfaces \& decisions}\\[2pt]dashboards, procurement,\\ workflows};

\node[
  font=\bfseries,
  anchor=east,
  text width=2.5cm, 
  align=right
] at ([xshift=-4mm]t1.west)
  {Epistemic \& algorithmic layer};

\node[proc, below=10mm of t1] (p1) {\textbf{Sensing}\\[-1pt]data capture};
\node[proc, below=10mm of t2] (p2) {\textbf{Assessment}\\[-1pt]prediction};
\node[proc, below=10mm of t3] (p3) {\textbf{Generation}\\[-1pt]simulation};

\draw[flow] (p1.east) -- (p2.west);
\draw[flow] (p2.east) -- (p3.west);

\node[card, below=10mm of p1] (b1) {\textbf{Built form}\\[2pt]streets, parks,\\ buildings};
\node[card, below=10mm of p2] (b2) {\textbf{Operations}\\[2pt]maintenance,\\ enforcement};
\node[card, below=10mm of p3] (b3) {\textbf{Everyday experience}\\[2pt]mobility, safety,\\ belonging};

\node[
  font=\bfseries,
  anchor=east,
  text width=2.5cm,
  align=right
] at ([xshift=-4mm]b1.west)
  {Material city};

\draw[link] (t1.south) -- (p1.north);
\draw[link] (t2.south) -- (p2.north);
\draw[link] (t3.south) -- (p3.north);
\draw[link] (p1.south) -- (b1.north);
\draw[link] (p2.south) -- (b2.north);
\draw[link] (p3.south) -- (b3.north);

\coordinate (in)  at ([xshift=12mm]L |- p1.west);   
\coordinate (out) at ([xshift=-15mm]R |- p3.east); 

\draw[flow] (in) -- (p1.west);
\draw[flow] (p3.east) -- (out);

\node[
  ext,
  font=\bfseries\footnotesize
] at ([xshift=-1.5mm]L |- p1.center)
  {observations\\complaints};

\node[
  ext,
  font=\bfseries\footnotesize
] at ([xshift= 1.5mm]R |- p3.center)
  {plans, priorities\\interventions};

\end{tikzpicture}

\caption[Urbanism as dual infrastructure]{Urbanism as dual infrastructure: the material city is coupled to an epistemic and algorithmic layer that produces representations and decision supports. The coupling is bidirectional: lived experience and material conditions produce data, while model outputs shape priorities, narratives, and interventions. \added{Source: author's synthesis.}}
\label{fig:dual-infrastructure}
\end{figure}

This dissertation therefore treats urbanism as \emph{dual infrastructure}. The first layer is the material city: built form, operational practice, and the everyday conditions of mobility, safety, and belonging. The second layer is the epistemic and algorithmic infrastructure through which the material city is rendered into data, metrics, and visual narratives that can be acted upon in municipal workflows \citep{LarteyLaw2025,Jasanoff2004,yigitcanlar2025editorial,Batty2024,Kitchin2023}. Figure~\ref{fig:dual-infrastructure} summarizes this framing and highlights coupling points where governance decisions move between layers.

Framing the algorithmic layer as infrastructure changes the governance question \citep{CostanzaChock2020,BowkerStar1999,star1999ethnography,Ostrom2009}. When AI systems are treated as optional tools, accountability is often limited to technical quality, procurement compliance, and project delivery. When they are treated as market products, governance is often reduced to vendor responsibility and consumer protection. When they are treated as infrastructure, governance concerns expand to include authorization, oversight, and contestability, because the layer shapes how collective life is represented and how resources and burdens are distributed through administrative action \citep{Crawford2021,Eubanks2018,Benjamin2019,Jasanoff2004}.

The dual-infrastructure framing is anchored empirically in systems that have become common in urban analytics and planning practice \citep{shelton2015actually,Kitchin2014b,Batty2024}. Computer vision and perception modeling have been used to infer perceived safety, wealth, and aesthetic judgments from street-level imagery at city scale \citep{Naik2014,Dubey2016}. Participatory platforms use imagery and structured annotation to document accessibility barriers and translate lived experience into actionable maps, often with the explicit aim of supporting municipal auditing and remediation \citep{SahaEtAl2019,Gabrys2016,LarteyLaw2025}. The practical appeal of these approaches is that they can produce comparable outputs across large geographies, but their governance stakes are that they require decisions about whose judgments count, what dimensions matter, and how uncertainty and disagreement are represented \citep{EspelandStevens1998,Porter1995}. The same stakes appear in generative systems, which are increasingly used to produce visual scenarios and prototypes in design and planning contexts. Such representations can influence problem framing and deliberation, and they can shift what stakeholders treat as feasible or appropriate, even when they are used as exploratory artifacts rather than as direct prescriptions \citep{Mattern2017a,Morozov2013,DubeyEtAl2024,zhang2024t2i-survey,yigitcanlar2025editorial}. Because text-to-image AI models and related systems are part of the technical substrate of this emerging practice, this dissertation treats generative systems not only as media tools but as governance-relevant infrastructures of representation \citep{podell2023sdxlimprovinglatentdiffusion,Mattern2017b}.

This dissertation therefore treats both evaluative systems, which measure and score streets, and generative systems, which produce alternative visual narratives, as components of an algorithmic layer that requires explicit governance. The remainder of this chapter specifies why that governance problem cannot be solved by universal standards alone and how pluralism changes what legitimate measurement and alignment require.

\section{Why universal standards fail}
\label{sec:pluralism}

Public space is a contested civic good. The right to the city tradition frames urban space not only as a physical container but as a political claim about participation, appropriation, and collective life \citep{Lefebvre1968,Harvey2003,Mitchell2003,LowSmith2006}. Within this tradition, inclusion cannot be reduced to technical compliance with any single code, because the relevant questions extend to who feels entitled to be present, who is protected or exposed, who is made visible or invisible, and whose practices are treated as legitimate in shared space \citep{Low2000,Mitchell1995,Mitchell2003,Harvey2003}. This understanding aligns with scholarship that treats publicness as negotiated rather than given and that documents how design and management practices can produce exclusion through surveillance, hostile architecture, privatization, and uneven enforcement \citep{Low2000,LowSmith2006,varna2010publicness,Mehta2014}.

In institutional practice, however, cities often operationalize public-space goals through universal standards and metrics, including design guidelines, levels-of-service measures, key performance indicators, and composite indices \citep{MitrasinovicMehta2021,low2020social}. These instruments enable budgeting, coordination, and accountability across agencies and contractors, and they produce legibility for evaluation and enforcement \citep{MitrasinovicMehta2021}. At the same time, they can convert contested values into commensurable quantities and then treat the resulting quantities as objective facts, with the result that disagreement is treated as error and minority perspectives are attenuated by aggregation \citep{EspelandStevens2008,Merry2016,Sorensen2024}. The governance risk is not simply that metrics are imperfect, but that they can reclassify political disagreement as a technical issue of measurement accuracy and thereby narrow the space in which value conflict is recognized and negotiated \citep{Mouffe2000}.

The rise of AI intensifies this risk because it expands the capacity to generate measurements and representations without necessarily expanding the legitimacy of the standards embedded in those outputs \citep{Morozov2013,Mattern2017a,Noble2018,Pasquale2015,shelton2015actually}. A city can produce large volumes of predictions from street imagery, but it cannot assume that heatmaps represent civic agreement about what counts as an inclusive street. A generative model can produce many design alternatives that appear plausible, but plausibility is not a substitute for legitimacy, and no single objective can encode the divergent values at stake in public space \citep{rawls1993political,Habermas1996,Sorensen2024}. In both evaluative and generative settings, the technical system can outpace the political process, particularly when procurement or platform adoption implicitly sets the measurement framework before civic negotiation has occurred \citep{Healey1997,Fishkin2009}.

This dissertation treats pluralism as a governance condition rather than as a nuisance parameter \citep{Berlin1969,Galston2002}. Pluralism is empirical because people evaluate the same streetscape differently and because differences are structured by social position and lived experience. In public-space contexts, social position is often intersectional, meaning that experiences of disability, age, gender, sexuality, and other axes combine and shape what a street affords and denies \citep{Crenshaw1989,low2020social}. Pluralism is also normative because claims about what makes a public space \enquote{good} or \enquote{inclusive} are political claims, even when expressed through design standards or technical indicators \citep{rawls1993political,Habermas1996,Mouffe2000}, and because communities should have meaningful authority over systems that shape their environments and daily mobility \citep{CostanzaChock2020}. These premises imply that legitimate governance cannot proceed by assuming that one universal score represents inclusion, but must instead represent heterogeneous evaluations in a form that enables comparison, negotiation, and revision \citep{Jasanoff2004}.

The empirical chapters of this dissertation provide concrete evidence for treating pluralism as structured rather than random. Chapter~5 reports a study in Montréal in which participants evaluated a shared set of streetscape images across multiple criteria, showing that agreement differs systematically by criterion and that socially contested dimensions exhibit lower agreement than dimensions with more direct visual referents. Chapter~5 also compares judgments produced individually with judgments produced after structured small-group discussion, showing that deliberation can increase convergence on selected criteria while preserving residual disagreement on contested judgments \citep{Habermas1996,Fishkin2009,Fung2004}. The same chapter demonstrates a measurement pipeline that scales co-produced criteria through a locally trained model and produces subgroup-aware outputs and citywide maps using street imagery, while documenting limits that arise from image quality and from aspects of public-space experience not captured by the input modality (Chapter 5 reports these studies in synthesized form; their complete methodological details and extended analyses are available in the corresponding papers on public perceptions of Montréal’s streets \citep{MushkaniJUM2025} and intersecting perspectives \citep{MushkaniHabitat2025}). Chapter~6 extends the analysis to generative systems by constructing a pluralistic alignment dataset for inclusive public spaces and evaluating preference-based tuning of a text-to-image model; it reports a substantial fraction of neutral or indeterminate comparisons in evaluation, which the dissertation interprets as a governance-relevant signal that some value conflicts remain unresolved by optimization and require procedure rather than forced aggregation \citep{ChristianoEtAl2017,ouyang2022,selbst2019fairness}.

Together, these findings motivate a second premise of the dissertation: \emph{measurement is a governance choice} \citep{EspelandStevens1998,Porter1995,Desrosieres1998,scott1998seeing}. Metrics do not merely describe streets; they reorganize what becomes governable by deciding what counts as evidence and which forms of disagreement are visible or suppressed \citep{BowkerStar1999,star1999ethnography}. When cities adopt universal standards and single-score dashboards, they decide to treat plural values as commensurable and to prioritize optimization over negotiation \citep{EspelandStevens2008,Power1997,Strathern2000}. When cities preserve pluralism through disaggregated outputs, documented disagreement, and deliberative procedures, they decide to treat value conflict as a legitimate part of public decision-making, and they enable accountability mechanisms that can respond to contestation rather than treating it as noise \citep{Habermas1996,Fishkin2009,EspelandStevens2008}.

This dissertation proposes a synthesis framework, \emph{Pluralistic-Alignment Urbanism (PAU)}, to describe what follows from treating pluralism as a governance condition and from treating the algorithmic layer as part of urban infrastructure. PAU has four commitments:

\begin{figure}[htbp]
\centering
\begin{tikzpicture}[font=\small, node distance=12mm]

\def\PAUwide{0.88\linewidth}
\def\PAUnarrow{0.28\linewidth}

\node[draw, very thick, align=center, minimum width=\PAUwide, minimum height=14mm] (pau) {%
  \textbf{Pluralistic-Alignment Urbanism (PAU)}\\
  A governance framework for cities operating through dual infrastructure};

\node[draw, align=center, minimum width=\PAUnarrow, minimum height=16mm, below=12mm of pau] (rights) {%
  \textbf{Right to AI}\\
  disclosure, participation\\
  contestation, recourse};

\node[draw, align=center, minimum width=\PAUnarrow, minimum height=16mm, left=3mm of rights] (dual) {%
  \textbf{Dual infrastructure}\\
  material city\\
  algorithmic layer};

\node[draw, align=center, minimum width=\PAUnarrow, minimum height=16mm, right=3mm of rights] (plural) {%
  \textbf{Pluralistic alignment}\\
  represent and compare\\
  heterogeneous values};

\node[draw, align=center, minimum width=\PAUwide, minimum height=18mm, below=16mm of rights] (inst) {%
  \textbf{Institutional mechanism: deliberative cells and lifecycle governance}\\
  co-framing to co-maintenance with checkpoints, artifacts, and documented disagreement};

\draw[-{Latex[length=3mm]}, thick] (pau) -- (dual);
\draw[-{Latex[length=3mm]}, thick] (pau) -- (rights);
\draw[-{Latex[length=3mm]}, thick] (pau) -- (plural);

\draw[-{Latex[length=3mm]}, thick] (dual) -- (inst);
\draw[-{Latex[length=3mm]}, thick] (rights) -- (inst);
\draw[-{Latex[length=3mm]}, thick] (plural) -- (inst);

\node[align=center, font=\small, text width=\PAUwide, inner sep=0pt, below=7mm of inst] (pipe) {%
  Methodological pipeline in this dissertation, beginning with co-produced descriptors and criteria,
  followed by subgroup-aware assessment and mapping (Street Review),
  pluralistic preference data and tuning (LIVS),
  and concluding with municipal procedures for procurement, audit, and recourse.};

\end{tikzpicture}
\caption[Pluralistic-Alignment Urbanism framework]{Pluralistic-Alignment Urbanism (PAU) as a framework for the dissertation. PAU treats AI in public space as part of an algorithmic layer that requires rights-based governance and pluralistic alignment. Deliberative cells and lifecycle governance provide an institutional mechanism that translates plural evaluations into standards. \added{Source: author's synthesis.}}
\label{fig:pau}
\label{fig:pau-framework}
\end{figure}

\section{Research objectives and research questions}
\label{sec:rqs}

The dissertation's overarching objective is to specify how cities can legitimately govern the algorithmic layer of public space while respecting pluralism. This objective is pursued through a research program that combines normative theory, participatory co-production, mixed-method empirical analysis, and technical modeling, with the aim of linking governance claims to documented empirical evidence and implementable municipal procedures \citep{Jasanoff2004}.

Three commitments guide the research design.

\begin{enumerate}[leftmargin=*]
\item \deleted{\textbf{Legitimacy as a design requirement.} Governance of public-space AI systems is not satisfied by technical performance alone, because performance does not establish democratic authorization, does not determine which values the system should represent, and does not provide standing for contestation. The dissertation therefore treats participation with meaningful authority, transparency about use and limits, and mechanisms for contestation and recourse as requirements of legitimate deployment \citep{Crawford2021,Eubanks2018,Benjamin2019,rawls1993political,Habermas1996,Fishkin2009}.}\par
\added{\textbf{Legitimacy as a design requirement.} \citet{rawls1993political} and \citet{Habermas1996} ground democratic authorization in public justification and in institutions through which people subject to collective decisions can contest their terms. \citet{Fishkin2009} contributes a deliberative model for informed public judgment. These accounts establish who requires standing and how authorization should be formed. Critical technology scholarship sets substantive limits on the procedure. \citet{Benjamin2019} shows how technical classification can reproduce racial hierarchy; \citet{Eubanks2018} documents punitive effects of automated public administration on poor and working-class people; and \citet{Crawford2021} traces AI systems to extractive relations involving data, labor, energy, and material resources. The dissertation therefore evaluates legitimacy through two linked tests: meaningful civic authority and substantive protection against discriminatory, punitive, and extractive arrangements. Chapter~3 expresses those protections through the \RightToAI{}. Chapter~7 attaches them to decision gates, prohibitions, contestation, and recourse.}
\label{rev:commitment-1}
\item \deleted{\textbf{Pluralism as data, not a nuisance parameter.} Disagreement is treated as a signal about contested values and differential experience rather than as an error to be eliminated through aggregation. Methods therefore measure agreement and disagreement explicitly and interpret residual disagreement as governance-relevant information \citep{EspelandStevens2008,Galston2002,Engin2020}.}\par
\added{\textbf{Pluralism as governance-relevant evidence.} \citet{Galston2002} supplies the normative premise that public life involves multiple goods whose conflicts may resist a common ranking. \citet{EspelandStevens2008} identify commensuration as an institutional process that converts unlike qualities into a shared metric and changes the relations among them. Observed variation may arise from substantive conflict among values, from the categories and scales used to measure them, or from both sources. The methods therefore measure agreement, report disaggregated results, and retain residual disagreement in the governance record.}
\label{rev:commitment-2}
\item \deleted{\textbf{Infrastructure over one-off interventions.} Because AI systems and public-space standards persist and evolve, governance must be lifecycle-oriented, with recurring checkpoints, documented artifacts, and recommissioning mechanisms that can respond to drift, complaint, and changing civic priorities \citep{Kitchin2016,Kitchin2014a,star1999ethnography,BowkerStar1999}.}\par
\added{\textbf{Lifecycle governance of infrastructure.} \citet{BowkerStar1999} show how classifications and standards become infrastructural: they stabilize practice, distribute consequences, and recede from view as they become routine. \citet{star1999ethnography} explains how breakdown makes those embedded relations visible. Kitchin's analyses of data-driven urbanism locate this dynamic in contemporary cities, where data systems shape administrative knowledge, coordination, and power \citep{Kitchin2014a,Kitchin2014b,Kitchin2016}. Recurring checkpoints, versioned artifacts, complaint pathways, and recommissioning rules make classifications and models reviewable as data, software, institutions, and civic priorities change.}
\label{rev:commitment-3}
\end{enumerate}
\label{rev:commitments}

These commitments lead to five research questions.

\begin{description}[leftmargin=0pt, itemsep=1.2ex]
\item[\textbf{RQ1 (Urban theory and governance)}] How should municipal governance conceptualize the algorithmic layer shaping public space as tools, markets, or infrastructure, and what institutional responsibilities follow?

\deleted{This question motivates the dual-infrastructure premise and frames AI systems that influence public space as governance objects requiring durable institutional commitments \citep{star1999ethnography,BowkerStar1999}. It draws on scholarship on data-driven urbanism and algorithmic governance to specify how computational representations become embedded in municipal justification and coordination \citep{Kitchin2014a,Kitchin2014b,Crawford2021,Batty2024,shelton2015actually,Mattern2017a}.}

\added{This question motivates the dual-infrastructure premise. \citet{BowkerStar1999} and \citet{star1999ethnography} explain how classifications and standards acquire durability and organize institutional practice. \citet{Batty2024} examines the computable city through simulation, scaling, and prediction. \citet{Kitchin2014a,Kitchin2014b} and \citet{shelton2015actually} locate data-driven urbanism within municipal institutions and uneven urban geographies. \citet{Mattern2017a} challenges the epistemic reduction of urban life to computational representation, while \citet{Crawford2021} identifies the labor, resource, and power relations that sustain AI systems. RQ1 uses these distinct lenses to determine which responsibilities follow when computation becomes part of municipal evidence and coordination.}
\label{rev:rq1}

\item[\textbf{RQ2 (Rights and legitimacy)}] What does a civic \RightToAI{} require in public-space contexts, including participation, transparency, and recourse, and what should it forbid?

\deleted{This question is developed primarily in Chapter~3 and asks how rights language can specify enforceable governance floors, define standing for contestation, and articulate prohibitions against unacceptable public-space uses \citep{rawls1993political,Habermas1996,Reisman2018,selbst2021institutional}. The emphasis on participation as power connects to planning scholarship that distinguishes between consultation and authority and that treats participation as an institutional design problem \citep{Arnstein1969,Healey1997,Fung2004,Jacobs1961}.}

\added{This question is developed primarily in Chapter~3. \citet{rawls1993political} and \citet{Habermas1996} provide normative accounts of democratic legitimacy and public justification. \citet{Reisman2018} translate accountability into administrative mechanisms, including impact assessment before acquisition and opportunities for public comment. \citet{selbst2021institutional} shows how the legal and institutional setting shapes an assessment's operation and limits. Planning scholarship specifies how participation distributes authority: \citet{Arnstein1969} defines participation through the redistribution of citizen power; \citet{Healey1997} develops collaborative planning through inclusionary argumentation; and \citet{Fung2004} examines empowered participation and accountable autonomy. \citet{Jacobs1961} grounds this discussion in the everyday social life of streets and the limits of top-down urban knowledge. Together, these lineages guide the governance floors, standing rules, prohibitions, and avenues of recourse.}
\label{rev:rq2}

\item[\textbf{RQ3 (Pluralism as an empirical condition)}] To what extent do perceptions of streetscapes diverge across social positions, and which dimensions are more or less amenable to convergence through deliberation? \deleted{This question is anchored in Chapters~5 and 6 and treats pluralism as measurable variation rather than as a residual \citep{Berlin1969,Galston2002}. It asks how disagreement varies by criterion and how structured discussion changes convergence while preserving residual contestation on judgments that remain politically contested \citep{Habermas1996,Fishkin2009,Mouffe2000}.}

\added{This question is anchored in Chapters~5 and 6. \citet{Berlin1969} and \citet{Galston2002} establish the possibility of enduring conflict among legitimate values. \citet{Habermas1996} and \citet{Fishkin2009} explain how reason-giving and structured deliberation can form considered public judgment. \citet{Mouffe2000} warns that consensus-oriented procedures can domesticate antagonism and obscure exclusions built into the terms of discussion. The empirical design therefore records movement toward agreement and residual disagreement as separate outcomes. A deliberative result supports collective action within its documented scope; unresolved reasons remain in disagreement logs and can trigger disaggregation, review, or limits on use.}
\label{rev:rq3}

\item[\textbf{RQ4 (Measurement-to-governance pipeline)}] How can cities elicit and represent heterogeneous values without collapsing them into a single score, and how do such representations change what becomes governable? \deleted{This question connects methodological design to governance implications by treating measurement as a choice about commensuration, representation, and administrative legibility \citep{EspelandStevens1998,scott1998seeing}. It is pursued through a pipeline that co-produces evaluative criteria and builds representations that preserve heterogeneity in outputs rather than collapsing it through a single objective \citep{EspelandStevens2008,Porter1995,Sorensen2024,ConitzerEtAl2024}.}

\added{This question connects methodological design to governance. \citet{Porter1995} explains mechanical objectivity as an institutional response to distrust. \citet{EspelandStevens1998,EspelandStevens2008} show how commensuration transforms the qualities and social relations it measures. \citet{scott1998seeing} identifies the administrative power created through legibility. \citet{Sorensen2024} formalize Overton, steerable, and distributional pluralism and identify corresponding benchmark classes. \citet{ConitzerEtAl2024} argue that social-choice theory should guide whose feedback is collected and how it is aggregated. The dissertation evaluates these technical proposals under the warnings supplied by the sociology of quantification. Category selection, annotation rules, aggregation, and optimization remain sites at which a pluralistic system can impose a single ordering. The pipeline therefore retains disaggregated outputs, neutral judgments, provenance, exclusions, and decision rules as auditable parts of the representation.}
\label{rev:rq4}

\item[\textbf{RQ5 (Implementation realism)}] How can co-produced lifecycle governance and procurement practices encode negotiated standards into models and streets over time, including mechanisms for recommissioning, audit, and redress? \deleted{This question motivates the synthesis in Chapter~7 and translates rights and empirical findings into municipal procedures that can be integrated into existing planning, commissioning, and maintenance workflows \citep{Power1997,Strathern2000,LarteyLaw2025}. It addresses the institutional conditions under which cities can adopt AI systems without ceding control over standards, documentation, and contestability \citep{Crawford2021,yigitcanlar2025editorial,Benjamin2019,Reisman2018,selbst2021institutional}.}

\added{This question motivates the synthesis in Chapter~7. \citet{LarteyLaw2025} examine the organizational and governance conditions of AI adoption in urban planning, while \citet{yigitcanlar2025editorial} situate adoption within contemporary urban modernization. \citet{Benjamin2019} and \citet{Crawford2021} identify discriminatory and extractive harms that procurement rules must address. \citet{Reisman2018} specify impact-assessment mechanisms for public agencies, and \citet{selbst2021institutional} shows how institutional context shapes their operation and limits. \citet{Power1997} and \citet{Strathern2000} warn that verification can become ritualized and displace substantive judgment. RQ5 therefore asks which artifacts carry consequences at municipal decision gates. A model card, impact assessment, audit report, or disagreement log becomes a governance instrument when it is linked to a named authority, a review trigger, and an available remedy.}
\label{rev:rq5}
\end{description}
\label{rev:rqs}

\section{Main argument}
\label{sec:claim-propositions}

\subsection{Central claim}

This dissertation advances the following central claim.

\begin{quote}
\noindent Contemporary urbanism must be governed as a dual infrastructure: the material city and an epistemic, algorithmic layer that shapes what can be perceived, evaluated, and acted upon.

Because public space is a contested civic good, this algorithmic layer requires a civic \RightToAI{} and an operational approach to pluralistic alignment through which cities elicit, compare, and legitimately negotiate divergent values, and encode negotiated standards into both physical designs and AI systems.
\end{quote}

Two implications follow for public-space governance. First, AI systems affecting public space must be treated as governance objects rather than as neutral tools, because they embed choices about evidence, classification, and acceptable trade-offs, and because those choices shape how decisions can be justified and contested \citep{Crawford2021,Eubanks2018,Benjamin2019,Noble2018,Pasquale2015}. Second, inclusion cannot be governed through universal metrics alone, because a single scalar score can erase structured disagreement and can reclassify political conflict as a technical issue of measurement, which can narrow the space in which value conflict is negotiated \citep{EspelandStevens2008,EspelandStevens1998,Mouffe2000}.

\subsection{Falsifiable propositions}

To make these implications testable, the dissertation develops four falsifiable propositions that connect the normative and theoretical claims to empirical evidence.

\begin{description}[leftmargin=0pt, itemsep=1.0ex]
\item[\textbf{P1 (Pluralism is structured, not noise)}] For the same streetscape stimuli, evaluations differ systematically across social positions, and \enquote{inclusivity} exhibits lower cross-participant agreement than more visually legible criteria (for example, aesthetics) \citep{Mehta2014,Galston2002}.
\item[\textbf{P2 (Deliberation changes the evidentiary object)}] Structured small-group discussion increases convergence on some criteria, but does not fully homogenize views on socially contested dimensions \citep{Habermas1996,Fishkin2009}. Residual disagreement is informative and should be documented rather than suppressed.
\item[\textbf{P3 (Co-produced scaling is feasible but bounded)}] A model trained on locally co-produced data can predict subgroup and group-level streetscape scores and produce spatial maps useful for governance, but it under-represents cultural and symbolic markers not captured in the input modality.
\item[\textbf{P4 (Neutrality is a governance signal)}] In pluralistic preference evaluation of public-space imagery, a substantial fraction of judgments remain neutral or indeterminate even when models improve on average, indicating contested or ambiguous values that should trigger deliberation rather than be forced into optimization \citep{rafailov2024,ChristianoEtAl2017,ouyang2022}.
\end{description}

Table~\ref{tab:propositions} maps each proposition to the dissertation chapters and the evidence sources that address it.

\begin{table}[htbp]
\centering
\caption{Falsifiable propositions and primary evidence sources in the dissertation.}
\label{tab:propositions}
\renewcommand{\arraystretch}{1.2}
\begin{tabularx}{\textwidth}{@{}p{0.07\textwidth} X X@{}}
\toprule
\textbf{Prop.} & \textbf{Statement (falsifiable)} & \textbf{Primary evidence sources in this dissertation} \\
\midrule
P1 &
Evaluations of street quality differ systematically across social positions, with inclusivity exhibiting lower inter-rater agreement than more visually legible criteria such as aesthetics or maintenance. &
Street Review descriptor elicitation and subgroup modeling; agreement and correlation analyses from the Montréal street perception study (see Chapter \ref{chap:streetreview}; \cite{MushkaniHabitat2025}). \\
\addlinespace
P2 &
Structured small-group discussion increases convergence on selected evaluative criteria, while persistent disagreement remains on normatively contested dimensions. &
Comparative analysis of individual versus group judgments in the Montréal street perception study; interpretive synthesis in Chapters~5 and~7. \\
\addlinespace
P3 &
Co-produced evaluative models can scale pluralistic street assessments to citywide representations, but remain constrained by data modality and participant representation. &
Model training, subgroup outputs, and citywide mapping in the Street Review study; cross-chapter limitations analysis. \\
\addlinespace
P4 &
A substantial share of neutral or indeterminate preferences persists even after preference tuning, indicating value conflict that cannot be resolved through optimization alone and instead requires procedural intervention. &
LIVS dataset construction and evaluation of preference tuning; governance-oriented interpretation in Chapters~6 and~7. \\
\bottomrule
\end{tabularx}
\end{table}

\subsection{Evidence strategy}

The dissertation's evidence strategy follows the logic of PAU by moving from premise to mechanisms, from mechanisms to demonstrations, and from demonstrations to governance implications. Chapters~3 and 4 specify the normative and procedural requirements for governing AI systems in public space, with Chapter~3 formalizing the \RightToAI{} and Chapter~4 operationalizing governance through an augmented lifecycle that attaches participation and documentation to recurrent stages of design, deployment, and maintenance \citep{star1999ethnography,BowkerStar1999,Haakman2021,Terzi2010}. Chapters~5 and 6 provide empirical and technical demonstrations of how pluralism appears in streetscape evaluation and how it persists in preference-based alignment for generative models, with particular attention to disagreement and neutrality as properties of the evidentiary object rather than as annotation error. Chapter~7 translates these constraints into a municipal governance architecture that specifies authorization, procurement requirements, auditable artifacts, deliberative oversight, and recourse mechanisms \citep{Reisman2018,selbst2021institutional}, and Chapter~8 consolidates the contributions and boundary conditions of the argument.

\section{Contributions and thesis structure}
\label{sec:contributions-structure}

\subsection{Contributions}

The dissertation's contributions are grouped into five categories.

\begin{enumerate}[leftmargin=*]
\item \textbf{Normative contribution:}
The dissertation articulates a \RightToAI{} that reframes AI systems affecting public space as governance objects requiring democratic authorization, participation as power, transparency about use and limits, and contestability through recourse mechanisms \citep{rawls1993political,Habermas1996,Reisman2018,selbst2021institutional}. The contribution specifies rights-based duties and prohibitions relevant to municipal use, emphasizing that technical performance and procedural compliance are not sufficient substitutes for legitimacy in contested public-space contexts \citep{AnttiroikoDeJong2020,Benjamin2019,low2020social}.

\item \textbf{Theoretical contribution:}
The dissertation proposes PAU as a synthesis framework that integrates dual infrastructure, rights-based governance, pluralistic alignment, and \DelibCells{} as an institutional unit for translating pluralism into revisable standards and documented residual contestation \citep{Habermas1996,Fishkin2009}.

\item \textbf{Methodological contribution:}
The dissertation develops and integrates methods for eliciting and consolidating evaluative descriptors, measuring pluralism and deliberation effects, scaling assessment through subgroup-aware modeling and mapping, and aligning generative models through pluralistic preference datasets and preference-based tuning \citep{ouyang2022,rafailov2024}.

\item \textbf{Empirical contribution:}
Using Montréal as a case, the dissertation provides evidence that agreement differs systematically across criteria and that structured group discussion can increase convergence on selected criteria while preserving residual disagreement on contested judgments. The dissertation also documents representational limits of image-based inputs and interprets these limits as governance constraints on how outputs should be used.

\item \textbf{Governance and implementation contribution:}
The dissertation provides an operational mapping from rights to municipal procedure through an augmented lifecycle, procurement and recommissioning rules, deliberative oversight design, and recourse mechanisms suitable for city agencies \citep{Power1997,Reisman2018,selbst2021institutional}. These procedures are derived from the empirical constraints documented in Chapters~5 and 6 and are structured to preserve pluralism as governance-relevant information rather than as error.
\end{enumerate}

\subsection{Thesis structure}
The dissertation is organized as a monograph with manuscript-based chapters. After this introduction, Chapter~2 presents the research methodology, including case selection, partnerships, data sources, analytic methods, and validity logic across manuscripts. Chapter~3 develops the normative foundation of the \RightToAI{} for public-space contexts, and Chapter~4 develops the procedural backbone through a lifecycle approach to participation, documentation, and maintenance. Chapter~5 presents Street Review as a participatory framework for assessing streetscape inclusivity that integrates qualitative elicitation, agreement analysis, subgroup-aware modeling, and citywide mapping, while Chapter~6 presents LIVS as a pluralistic alignment dataset for inclusive public spaces and evaluates preference-based tuning of a text-to-image model, with neutrality treated as governance-relevant information. Chapter~7 synthesizes the dissertation into a municipal governance and implementation architecture, and Chapter~8 concludes by answering the research questions, summarizing contributions, and outlining limits and future research.

\section{Definitions and scope}
\label{sec:definitions-scope-ethics}
This section defines key terms used throughout the dissertation and states the scope conditions under which the claims are made.

\subsection{Key definitions}

\paragraph{\textbf{Public space and streetscapes.}} In this dissertation, \emph{public space} refers to spaces that are publicly accessible and socially consequential, including streets, sidewalks, plazas, and parks \citep{Low2000,LowSmith2006,Mitchell2003}. The dissertation focuses empirically on \emph{streetscapes}, defined as the spatial and experiential configuration of streets and adjacent pedestrian environments as represented through street-level imagery and described by residents. This focus is motivated by the centrality of streets to everyday mobility and by the institutional reality that streets are frequent sites of municipal investment, maintenance, and conflict.

\medskip

\paragraph{\textbf{AI systems in scope.}} The dissertation focuses on AI systems that materially influence public-space decision-making through \emph{representation} and \emph{generation}. Representational systems include models that infer or score street qualities from imagery and that produce spatial summaries such as maps and dashboards \citep{Naik2014,Dubey2016}. Generative systems include text-to-image models used to create visual scenarios, prototypes, and alternatives \citep{DubeyEtAl2024,zhang2024t2i-survey}, with Stable Diffusion XL used as an implementation reference point in the dissertation's generative experiments \citep{podell2023sdxlimprovinglatentdiffusion}. Although surveillance and enforcement systems are not the main empirical focus, the \RightToAI{} developed here is motivated by the broader class of AI systems that can affect public-space access and treatment, and the governance architecture in Chapter~7 explicitly addresses risks of repurposing evaluative outputs for punitive enforcement \citep{Crawford2021,Eubanks2018,Benjamin2019,Noble2018,Pasquale2015}.

\medskip

\paragraph{\textbf{Algorithmic layer and epistemic infrastructure.}} The \emph{algorithmic layer} is the socio-technical infrastructure that produces computational representations of public space, including data collection, curation, modeling, evaluation, interfaces, and organizational workflows \citep{star1999ethnography,lazar2024lectureigoverningalgorithmic}. It is epistemic because it shapes what the city knows and can justify \citep{Jasanoff2004,scott1998seeing}, and algorithmic because it increasingly relies on machine learning models and automated decision supports \citep{Kitchin2016,LarteyLaw2025,lazar2024lectureigoverningalgorithmic}.

\medskip

\paragraph{\textbf{Pluralism.}} \emph{Pluralism} refers to the condition that individuals and groups hold divergent, sometimes incommensurable, values and interpretations about what constitutes an inclusive and desirable public space \citep{Berlin1969,Galston2002,Mouffe2000}. In this dissertation, pluralism is treated as both empirical, in the form of observable disagreement and neutrality, and normative, in the sense that value conflict is a legitimate feature of public decision-making in contested civic domains \citep{rawls1993political,Habermas1996,Crenshaw1989,Low2000,varna2010publicness}.

\medskip

\paragraph{\textbf{Pluralistic alignment.}} \emph{Pluralistic alignment} is the operational approach proposed here for governing AI systems under pluralism. Rather than treating aggregated preferences as ground truth for optimization, pluralistic alignment emphasizes eliciting heterogeneous values, representing them without collapsing them into a single target, negotiating and documenting standards through legitimate procedures, and encoding those negotiated standards into models and material interventions \citep{Sorensen2024,Habermas1996,Jasanoff2004}.

\medskip

\paragraph{\textbf{Right to AI}} The \RightToAI{} is a civic rights-based framing that specifies minimum guarantees and duties for AI systems that materially affect people, especially in public governance contexts \citep{Reisman2018,selbst2021institutional}. In this dissertation, it includes rights to disclosure, participation as power, contestation and recourse, and documentation, along with prohibitions against deploying high-impact systems without democratic authorization and enforceable accountability mechanisms \citep{Crawford2021,Turchin2019-valueDoNOtExist,Mattern2017a,Jacobs1961,rawls1993political,Habermas1996}.

\medskip

\paragraph{\textbf{Deliberative cells.}} Deliberative cells are small, structured, recurring deliberative bodies that serve as the institutional unit for translating plural evaluations into legitimate standards \citep{Healey1997,Fishkin2009}. They are designed to support convergence through structured justification while preserving residual disagreement as documented governance information that can trigger review and revision rather than being suppressed through aggregation \citep{Mouffe2000}.

\medskip

\paragraph{\textbf{Commissioning and recommissioning.}} \emph{Commissioning} refers to the institutional process through which a city authorizes, procures, and deploys both physical interventions and AI systems. \emph{Recommissioning} refers to the recurring process of revisiting these authorizations as conditions change, including demographic shifts, policy updates, model drift, and documented harms. The dissertation argues that governance should treat streets and models as objects requiring symmetric commissioning logic because both persist, both condition future decisions, and both require maintenance and revision \citep{BowkerStar1999,Harvey2012}.

\medskip

\paragraph{\textbf{Neutrality.}} In preference annotation, \emph{neutrality} refers to judgments where neither of two alternatives is preferred, or where preference is indeterminate given the prompt and criteria. In this dissertation, neutrality is treated as a governance-relevant outcome state that can indicate contested or context-dependent values and can therefore serve as a trigger for deliberation, additional contextual information, or constraint-based decision rules \citep{Sorensen2024,ChristianoEtAl2017,ouyang2022,ConitzerEtAl2024}.

\subsection{Scope conditions}

The dissertation makes three scope commitments. First, the empirical studies are grounded in Montréal and in partnerships with community organizations; the dissertation does not claim that the distribution of values observed in Montréal is universal, but it claims that pluralism is a predictable condition of public-space governance and that the methods and governance architecture provide a transferable pattern for eliciting and governing pluralism under local adaptation \citep{Galston2002}. Second, both Street Review and LIVS use imagery as a primary modality because it is scalable and legible for participatory tasks; the dissertation therefore treats image-based AI as partial measurement infrastructure and does not treat image-derived outputs as substitutes for lived experience, on-the-ground engagement, or multimodal evidence \citep{SahaEtAl2019,Gabrys2016}. Third, pluralistic alignment is not framed as an argument for indecision; the dissertation argues that cities can decide under pluralism by distinguishing between dimensions that can be benchmarked through negotiated standards and dimensions that remain contested and therefore require documentation, deliberation, and recourse rather than forced optimization \citep{EspelandStevens2008}.

\subsection{Ethical stance}

The dissertation's ethical stance follows from its central claim that the algorithmic layer is infrastructure \citep{star1999ethnography,lazar2024lectureigoverningalgorithmic}. If models, datasets, and interfaces influence how public space is evaluated and governed, then research that constructs these artifacts raises ethical questions that extend beyond consent and privacy to include extractive participation, benefit sharing, documentation, and downstream repurposing risks \citep{CostanzaChock2020,Noble2018,Pasquale2015}. Across Street Review and LIVS, participation is treated as knowledge production rather than as post hoc validation, and the dissertation emphasizes the labor and situated expertise involved in producing training and evaluation data \citep{Strathern2000,sloane2022}. Models and datasets are documented to clarify intended uses, disallowed uses, representational limits, and uncertainty, aligning with broader work on dataset documentation and accountability \citep{gebru2021}. Because public-space AI outputs can be repurposed for punitive enforcement or exclusionary practice, the dissertation argues for explicit prohibitions against secondary uses that bypass due process and for recourse mechanisms through which affected groups can contest outputs and trigger review \citep{Eubanks2018,Benjamin2019,Reisman2018,selbst2021institutional}.

\medskip

The remainder of the dissertation develops PAU through a sequence of governance, empirical, and synthesis chapters. Next, Chapter 2 presents the unified research methodology.

\chapter{Research Methodology}
\label{chap:methodology}

This dissertation adopts a methodological premise that is also a governance claim: pluralism in public-space values is an empirical condition with institutional consequences. When publics disagree about whether a street feels accessible, safe, or inclusive, the divergence is treated as evidence about contested criteria, ambiguous stimuli, and heterogeneous experiences that must be represented and negotiated through legitimate procedures.

That premise constrains research design. Conventional evaluation pipelines often treat divergence as measurement error and respond by tightening instruments or aggregating responses into a single score. This dissertation instead treats disagreement and neutrality as empirical signals that require explanation, documentation, and procedural handling. The methodological objective is therefore not only to measure perceptions of public space, but also to make the conditions of agreement, contestation, and indeterminacy explicit in ways that can be audited and used as governance inputs.

This chapter provides a unified account of methods across the dissertation's manuscript chapters. It explains how normative analysis, participatory co-production, mixed-method empirical studies, and machine learning and preference-learning experiments are integrated into one auditable research program.

The chapter has seven sections. Section~\ref{sec:strategy} describes the program-level research strategy and design logic, including how the dissertation links governance claims to empirical demonstrations and technical artifacts. Section~\ref{sec:cases} justifies case selection and describes partnerships in Montréal that made co-production feasible. Section~\ref{sec:data} summarizes data sources, units of analysis, and instruments (with full protocols in the appendices). Section~\ref{sec:analysis} details analytic methods: qualitative synthesis, agreement analysis, the Street Review scaling pipeline, and pluralistic alignment through Direct Preference Optimization (DPO). Section~\ref{sec:validity} addresses validity, reliability, and methodological limits, with attention to the risk of misreading pluralism as noise. Section~\ref{sec:ethics} documents research ethics, consent, data governance, and reciprocity practices. Section~\ref{sec:summary} summarizes the chapter and previews how the methodological commitments motivate the rights-based argument in Chapter Three.

\section{Research strategy}
\label{sec:strategy}

\subsection{A programmatic, multi-method design}
\label{subsec:programmatic}

The dissertation is organized as a program of research rather than as a sequence of independent studies. Its methods were designed to support five linked functions:

\begin{enumerate}
\item \textbf{Conceptualization and governance framing:} specify AI systems affecting public space as governance objects and articulate institutional responsibilities for legitimacy (developed in Chapter Three and the thesis synthesis).
\item \textbf{Procedural operationalization:} translate participation and accountability from principles into lifecycle checkpoints, artifacts, and roles (developed in Chapter Four and used as a governance scaffold across empirical chapters).
\item \textbf{Elicitation of situated values:} elicit context-specific public-space descriptors and consolidate them into criteria usable for evaluation tasks (reported in the Street Review chapters and appendices).
\item \textbf{Empirical measurement of pluralism:} quantify where perceptions converge or diverge and how structured discussion changes convergence, while preserving residual disagreement as an empirical outcome (reported in the Montréal perceptions study; \cite{MushkaniJUM2025}).
\item \textbf{Scaling and alignment:} demonstrate how co-produced judgments can support subgroup-aware prediction and mapping, and how community-defined criteria can guide generative models while preserving neutrality as an informative outcome (reported in the Street Review chapter and the LIVS chapter).
\end{enumerate}

Methodologically, this program blends qualitative and quantitative approaches in a mixed-method logic that prioritizes interpretive validity and governance relevance over a single ideal of statistical generalization \citep{CreswellCreswell2022,Bryman2012}. Qualitative methods are used to produce categories, meanings, and descriptors grounded in participant language and interpretation practices \citep{BraunClarke2006}. Quantitative agreement analysis is used to characterize convergence and contestation and to evaluate how structured discussion changes collective judgments \citep{Bartko1966,KooLi2016,PuthEtAl2015}. Machine learning is used to scale measurements and to produce artifacts (predictions, subgroup outputs, heatmaps, documentation) that can be treated as decision-support evidence subject to audit and contestation, rather than as automated decisions \citep{Mitchell2019,gebru2021}. Preference learning and DPO are used to examine how community-defined criteria can be operationalized for generative models under documented constraints of binary preference signals and persistent neutrality \citep{wallace2023,rafailov2024}.

\subsection{Design logic}
\label{subsec:values-to-infra}

Across the empirical and technical manuscripts, the dissertation instantiates a \enquote{values-to-infrastructure} pipeline. The pipeline is not treated as a linear sequence in practice, but as an audit structure that makes dependencies between stages explicit and revisable.

\begin{itemize}
\item \textbf{Values elicitation and consolidation:} generate a descriptor set through interviews and focus groups, then consolidate descriptors into criteria that participants recognize as meaningful and usable for evaluation.
\item \textbf{Pluralistic measurement:} use structured rating, ranking, and preference tasks to measure within-participant coherence, between-participant agreement, and neutrality; distinguish criteria that are visually legible from criteria that remain socially contested.
\item \textbf{Scaling with accountability:} train predictive models on locally co-produced labels and report subgroup outputs and spatial patterns as decision-support artifacts, accompanied by documentation of intended use, disallowed use, and modality limits \citep{Mitchell2019,gebru2021}.
\item \textbf{Pluralistic alignment for generation:} co-produce criteria and preference annotations for text-to-image outputs, apply DPO to fine-tune a model, and treat residual neutrality as a governance signal rather than as missing data \citep{wallace2023,rafailov2024}.
\end{itemize}

\begin{figure}[htbp!]
\centering
\includegraphics[width=1\textwidth]{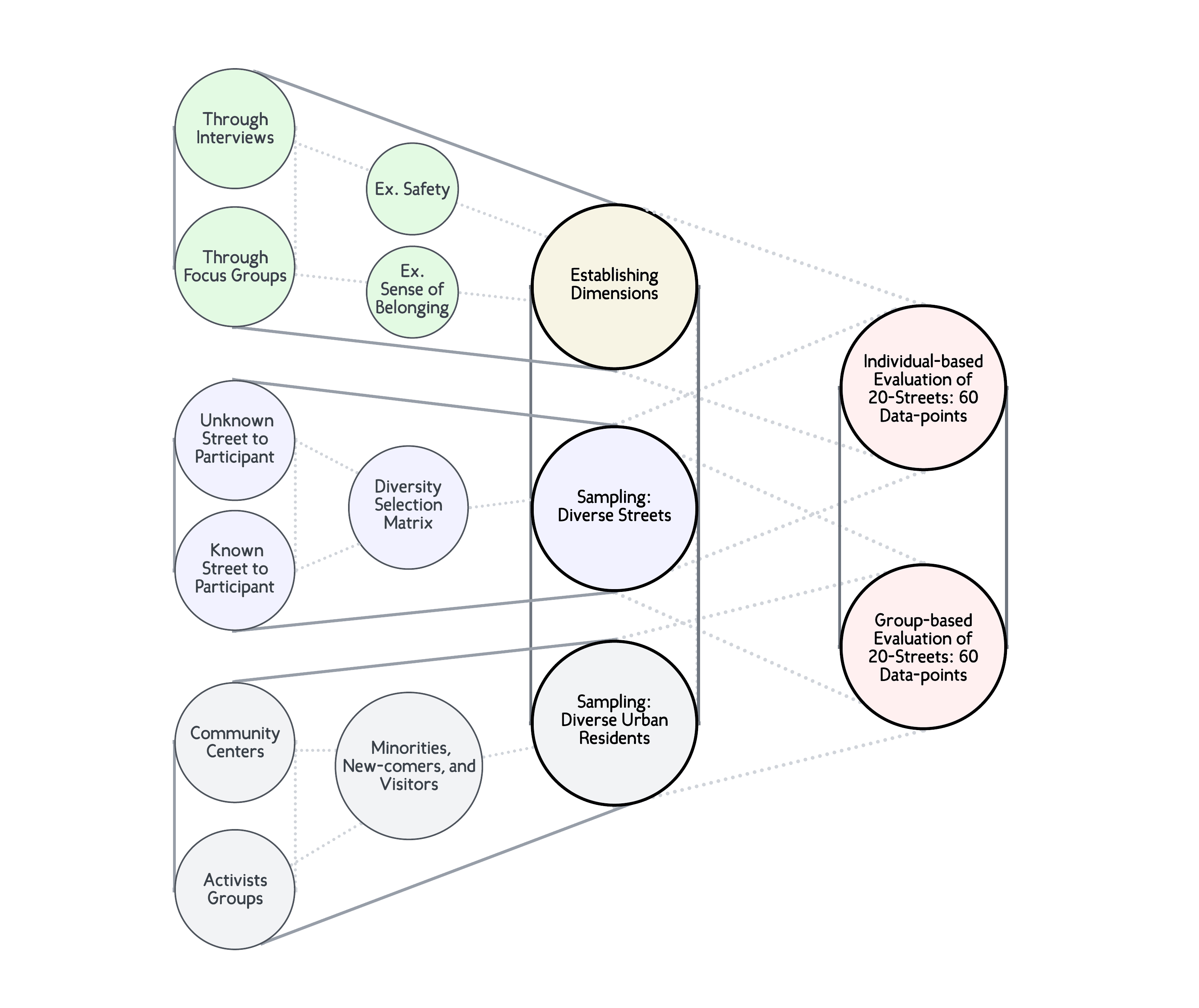}
\caption[Integrated Street Review workflow]{Integrated Street Review study workflow. Recruitment through community organizations supported a diverse participant pool; interviews and focus groups elicited descriptors (e.g., safety, sense of belonging) that were consolidated into evaluative dimensions; a diversity selection matrix guided the sampling of heterogeneous streets; and participants completed both individual and group-based evaluations of the same 20 streets represented through 60 data points. See \cite{MushkaniJUM2025} and \cite{MushkaniHabitat2025} for full protocols. \added{Source: Author's work, originally published in \citet{MushkaniJUM2025} and \citet{MushkaniHabitat2025}.}}
\label{fig:method_streetreview_workflow}
\label{fig:street-review-workflow}
\end{figure}

Figure~\ref{fig:method_streetreview_workflow} provides a concrete example of how the values-to-infrastructure pipeline was operationalized in the Street Review module, integrating recruitment through community partners, interviews and focus groups, sampling of diverse streets, and individual and group-based rating sessions on shared Montréal streetscape stimuli.

The design logic also reflects the participatory lifecycle developed in Chapter Four. In methodological terms, participation is distributed across framing, instrument development, iterative feedback on artifacts, and documentation practices, rather than being confined to early consultation. This distribution aligns participation with decision points in the production of datasets, models, and evaluation outputs, and supports traceability between community-defined criteria and technical artifacts \citep{Arnstein1969,CostanzaChock2020}.

\subsection{Units of analysis}
\label{subsec:units}

The dissertation uses multiple units of analysis because the research questions span governance concepts, social perceptions, and technical systems. The core units include:

\begin{itemize}
\item \deleted{\textbf{Participants and identity markers:} self-declared identity markers are used to describe heterogeneity and to support subgroup analyses, with explicit attention to the limits and risks of categorical representation.}\par
\added{\textbf{Participant descriptors and analytic categories:} Self-declared descriptors characterize the Street Review sample and structure within-sample outputs; the descriptors may overlap. Recruitment records document how community partners circulated the call. In Chapter~4, primary professional roles describe the purposively recruited workshop sample, while workshop transcripts were coded separately by theme. These categories support descriptive, within-sample analysis and no claim about population prevalence or sector-wide views.}
\label{rev:participant-units}
\item \textbf{Streetscape stimuli:} street-level images and vantage points are used as standardized stimuli for comparative evaluation, enabling agreement analysis and scalable prediction.
\item \textbf{Evaluative judgments:} ratings, rankings, pairwise preferences, and neutral responses are treated as empirical signals of convergence, contestation, and ambiguity.
\item \textbf{Model artifacts:} prediction outputs, subgroup distributions, maps, and documentation are treated as governance-relevant artifacts that can be audited, contested, and revised \citep{Mitchell2019,gebru2021}.
\item \textbf{Governance arguments and procedural designs:} normative claims and lifecycle proposals are supported by conceptual analysis and synthesis of prior work, including participatory and design-justice traditions \citep{Jacobs1961,Arnstein1969,CostanzaChock2020}.
\end{itemize}

This multi-unit structure enables traceability between governance claims and empirical and technical outputs. It also makes failure points explicit. If pluralism were not empirically structured, agreement would be high across criteria and subgroups for shared stimuli. If deliberation did not change convergence, structured discussion would not measurably alter agreement. If scaling were infeasible, prediction performance and mapping would fail to generalize beyond training conditions. If alignment could be reduced to a single objective, neutrality would be expected to collapse under tuning rather than persist as an outcome requiring procedural interpretation.

\section{Cases and partnerships}
\label{sec:cases}

\subsection{Case selection}
\label{subsec:montreal}

\deleted{Montréal, Québec is the principal empirical site for this dissertation. The Street Review studies and the Montréal perceptions study use Montréal because its socio-cultural diversity and bilingual governance context provide a setting for examining heterogeneous perceptions of streetscapes and for operationalizing inclusion as a criterion that may not admit a single stable interpretation. The case logic is not that Montréal is representative of all cities. The rationale is that the research questions require a context where pluralism is observable and where participatory partnerships are feasible.}

\deleted{Accordingly, the dissertation emphasizes transferability rather than statistical representativeness. Transferability is treated as an outcome of methodological design: criteria elicitation, annotation protocols, and documentation practices are specified to support adaptation to other settings, with scope conditions and limits stated alongside results \citep{CreswellCreswell2022,Bryman2012}.}

\added{Montréal's French-language municipal setting, multilingual population, established community partnerships, and heterogeneous street types made it possible to examine how people interpreted common streetscape stimuli through different histories and everyday constraints. The shared-image protocol held the stimuli and evaluation procedure constant \citep{MushkaniJUM2025,MushkaniKoseki2026}.}

\added{The evidentiary claims remain bounded to the participating residents, sampled streets, image dates, and Montréal's institutional setting. Application in another city requires renewed criteria elicitation, recruitment through trusted local organizations, sampling across local street types, and fresh model validation before municipal use. Transferability therefore resides in the documented procedure and its local revalidation \citep{CreswellCreswell2022,Bryman2012}.}
\label{rev:montreal-case}

\subsection{Partnerships}
\label{subsec:partnerships}

Across the empirical studies, recruitment relied on partnerships with community organizations serving diverse populations. This strategy was adopted to reduce barriers to participation, align study protocols with community context, and mitigate reliance on convenience samples. In the Street Review study, over one hundred community organizations were contacted in spring and summer 2023; thirty-five citizens expressed interest; twenty-eight individuals participated in interviews and focus groups; and twelve contributed to the structured rating exercise. To scale the analysis, the study used the same participatory base to produce local labels and integrated those evaluations with large-scale street-view imagery to enable citywide inference. The LIVS dataset was developed over a two-year period in collaboration with thirty community organizations through workshops and interviews used to elicit and refine criteria and to develop an annotation protocol for text-to-image outputs.

\added{Recruitment sources, participant descriptors, and analytic coding were recorded separately. Community organizations circulated the call and supported access. Street Review used self-declared identity markers to describe the sample and organize subgroup outputs; markers could overlap, and findings are interpreted as patterns within the sample. The Chapter~4 workshops used purposive recruitment to obtain disciplinary and institutional breadth. Participants self-identified a primary professional role for sample description, while workshop transcripts were coded inductively and deductively by theme. The professional-role labels were not used to claim representative sector positions.}

\added{Digital-participation scholarship distinguishes data contribution from authority over problem definition and outcomes. \citet{Goodchild2007} establishes how residents can produce volunteered geographic information at scale. \citet{Sieber2006} shows that public-participation GIS is shaped by place, participants, technology, process, and institutional outcomes. \citet{Haklay2013} distinguishes forms of citizen science according to participants' involvement in problem definition, data collection, and analysis. \citet{Brabham2009} shows that crowdsourcing can widen participation while the commissioning institution may retain selection rules and final authority. These distinctions guide the allocation of roles here: community partners supported recruitment and protocol adaptation; participants generated descriptors and evaluated shared stimuli; and the research team documented criteria consolidation, model construction, scaling, and final analytic decisions.}
\label{rev:participation-sources}

Two partnership commitments were carried across projects. \deleted{First, co-production was treated as a knowledge-production method in which community members contribute to defining criteria and interpreting artifacts, not as a data-extraction practice. This commitment aligns with participatory traditions that treat participation as a question of authority and decision power \citep{Arnstein1969}.}
\added{First, co-production was organized as shared knowledge production: community members contributed to criteria formation and artifact interpretation, and the research record identifies the later stages controlled by the research team. This allocation follows the account of participation as decision power developed by \citet{Arnstein1969} and keeps the limits of participant authority visible.} Second, studies were designed with iterative feedback loops and role clarity, including attention to resourcing and logistics that condition feasibility and equity of participation.
\label{rev:participant-coding}

\subsection{Why streets, images, and generative models}
\label{subsec:why-streets}

The dissertation focuses on streets because streets are a primary unit of public-space experience and a routine unit of municipal planning, maintenance, and investment. Streets also operationalize the dissertation's dual infrastructure claim: streets are materially designed and managed, and are increasingly mediated by evaluative systems, maps, and image-based decision support.

Street-level images are used as standardized stimuli for two reasons. First, images enable participants to evaluate shared stimuli, which makes pluralism observable in comparative tasks and supports agreement analysis. Second, image-based datasets enable scaling through computer vision, which makes it possible to examine citywide patterns while documenting what remains outside the modality's representational \citep{Naik2014}.

Generative text-to-image models are included because they increasingly shape how urban futures are visualized and communicated in planning and design contexts \citep{vonBrackelSchmidtEtAl2024,valenca2025visualizations}. In this dissertation, generative systems are treated as part of the algorithmic layer through which public-space imaginaries are produced and circulated. The methodological objective is to test whether community-defined criteria can be operationalized in preference data and whether residual neutrality persists as a measurable outcome that requires governance interpretation rather than technical elimination \citep{wallace2023,rafailov2024}.

\section{Data sources and instruments}
\label{sec:data}

\subsection{Overview of data sources}
\label{subsec:data-overview}

\Cref{tab:data-sources} summarizes primary data sources across the dissertation and the unit of analysis for each. Detailed protocols, recruitment materials, interview guides, and annotation interfaces are documented in the dissertation appendices. This chapter consolidates shared methods while manuscript chapters report chapter-specific operational details and results.
\begin{table}[tbp]
\centering
\caption{Data sources and units of analysis across the dissertation.}
\label{tab:data-sources}
\small
\renewcommand{\arraystretch}{1.1}
\setlength{\tabcolsep}{6pt}

\begin{tabularx}{\textwidth}{@{}
>{\raggedright\arraybackslash}X
>{\raggedright\arraybackslash}X
>{\raggedright\arraybackslash}X
@{}}
\toprule
\textbf{Source type} &
\textbf{Examples in this dissertation} &
\textbf{Primary unit(s) of analysis} \\
\midrule

Interviews and focus groups &
Semi-structured interviews and focus groups used to elicit descriptors and interpret streetscape stimuli \citep{Krueger2002} &
Transcript segments; elicited descriptors; participant narratives \\

\addlinespace[0.4em]
Structured rating and ranking tasks &
Individual and group-based evaluations of street images across criteria, including structured discussion and re-evaluation &
Ratings, rankings, pairwise choices, and within-group convergence measures \\

\addlinespace[0.4em]
Street imagery datasets &
Locally collected street-view images and large-scale street-view imagery for citywide inference &
Image frames; vantage points; street segments \\

\addlinespace[0.4em]
Workshop data (participatory AI governance) &
Multidisciplinary workshops supporting the augmented AI lifecycle synthesis &
Workshop transcripts; coded themes; lifecycle artifacts \\

\addlinespace[0.4em]
Preference annotations for generative models &
Pairwise comparisons of text-to-image outputs under one to three criteria, including a neutral option &
Pairwise preference labels; neutral judgments; prompt metadata \\

\addlinespace[0.4em]
Model artifacts and documentation &
Prediction outputs, subgroup scores, maps; dataset and model documentation artifacts \citep{Mitchell2019,gebru2021} &
Model outputs; error analyses; documentation entries \\

\bottomrule
\end{tabularx}
\end{table}

\subsection{Interviews and focus groups}
\label{subsec:interviews}

The Street Review studies used a multi-phase design of interviews, focus groups, and structured evaluation exercises to elicit how streetscapes are interpreted across multiple evaluative dimensions. Semi-structured interviews and focus groups were used to surface descriptors, interpret visible cues in street images, and articulate how built features relate to experiences of access and inclusion. Focus groups used images as prompts and invited participants to generate written and verbal descriptors to surface tacit categories and context-specific meanings \citep{Krueger2002,BraunClarke2006}.

Descriptor consolidation proceeded iteratively. The analytic objective was to reduce a large pool of participant-generated descriptors into a smaller set of criteria usable for structured evaluation, while preserving fidelity to participant language and interpretive distinctions. Consolidation combined thematic analysis and clustering as complementary procedures: thematic coding supported semantic consolidation and traceability to transcript segments, while clustering supported reduction of redundancy and identification of descriptor families. Reporting follows established guidance for qualitative transparency, including structured description of recruitment, data collection, and analytic procedures \citep{tong2007coreq,CreswellCreswell2022}.

\subsection{Street image stimuli and sampling}
\label{subsec:stimuli}

Street Review used street-level image stimuli to enable comparative evaluation of shared scenes. Streets were selected using a diversity sampling matrix that considered land use, history, socio-economic status, density, and pre-occupancy and post-occupancy conditions. The resulting sample included twenty streets and sixty vantage points in the Montréal metropolitan region. In the rating exercise, participants evaluated one hundred twenty images representing those streets, with multiple images per vantage point.

Street Review extends this design for citywide inference. Locally collected imagery was captured in a three-hundred sixty degree rotation for each vantage point to increase coverage of streetscape features visible from a single location. Large-scale inference was conducted on street-view imagery to produce citywide assessment outputs. Throughout, images are treated as partial proxies for public-space experience: they support standardization for comparative tasks, but they cannot represent determinants of inclusion that are non-visual, historical, or interactional.

\subsection{Rating and ranking instruments}
\label{subsec:ratings}

The empirical studies use structured evaluation tasks designed to support individual-level measurement and deliberation-aware analysis. In Street Review, structured rating exercises asked participants to assign scores across evaluative dimensions for each vantage point based on visible cues and participant-defined meanings. The Montréal perceptions study similarly examines subjective assessments and compares individual and group-based evaluations to measure convergence and the persistence of discrepancy for socially contested criteria \citep{MushkaniJUM2025}.

Because these tasks rely on ordinal response formats, analysis decisions are documented as part of validity logic. Where Likert-type responses are treated as approximately interval for purposes of agreement, correlation, and predictive modeling, this is justified as a pragmatic convention in applied social research, while recognizing that interpretation must remain cautious and triangulated \citep{CarifioPerla2008,Norman2010}. The analytic strategy therefore emphasizes agreement patterns, distributional differences, and sensitivity to deliberation, rather than inference to a single latent ground-truth score.

\subsection{LIVS instruments}
\label{subsec:livs-instruments}

The LIVS chapter reports a community-based participatory process used to develop a pluralistic alignment dataset for inclusive public spaces. Workshops and interviews were used to elicit a large pool of concepts that participants associated with equitable public-space design. Through validation activities, concepts were consolidated into intermediate criteria and then distilled into six criteria used throughout the dataset: Accessibility, Safety, Comfort, Invitingness, Inclusivity, and Diversity. Prompting workshops produced prompts that represent public-space scenarios and features, and the annotation platform was tested through tutorial and feedback sessions to assess usability and representation quality. Generated images were produced using Stable Diffusion XL \citep{podell2023sdxlimprovinglatentdiffusion}.

The dataset comprises more than one hundred thousand pairwise comparisons spanning thousands of generated images, with preference annotations structured to cover one to three criteria per comparison. The annotation interface included an option to record neutral or indeterminate judgments. This instrument design operationalizes the dissertation's pluralism premise: neutrality is treated as an observable outcome that may reflect ambiguity, multi-criteria trade-offs, or value conflict, and is therefore preserved for analysis rather than discarded.

\section{Analytic methods}
\label{sec:analysis}

\subsection{Qualitative analysis}
\label{subsec:qual}

Qualitative data across the Street Review studies and the lifecycle workshops were analyzed with thematic coding approaches that combine inductive emergence with deductive scaffolding \citep{BraunClarke2006}. In Street Review, interviews and focus groups were analyzed to elicit descriptors and to interpret how participants relate built features to experiences of access, comfort, safety, and inclusion. The analysis proceeded iteratively: early coding and consolidation informed subsequent data collection, and consolidated criteria then guided subsequent structured evaluation exercises.

In the chapter on the co-production of the AI lifecycle (Chapter~\ref{chap:coproducing_ai_lifecycle}), the literature review and workshop transcripts were coded using an inductive-deductive approach. Codes were initially informed by design-justice and expansive-learning theory, then expanded as categories emerged and stabilized through iterative review and peer debriefing. The synthesis triangulated literature and workshop data to produce a five-stage augmented lifecycle intended to map participation to risks and artifacts across phases \citep{CostanzaChock2020,Engestrm2010,Engestrom2014}.

\subsection{Quantitative agreement and convergence analysis}
\label{subsec:agreement}

Agreement and correlation analysis are used to distinguish convergence from contestation across evaluative dimensions and to measure the impact of structured discussion on collective judgments. Agreement is evaluated using complementary metrics, including intraclass correlation coefficients (ICC) and rank or linear correlations, to reduce reliance on any single statistic \citep{Bartko1966,KooLi2016,PuthEtAl2015}. The analytic objective is not to validate a single collective truth, but to identify where judgments are stable across participants, where they diverge, and whether structured discussion changes the evidentiary object by producing different patterns of convergence.

Two methodological decisions shape interpretation. First, agreement metrics are used diagnostically rather than as proof of a single correct assessment. Low agreement is interpreted as a candidate signal of contestation or context dependence, while remaining open to the possibility of instrument ambiguity. Second, individual and group-based judgments are compared to characterize the effects of structured discussion on convergence, and to identify where disagreement persists after deliberation as an empirical outcome that requires documentation.

\subsection{Machine learning pipeline}
\label{subsec:ml}

The Street Review chapter develops a supervised multi-output regression pipeline that predicts subgroup and group-level ratings from street-view imagery and produces citywide maps as decision-support artifacts. The pipeline uses semantic segmentation to extract pixel-level features and an attention-based multi-layer perceptron (MLP) to map those features to predicted scores across criteria and identity marker groups. Training uses mean squared error loss and a stratified split of labeled data into training, validation, and test sets to reduce leakage across closely related vantage points \citep{Bishop2006}.

Model evaluation emphasizes predictive performance and interpretability-oriented diagnostics rather than deployment claims. Performance is reported using coefficient of determination ($R^{2}$) on validation and held-out test sets, and permutation importance is used to estimate the influence of input features on outputs \citep{Molnar2025}. The methodological stance is that prediction outputs and maps are governance artifacts that require documentation of scope, assumptions, and limits. Accordingly, model reporting and dataset documentation are treated as part of the analytic method, enabling contestability of outputs and constraints on use \citep{Mitchell2019,gebru2021}.

\subsection{Preference learning and pluralistic alignment}
\label{subsec:dpo}

The LIVS chapter operationalizes pluralistic alignment for generative models by combining community-defined criteria with pairwise preference data and DPO fine-tuning of a Stable Diffusion XL model \citep{podell2023sdxlimprovinglatentdiffusion,wallace2023,rafailov2024}. The dataset includes multi-criteria annotations and a neutral option. Because DPO requires binary preference signals, training requires a documented reduction from multi-criteria feedback to a single preference label for each pair. This reduction is treated as a methodological constraint that motivates careful interpretation of both gains in preference match and persistence of neutrality.

Evaluation compares a DPO-tuned model to a baseline and reports the distribution of preferred, dispreferred, and neutral judgments on held-out comparisons. Within the dissertation's logic, the methodological objective is not to eliminate pluralism through optimization, but to use preference learning as a measurable lens on contested criteria, prompt sensitivity, and residual indeterminacy. Neutral judgments are treated as governance-relevant outcomes that can motivate additional deliberation, re-specification of criteria, or constraints on model use.

\subsection{Normative and governance analysis methods}
\label{subsec:normative}

The governance chapters rely on interpretive and conceptual analysis grounded in urban rights traditions and participatory governance literatures, and operationalized through procedural synthesis. The methodological output of this work is a set of institutional requirements and process designs that connect empirical findings from Street Review and LIVS to municipal procedures. In Chapter Four, the lifecycle synthesis method combines a scoping review with multidisciplinary workshops and thematic coding to generate a five-stage augmented lifecycle, and to map participation activities to risks and artifacts across phases \citep{CostanzaChock2020,Engestrm2010,Engestrom2014}.

Across the dissertation, technical artifacts (datasets, model cards, evaluation logs, subgroup outputs, maps) are treated as governance objects whose production and interpretation must be institutionally authorized and contestable \citep{kaminski2021right}. Documentation practices are therefore treated as methodological instruments that specify provenance, intended use, and disallowed use, and that support audit and recourse \citep{kaminski2018right,zhang2024right}. Privacy and legitimacy claims are similarly operationalized through a contextual integrity approach that treats data use as context-dependent and institutionally bounded \citep{Nissenbaum2010,zhang2024right}.

\section{Methodological limits}
\label{sec:validity}

\subsection{Validity logic}
\label{subsec:pluralism-validity}

A primary validity risk in this dissertation is interpretive: pluralism can be misread as noise and therefore treated as a defect to be eliminated through aggregation. The research design addresses this risk by separating three questions that are often conflated.

First, do participants understand and apply criteria coherently within themselves across tasks and stimuli. This is addressed through iterative descriptor consolidation and participant validation during instrument development, and by using shared stimuli to reduce ambiguity in what is being evaluated.

Second, do participants agree with one another on shared stimuli. This is addressed through agreement metrics and correlation analyses, interpreted as evidence about convergence and contestation rather than as a reliability defect by default \citep{Bartko1966,KooLi2016,PuthEtAl2015}.

Third, can models replicate co-produced judgments at scale under documented assumptions. This is addressed through held-out evaluation, subgroup reporting, and interpretability diagnostics, interpreted as bounded feasibility under image-based proxies \citep{Molnar2025,Mitchell2019}.

This logic does not deny conventional measurement concerns. Instruments can be underspecified, and data quality limitations can reduce reliability. The methodological commitment is that high agreement is not treated as an unconditional goal for governance questions. The relevant distinction is between disagreement driven by instrument ambiguity and disagreement driven by value conflict or context dependence. The dissertation therefore treats residual disagreement and neutrality as findings that motivate procedural responses, not as artifacts to be statistically eliminated.

\subsection{Reliability}
\label{subsec:reliability}

Reliability is addressed through triangulation across methods and through complementary metrics. In Street Review, qualitative elicitation informs quantitative scoring, and scores are interpreted alongside narratives that explain how judgments are produced and justified \citep{BraunClarke2006}. In the Montréal perceptions study, multiple agreement metrics are used to reduce dependence on a single statistical lens \citep{Bartko1966,KooLi2016,PuthEtAl2015}. In Street Review scaling, predictive performance is reported on validation and test splits and complemented with feature-importance analysis to support interpretability and to surface potential proxy risks \citep{Molnar2025}.

For LIVS, reliability concerns arise because preference judgments are context-dependent and because DPO requires binary labels. The dataset design mitigates cognitive fatigue by limiting each comparison to one to three criteria and by structuring annotation batches to provide coverage across criteria. Persistence of neutral responses is interpreted conservatively: neutrality can reflect ambiguous images, conflicting criteria, prompt sensitivity, or value pluralism. The dissertation treats these possibilities as analytically separable hypotheses and as governance-relevant outcomes, rather than as a single category of annotation error.

\subsection{Limits}
\label{subsec:limits}

\deleted{Three methodological limits are carried explicitly into the dissertation's claims.}

\deleted{First, \textbf{representational limits of identity markers}. Subgroup analyses rely on self-declared categories that are necessarily coarse. They are used to reveal structured differences and to reduce majority averaging, but they risk reifying categories and obscuring within-group heterogeneity. The dissertation therefore treats subgroup outputs as prompts for deliberation and safeguards, not as definitive statements about groups.}

\deleted{Second, \textbf{image-based proxy limits}. Both Street Review scaling and LIVS alignment rely on visual modalities. Images can represent some features relevant to accessibility and design, but they may omit determinants of inclusion linked to history, policing, stigma, or symbolic cues. These limits are treated as governance constraints on permissible use of image-based outputs, and are documented through model reporting \citep{Mitchell2019,gebru2021}.}

\deleted{Third, \textbf{local grounding and transferability}. Empirical work is grounded in Montréal and in partnerships with specific organizations. The dissertation's claims are therefore about governance patterns and methodological design for adaptation, not about universal citywide truths. Transferability requires re-elicitation of criteria, localized annotation, and institutional design appropriate to the local context \citep{CreswellCreswell2022,Bryman2012}.}

\added{Recruitment and participant coding define the first limit. Partner-based purposive recruitment reached participants through community organizations serving people with distinct relationships to public space. The resulting samples remain shaped by partner networks, willingness, availability, and the accessibility of the research format. Residents beyond those networks may be absent. Self-declared markers overlap, and several analytic cells are small. Small cells restrict subgroup findings to descriptive, within-sample contrasts. The categories can reveal majority averaging and can also stabilize identities that are fluid, contextual, and internally heterogeneous.}

\added{The visual modality defines a second limit. Static imagery standardizes the stimulus and leaves sound, time of day, weather, maintenance cycles, enforcement, social interaction, sensory conditions, and symbolic histories outside the frame. Mapillary adds uneven spatial coverage and variation in lighting, blur, distortion, and date of capture. Authorized use of visual outputs is consequently limited to exploratory assessment of visible features relevant to the named criteria, accompanied by field verification and other forms of local evidence wherever decisions concern lived experience.}

\added{The structure of the Street Review dataset defines a third limit. The 15{,}000 local frames inherit labels from 60 street-level data-point groups across 20 streets and remain correlated observations. Grouped train, validation, and test splits keep frames from the same data-point group together, with street-level grouping used as a stronger test. The reported validation and held-out test results establish bounded predictive feasibility under the sampled streets and image conditions. The approximately 45{,}000 Mapillary images used for inference add spatial coverage and no independent human labels. External application therefore requires renewed local sampling, annotation, grouped validation, and error analysis.}

\added{The deliberative procedure defines a fourth limit. Changes in agreement after small-group discussion can reflect reason-giving, facilitation, group composition, compromise, fatigue, or silence. Convergence is interpreted as an outcome of the documented procedure. Disagreement logs and qualitative records are needed to preserve the reasons, exclusions, and unresolved claims hidden by a summary agreement statistic.}

\added{LIVS defines a fifth limit. Neutral judgments can register value indeterminacy, ambiguous images, limited visual difference, prompt sensitivity, or fatigue. In the resident evaluation workshop, 1{,}100 of 2{,}100 comparisons were neutral. Preference tuning therefore supports claims about the comparisons on which participants expressed a direction and under the six elicited criteria. The neutral region constrains any broader claim of community preference, and scalar comparisons leave harm, stereotyping, and symbolic exclusion requiring separate review.}

\added{Local grounding bounds all five limits. The descriptors, criteria, samples, and institutional relationships arise from Montréal and its participating organizations. Transferability resides in a documented procedure for renewed elicitation, sampling, annotation, deliberation, and validation \citep{CreswellCreswell2022,Bryman2012}. Each adopting city must also specify the authority through which local evidence can revise, pause, or end a municipal use.}
\label{rev:method-limits}

\Cref{tab:validity-mitigations} summarizes selected validity threats and mitigations, with attention to pluralism and proxy risks.
\begin{table}[htbp]
\centering
\caption{Selected validity threats and mitigations, with particular focus on pluralism and proxy risks.}
\label{tab:validity-mitigations}

\renewcommand{\arraystretch}{1.08}
\setlength{\tabcolsep}{6pt}
\footnotesize
\begin{tabularx}{\textwidth}{@{}P{0.23\textwidth}P{0.34\textwidth}L@{}}
\toprule
\textbf{Threat} & \textbf{Why it matters here} & \textbf{Mitigation in this dissertation} \\
\midrule
Pluralism misread as noise &
Disagreement could be treated as measurement error, motivating unjustified aggregation &
Use agreement metrics diagnostically; interpret residual disagreement and neutrality as outcomes; triangulate quantitative patterns with qualitative explanations \citep{Bartko1966,KooLi2016,PuthEtAl2015,BraunClarke2006} \\
\addlinespace
Instrument drift and criterion ambiguity &
Criteria such as inclusivity can be socially contested and context-dependent &
Iterative elicitation and consolidation; participant validation during instrument development; structured deliberation and re-evaluation \citep{Krueger2002,BraunClarke2006,tong2007coreq,CreswellCreswell2022} \\
\addlinespace
Image proxy limitations &
Visual stimuli may omit non-visual or historical determinants of inclusion &
Explicit scope limits; treat outputs as audit and decision-support artifacts; document disallowed uses in model reporting \citep{BarocasSelbst2016,gebru2021} \\
\addlinespace
Category reification in subgroup modeling &
Identity categories can obscure within-group heterogeneity and reinforce stereotypes &
Use self-declaration; report subgroup outputs as governance inputs; emphasize deliberation and documentation; avoid normative claims about groups \citep{Dwork2012,CostanzaChock2020,BarocasHardtNarayanan2023} \\
\addlinespace
Binary simplification in DPO &
Collapsing multi-criteria feedback to binary labels can erase trade-offs and internal conflict &
Document as methodological constraint; interpret neutrality as governance signal; treat preference-learning outputs as diagnostic rather than dispositive \citep{Sorensen2024,wallace2023,rafailov2024} \\
\bottomrule
\end{tabularx}
\end{table}

\section{Ethics, consent, and data governance}
\label{sec:ethics}

The dissertation's ethics approach follows the substantive commitments of the work. If AI systems and their data practices must be governed with transparency, participation, and recourse, then research that produces datasets and models for urban governance must also be accountable to participants and communities. Consent, privacy, and reciprocity are treated as methodological requirements integrated into study design and documentation.

\subsection{Consent}
\label{subsec:consent}

Across studies involving interviews, focus groups, workshops, and evaluation tasks, participants provided informed consent. Interview and workshop data were audio recorded where consented, transcribed for analysis, and anonymized for reporting. The Montréal perceptions study was reviewed by a university research ethics committee. Protections were implemented through recruitment via community organizations, anonymization of transcripts, and careful handling of sensitive judgments about inclusion and representation. Reporting practices follow established guidance for transparency in qualitative research \citep{tong2007coreq,CreswellCreswell2022}.

\subsection{Privacy}
\label{subsec:privacy}

Two privacy commitments are emphasized. First, data minimization and contextual integrity principles guide what is collected and how it is reused, treating privacy as context-dependent and institutionally bounded \citep{Nissenbaum2010}. Second, documentation is treated as an ethical instrument. Dataset and model reporting practices are used to make provenance, intended uses, and disallowed uses visible to technical and non-technical stakeholders \citep{gebru2021,Mitchell2019}.

Street imagery is used because it enables standardized stimuli and scalable inference, but street imagery and derived maps can raise ethical concerns when combined with scoring and spatial visualization. The dissertation therefore treats such outputs as governance artifacts requiring explicit scope limits and documentation of disallowed uses. For LIVS, generated images and preference annotations concern inclusion and representation; the dataset is framed as a research benchmark for pluralistic alignment and is interpreted in light of local grounding and the limits of preference-based optimization \citep{wallace2023,rafailov2024}.

\subsection{Reciprocity}
\label{subsec:reciprocity}

The dissertation adopts a non-extractive stance in which participants' time, labor, and expertise are treated as inputs requiring both material resourcing and role clarity. Reciprocity is enacted through appropriate financial compensation, iterative feedback processes, and accessible sharing of study outputs. In addition, the research design includes opportunities for participant learning, such as workshops and hands-on sessions focused on understanding and critically engaging with AI systems, thereby supporting capacity-building alongside knowledge production. 

Reciprocity is further expressed through the framing of datasets and models as civic audit infrastructure rather than as proprietary extraction of local knowledge. This commitment aligns with design-justice principles that center decision-making authority with those most affected by technological outcomes \citep{CostanzaChock2020,Arnstein1969}. Where research artifacts are made public, documentation practices are used to clarify intended uses, governance constraints, and potential downstream impacts \citep{Mitchell2019,gebru2021}.

\section{Summary}
\label{sec:summary}

This chapter consolidated the dissertation's methodological commitments into a program-level account. The research is designed to treat pluralism as an empirical condition and a governance requirement, integrating participatory co-production with mixed-method measurement, scalable modeling, and preference learning. The chapter specified a values-to-infrastructure design logic in which qualitative elicitation and consolidation feed structured evaluation tasks, agreement analysis distinguishes convergence from contestation, and technical artifacts are produced as auditable governance inputs. It articulated a validity logic that separates within-participant coherence, between-participant agreement, and model-based scaling feasibility, and it treated neutrality and residual disagreement as outcomes requiring documentation and procedural handling rather than elimination.

The next chapter builds directly on these methods by addressing the institutional question that follows from the empirical and technical findings: what rights, duties, and participatory powers are required for legitimacy when AI mediates public space.


\chapter{The Right to AI}
\label{chap:right-to-ai}

This chapter provides the normative foundation for the dissertation by developing a civic \emph{Right to AI} for public-space contexts.
As AI systems increasingly shape the conditions of urban life, individuals and communities require not only protection from harms, but ongoing, institutionally recognized capacity to co-govern the AI infrastructures that mediate public decisions.

The dissertation's core premise is that twenty-first-century cities are governed through a dual infrastructure: the material city (streets, buildings, services) and an epistemic layer that makes places legible through data, models, and algorithmic representations.
When the epistemic layer influences planning, maintenance, enforcement, and design, it becomes a governance object rather than a neutral tool.
The Right to AI specifies what democratic legitimacy requires for this layer, especially where AI systems affect the public realm.

The chapter proceeds in six steps.

\begin{enumerate}[leftmargin=*, itemsep=0.4em]
\item \cref{sec:ch3-introduction} motivates why a rights-based approach is needed when AI becomes civic infrastructure.
\item \cref{sec:ch3-rights-framing} grounds the Right to AI in the urban-right tradition and in the broader human-rights lineage concerning technology, participation, and collective goods.
\item \cref{sec:ch3-right-to-ai-requirements} specifies the scope of the Right to AI in public-space governance, clarifying the difference between individual procedural rights and collective power over infrastructural design choices.
\item \cref{sec:ch3-participation-power} operationalizes participation as power using Arnstein's ladder and a four-tier model that characterizes prevailing governance regimes and aspirational endpoints.
\item \cref{sec:ch3-municipal-implications} translates the Right to AI into municipal responsibilities and procurement implications, anticipating the procedural lifecycle developed in Chapter Four.
\item \cref{sec:ch3-discussion} addresses market-led and state-centric alternatives, common objections to participatory governance, and boundary conditions for public-space applications.
\end{enumerate}

The chapter advances a governance claim:
if AI systems become part of the epistemic infrastructure through which cities see and act, then democratic urbanism requires enforceable participation rights and institutions that treat AI as collectively governable infrastructure rather than as a vendor product or an expert-only domain \citep{Lefebvre1968,Arnstein1969,plantin2018infrastructure,Sun2020,Wenar2023}.

\section{Introduction}
\label{sec:ch3-introduction}

AI systems increasingly mediate how cities represent and manage public  \citep{LarteyLaw2025,yigitcanlar2025editorial,Ibrahim2020}.
In contemporary municipal practice, the relevant systems are not limited to narrow ``smart city'' sensing deployments.
They include predictive models used in service allocation and risk assessment, computer-vision systems that classify street-level imagery, large language models that summarize citizen input, and generative systems that produce design alternatives and visualizations for planning workflows \citep{Kitchin2023,LarteyLaw2025,Ibrahim2020,Lepri2018,Taeih2021,BommasaniEtAl2022}.
Even when these systems are described as decision support, their representations shape what becomes legible, measurable, and therefore governable \citep{plantin2018infrastructure,Kitchin2023}.
Such representational infrastructures can also produce epistemic injustice by systematically discounting some speakers and some forms of local knowledge \citep{Fricker2007,GowaikarEtAl2024}.

Public spaces are shared and contested civic goods whose ``quality'' cannot be reduced to a single technical objective \citep{Gehl2011,low2020social}.
Streets, parks, and public facilities organize access, mobility, visibility, and safety.
They also carry symbolic meanings and govern social recognition.
Because the public realm is a site of ongoing contestation, the legitimacy of any AI-mediated representation depends not only on accuracy, but on whether the underlying objectives, data practices, and trade-offs have been democratically authorized \citep{Lefebvre1968,Harvey2012,Purcell2014}.

This chapter argues that a rights-based framing is the appropriate starting point for that authorization.
A Right to AI asserts that individuals and communities affected by AI systems should have meaningful, continuing capacity to shape, critique, and govern them. The claim is not that every resident must become a machine learning specialist. Rather, it is that participation must be defined as power over consequential choices: what a system is for, what it is not for, what data may be extracted, which risks are acceptable, who can audit, and how disagreement and contestation are represented \citep{Arnstein1969,CostanzaChock2020,Birhane2022-power,sloane2022}.

The Right to AI is motivated by three structural facts about contemporary AI governance.

\paragraph{First, AI systems are infrastructural.} Infrastructure is durable, interconnected, and path-dependent \citep{star1999ethnography,BowkerStar1999}. When AI models and datasets become embedded in workflows, they shape downstream possibilities and redistribute agency \citep{plantin2018infrastructure,Graham2001,Kitchin2016}. The infrastructure framing therefore shifts the normative question. Instead of asking only whether a model is ``ethical,'' the governance question becomes: who has standing to co-govern the infrastructural layer that mediates public decisions and public knowledge?

\paragraph{Second, AI depends on socially produced data.}
Data is created through collective life and institutional practice, including public administration and everyday participation in digital platforms.
When this data is enclosed through proprietary ownership and opaque licensing, the communities that produced the data lose agency over how it is used and monetized \citep{Benkler2006,Beer2016,Kitchin2016,NuceraOnuoha2018}.
This applies in public-space contexts where street imagery, mobility traces, and administrative records can be repurposed for surveillance and exclusion if governance is weak \citep{Brayne2017,CohenSuzor2024}.

\paragraph{Third, value pluralism is an empirical and normative condition.}
Urban publics differ in what they need and what they consider  \citep{rawls1993political,Mouffe2000,Mehta2019}.
Pluralism cannot be treated as noise to be averaged away, because doing so can erase minority harms and foreclose legitimate disagreement \citep{Habermas1996,Fraser1995,CostanzaChock2020,Jain_2024}.
A Right to AI therefore requires institutions that can register and negotiate plural values rather than collapsing them into a single metric.

The remainder of the chapter formalizes this rights-based approach and ties it to municipal practice.
The next sections ground the Right to AI in urban-right traditions, specify its scope for public-space governance, and operationalize participation as power.

\section{Rights-based framing for AI in public space}
\label{sec:ch3-rights-framing}

\subsection{From the Right to the City to the Right to AI}
\label{sec:ch3-right-to-ai-public-space}

A conceptual bridge for a civic Right to AI is Henri Lefebvre's \emph{Right to the City}.
Lefebvre argued that urban life should not be governed solely through state technocracy or market exchange, but through collective participation in the production of the city as an oeuvre, a shared work that expresses and sustains social life \citep{Lefebvre1968,Purcell2014}.
Subsequent right-to-the-city scholarship has clarified the political stakes of this claim: who can shape urban space, whose needs count, and how the distribution of urban resources and recognition is contested and renegotiated \citep{Harvey2012,MaddenMarcuse2017}.

The Right to AI extends this lineage to the epistemic layer of contemporary urbanism.
If the city is increasingly governed through algorithmic representations of neighborhoods, risk, accessibility, and ``quality,'' then the right to collectively shape urban life must include the right to contest and co-govern the infrastructures that produce those representations.
The extension is not metaphorical.
Urban residents already encounter AI systems as part of housing access, mobility services, public communication, and policing.
These systems change the terms under which residents can participate in and appropriate public space, sometimes by shaping what institutions can perceive and sometimes by reshaping the incentives that structure everyday movement and presence \citep{Kitchin2016,Taeih2021,Brayne2017}.

Jane Jacobs' critique of top-down planning provides a complementary anchor.
Jacobs emphasized that urban knowledge is situated and that street-level lived experience can reveal harms that are not legible to centralized planning and abstract indicators \citep{Jacobs1961}.
The Right to AI imports this epistemic critique into AI governance.
If AI systems are trained and evaluated only through distant datasets and expert benchmarks, then they risk encoding partial perceptions into the city's epistemic infrastructure.
A rights-based approach requires that those who live with the consequences have institutionalized capacity to reshape system objectives, data, and constraints.

\subsection{AI as civic infrastructure}
\label{sec:ch3-ai-infrastructure}

The Right to AI reframes AI from \emph{product} to \emph{infrastructure}.
Infrastructure studies emphasize that infrastructures are relational systems that organize access to resources and coordinate social life \citep{star1999ethnography}. They tend to become invisible when they function, yet they structure what is possible \citep{plantin2018infrastructure,Graham2001}.
In cities, infrastructures such as streets, sewers, and transit networks are governed through public law and public accountability because their benefits and burdens are collectively experienced.

AI systems increasingly share these infrastructural characteristics. They are integrated across organizational boundaries, depend on shared datasets and model components, update over time, and mediate access to opportunities and services \citep{BommasaniEtAl2022,Taeih2021,Batty2024}.
They also produce dependencies that are difficult to reverse, especially when public agencies rely on vendor-controlled systems or when model outputs become embedded in policy routines \citep{Dignam2020,Kitchin2023}.

The infrastructure characterization can be specified through three features that matter for governance.
First, AI systems can have broad societal impact because they are embedded across policy domains and can reorganize how institutions allocate resources and manage risk \citep{Ulnicane2024,Taeih2021}.
Second, AI systems can become embedded in administrative and professional practice when they are integrated into workflows for planning, service delivery, and organizational decision support \citep{plantin2018infrastructure,Kitchin2023}.
Third, AI systems often require collective management because they operate through shared data resources and create externalities that are not internalized by individual users, agencies, or vendors \citep{Ostrom2009,Murray2017}.

Treating AI as infrastructure clarifies why participation is not optional.
Infrastructural governance is not only about technical choices.
It is about defining collective goods, acceptable risks, and legitimate authority.
These decisions cannot be outsourced to private firms or confined to professional expertise without eroding democratic legitimacy \citep{Dahl1971,Habermas1996,Sun2020}.

\subsection{Why rights?}
\label{sec:ch3-rights-not-only-ethics}

Contemporary AI governance is often framed through ethical principles, voluntary guidelines, or risk-based regulation.
These approaches remain relevant, and recent policy activity, including the European Union's AI Act, has created enforceable obligations in some jurisdictions \citep{Schiff2021b,Saheb2024,EuropeanUnion2024}.
However, principles and regulation alone often leave under-specified a question: who has legitimate authority to define objectives and trade-offs in contested civic domains? Rights language is used here for two reasons.

First, rights specify a \emph{floor}.
They define minimum entitlements that should not depend on organizational goodwill, consumer power, or ad hoc consultation \citep{Sun2020,Wenar2023}.
In the public realm, where residents cannot easily opt out of infrastructural systems, a rights floor constrains discretion.

Second, rights create \emph{standing}.
They identify right-holders and duty-bearers, making contestation and recourse institutionally legible rather than discretionary \citep{Wenar2023,kaminski2021right}. This standing is especially relevant for AI systems whose harms may be diffuse, probabilistic, or mediated through organizational procedures \citep{rawls1993political,Berlin1969}.

The Right to AI does not replace existing procedural rights in AI governance, such as the right to contest automated decisions or the right to obtain explanations. Instead, it complements them by treating collective participation in AI design and governance as a power right rather than merely an informational right \citep{kaminski2018right,kaminski2021right,zhang2024right}. The core normative shift is from after-the-fact redress to ex ante and ongoing co-governance.

\medskip

\paragraph{\textbf{Privilege rights and power rights.}} Much of the AI governance debate is framed as a question of access and protection: can people use beneficial systems, and can they be shielded from harmful ones.
In rights-theoretic terms, these are often \emph{privilege rights}.
A Right to AI, as developed in this dissertation, is a \emph{power right}: it grants communities authority to reshape the AI systems that structure their options, including by changing objectives, constraining uses, and withdrawing authorization \citep{Sun2020,Wenar2023,Power1997,Galston2002}.

\medskip

\paragraph{Definition of the \textbf{Right to AI}}
The Right to AI is the civic entitlement of individuals and communities to meaningfully participate in, and exercise institutional power over, the design, deployment, and ongoing governance of AI systems that materially shape their lives, including the authority to contest, revise, and decommission such systems.

\subsection{What the Right to AI requires in public-space contexts}
\label{sec:ch3-right-to-ai-requirements}

To apply the Right to AI to public space, this dissertation distinguishes between \emph{individual procedural protections} and \emph{collective infrastructural governance}.
Both matter, but they operate at different levels.

\medskip

\paragraph{\textbf{Individual procedural protections.}}
When a public agency uses an AI system in ways that can affect a person's rights or access, procedural protections include notice, explanation where feasible, and mechanisms for contestation and review \citep{kaminski2018right,kaminski2021right}.
These protections are often discussed in relation to automated decision-making systems.

\medskip

\paragraph{\textbf{Collective infrastructural governance.}}
Public-space AI systems also operate by shaping how the city sees and evaluates places and populations, including through aggregated maps, scores, forecasts, and generated design alternatives.
These systems affect publics collectively, even when they do not target identifiable individuals.
Collective governance therefore requires participation and oversight at the level of objectives, data practices, and permissible uses.

\medskip

\paragraph{\textbf{Who holds the right in the public realm.}}
Public-space AI systems do not affect only identifiable ``users.''
They shape the conditions under which publics form and act, including residents, workers, and visitors who experience the indirect consequences of public decisions \citep{Dewey1927,Lefebvre1968}.
A Right to AI therefore treats the relevant right-holders as those impacted by the system's representations and uses, with particular attention to groups exposed to disproportionate burdens or exclusion \citep{Fraser1995,Benjamin2019}.

Building on the broader human-right-to-technology tradition \citep{Sun2020}, this chapter specifies six core requirements for public-space contexts.

\begin{enumerate}[leftmargin=*, itemsep=0.4em]
\item \textbf{Right to disclosure and intelligibility.} Residents should be informed when AI systems materially influence public-space planning, management, or representation.
Disclosure must be intelligible to non-experts and include the system's intended uses, disallowed uses, and known limitations \citep{Schiff2021b,ZaidanIbrahim2024}; see also \citep{burrell_how_2016,miller_explanation_2019}.

\item \textbf{Right to participate as power.} Participation must include influence over problem framing, objectives, success criteria, and trade-offs, not only downstream feedback on outputs \citep{Arnstein1969,CostanzaChock2020,Birhane2022-power,Sieber2024PublicsEngaging}.

\item \textbf{Right to collective data stewardship.} Where public-space AI depends on data produced through collective life, governance should support communal stewardship mechanisms such as data trusts, co-governed repositories, or Indigenous data sovereignty protocols \citep{Ostrom1996,KukutaiTaylor2016,Lewis2020}; see also \citep{Ostrom2009}.

\item \textbf{Right to documentation and auditability.} Public agencies should require documentation and technical access sufficient for independent review proportional to the system's public-space impacts, including traceability of updates, evaluation protocols, and institutional decision rules.
Where legal or technical constraints limit disclosure, governance should provide alternative mechanisms for independent oversight and verifiable claims about performance and compliance \citep{Mitchell2019,gebru2021,ZaidanIbrahim2024,Reisman2018}.

\item \textbf{Right to contestation and recourse.} Individuals and communities should be able to challenge system outputs and uses, trigger review, and obtain remedies when systems are misused or harmful \citep{kaminski2021right,CohenSuzor2024}.

\item \textbf{Right to pluralistic representation.} The epistemic layer should be designed to preserve and make visible relevant disagreements, especially where values are contested.
In public-space governance, legitimacy requires not only aggregation but also mechanisms to recognize minority perspectives and contestation \citep{Fraser1995,Jain_2024,sloane2022}; see also \citep{Sorensen2024}.
\end{enumerate}

These requirements are design constraints on municipal institutions and on the technical artifacts produced within AI lifecycles, including datasets, documentation, evaluation protocols, and audit processes \citep{gebru2021}; see also \citep{Mitchell2019,Reisman2018,ZaidanIbrahim2024}.

\section{Participation as power}
\label{sec:ch3-participation-power}

\subsection{The participation ladder and the problem of tokenism}
\label{sec:ch3-ladder}

Participation is often invoked in both planning and AI governance, yet its meaning varies.
Arnstein's ladder remains influential because it makes explicit that participation is a distribution of power, not a count of attendees \citep{Arnstein1969}.
At the bottom are forms of non-participation that manage dissent rather than share authority.
In the middle are tokenistic modes that invite feedback without granting decision power.
At the top are forms of partnership and citizen control that institutionalize shared authority.

\begin{figure}[H]
\centering
\includegraphics[width=\textwidth]{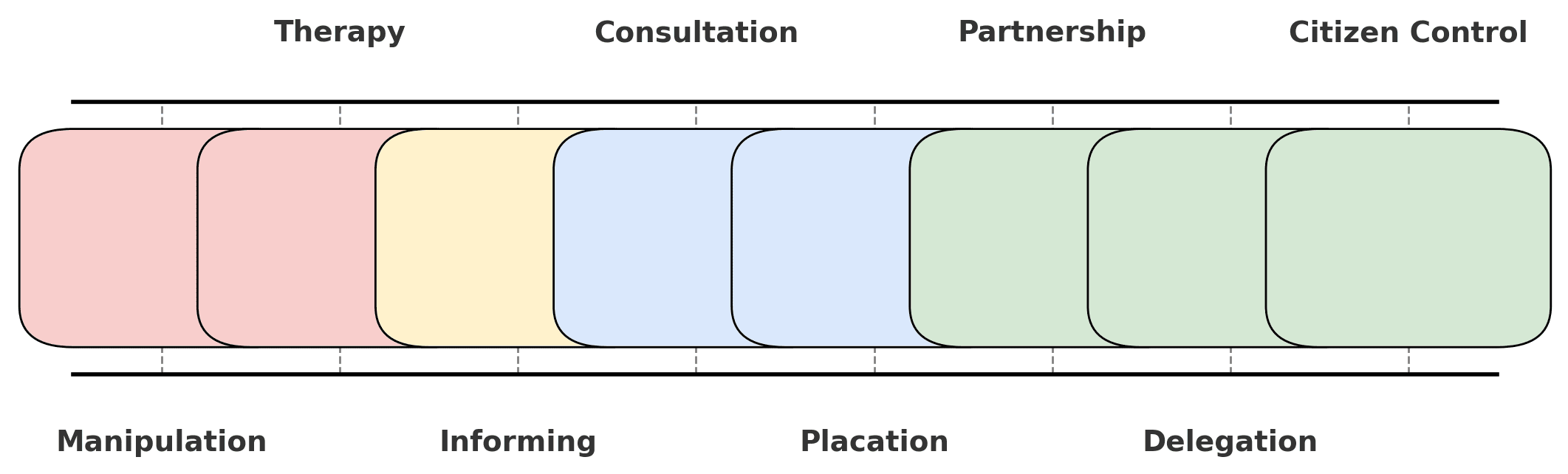}
\caption[Arnstein's ladder of citizen participation]{\deleted{Arnstein's ladder of participation, adapted for AI governance contexts. The ladder distinguishes non-participation (manipulation, therapy), tokenism (informing, consultation, placation), and power-sharing (partnership, delegation, citizen control) \citep{Arnstein1969}.} \added{Arnstein's ladder of citizen participation, adapted for AI governance. Source: adapted by the author from \citet{Arnstein1969}.}}
\label{fig:participation-ladder}
\label{fig:arnstein-ladder}
\end{figure}

The relevance for AI governance is that many ``participatory'' initiatives remain within the tokenistic middle of the ladder, for example by collecting user surveys, running one-off consultations, or publishing transparency reports without creating enforceable rights of influence \citep{CostanzaChock2020,sloane2022}.

In urban planning, the ladder also clarifies why consultation alone does not resolve legitimacy problems.
Consultation can be necessary but insufficient when agenda-setting, data access, and evaluative criteria remain centralized. In public-space contexts, this matters because the epistemic layer can embed value judgments in technical artifacts: what gets labeled as `safety,'' what counts as `accessibility,'' which trade-offs are coded into optimization, and which populations are treated as normative \citep{Kitchin2016,Benjamin2019}. A rights-based approach therefore requires participation in the upstream stages where these judgments are made.

This positioning also aligns the Right to AI with several planning-theory strands beyond Arnstein. Communicative and collaborative planning emphasize deliberation as situated practice and institutional design as a mediator of power and workable agreement \citep{Forester1988,Forester1999,InnesBooher2010}. Advocacy and equity planning highlight that inclusion requires safeguards against structural inequality and that distributive and recognitional justice cannot be assumed to follow from consultation alone \citep{Davidoff1965,KrumholzForester1990,Fainstein2010,young2000inclusion}. Power-oriented planning scholarship further cautions that participatory forms can be captured without enforceable decision authority, motivating audit rights, pause triggers, and contestation as governance primitives rather than optional add-ons \citep{Flyvbjerg1998}.

\subsection{A four-tier model of the Right to AI}
\label{sec:ch3-four-tier}

To operationalize participation as power for AI governance, this dissertation adapts Arnstein's ladder into a four-tier model. The tiers distinguish prevailing governance regimes by their levels of transparency, agency, and locus of authority. They also identify characteristic risks that matter for public-space applications.

\begin{figure}[!htbp]
\centering
\includegraphics[width=0.5\textwidth]{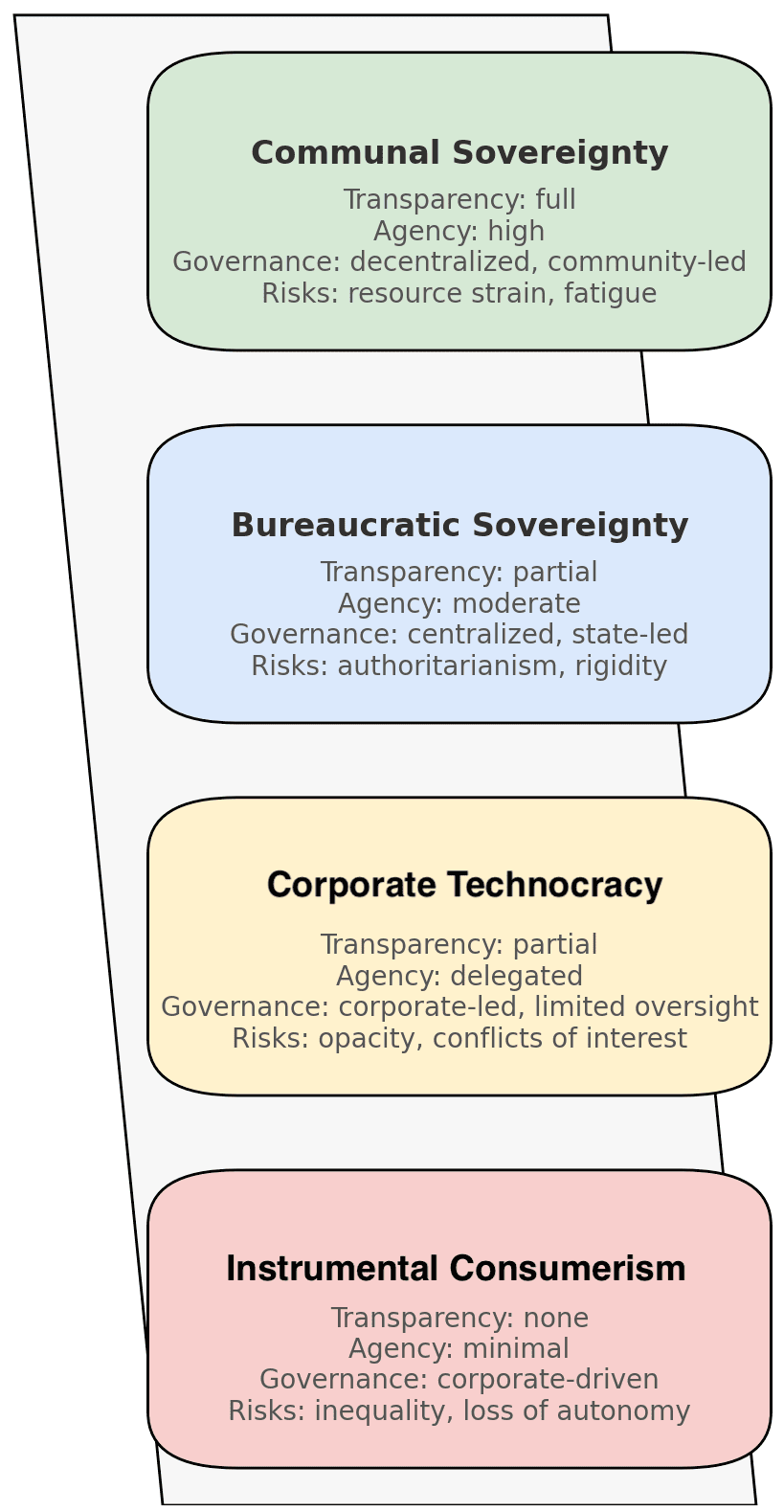}
\caption[Four-tier model of the Right to AI]{\deleted{A four-tier model of the Right to AI: instrumental consumerism, corporate technocracy, bureaucratic sovereignty, and communal sovereignty. Each tier summarizes typical levels of transparency and agency, prevailing governance structures, and characteristic risks.} \added{Four-tier model of the Right to AI. Source: author's work.}}
\label{fig:four-tier-model}
\label{fig:right-to-ai-tiers}
\end{figure}

\paragraph{\textbf{Tier 1: Consumer-based (instrumental consumerism).}}
In this tier, the public primarily encounters AI as a service, tool, or product.
Participation is limited to passive acceptance, intermittent feedback, or individual complaint mechanisms.
Governance remains largely corporate-driven or expert-led, and transparency is minimal.
For public-space systems, this tier corresponds to arrangements where residents are treated as end-users of an AI-mediated city rather than as co-authors of the epistemic infrastructure that shapes collective life \citep{Baudrillard1970,Kitchin2023}.

\medskip

\paragraph{\textbf{Tier 2: Private organization-led (corporate technocracy).}}
In this tier, private organizations incorporate some feedback into internal governance, for example through user studies, feedback dashboards, or advisory boards.
Decision authority remains private, and conflicts of interest can persist when public input is treated as product research rather than as democratic authorization \citep{Dignam2020,Huang2024,kirk2024prism}.
In urban contexts, corporate technocracy can take the form of vendor-controlled systems in procurement pipelines where the city cannot audit components that determine system behavior or meaningfully negotiate design objectives.

\medskip

\paragraph{\textbf{Tier 3: Government-controlled (bureaucratic sovereignty).}}
In this tier, public agencies set rules and standards, and may require transparency, privacy protections, and anti-discrimination constraints.
Government control can improve accountability, but risks include rigidity, limited local knowledge, and governance arrangements that remain expert-driven with weak mechanisms for community agenda-setting \citep{Fischer2000,Morison2020,EuropeanUnion2024}.
For public-space AI, bureaucratic sovereignty corresponds to a city that treats AI as part of administrative infrastructure but construes participation primarily as consultation rather than shared authority.

\medskip

\paragraph{\textbf{Tier 4: Citizen-controlled (communal sovereignty).}}
In this tier, communities hold formal authority over AI governance through mechanisms such as data trusts, cooperative ownership of datasets, citizen assemblies for oversight, and enforceable audit rights \citep{Ostrom1996,KukutaiTaylor2016,Lewis2020,Birhane2022-power}.
This tier does not imply that technical expertise is excluded.
It implies that experts operate within democratically defined objectives and constraints, with communities holding standing to shape and contest those constraints.

The four tiers are ideal types.
In practice, municipal governance may combine elements of bureaucratic and communal sovereignty, especially where high-stakes domains require expert judgment.
The model is intended to be diagnostic and directional: it clarifies that many current arrangements remain at low-power tiers, and it provides a vocabulary for specifying what institutional changes are required to move toward democratic legitimacy.

\subsection{Lessons from participatory AI practices}
\label{sec:ch3-lessons}

Participation is feasible, but it is not automatic.
A persistent concern is that participatory rhetoric can be used to legitimize decisions without redistributing power.
To ground the Right to AI in practice, the chapter draws on participatory AI initiatives across domains.
\cref{tab:cases} synthesizes examples that illustrate how participation has been implemented, who is involved, and what kinds of outcomes have been reported.

\begingroup
\scriptsize
\renewcommand{\arraystretch}{1.08}
\setlength{\tabcolsep}{3pt}

\begin{longtable}{@{}%
    P{0.17\linewidth}%
    P{0.12\linewidth}%
    P{0.17\linewidth}%
    P{0.15\linewidth}%
    P{0.125\linewidth}%
    P{0.185\linewidth}%
@{}}
\caption{Eight examples of participatory AI initiatives. The cases illustrate a range of domains and participation designs, from value elicitation for alignment datasets to community-led data governance.}
\label{tab:cases}\\

\toprule
\textbf{Project} &
\textbf{Why it was done} &
\textbf{How it was implemented} &
\textbf{Stakeholder involvement} &
\textbf{Domain / application} &
\textbf{Key outcomes and impact} \\
\midrule
\endfirsthead

\caption[]{Eight examples of participatory AI initiatives (continued).}\\
\toprule
\textbf{Project} &
\textbf{Why it was done} &
\textbf{How it was implemented} &
\textbf{Stakeholder involvement} &
\textbf{Domain / application} &
\textbf{Key outcomes and impact} \\
\midrule
\endhead

\midrule
\multicolumn{6}{r}{\itshape Continued on next page.}\\
\endfoot

\bottomrule
\endlastfoot

\textbf{Anthropic's Collective Constitutional AI \citep{Huang2024}} &
Align AI with shared values &
Ethical `constitution,'' iterative feedback &
AI researchers, end-users, ethicists &
AI alignment &
Surfaced tensions in ethical frameworks and value trade-offs \\

\addlinespace[0.35em]

\textbf{PRISM alignment dataset \citep{kirk2024prism}} &
Investigate cross-cultural alignment &
Surveys with diverse participants &
International participants, researchers &
AI ethics &
Documented cross-cultural disagreement about alignment criteria \\

\addlinespace[0.35em]

\textbf{MID-Space \citep{nayak2024midspace}} &
Democratize design visualization &
Community-based annotation and mediation &
Marginalized groups, planners &
Urban planning &
Incorporated localized perspectives into design representations \\

\addlinespace[0.35em]

\textbf{Participatory modelling for agro-pastoral restoration \citep{Eitzel2021}} &
Include Indigenous and local knowledge &
Co-created computational models &
Farmers, modelers &
Environmental sustainability &
Supported context-driven land management strategies \\

\addlinespace[0.35em]

\textbf{Co-design of trustworthy AI in healthcare \citep{Zicari2021}} &
Address bias in medical AI &
Iterative design with patients and clinicians &
Patients, clinicians, ethicists &
Healthcare &
Identified bias risks and improved accountability, with resource constraints \\

\addlinespace[0.35em]

\textbf{WeBuildAI: participatory algorithmic governance \citep{Lee2019}} &
Develop collaborative governance &
Workshops with civic groups &
Civic groups, public officials &
Civic algorithms &
Prototyped participatory design of allocation mechanisms \\

\addlinespace[0.35em]

\textbf{Participatory research for low-resourced machine translation \citep{Nekoto2020}} &
Scale NLP for low-resource African languages &
Community-driven data collection and annotation &
Language communities, researchers, linguists &
Machine translation, NLP &
Enabled new datasets and benchmarks through community contributions \\

\addlinespace[0.35em]

\textbf{Māori data sovereignty initiative \citep{KukutaiTaylor2016}} &
Protect Māori language data and ensure community benefits &
Community protocols and governance structures &
Māori community, linguists, Indigenous organizations &
Language technology, data sovereignty &
Established governance principles for data use and benefit sharing \\

\end{longtable}
\endgroup

Across cases, several recurring lessons are relevant for municipal public-space governance.

\medskip

\paragraph{\textbf{Early involvement matters.}} When participation begins after objectives and datasets are fixed, it tends to be experienced as tokenistic and has limited power to change foundational choices \citep{Arnstein1969,Birhane2022-power,sloane2022}.
In municipal contexts, this implies that Right to AI participation must attach to commissioning and procurement, not only to post-deployment feedback.

\medskip

\paragraph{\textbf{Pluralism requires conflict mediation.}} When values differ, participation must include processes for surfacing disagreement, negotiating trade-offs, and documenting contested issues rather than suppressing them \citep{Bohman2000,Habermas1996,kirk2024prism}.
For public-space AI, this follows because the relevant values include social recognition and inclusion, not only technical performance.

\medskip

\paragraph{\textbf{Participation requires resources and capacity.}}
Many participatory initiatives depend on training, facilitation, translation, and sustained compensation.
Without institutional support, participation can become extractive or limited to those with time and expertise \citep{KukutaiTaylor2016,Birhane2022-power}. For cities, this implies that participation cannot be treated as a volunteer add-on; it must be budgeted and institutionalized.

\medskip

\paragraph{\textbf{Cooptation is a real risk.}} Organizations can selectively incorporate community input while retaining control over agenda-setting and ownership. Avoiding cooptation requires explicit governance structures and rights of audit, refusal, and revision \citep{MikalefEtAl2022,CohenSuzor2024}.
The Right to AI is designed to create those rights as enforceable standing rather than optional engagement.

\section{Implications}
\label{sec:ch3-municipal-implications}

The Right to AI becomes meaningful only when translated into municipal procedures and institutional responsibilities. Cities are not sovereign states, but they possess concrete levers over public-space AI because they commission infrastructure, procure services, regulate use in public facilities, and manage data produced through public administration. This section translates the chapter's rights-based framing into a set of municipal implications that set up the procedural lifecycle developed in Chapter Four.

\subsection{Municipal levers over the AI layer}
\label{sec:ch3-municipal-levers}

Municipal agencies can govern AI systems affecting public space through at least four levers.

\paragraph{\textbf{Commissioning and procurement.}} Contracts can define scope, permissible uses, audit rights, documentation requirements, and participation obligations. Because many AI systems enter cities through procurement, procurement is a site of democratic authorization \citep{Dignam2020,Taeih2021}.

\medskip

\paragraph{\textbf{Data stewardship.}} Cities collect and manage administrative data (permits, maintenance logs, service requests) and may oversee or facilitate access to street imagery and environmental data.
Data governance decisions therefore shape what can be modeled and by whom. Mechanisms such as data trusts and community protocols offer pathways for collective stewardship \citep{Ostrom1996,KukutaiTaylor2016,Lewis2020}.

\medskip

\paragraph{\textbf{Oversight and accountability.}}
Municipalities can establish internal audit functions, independent review boards, or citizen assemblies that monitor deployments, evaluate impacts, and authorize updates or decommissioning \citep{ZaidanIbrahim2024}. Algorithmic impact assessment frameworks provide additional procedural templates for public agencies to document objectives, risks, and accountability mechanisms \citep{Reisman2018}.

\medskip

\paragraph{\textbf{Public communication and recourse.}} Cities can create channels for notice, contestation, and remedy, including public-facing documentation and complaint mechanisms with standing for affected groups \citep{kaminski2021right,CohenSuzor2024}. 

These levers are not purely administrative. They instantiate the difference between low-power tiers (where residents are treated as consumers of AI-mediated urbanism) and higher-power tiers (where residents hold governance authority).

\subsection{From rights to duties}
\label{sec:ch3-rights-to-duties}

\cref{tab:right-to-ai-duties} maps the Right to AI requirements to concrete municipal duties and deliverables.
The table is a governance design template intended to be adapted to local legal contexts and capacity constraints, while preserving the premise that participation is power.

\begingroup
\small
\renewcommand{\arraystretch}{1.08}
\setlength{\tabcolsep}{4pt}

\begin{longtable}{@{}%
    P{0.19\linewidth}%
    P{0.44\linewidth}%
    P{0.29\linewidth}%
@{}}
\caption{A municipal translation of the Right to AI: core rights, institutional duties, and typical governance artifacts for public-space AI systems. The mapping builds on the Right to AI and on participatory and documentation practices discussed in the broader governance literature \citep{Arnstein1969,Ostrom1996,gebru2021,ZaidanIbrahim2024}.}
\label{tab:right-to-ai-duties}\\

\toprule
\textbf{Right to AI requirement} &
\textbf{Municipal duty} &
\textbf{Artifacts and checks} \\
\midrule
\endfirsthead

\caption[]{A municipal translation of the Right to AI (continued).}\\
\toprule
\textbf{Right to AI requirement} &
\textbf{Municipal duty} &
\textbf{Artifacts and checks} \\
\midrule
\endhead

\midrule
\multicolumn{3}{r}{\itshape Continued on next page.}\\
\endfoot

\bottomrule
\endlastfoot

Disclosure and intelligibility &
Provide timely, plain-language notice when AI systems materially shape public-space representations or decisions; publish intended use and disallowed use. &
Public AI register; plain-language system brief; signage or digital notice; ``scope and limits'' statement. \\

\addlinespace[0.35em]

Participation as power &
Attach participation to agenda-setting and commissioning; define who has decision authority at key gates; resource participation (compensation, accessibility, facilitation). &
Co-framing charter; deliberation protocol; documented decision memos; participation budget and accessibility plan. \\

\addlinespace[0.35em]

Collective data stewardship &
Ensure lawful and legitimate data acquisition; prevent extractive reuse; support collective governance mechanisms over locally produced data. &
Data governance plan; consent and retention rules; data trust or stewardship board; benefit-sharing commitments. \\

\addlinespace[0.35em]

Documentation and auditability &
Require documentation sufficient for independent review; guarantee audit rights and technical access appropriate to risk. &
Dataset documentation (datasheets); model documentation; evaluation reports; access provisions for auditors; versioned change logs. \\

\addlinespace[0.35em]

Contestation and recourse &
Provide accessible mechanisms to challenge outputs and uses; create remedies, including suspension, revision, or decommissioning when warranted. &
Complaint intake and tracking; escalation pathway; independent review option; remediation plan; sunset or pause clauses. \\

\addlinespace[0.35em]

Pluralistic representation &
Prevent majority averaging from erasing minority harms; require reporting that preserves heterogeneity where relevant. &
Disaggregated reporting; documentation of contested dimensions; triggers for deliberation when disagreement or indeterminacy is high. \\

\end{longtable}
\endgroup

Two design implications follow directly.

\medskip

First, the Right to AI shifts attention from single moments of participation to a lifecycle of governance.
Participation must recur at points where the system is framed, updated, and maintained, not only at initial deployment. This motivates the augmented participatory lifecycle introduced in Chapter Four.

\medskip

Second, the Right to AI requires a change in procurement norms. If a city cannot audit a system, cannot constrain its uses, and cannot renegotiate objectives as public values evolve, then the system functions as a private infrastructure operating inside public governance.
That condition is incompatible with democratic legitimacy for public-space AI systems, especially when systems are difficult to reverse or when they create new surveillance capacities \citep{Brayne2017,CohenSuzor2024}.

\subsection{Procurement implications}
\label{sec:ch3-procurement}

Municipal procurement is often treated as an administrative process, but in public-space AI it is also a moment of democratic authorization. A Right to AI perspective implies that procurement should include at least four categories of requirements.

\medskip

\paragraph{\textbf{Scope control and disallowed uses.}} Contracts should specify not only what the system is intended to do, but what it is forbidden to do. For public-space AI, disallowed uses include repurposing evaluative tools for punitive enforcement without explicit due process authorization, or using public imagery and data for unrelated commercial objectives \citep{CohenSuzor2024,EuropeanUnion2024}.

\medskip

\paragraph{\textbf{Audit rights and access.}}
Procurement can require auditability proportional to risk, including access to documentation, evaluation protocols, and system logs, and where feasible access to model interfaces and data pipelines for independent review \citep{ZaidanIbrahim2024,Mitchell2019,gebru2021}.
Algorithmic impact assessment processes can be incorporated as contractual deliverables and review gates \citep{Reisman2018}.

\medskip

\paragraph{\textbf{Participation and resourcing.}}
Participation obligations can be written into procurement by specifying co-framing requirements, stakeholder recruitment criteria, facilitation standards, and compensation.
This moves participation from an informal consultation norm to a contractual deliverable \citep{Arnstein1969,CostanzaChock2020}.

\medskip

\paragraph{\textbf{Update governance and re-commissioning.}}
Because AI systems change, procurement should address updates, drift, and re-commissioning.
A rights-based procurement approach treats models as maintainable infrastructures with versioning, periodic review, and decommissioning triggers, not as static products delivered once \citep{Taeih2021}.

These implications are not primarily technical. They are institutional design choices that determine whether cities can govern AI systems as part of the public realm.

\section{Discussion: alternatives and boundary conditions}
\label{sec:ch3-discussion}

The Right to AI is not the only possible approach to AI governance.
This section considers two alternatives and then addresses common objections to participatory governance, with attention to municipal public-space contexts.

\subsection{Market-led governance}
\label{sec:ch3-market-led}

A market-led view argues that competition, consumer choice, and reputational incentives can drive responsible AI. Under this view, firms that deploy harmful or biased systems will lose trust and market share, and self-regulation may be more flexible than public regulation \citep{Dignam2020,demarcellis-warin2022ai,HadfieldClark2023}. Market approaches can generate innovation, but they face limits in public-space contexts.

First, residents often cannot opt out of AI-mediated public services. Public infrastructure is not a consumer good in the relevant sense, and the burden of exit is unequally distributed.

Second, markets tend to reflect purchasing power rather than democratic legitimacy. Marginalized communities frequently have limited market influence, yet may experience disproportionate harms \citep{Eubanks2018,CohenSuzor2024}.

Third, vendor-controlled systems can create lock-in. Once embedded in municipal workflows, proprietary systems can become difficult to audit or replace, weakening public accountability \citep{Kitchin2023,Dignam2020}.

These limits do not imply that markets are irrelevant. They imply that market incentives are insufficient as a legitimacy mechanism for AI systems that mediate public space.

\subsection{State-centric governance}
\label{sec:ch3-state-centric}

A state-centric view emphasizes strong regulation, central standards, and expert-led oversight to ensure safety, consistency, and enforceability \citep{Taeih2021,EuropeanUnion2024}.
This approach can address some market failures, but it also has limits.

Central regulation can be slow relative to technological change, can under-attend to local knowledge, and can reproduce technocratic assumptions about what counts as harm or benefit \citep{Fischer2000,Morison2020}.
In public-space contexts, where values are contested and where the meaning of inclusion is context-dependent, purely centralized governance risks imposing universal metrics without legitimate local negotiation.

The Right to AI is compatible with state-centric regulation, but it adds a missing dimension: participatory institutions that enable local authorization and contestation within broader regulatory floors.
Regulation can specify minimum protections and risk categories, while the Right to AI specifies participation and governance structures that make the epistemic layer democratically negotiable.

\subsection{Objections to participatory governance}
\label{sec:ch3-objections}

Three objections recur in critiques of participatory AI and participatory planning.

\medskip

\paragraph{\textbf{Objection 1: participation is infeasible and too costly.}}
Participatory processes require time, facilitation, and compensation.
The response is to treat these requirements as part of legitimate infrastructural governance.
Cities already invest in public consultation for physical infrastructure, and high-impact AI systems warrant comparable investment, scaled by risk \citep{Arnstein1969,Taeih2021}.
The cost of participation should be compared to the cost of illegitimate or harmful deployments, including loss of public trust and downstream remediation.

\medskip

\paragraph{\textbf{Objection 2: citizens lack expertise.}}
Public-space AI systems can be technically complex, and some domains are safety-critical.
The Right to AI does not require that residents directly set technical parameters.
It requires that residents have power over objectives, constraints, and acceptable risks, while experts contribute within those democratically set boundaries \citep{Habermas1996,Fischer2000}.
The four-tier model allows hybrid forms of citizen authority that incorporate expert advisory roles, especially in high-stakes settings.

\medskip

\paragraph{\textbf{Objection 3: participatory bodies can be captured or parochial.}}
Participation can be captured by organized interests, and local deliberation can reproduce exclusion.
This is a risk.
The response is institutional design: recruitment that centers underrepresented groups, rotation, compensation, independent facilitation, transparency about decision rules, and conflict mediation mechanisms \citep{Bohman2000,CostanzaChock2020,Birhane2022-power}.
The participatory cases surveyed in \cref{tab:cases} also suggest that capacity-building and explicit governance protocols are necessary to avoid tokenism and cooptation.

\subsection{Boundary conditions for public-space AI}
\label{sec:ch3-boundary}

Applying the Right to AI in public-space contexts requires clarity about boundary conditions.

\medskip

\paragraph{\textbf{Surveillance and enforcement.}}
Public-space data and models can be repurposed for surveillance or punitive enforcement.
A rights-based approach must therefore include prohibitions and due process requirements for repurposing, especially when systems create new capacities for monitoring presence and movement \citep{Brayne2017,CohenSuzor2024}.

\medskip

\paragraph{\textbf{Pluralism and non-resolution.}}
Participation does not ensure consensus; for contested public-space values, legitimacy may require preserving documented disagreement and plural representations rather than enforcing a single metric \citep{Fraser1995,Jain_2024}.

\medskip

\paragraph{\textbf{Local adaptation and transferability.}} The Right to AI is formulated as a civic governance principle that can be adapted across contexts, but its institutional realizations must be local.
Data governance norms, legal authorities, and participatory institutions vary by jurisdiction.
The chapter therefore  positions the Right to AI as a design and legitimacy standard, not an institutional blueprint.

\section{Conclusions}
\label{sec:ch3-conclusion}

The chapter contends that when AI mediates urban perception and governance, democratic legitimacy demands more than ethical principles or technical performance. A civic Right to AI provides a normative floor and a governance claim: individuals and communities affected by AI infrastructures must have meaningful, ongoing power to co-govern objectives, data practices, permissible uses, and mechanisms of accountability. Operationally, the chapter clarified participation as power through Arnstein's ladder (\cref{fig:participation-ladder}) and through a four-tier model of AI governance regimes (\cref{fig:four-tier-model}). It also translated the Right to AI into municipal duties and governance artifacts (\cref{tab:right-to-ai-duties}), emphasizing procurement, data stewardship, auditability, and recourse as practical levers for cities. Rights are not self-executing. They require procedures, roles, and lifecycle checkpoints that attach participation to the full duration of an AI system's existence in the public realm. Chapter Four therefore develops an augmented, participatory AI lifecycle that operationalizes these rights into a co-produced municipal workflow.


\chapter{Co-Producing AI: Toward an Augmented, Participatory AI Lifecycle}
\label{chap:coproducing_ai_lifecycle}

This chapter specifies an operational procedure for governing AI systems whose outputs shape civic life: an augmented AI lifecycle that treats co-production as a recurring governance practice rather than an upstream consultation. The chapter is adapted and extended from the paper \textit{Co-Producing AI: Toward an Augmented, Participatory Lifecycle} \citep{mushkani2025position}. In the dissertation argument, it operationalizes the normative requirements developed in Chapter Three (the Right to AI) and provides a procedural link to the empirical and technical instantiations in Chapters Five and Six, where co-produced standards are encoded into measurement and into generative-model alignment.

This chapter argues that dominant approaches to AI governance fail not because of missing principles or tools, but because they confine participation to narrow moments while leaving authority concentrated in expert and vendor hierarchies. It first clarifies why ethical principles, fairness metrics, and accountability toolkits remain operationally incomplete under these conditions \citep{Jobin2019,NIST2023,wong2023toolkit}. It then synthesizes lifecycle thinking from product development, software engineering, and machine learning pipelines with participatory design and design-justice traditions, reframing participation as an allocation of decision-making authority rather than a method of information gathering \citep{KiritsisEtAl2003,BeckEtAl2001,Sculley2015,CostanzaChock2020}. Finally, it presents empirical findings from a scoping review and four multidisciplinary workshops conducted in Montréal, Canada (January to May 2024), which ground an augmented five-phase lifecycle: \textit{co-framing}, \textit{co-design}, \textit{co-implementation}, \textit{co-deployment}, and \textit{co-maintenance}.

The remainder of the chapter is structured as follows. Section~\ref{sec:ch4_intro} motivates lifecycle co-production as an institutional response to recurring AI harms and governance failures. Section~\ref{sec:ch4_lit} reviews lifecycle models, participatory AI, and theoretical orientations that justify co-production as a governance requirement. Section~\ref{sec:ch4_methods} describes the scoping review and workshop methods used to develop the lifecycle. Section~\ref{sec:ch4_findings} presents workshop themes and details each lifecycle phase with its core tasks, artifacts, and participation checkpoints, alongside a risk and ethics mapping. Section~\ref{sec:ch4_discussion} extends the lifecycle into municipal settings by mapping phases to commissioning, procurement, and maintenance practices. Section~\ref{sec:ch4_conclusion} concludes with limitations.

\section{Introduction}
\label{sec:ch4_intro}
AI systems are increasingly embedded in decision-critical infrastructures, including systems used to allocate resources, enforce rules, and produce authoritative representations of social reality \citep{Dwivedi2021,GalazEtAl2021}. In urban contexts, such systems can shape what becomes visible to government, what is counted as a problem, and which interventions appear justified or feasible \citep{Koseki2022}. Data-driven urban governance has also expanded the range of actors and artifacts that mediate public decisions, including platform infrastructures and proprietary analytics that can restructure municipal capacities and accountabilities \citep{Kitchin2014a,Kitchin2016}. As Chapter Three argues, when AI systems materially shape civic life, they should be treated as governance objects requiring democratic authorization, contestability, and participatory power rather than as discretionary technical tools \citep{airight2025}. A rights-based claim, however, is not sufficient without a procedure through which rights are exercised over time. Participation confined to initial consultation cannot address downstream harms that emerge during training, deployment, monitoring, updates, and decommissioning \citep{Gerdes2022,sloane2022}.

Two recurrent patterns motivate a lifecycle approach.

First, many harms are not one-time design errors but lifecycle phenomena. Predictive policing and criminal-justice risk scoring can reproduce structural inequities through feedback loops \citep{AngwinEtAl2016,Brayne2017}. Facial-recognition systems can exhibit disparate error rates across demographic groups, and their impacts can intensify when deployed at scale without meaningful recourse \citep{BuolamwiniGebru2018}. Automated hiring systems and recommender systems can codify and amplify exclusionary patterns, sometimes in ways that are difficult to detect when only aggregate performance is measured \citep{Dastin2022}. Comparable dynamics are documented in automated decision systems used in public-service contexts, where technical choices can be coupled to administrative routines that affect eligibility, burden, and recourse \citep{Eubanks2018}. These cases support a governance implication: oversight must attach to post-deployment behavior, not only to pre-deployment design intentions.

Second, existing governance instruments tend to be aspirational or expert-centric. Policy frameworks and professional guidelines, including the Montréal Declaration for Responsible AI, IEEE guidance, European Commission ethical guidance, and the NIST AI Risk Management Framework, articulate principles such as transparency, accountability, privacy, and fairness \citep{montreal2018declaration,ieee2019ethically,AIHLEG2018,NIST2023}. Yet they often under-specify the institutional mechanisms by which affected publics can participate in the definition of objectives, the negotiation of trade-offs, and the oversight of ongoing system behavior \citep{Jobin2019,Schiff2021b,wong2023toolkit}. In parallel, technical work offers a growing arsenal of fairness and interpretability methods, but fairness metrics are plural and sometimes incompatible, and their selection is itself a normative choice \citep{HardtEtAl2016,Chouldechova2017,Binns2020,BarredoArrieta2020}. Treating such choices as purely technical can displace political disagreement into design decisions that remain unaccountable to affected publics.

This chapter responds to these patterns by proposing an augmented AI lifecycle in which co-production is continuous and phase-specific. The central question is the one posed in the originating paper: \textit{How can citizen participation be integrated throughout the AI lifecycle to balance process-oriented and outcome-oriented considerations and to produce systems that are both effective and just?}. The answer developed here is not that participation is universally required in the same form for all systems. Participation becomes implementable when it is mapped to lifecycle tasks, risks, and decision rights, with explicit artifacts that preserve institutional memory and enable contestation across time \citep{Mitchell2019,sloane2022}.

\section{Literature Review}
\label{sec:ch4_lit}
The chapter uses \emph{co-production} to denote shared authority over problem definition, evidentiary standards, and ongoing system governance across the lifecycle. This use is compatible with participatory design traditions in HCI and with public-administration accounts of co-production as joint work between institutions and publics in producing public outcomes \citep{asaro_transforming_2000,Ostrom1996}. Within the dissertation, co-production is treated as a procedural counterpart to the Right to AI: a specification of when decision rights are exercised, by whom, and with what artifacts and recourse pathways \citep{airight2025}.

\subsection{Lifecycle models across domains}
Lifecycle governance is not unique to AI. Product development research describes beginning-of-life, middle-of-life, and end-of-life phases through which artifacts progress, emphasizing continuity of knowledge and accountability across handoffs \citep{KiritsisEtAl2003}. Product lifecycle management (PLM) extends this concern by structuring how design decisions, requirements, and performance evidence remain accessible as products evolve \citep{AmeriDutta2005}. In software engineering, the software development lifecycle (SDLC) and related process models formalize sequences of planning, analysis, design, implementation, testing, deployment, and maintenance, with iterative variations such as agile methods \citep{Deming1986,BeckEtAl2001,Mohammed2017,AssalChiasson2018}.

Machine learning systems inherit and complicate these lifecycles. Data acquisition, labeling, and preprocessing become central governance points; model selection and training introduce statistical dependencies and validation regimes; and deployment occurs in non-stationary environments where concept drift and feedback loops can emerge \citep{Sculley2015,Haakman2021,DeSilvaAlahakoon2022}. Pre-trained models and fine-tuning practices add additional loops devoted to adapting foundation models to local contexts . Across these domains, a recurrent governance problem is the loss of context during phase transitions: assumptions made at problem formulation may not remain visible during deployment, and post-deployment failures may not be legible to those who framed the objectives \citep{Haakman2021,DeSilvaAlahakoon2022}. Lifecycle thinking provides a structural vocabulary for continuity, but it does not, by itself, specify who gets to decide and with what standing to intervene.

\subsection{Participation in the AI lifecycle}
Participatory design traditions emphasize that socio-technical systems should be designed \textit{with} affected people, not merely \textit{for} them \citep{asaro_transforming_2000}. In AI, participatory approaches have been proposed and implemented in varied forms: participatory frameworks for algorithmic governance \citep{Lee2019}, participatory AI for humanitarian innovation \citep{BerditchevskaiaEtAl2021}, and participatory practices in public-sector applications and clinical decision support \citep{Donia2021,Zicari2021}. Recent work in alignment and evaluation also points toward methods that incorporate public input, including collective or constitutional-style approaches and datasets explicitly designed to represent plural perspectives \citep{Huang2024,kirk2024prism}.

Despite these advances, the literature repeatedly identifies a gap between the aspiration of participation and its operationalization across time. Many initiatives involve short-term engagement focused on a single phase of the pipeline, typically after objectives and constraints have already been set, and rarely include post-deployment governance rights such as contestation, audits, and re-commissioning \citep{Gerdes2022,Birhane2022-power,Aizenberg2020}. This gap is not merely procedural. It reflects power asymmetries in which expert institutions define the problem, select the metrics, and retain control over updates, while publics are invited to provide feedback without decision rights \citep{sloane2022,wong2023toolkit}. Participatory AI is therefore often implemented as consultation rather than shared authority.

\subsection{Design justice, DEI, and the distribution of decision rights}
Design justice argues that those most affected by design outcomes should have decision-making power, and it foregrounds how technology can reproduce structural inequities when designed within narrow institutional and cultural frames \citep{CostanzaChock2020}. Diversity, equity, and inclusion frameworks similarly emphasize that individual identities and social positions shape access to resources, exposure to harm, and capacity to participate in governance \citep{calabrese_barton_beyond_2020,DeHond2022}. Within AI practice, calls for DEI are linked not only to representation in datasets but also to who participates in setting objectives, defining harms, and authorizing deployment \citep{cachat-rosset_diversity_2023}.

A key implication for lifecycle governance is that participation must be specified as power. Arnstein's ladder remains a diagnostic: many engagements remain at the rungs of informing or consultation rather than partnership or citizen control \citep{Arnstein1969}. The Right to AI, developed in Chapter Three, builds on this insight by framing participation as a progression of authority and by treating AI systems that shape civic life as societal infrastructure rather than discretionary tools. The lifecycle developed in this chapter is the procedural counterpart: it distributes opportunities for participation across phases where commitments are made, altered, or potentially violated.

\subsection{The operational limits of AI ethics frameworks}
The proliferation of ethics principles and responsible AI frameworks demonstrates recognition of AI risks. Comparative reviews show that many frameworks remain high-level, leaving unresolved how principles become enforceable practices and how stakeholders can intervene when trade-offs emerge \citep{Jobin2019,Schiff2021b}. Practitioner-facing toolkits can also encode a particular vision of what ethical work looks like, often framing ethics as an internal technical task rather than a public governance practice \citep{wong2023toolkit}. Meanwhile, the multiplicity of fairness definitions creates room for strategic selection of metrics that align with institutional incentives rather than with affected publics' priorities \citep{Chouldechova2017,Binns2020}.

These critiques motivate an alternative framing: rather than asking only \textit{what} principles an AI system should satisfy, lifecycle governance asks \textit{who} decides, \textit{when} they decide, \textit{with what evidence and artifacts}, and \textit{with what recourse} \citep{Mitchell2019,NIST2023}. This reframing is consistent with the chapter's theoretical orientation. Co-production is not treated as a moral supplement but as a governance mechanism that can redistribute epistemic authority and reduce the risk that value conflicts are collapsed into hidden technical choices \citep{CostanzaChock2020,Engestrom2014}.

\subsection{Expansive learning and co-production}
Expansive learning and activity theory conceptualize learning as collective transformation across interacting activity systems, emphasizing boundary crossing between communities with distinct norms, languages, and accountabilities \citep{Engestrm2010,Engestrom2014}. In the AI governance context, this perspective helps clarify why one-off engagement is insufficient: publics and experts do not simply exchange preferences; they iteratively construct shared problem representations, contest assumptions, and revise artifacts that mediate future decisions \citep{Engestrom2014}. Co-production, from this view, is a structured process of iterative knowledge exchange, not a single event.

Combined with design justice, this perspective motivates the chapter's design move: the AI lifecycle itself must be re-architected to embed repeated opportunities for boundary crossing, negotiation, and revision, with explicit attention to power and inclusion \citep{CostanzaChock2020}.

\section{Methodology}
\label{sec:ch4_methods}
The empirical work aimed to develop a grounded procedural model for ethical AI practice, rather than to evaluate a deployed system. Insights from a scoping review and multidisciplinary deliberation were synthesized to inform a lifecycle-oriented approach.

\subsection{Scoping review}
Between October 2023 and May 2024, a scoping review was conducted of academic and gray literature published from January 2013 to May 2024. Searches were conducted across Scopus, PubMed, Web of Science, and Google Scholar, spanning computer science, social sciences, and humanities. A Boolean search expression targeted work on ethical AI, fairness, transparency, accountability, bias, participatory methods, co-production, lifecycle models, and ethics-by-design approaches.

The initial search returned 330 documents. After duplicate removal and title and abstract screening for relevance to AI ethics, inclusivity, and lifecycle processes, 147 records remained. Full-text inclusion required explicit engagement with ethics and AI, co-creation or co-production, lifecycle or process models, and design-justice perspectives. Targeted searches of organizational repositories were also used to include standards and policy frameworks. In total, 76 sources were synthesized to prepare workshop materials.

\subsection{Workshop design and participants}
Four three-hour workshops were conducted in Montréal, Canada between January and May 2024. Recruitment used purposive sampling intended to ensure disciplinary breadth and institutional diversity. Across sessions, twenty participants took part (5--9 per workshop), affiliated with organizations including Mila--Quebec AI Institute, Université de Montréal, IVADO, INRS, and OBVIA. Participants' disciplinary backgrounds spanned computer science, social science, law, philosophy, and DEI practice. Nine participants identified primarily as researchers, six as industry practitioners, and five as civil-society advocates.

\added{Organizational affiliations document the institutional breadth sought through purposive recruitment. Each participant's self-described primary professional role was used to characterize the sample and preserve anonymity in reporting. Workshop transcripts were coded inductively and deductively by theme. The three broad role labels were not used to estimate sector-wide views or to support a role-stratified thematic comparison.}
\label{rev:workshop-coding}

Each workshop began with a synthesis of scoping review themes, followed by moderated discussion around three questions: how to address ethical challenges in AI design and use, which methods have demonstrated efficacy, and which methodological scenarios could guide future development. Audio was recorded and supplemented with detailed notes; transcripts were anonymized for analysis.

\subsection{Analysis approach and theoretical orientation}
Transcripts were coded using an inductive-deductive thematic approach. Initial codes were informed by design justice and expansive learning theory, and new categories were added iteratively until saturation was reached; disagreements were resolved through peer debriefing. The resulting themes inform the lifecycle phases, the cross-cutting conditions for co-production, and the mapping between participation and risk mitigation.

The theoretical orientation treats co-production as a normative requirement grounded in design justice, which centers decision-making authority with those most affected, and expansive learning, which frames participatory engagements as boundary-crossing interventions that enable shared problem reconstruction across communities \citep{CostanzaChock2020,Engestrom2014}.

\vspace{-0.5cm}

\section{Findings}
\label{sec:ch4_findings}
The findings include cross-workshop themes that motivate co-production, the augmented five-phase lifecycle, and a mapping of lifecycle co-production to risk categories and to major ethical frameworks.

\subsection{Key themes} 
Thematic analysis of the four workshops produced four interdependent themes that shaped the lifecycle design. These themes provide empirical grounding for why co-production must be continuous and why participation requires explicit support structures.

\subsubsection{Distributed authority}
Participants argued that decision rights should reside closer to the communities who bear consequences of an AI system. The lifecycle therefore embeds shared governance checkpoints and, where appropriate, community veto rights at multiple stages rather than limiting influence to initial requirements gathering \citep{CostanzaChock2020,Benjamin2019}. This theme aligns with the dissertation's rights-based framing in Chapter Three.

\subsubsection{Iterative knowledge exchange}
Workshops emphasized that participation is itself a learning process: publics acquire AI literacy and experts acquire contextual knowledge through repeated interaction. The lifecycle therefore specifies cyclic feedback mechanisms and shared artifact repositories to maintain context as teams evolve and as projects transition across phases \citep{Engestrom2014}.

\subsubsection{Contextual privacy}
Participants framed privacy practices as inseparable from local norms and from the social meaning of data use. This theme supports layered privacy strategies calibrated to context, consistent with contextual integrity \citep{Nissenbaum2010}. In municipal settings, this implies that privacy governance is not a one-time compliance check but a continuing negotiation tied to changing uses and public expectations.

\subsubsection{Resource constraints}
Participants highlighted that sustained engagement is feasible only when budgets address the tangible costs borne by community members, including time, travel, accessibility supports, and care obligations. The lifecycle therefore treats resourcing as a governance requirement: without compensation and logistical support, participation risks becoming extractive or tokenistic \citep{sloane2022}.

Together, these themes motivate a shift from participation-as-event to participation-as-infrastructure: sustained, resourced, and embedded in phase-specific decision-making.

\subsection{The augmented AI lifecycle}
\label{sec:augmented_lifecycle_definition}
The augmented lifecycle operationalizes co-production across five interdependent phases: \textit{co-framing}, \textit{co-design}, \textit{co-implementation}, \textit{co-deployment}, and \textit{co-maintenance}. The lifecycle is grounded in design justice, expansive learning theory, and DEI scholarship, and it positions citizens, domain specialists, and technologists as joint decision-makers rather than as sequential inputs \citep{CostanzaChock2020,Engestrom2014,DeHond2022}.

\begin{figure}[t]
\centering
\includegraphics[width=\textwidth]{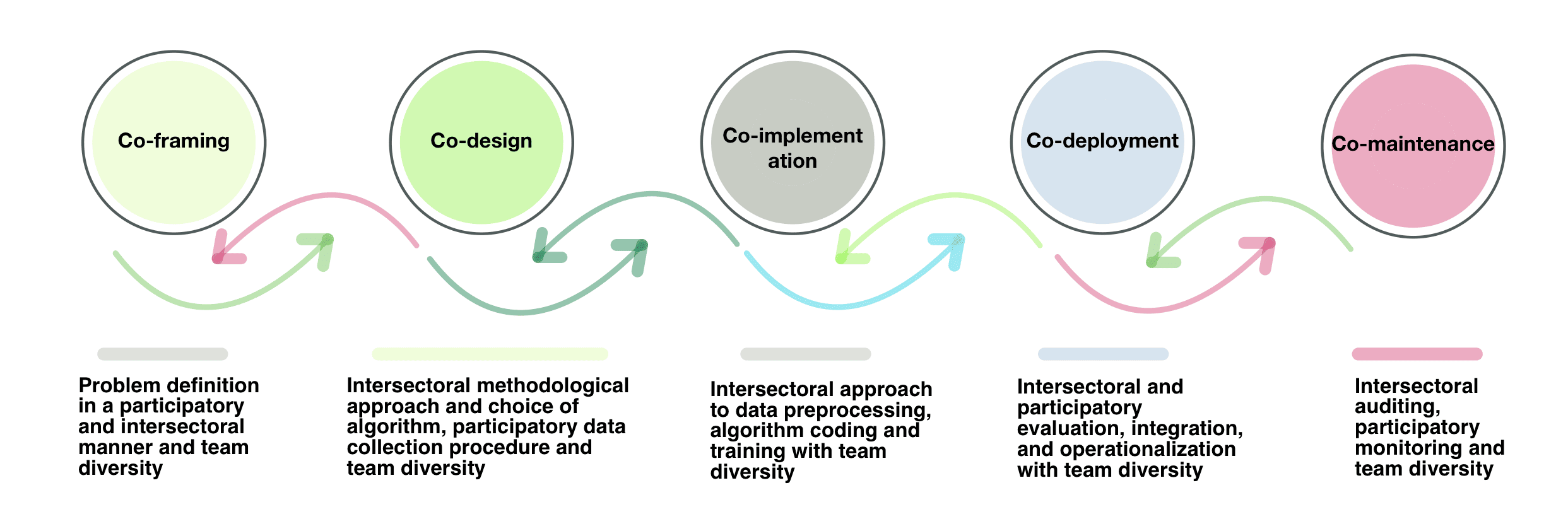}
\caption[Augmented AI lifecycle]{The augmented AI lifecycle with five co-production phases: co-framing, co-design, co-implementation, co-deployment, and co-maintenance. \added{Source: author's synthesis; see \citet{mushkani2025position}.}}
\label{fig:augmented_ai_lifecycle_ch4}
\label{fig:augmented-lifecycle}
\end{figure}

Figure~\ref{fig:augmented_ai_lifecycle_ch4} visualizes the augmented AI lifecycle. The diagram emphasizes continuous knowledge exchange and shared accountability across phases, which addresses the loss of context that often occurs as projects move from planning to implementation and into operations.

\subsubsection{Co-framing}
Co-framing establishes a shared problem definition, an initial risk register, and a participation plan. It draws on participation theory as a way to diagnose power, and on expansive learning as a way to create boundary spaces where different activity systems can jointly reconstruct the problem \citep{Arnstein1969,Engestrm2010}. In this phase, affected communities are not asked only to react to a pre-defined solution. They participate in defining what the system is for, what it is not for, which harms are plausible, and what evidence would count as acceptable performance.

Core artifacts include a problem definition charter and scope statement that record objectives, non-objectives, decision boundaries, and anticipated coupling points with existing institutions. Co-framing also produces an initial risk register that includes distributional harms, privacy risks, transparency failures, and misuse scenarios \citep{Koseki2022}. It further specifies a participation plan that identifies who participates, how representation is achieved, what decision rights exist at each phase, and how participation is resourced \citep{sloane2022}. Guiding questions include identifying affected communities and their geographic distribution, documenting comparable systems and known failures, selecting engagement methods aligned with DEI objectives, and recording how citizen perspectives reshape the problem statement.

In municipal settings, co-framing aligns with upstream moments of authorization: deciding whether a system should exist, what public purpose it serves, and what governance conditions should attach before procurement proceeds.

\subsubsection{Co-design}
In co-design, participants select data sources, model families, and interface concepts aligned with co-framed objectives. Participatory prototyping and scenario walk-throughs translate contextual knowledge into technical specifications, while comparative risk assessments evaluate trade-offs among pipeline options \citep{sloane2022}. This phase is where governance commitments become embedded, including whether a dataset adequately represents marginalized groups, which proxies are acceptable, and what forms of interpretability are meaningful to non-technical stakeholders \citep{cachat-rosset_diversity_2023}. Co-design also clarifies which documentation artifacts are required for later contestation and public accountability \citep{Mitchell2019}.

\begin{figure}[t]
\centering
\includegraphics[width=\textwidth]{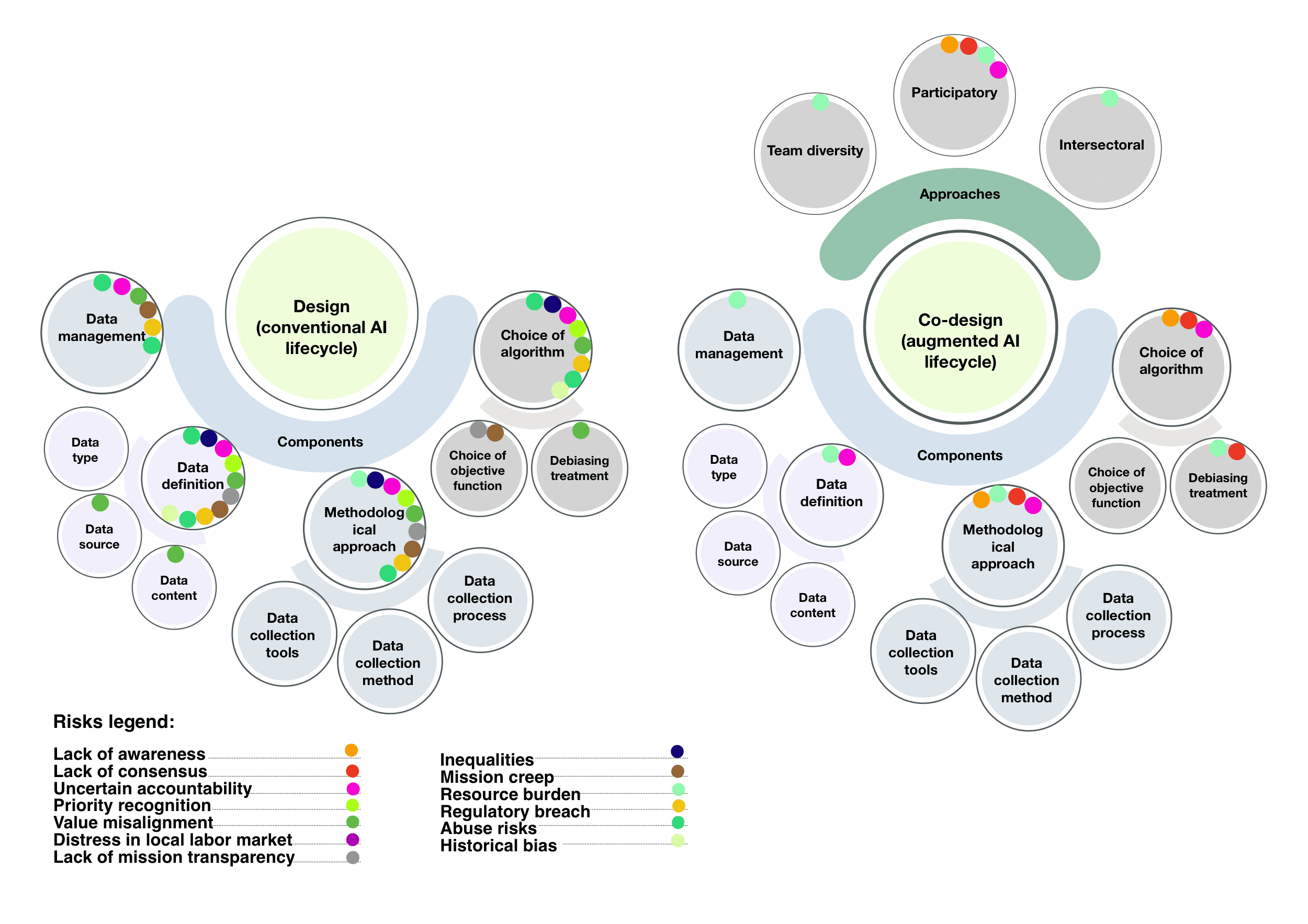}
\caption[Lifecycle risks and co-design responses]{Phase-specific risks in a conventional AI lifecycle and corresponding mitigation via co-design. \added{Source: author's synthesis; see \citet{mushkani2025position}.}}
\label{fig:risk_lifecycle_ch4}
\label{fig:lifecycle-risk-mitigation}
\end{figure}

Figure~\ref{fig:risk_lifecycle_ch4} reproduces the study's risk mapping that contrasts phase-specific risks in conventional lifecycles with mitigation via co-design. The figure supports a governance claim that runs through the dissertation: co-production functions as a form of risk mitigation by shifting which harms are detectable early enough to change system decisions.

\subsubsection{Co-implementation}
Co-implementation includes data acquisition, feature engineering, model training or fine-tuning, and iterative validation. The governance emphasis is traceability: decisions should be recorded in version-controlled artifacts and model documentation so that later disputes can be adjudicated with reference to what was agreed and what changed \citep{Mitchell2019}. Documentation practices that bind datasets and model behavior to explicit intended-use statements can support this traceability, including dataset documentation norms such as datasheets and closely related artifacts \citep{gebru2021}. Citizen partners review intermediate outputs such as data summaries, error reports, and subgroup performance breakdowns to check whether implementation remains aligned with co-designed commitments \citep{cachat-rosset_diversity_2023}.

The phase foregrounds privacy and accountability. Workshop discussions supported a layered privacy strategy calibrated to data sensitivity, spanning informed consent and anonymization practices through differential privacy where appropriate \citep{PawarEtAl2018,Vliz2021,Nissenbaum2010}. Co-implementation therefore treats privacy not as a one-time compliance artifact but as an evolving set of safeguards tied to concrete implementation choices.

\subsubsection{Co-deployment}
Deployment introduces the system into non-stationary contexts where latent risks can emerge, including mission creep, secondary use, and feedback loops that were not visible in controlled testing \citep{Koseki2022}. In co-deployment, citizen representatives and domain experts conduct user-acceptance testing, while a governance charter assigns responsibility for emergent harms and defines pathways for pause, rollback, redesign, or decommissioning. Performance dashboards and recourse mechanisms are central: affected individuals must be able to contest outputs, request review, and trigger investigation \citep{GalazEtAl2021}.

This phase shifts participation from design input to operational oversight. A core question is whether evaluation metrics capture distributional impacts across stakeholder groups, not only aggregate accuracy. Another is what safeguards prevent unauthorized secondary use. In municipal contexts, this is salient when evaluative tools can be repurposed for enforcement, surveillance, or exclusionary policy.

\subsubsection{Co-maintenance}
Co-maintenance treats the system as a living artifact subject to concept drift, regulatory change, and evolving norms \citep{DeSilvaAlahakoon2022}. Periodic audits, technical and participatory, evaluate whether the system remains aligned with its original objectives and whether its impacts have shifted across groups. Citizen assemblies or standing committees may recommend updates, suspension, or decommissioning, and they play a role in preserving institutional memory as team membership changes \citep{HelbingEtAl2023}.

In municipal settings, co-maintenance parallels the reality that public infrastructure is maintained through recurring cycles: budgets, inspections, complaints, and periodic redesigns. The lifecycle suggests that AI governance should adopt an analogous cadence, including explicit funding for monitoring and audit.

\subsubsection{Cross-cutting considerations}
Across phases, co-production depends on three cross-cutting conditions emphasized in the originating work: recruitment practices that meet DEI goals, anonymization protocols that uphold contextual integrity, and facilitation methods that reduce power asymmetries \citep{calabrese_barton_beyond_2020,Nissenbaum2010,turnhout2010participation}. The workshops further highlighted multidisciplinary collaboration challenges: while diverse teams can increase reflexivity, they can also face epistemic divergence and communication barriers, which may require rotating facilitation, shared glossaries, and just-in-time expert input \citep{BiscontiEtAl2023,sloane2022}.

\subsection{Risk categories motivating lifecycle co-production}
A central claim of the lifecycle proposal is that participation functions as risk mitigation, not only as a moral add-on. Table~\ref{tab:ai_risks_ethics_ch4} reproduces the paper's consolidated list of risk categories and ethical considerations that exceed the remit of conventional expert-centric lifecycles. The table is used here as an interpretive device: it identifies harms that should be surfaced during co-framing, operationalized during co-design and co-implementation, monitored during co-deployment, and revisited during co-maintenance.

\begin{table}[!htbp]
\centering
\footnotesize
\setlength{\abovecaptionskip}{4pt}
\setlength{\belowcaptionskip}{6pt}
\caption{Risks and ethical considerations in AI development and use.}
\label{tab:ai_risks_ethics_ch4}
\renewcommand{\arraystretch}{1.2}

\begin{tabularx}{0.95\linewidth}{@{}p{4cm}X@{}}
\toprule
\textbf{Risk Category} & \textbf{Description} \\
\midrule
Social Responsibility & Ethical duty to consider AI's societal impacts and promote public benefit. \\
Vulnerable Populations & Risk of exacerbating harm to marginalized or disadvantaged groups. \\
Transparency & Need for clear, understandable AI processes and decisions. \\
Misuse and Hostile Use & Potential for malicious or unethical applications of AI. \\
Human Rights & Threats to privacy, expression, and freedom from discrimination. \\
Deception and Manipulation & Use of AI to mislead or covertly influence behavior. \\
Inequalities & Risk of deepening social and economic divides. \\
Workforce Diversity & Importance of inclusive development to avoid narrow perspectives. \\
AGI and Existential Risk & Long-term concerns over uncontrolled advanced AI. \\
Cultural Sensitivity & Need to respect diverse norms in AI design and use. \\
Trust & Building confidence in AI's reliability and ethical use. \\
Psychological Impacts & Effects on mental health, identity, and human purpose. \\
Unemployment & Job loss risks from automation-driven displacement. \\
Misinformation & Generation or spread of false information via AI. \\
Economic Growth & AI's role in boosting innovation and productivity. \\
Explainability & Clarity on how AI makes decisions. \\
Privacy & Safeguarding personal data from misuse or leaks. \\
Safety and Reliability & Ensuring systems function correctly and safely. \\
Human Control & Maintaining human oversight of AI actions. \\
Accountability & Assigning responsibility for AI outcomes. \\
Consent and Autonomy & Respecting individuals' control over data and impact. \\
\bottomrule
\end{tabularx}
\end{table}

These risks connect directly to the dissertation's urban context. Public-space AI systems can affect vulnerable populations through uneven accessibility or surveillance burdens; can invite misuse through secondary enforcement uses; and can generate legitimacy crises when transparency and contestability are weak. A lifecycle approach is therefore justified not only by technical complexity but by the distributional and political stakes of AI in civic infrastructures \citep{Koseki2022}.

\subsection{Mapping the lifecycle to prevailing ethical frameworks}
The originating paper compares the augmented lifecycle to multiple ethics and governance frameworks, highlighting where they correspond to co-production dimensions and where participatory practice remains under-specified. Table~\ref{tab:guideline_alignment_ch4} reproduces that alignment analysis. In the dissertation framing, the table supports a claim that appears throughout responsible AI debates: high-level ethical principles often assume participation and accountability, but they rarely specify operational artifacts and decision rights required to implement these commitments across time.

\begingroup
\scriptsize
\renewcommand{\arraystretch}{1.12}
\setlength{\tabcolsep}{4pt}

\begin{longtable}{@{}p{3.38cm}p{4.92cm}%
  >{\centering\arraybackslash}p{2.12cm}%
  >{\centering\arraybackslash}p{2.12cm}%
  >{\centering\arraybackslash}p{2.12cm}@{}}

\caption{Alignment of selected AI ethics guidelines with dimensions of co-production.}
\label{tab:guideline_alignment_ch4}\\

\toprule
\textbf{Title} &
\textbf{Principles} &
\textbf{\shortstack{DEI\\Participation}} &
\textbf{\shortstack{Multi-\\disciplinarity}} &
\textbf{\shortstack{Team\\Diversity}} \\
\midrule
\endfirsthead

\caption[]{Alignment of selected AI ethics guidelines with dimensions of co-production (continued).}\\
\toprule
\textbf{Title} &
\textbf{Principles} &
\textbf{\shortstack{DEI\\Participation}} &
\textbf{\shortstack{Multi-\\disciplinarity}} &
\textbf{\shortstack{Team\\Diversity}} \\
\midrule
\endhead

\midrule
\multicolumn{5}{r}{\itshape Continued on next page.}\\
\endfoot

\bottomrule
\endlastfoot

\multicolumn{5}{@{}l}{\textsc{The Montr\'eal Declaration for Responsible AI}} \\
\addlinespace[0.2em]
& Well-being & \textit{High} & \textit{High} & \textit{Medium} \\
& Respect for autonomy & \textit{High} & \textit{High} & \textit{High} \\
& Protection of privacy and intimacy & \textit{High} & \textit{High} & \textit{High} \\
& Solidarity & \textit{High} & \textit{High} & \textit{High} \\
& Democratic participation & \textit{Medium} & \textit{High} & \textit{Low} \\
& Equity & \textit{Medium} & \textit{High} & \textit{Low} \\
& Diversity inclusion & \textit{Low} & \textit{High} & \textit{Low} \\
& Caution & \textit{High} & \textit{High} & \textit{High} \\
& Responsibility & \textit{High} & \textit{High} & \textit{High} \\
& Sustainable development & \textit{High} & \textit{High} & \textit{High} \\

\addlinespace[0.5em]
\multicolumn{5}{@{}l}{\textsc{IEEE Global Initiative on Ethics of Autonomous and Intelligent Systems}} \\
\addlinespace[0.2em]
& Human rights & \textit{High} & \textit{High} & \textit{High} \\
& Well-being & \textit{Low} & \textit{Low} & \textit{Low} \\
& Data agency & \textit{Low} & \textit{Low} & \textit{Low} \\
& Effectiveness & \textit{Low} & \textit{Low} & \textit{Low} \\
& Transparency & \textit{High} & \textit{High} & \textit{High} \\
& Accountability & \textit{High} & \textit{High} & \textit{High} \\
& Awareness of misuse & \textit{High} & \textit{High} & \textit{High} \\
& Competence & \textit{Low} & \textit{Low} & \textit{Low} \\

\addlinespace[0.5em]
\multicolumn{5}{@{}l}{\textsc{NIST AI Risk Management Framework}} \\
\addlinespace[0.2em]
& Valid and reliable & \textit{Low} & \textit{Low} & \textit{Low} \\
& Safe & \textit{Low} & \textit{Low} & \textit{Low} \\
& Secure and resilient & \textit{Low} & \textit{Low} & \textit{Low} \\
& Accountable and transparent & \textit{High} & \textit{High} & \textit{High} \\
& Explainable and interpretable & \textit{Low} & \textit{Low} & \textit{Low} \\
& Privacy-enhanced & \textit{Low} & \textit{Low} & \textit{Low} \\
& Fair with harmful bias managed & \textit{Low} & \textit{Low} & \textit{Low} \\

\addlinespace[0.5em]
\multicolumn{5}{@{}l}{\textsc{European Commission Ethics Guidelines for Trustworthy AI}} \\
\addlinespace[0.2em]
& Human agency and oversight & \textit{High} & \textit{High} & \textit{High} \\
& Technical robustness and safety & \textit{Low} & \textit{Low} & \textit{Low} \\
& Privacy and data governance & \textit{Low} & \textit{Low} & \textit{Low} \\
& Transparency & \textit{High} & \textit{High} & \textit{High} \\
& Diversity, non-discrimination, and fairness & \textit{Low} & \textit{Low} & \textit{Low} \\
& Environmental and societal well-being & \textit{Low} & \textit{Low} & \textit{Low} \\
& Accountability & \textit{High} & \textit{High} & \textit{High} \\

\end{longtable}
\endgroup

The table should not be read as a definitive scoring of frameworks. Its purpose is analytic: it shows that many frameworks articulate values consistent with participation and inclusion, but the translation into enforceable roles, checkpoints, and artifacts remains a governance challenge \citep{Schiff2021b,NIST2023}. This observation supports a thesis-level claim developed across Chapters Three and Seven: rights require procedures and institutions, not only principles.

\subsection{Impacts and trade-offs}
Workshop participants described multiple anticipated impacts of lifecycle co-production, summarized in Table~\ref{tab:augmented_ai_impact_ch4}. These impacts are framed as plausible consequences grounded in workshop synthesis, not as empirically validated causal effects.

\begin{table}[t]
\centering
\small
\caption{Impacts of the augmented AI lifecycle.}
\label{tab:augmented_ai_impact_ch4}
\renewcommand{\arraystretch}{1.15}
\begin{tabular}{p{4.6cm}p{10cm}}
\toprule
\textbf{Impact Category} & \textbf{Description} \\
\midrule
Enhanced AI Reliability & Diverse perspectives across phases can increase robustness and reliability. \\
Minimized Biases & Involving marginalized communities can surface and mitigate bias and exclusion. \\
Contextual Solutions & Co-produced systems are more likely to address local challenges and norms. \\
Empowered Communities & The process democratizes AI development and increases civic agency. \\
\bottomrule
\end{tabular}
\end{table}

The same synthesis highlights trade-offs and challenges. Diverse perspectives can complicate consensus formation and increase resource demands \citep{sloane2022,Varanasi2023}. Multidisciplinary collaboration may be burdened by epistemic divergence and differences in disciplinary language \citep{BiscontiEtAl2023}. Ambiguities in role definitions can produce participation washing, in which engagement is formal rather than substantive \citep{sloane2022}. These challenges motivate the chapter's emphasis on explicit role charters, resourcing plans, and phase-specific decision rights \citep{Birhane2022-power}.

\section{Discussion}
\label{sec:ch4_discussion}
The augmented lifecycle was developed as a general governance framework, but this dissertation's stakes are municipal and urban: AI systems increasingly mediate how public spaces are evaluated, designed, and governed \citep{LarteyLaw2025}. This section extends the lifecycle into municipal procedure by specifying how each phase can be anchored in public-sector commissioning and maintenance practices. The goal is not to assume that all cities can adopt full-scale co-production in a uniform form. The aim is to show how co-production becomes implementable when mapped onto existing decision points and accountability structures.

\subsection{Why lifecycle co-production matters for municipal AI}
Municipalities face distinctive constraints. They procure many systems through vendor contracts; they operate under public law and administrative accountability; and they manage infrastructures that persist and evolve over long time horizons. These characteristics make the lifecycle perspective salient. A city can rarely treat an AI system as a one-off pilot, because systems become entangled with budgets, workflows, political commitments, and public expectations. Post-deployment drift and repurposing are therefore predictable governance risks \citep{DeSilvaAlahakoon2022,Koseki2022}. These dynamics occur within broader political economies of data and automation that can shape what is procured, what is measured, and which forms of contestation are feasible \citep{Crawford2021,Kitchin2016}.

At the same time, municipalities possess governance levers that many private organizations do not: public notice requirements, procurement and contracting rules, routine audit practices, and formal channels for complaint and appeal. The augmented lifecycle can be understood as a way to redesign these levers so that they operationalize the Right to AI's requirements for participation, transparency, and recourse.

\subsection{Mapping lifecycle phases onto municipal process stages}
Table~\ref{tab:municipal_mapping_ch4} provides a mapping between the five lifecycle phases and typical municipal process stages. The mapping is a synthesis intended to support implementation planning. It does not assume a single universal municipal workflow; it identifies common governance touchpoints where participation and documentation can be institutionalized.

\begin{table}[t]
\centering
\small
\caption{Synthesis mapping from the augmented AI lifecycle to common municipal process stages for commissioned systems.}
\label{tab:municipal_mapping_ch4}
\renewcommand{\arraystretch}{1.15}
\begin{tabular}{p{4.2cm}p{11.0cm}}
\toprule
\textbf{Lifecycle phase} & \textbf{Municipal anchoring points} \\
\midrule
Co-framing & Council or departmental authorization to explore an AI-supported intervention; early engagement with affected publics; creation of a public problem charter, initial risk register, and participation and resourcing plan. \\
Co-design & Drafting requirements for procurement or internal build; data governance plan; decisions about modalities (images, sensors, administrative data) and acceptable proxies; accessibility and equity requirements; documentation plan aligned with public reporting norms. \\
Co-implementation & Vendor development or internal build; version-controlled documentation and model cards; interim reviews with community representatives; privacy safeguards calibrated to context; testing protocols that include distributional evaluation. \\
Co-deployment & Pilot authorization rather than full rollout; public notice and communication; operational dashboards and logging; recourse mechanisms and complaint intake; safeguards against secondary use and mission creep. \\
Co-maintenance & Ongoing operations and budget cycles; periodic technical and participatory audits; change logs for updates; re-authorization or re-commissioning triggers; decommissioning or replacement when harms outweigh benefits. \\
\bottomrule
\end{tabular}
\end{table}

Three implications follow from this mapping. Municipal co-production must begin before procurement. If objectives and constraints are vendor-defined, later participation is structurally limited. Co-framing therefore functions as an authorization step that can block procurement until a problem charter, risk register, and participation plan exist. Co-production must also persist into operational phases where harms surface. Municipalities already have maintenance cycles for physical infrastructure; co-maintenance proposes an analogous cadence for AI systems, including audits, updates, and sunset decisions \citep{DeSilvaAlahakoon2022}. Finally, the lifecycle implies a shift in what counts as a governance artifact. Public-sector legitimacy depends on documentation that is accessible and contestable, not only on technical performance. Model cards and versioned documentation provide one mechanism for maintaining institutional memory and enabling later contestation \citep{Mitchell2019}.

\subsection{Participation as risk mitigation}
A municipal interpretation of the lifecycle emphasizes proportionality: participation should be scaled to the impact and risk profile of the system, but high-impact public-space systems should not be governed through minimal consultation. The lifecycle clarifies how participation attaches to distinct risks. Representation and exclusion risks are most actionable in co-framing and co-design, when data sources and evaluation criteria are selected \citep{cachat-rosset_diversity_2023,CostanzaChock2020}. Privacy and secondary-use risks become salient at co-implementation and co-deployment, when data flows and operational uses are concretized \citep{Nissenbaum2010,Vliz2021}. Drift, legitimacy, and mission creep risks concentrate in co-maintenance, which requires standing oversight and periodic re-authorization \citep{DeSilvaAlahakoon2022,Koseki2022}.

This risk-based view supports a thesis-level argument developed later: pluralism and contested values cannot be resolved by selecting a single metric once. They require recurring opportunities for negotiation and revision supported by stable artifacts and oversight institutions.

\subsection{Feasibility constraints}
The chapter's empirical synthesis foregrounds feasibility constraints: sustained participation requires resources, and multidisciplinary collaboration requires infrastructure for communication and power balancing \citep{sloane2022,BiscontiEtAl2023}. Municipal adoption therefore requires institutional design responses rather than exhortations.

\medskip

\paragraph{\textbf{Participation must be budgeted.}} If municipalities do not fund compensation, accessibility supports, and facilitation, participation risks becoming extractive and unrepresentative \citep{sloane2022}. 

\medskip

\paragraph{\textbf{Decision rights must be explicit.}} Role charters should specify who can approve progression between phases, who can request additional evidence, and who has standing to trigger review. Without explicit decision rights, participation can revert to symbolic input \citep{Arnstein1969,sloane2022}.

\medskip

\paragraph{\textbf{Documentation must be designed for contestability.}} Transparency in municipal contexts is not only disclosure; it is the creation of usable records that enable affected publics to understand and contest system behavior over time \citep{Mitchell2019,NIST2023}. This includes documenting non-objectives and disallowed uses, which is critical for limiting repurposing in public space domains.

These implementation considerations prepare the dissertation's later governance architecture. Chapter Seven returns to procurement and oversight design, including deliberative cells as a recurring institutional unit that can exercise lifecycle decision rights.

\section{Conclusions}
\label{sec:ch4_conclusion}
This chapter operationalizes the dissertation's normative commitment to a civic Right to AI by providing a procedure through which participation becomes continuous, phase-specific, and enforceable. Drawing on a scoping review (76 synthesized sources) and four multidisciplinary workshops conducted in Montréal (20 participants), the chapter presents an augmented AI lifecycle with five co-produced phases: co-framing, co-design, co-implementation, co-deployment, and co-maintenance. The lifecycle treats participation as a risk-mitigation practice and a governance mechanism that redistributes epistemic authority, preserves institutional memory, and enables contestation across time.

\deleted{The chapter also clarifies limitations. The lifecycle remains a conceptual framework and requires validation through longitudinal field trials across multiple jurisdictions and institutional settings. The workshops were geographically and professionally concentrated, and they did not include lay citizen cohorts outside the local AI ecosystem. As a result, the lifecycle should be read as an implementable design pattern grounded in multidisciplinary synthesis rather than as a proven intervention with quantified causal effects.}

\added{The four workshops support a design hypothesis about lifecycle governance. Their 20 participants were concentrated in Montréal's AI, research, industry, and civil-society ecosystem; lay resident cohorts outside that ecosystem fell beyond the sample. Primary-role descriptions also compress professional identities that may span several sectors. The findings establish themes present in this deliberative sample and provide a basis for field testing. Resident authorization, causal effects, implementation costs, and durability across institutions remain unresolved. Applied validation should examine whether participation changes decisions at named gates, disagreement remains traceable, community members can trigger review, and municipalities can maintain the required artifacts under procurement, staffing, vendor-access, and budget constraints. Longitudinal trials should record participant retention, compensation, decision changes, disputes, remedies, and recommissioning events. These observations would permit evaluation of institutional authority and effects over time.}
\label{rev:ch4-limits}

Lifecycle governance requires domain instantiation to become concrete. Chapter Five provides that instantiation in the context of public-space evaluation through Street Review, showing what co-framing, co-design, and co-assessment look like when plural values about streetscapes are elicited, represented, and scaled. Chapter Six then extends pluralistic co-production into generative-model alignment, where the algorithmic layer does not only measure public space but also generates candidate representations and alternatives that can shape planning imaginaries and decisions.


\chapter{Street Review: A Participatory AI-Based Framework for Assessing Streetscape Inclusivity}
\label{chap:streetreview}

This chapter develops and evaluates \textit{Street Review}, a participatory, AI-supported framework for assessing streetscape inclusivity. It provides an empirical account of how a municipality can commission a measurement pipeline grounded in resident-defined criteria while representing heterogeneity across social positions. Street Review is situated in public-space governance contexts in which streets are both material infrastructures and settings of public life, and in which inclusion is negotiated rather than reducible to a single indicator set \citep{AnttiroikoDeJong2020,Gehl2011,low2020social,Whyte2001}. The framework is designed to support accountability and public reasoning by making explicit what is being measured, how it is measured, and where evidence converges or diverges.

Street Review was developed in Montréal between 2023 and 2025 and integrates three linked research components. Semi-directed interviews with residents recruited through community organizations elicited grounded descriptors of inclusive and exclusionary streets, which were consolidated into four criteria: Accessibility, Aesthetics, Practicality, and Inclusivity. Structured image evaluation sessions combined independent scoring with moderated small-group deliberation to enable empirical analysis of agreement, convergence, and persistent contestation across criteria and participant groups \citep{MushkaniJUM2025}. A computer-vision pipeline then scaled co-produced evaluations by segmenting street-view imagery, learning a multi-output regression model, and producing subgroup-aware predictions and heatmaps across approximately 45{,}000 images.

Across these components, the chapter contributes evidence for three dissertation propositions introduced in Chapter One. First, pluralism is structured rather than annotation error: evaluations of the same streetscapes vary systematically by criterion and by social position, and socially mediated dimensions such as inclusivity exhibit lower agreement than more visually legible dimensions. Second, deliberation alters the evidentiary object: small-group discussion increases convergence on selected criteria while leaving residual contestation on dimensions where participants draw on different experiences or interpretive frames. Third, scaling is feasible but bounded: a model trained on locally co-produced labels can reproduce group and subgroup patterns and support citywide audits, while remaining constrained by what is visible in imagery and by the quality and coverage of street-view data.

The chapter is organized as follows. Section~\ref{sec:ch5_intro} motivates the governance problem of measuring inclusivity at scale without collapsing heterogeneity. Section~\ref{sec:ch5_lit} reviews work on public-space evaluation, GeoAI perception modeling, and intersectional approaches to inclusion. Section~\ref{sec:ch5_methods} details Street Review as a values-to-infrastructure pipeline from participatory elicitation to modeling and mapping. Section~\ref{sec:ch5_findings} reports findings on criteria consolidation, convergence and contestation, model performance and interpretability, and citywide spatial patterns. Section~\ref{sec:ch5_discussion} specifies what Street Review makes measurable, what it cannot represent, and the safeguards required for legitimate use. Section~\ref{sec:ch5_conclusion} summarizes the chapter’s contributions.

\section{Introduction}
\label{sec:ch5_intro}
Urban environments continue to undergo changes in demographic composition and cultural norms due to shifting migration patterns, economic developments, and mobility preferences \citep{AnttiroikoDeJong2020,Broderick2022,Youngbloom2023}. City streets, sidewalks, and public areas are primary interaction points among residents, commuters, and visitors \citep{Gehl2011}. These spaces carry social, economic, and cultural significance that influences navigation, use, and perceived belonging \citep{MitrasinovicMehta2021}. As a result, municipal claims about inclusive public space depend not only on physical provision but also on how diverse residents interpret and experience streetscapes in daily life \citep{Whyte2001}.

Municipal governments and planning agencies frequently state commitments to inclusive public space, yet they often lack operational infrastructures to evaluate inclusion in ways that remain accountable to lived experience. In practice, inclusion is often proxied by a narrow set of measurable features such as curb cuts, sidewalk width, or proximity to amenities. These indicators matter, but they only partially capture how different residents experience welcome, safety, dignity, and belonging in streetscapes \citep{Gehl2011,low2020social,Whyte2001,Mehta2014}. When contested values are compressed into a universal index, the index can become the de facto representation of ``inclusion'' in planning, procurement, and resource allocation.

This chapter responds to that dilemma with Street Review, an approach that treats inclusive streetscape evaluation as a civic measurement problem that requires democratic grounding and pluralistic representation. Street Review is designed to translate resident-defined descriptors into a documented measurement pipeline without forcing a single score to stand in for a contested concept. The core design choice is representational. Rather than outputting one scalar value that purports to capture inclusive quality, Street Review yields a structured set of criteria and subgroup-aware layers intended to support interpretation, contestation, and revision. In this sense, Street Review is not a substitute for political judgment. It is a decision-support infrastructure that clarifies where evidence converges, where it diverges, and what remains beyond the reach of image-based proxies.

\medskip

Street Review is developed and tested in Montréal, where linguistic diversity, varied street typologies, and ongoing debates about mobility, safety, and publicness provide a context in which inclusion cannot be assumed to be a shared construct \citep{Litman2024}. The chapter addresses four operational questions: how residents from different backgrounds describe inclusive and exclusionary streets and which descriptors recur across groups; which evaluative dimensions exhibit convergence and which remain contested even under structured discussion; whether a model trained on locally co-produced evaluations can predict group and subgroup judgments with sufficient fidelity to support citywide audits while retaining subgroup differences; and what interpretive safeguards are required to prevent metric substitution and preserve contestation as a legitimate state.

\medskip

The chapter positions Street Review as a values-to-infrastructure pipeline. Values are elicited through participatory methods; evaluation instruments are constructed to preserve heterogeneity; models are trained to scale co-produced targets rather than vendor-defined metrics; and outputs are produced as civic artifacts that can be interrogated, revised, and recommissioned. The remainder of the chapter specifies how Street Review is constructed, what it reveals about pluralism in street evaluation, and how it can be used as part of legitimate public-space governance.

\section{Literature review}
\label{sec:ch5_lit}

\subsection{From indicators to lived experience}
Public space scholarship has emphasized that streets are not only transport corridors but settings of public life in which daily encounters and civic presence unfold \citep{Mehta2019,Gehl2011,Whyte2001}. This insight has motivated evaluative frameworks that translate qualitative aims into operational indicators, including publicness typologies, place-quality toolkits, and audits of comfort, walkability, and amenity provision \citep{Mehta2014,varna2010publicness,Zamanifard2019}. In practice, these frameworks often combine measurable physical features with interpretive qualities such as legibility, sociability, or perceived safety.

Standardized indicator sets also produce a reduction problem. When public space is evaluated through a fixed set of indicators, contested concepts such as inclusion, safety, and belonging can be collapsed into metrics that are legible to agencies while misaligned with lived experience \citep{low2020social,AnttiroikoDeJong2020}. Once a metric is installed in governance, it can shape problem definitions and intervention priorities by functioning as an institutional proxy for the underlying concept \citep{EspelandStevens2008,CarnemollaEtAl2021}. This is both a technical and an institutional risk: metric substitution can obscure disagreement about what inclusion means and can make the proxy appear as an objective representation of a contested value.

Street Review is situated in this gap. It treats inclusive evaluation as requiring resident-grounded descriptors, criteria that remain interpretable to participants and actionable for agencies, and explicit representation of disagreement as a measurable outcome rather than an error state.

\subsection{GeoAI and perception-based modeling of streetscapes}
Advances in computer vision and the availability of street-view imagery have enabled scalable characterization of urban form and perception-related attributes \citep{Danish2025,Huang2023,Zhu2025}. Early perception datasets such as Place Pulse used crowdsourced comparisons to label streetscapes along attributes such as perceived safety or beauty \citep{Dubey2016}. Building on these data, StreetScore and related systems learned models that predict perceived safety at scale \citep{Naik2014}. Project Sidewalk extended perception modeling toward accessibility by combining imagery, human annotation, and computer vision to map features relevant to disability access \citep{NajafizadehFroehlich2018,SahaEtAl2019}.

These systems demonstrate the feasibility of combining human judgment and computational scaling, but they also illustrate limitations relevant to governance. Approaches that rely on predefined criteria or unexamined labeler demographics risk encoding systematic bias and misrecognition \citep{BuolamwiniGebru2018,Mehrabi2021}. Model outputs can diverge from local survey responses and lived experience, indicating that perceptual constructs are context-dependent and may not transfer across settings \citep{SahaEtAl2019}. Many perception modeling approaches also assume a single target construct, which can mask plural interpretations and value conflict. Street Review builds on GeoAI methods but reframes the problem as one of legitimate commissioning: whether a city can authorize an assessment pipeline whose criteria are co-produced and whose outputs preserve pluralism, while remaining explicit about proxy limits and permissible uses.

\subsection{Intersectionality}
Intersectionality foregrounds how overlapping identities and social positions shape experience, including experience of public space \citep{Crenshaw1989}. In streetscapes, exclusion can be produced through material barriers but also through social and symbolic cues, including policing, cultural representation, lighting, and the presence of venues associated with particular communities \citep{low2020social}. Urban design interventions that improve physical access may not address other forms of exclusion, and different groups may disagree about what constitutes a ``good'' street or who a street is for.

\medskip
\vspace{-0.5cm}

This creates a measurement dilemma. If evaluation is collapsed to a single score, aggregation can erase minority experiences. If evaluation attempts to represent all identity combinations, measurement becomes infeasible and can reify categories as fixed groups. Street Review adopts a pragmatic stance in which demographic groupings are used as initial heuristics to test whether perceptions differ systematically and to avoid majority averaging, while treating group boundaries as coarse and revisable. This stance aligns with design justice and participatory HCI critiques that warn against extractive data practices and misrecognition while still seeking actionable accountability mechanisms \citep{CostanzaChock2020}.

\subsection{Participatory AI in urban evaluation}
Co-production and participatory planning approaches seek to move beyond extractive consultation by involving residents in defining problems, generating knowledge, and shaping outputs \citep{McKercher2020,Arnstein1969}. In AI contexts, documentation and auditing practices remain uneven across deployments, and systems trained on unrepresentative data can reproduce social biases \citep{gebru2021,Mehrabi2021}. Participatory AI approaches respond by involving stakeholders in dataset creation, model evaluation, and governance artifacts such as documentation, evaluation protocols, and accountability \citep{Sloane2024,sloane2022}.

\medskip
\vspace{-0.5cm}

Street Review contributes to this agenda by treating co-produced criteria, labels, and outputs as governance artifacts rather than as training signals alone. The framework connects qualitative elicitation, structured evaluation, model training, and spatial visualization into a documented pipeline designed to support municipal accountability. It provides an empirical setting for assessing how participatory evaluation can be operationalized without collapsing pluralism into a single objective.

\vspace{2cm}

\section{Methodology}
\label{sec:ch5_methods}

\subsection{Research design and case context}
Street Review is implemented as a multi-stage, community-centered research program in Montréal. The design is iterative. Qualitative elicitation informs the construction of evaluation instruments, and quantitative outputs are interpreted against participant accounts rather than treated as self-justifying metrics. This chapter synthesizes the stages into a methodological arc: co-production of criteria through interviews and descriptor consolidation; structured evaluation to measure convergence, contestation, and deliberation effects; and model-based scaling and mapping for citywide auditing. Table~\ref{tab:study_design_overview} summarizes the modules and data flows that underpin the chapter.

\begin{table}[htbp]
\centering
\caption{Street Review study design overview across integrated modules in Montréal. Source papers: \citet{MushkaniHabitat2025}, \citet{MushkaniJUM2025}, and \citet{MushkaniKoseki2026}.}
\label{tab:study_design_overview}
\renewcommand{\arraystretch}{1.15}
\setlength{\tabcolsep}{4pt}

\begin{tabular}{
>{\raggedright\arraybackslash}p{0.15\textwidth}
>{\raggedright\arraybackslash}p{0.25\textwidth}
>{\raggedright\arraybackslash}p{0.26\textwidth}
>{\raggedright\arraybackslash}p{0.26\textwidth}
}

\toprule
\textbf{Module} & \textbf{Participants} & \textbf{Inputs and tasks} & \textbf{Primary outputs} \\
\midrule
Interviews and descriptor elicitation & 28 interviewees recruited via community organizations & Semi-directed interviews on inclusive streets; descriptor generation and thematic analysis & Resident-derived descriptors; consolidated criteria; theme maps \\
\addlinespace
Focus groups and deliberative evaluation & 12 participants in 3 focus groups; additional ranking participants (N=17) & Individual scoring and group scoring of street images; ranking experiment across 12 criteria; agreement analysis & Convergence and contestation measures (ICC, Pearson, Kendall); evidence on deliberation effects \\
\addlinespace
Scaling and citywide mapping & 12 evaluators for rating; model trained on co-produced labels & 120 rated images; label propagation across 15{,}000 local images; inference on 45{,}000 Mapillary images & Subgroup-aware predictions; interpretability analyses; citywide heatmaps \\
\bottomrule
\end{tabular}
\end{table}

\subsection{Resident-derived descriptors}
Street Review begins with semi-directed interviews designed to elicit resident descriptors of what makes streets inclusive or exclusionary from the standpoint of people who navigate the city under different constraints. Participants were recruited through collaborations with over one hundred community organizations, enabling engagement with older adults, mobility-impaired individuals, LGBTQ2+ residents, and younger adults.

Across interviews, participants produced a heterogeneous descriptor set. Participants collectively generated more than 600 adjectives describing streetscapes. These descriptors were consolidated through thematic analysis into a smaller set and refined into four criteria that remained interpretable across groups and actionable for planning: Accessibility, Aesthetics, Practicality, and Inclusivity. Thematic analysis foregrounded recurring concerns and supported refinement of the criteria for subsequent evaluation and modeling.

To keep criteria interpretable, Street Review uses a descriptor-to-criterion logic in which criteria are defined as clusters of resident-described conditions rather than as imported planning categories. Table~\ref{tab:criteria_definitions} summarizes the criteria and the descriptors participants associated with them.

\begin{table}[htbp]
\centering
\caption{Street Review criteria and indicative descriptors derived from participatory elicitation and thematic consolidation.}
\label{tab:criteria_definitions}
\renewcommand{\arraystretch}{1.15}
\begin{tabular}{p{0.17\textwidth}p{0.78\textwidth}}
\toprule
\textbf{Criterion} & \textbf{Indicative descriptors and cues discussed by participants} \\
\midrule
Accessibility & Barrier-free routes, sidewalk continuity, manageable crossings, navigation ease, and conditions that enable mobility for people with varied physical abilities. \\
Aesthetics & Visual appeal and perceived care: cleanliness, coherent streetscape composition, greenery presence, and perceived pleasantness of the environment. \\
Practicality & Functional usability and everyday affordances: signage and legibility, maintenance and order, availability of amenities, and perceived ease of accomplishing daily tasks. \\
Inclusivity & Perceived welcome and social openness: safety and security cues, sense of acceptance, cultural representation, and the extent to which different people are perceived as belonging in the space. \\
\bottomrule
\end{tabular}
\end{table}

\subsection{Sampling streetscapes}
To curate evaluation stimuli that cover diverse urban contexts, Street Review uses a sampling matrix that stratifies street environments by land use, neighborhood history, urbanization, socioeconomic context, density, greenery, and activity affordances \citep{talen2012design,Ye2019}. The matrix is a stimulus selection tool rather than a causal model. Its purpose is to reduce the likelihood that evaluation results are artifacts of a narrow set of streetscapes.

\begin{table}[htbp]
\centering
\caption{Street image selection matrix used to sample diverse Montr\'eal streetscapes for evaluation exercises.}
\label{tab:street_selection_matrix}
\begin{tabular}{>{\raggedright\arraybackslash}p{0.34\textwidth}p{0.19\textwidth}p{0.19\textwidth}p{0.19\textwidth}}
\toprule
\textbf{Characteristic} & \textbf{Type I} & \textbf{Type II} & \textbf{Type III} \\
\midrule
Land use & Predominantly residential & Mixed use & Predominantly commercial \\
History & Historic neighborhoods (1920s) & Modern neighborhoods (1970s) & Post-modern neighborhoods (2010s) \\
Urbanization spectrum & Suburban & Urban & City center \\
Socioeconomic status (income) & Low & Medium & High \\
Density & Low & Medium & High \\
Space-to-user relationship & Not occupied & Occupied & Well known place \\
Greenery & Minimal & Moderate & Abundant \\
Affordance (activities and amenities) & Limited & Basic & Diverse \\
\bottomrule
\end{tabular}
\end{table}

\begin{figure}[htbp]
\centering
\includegraphics[width=0.95\textwidth]{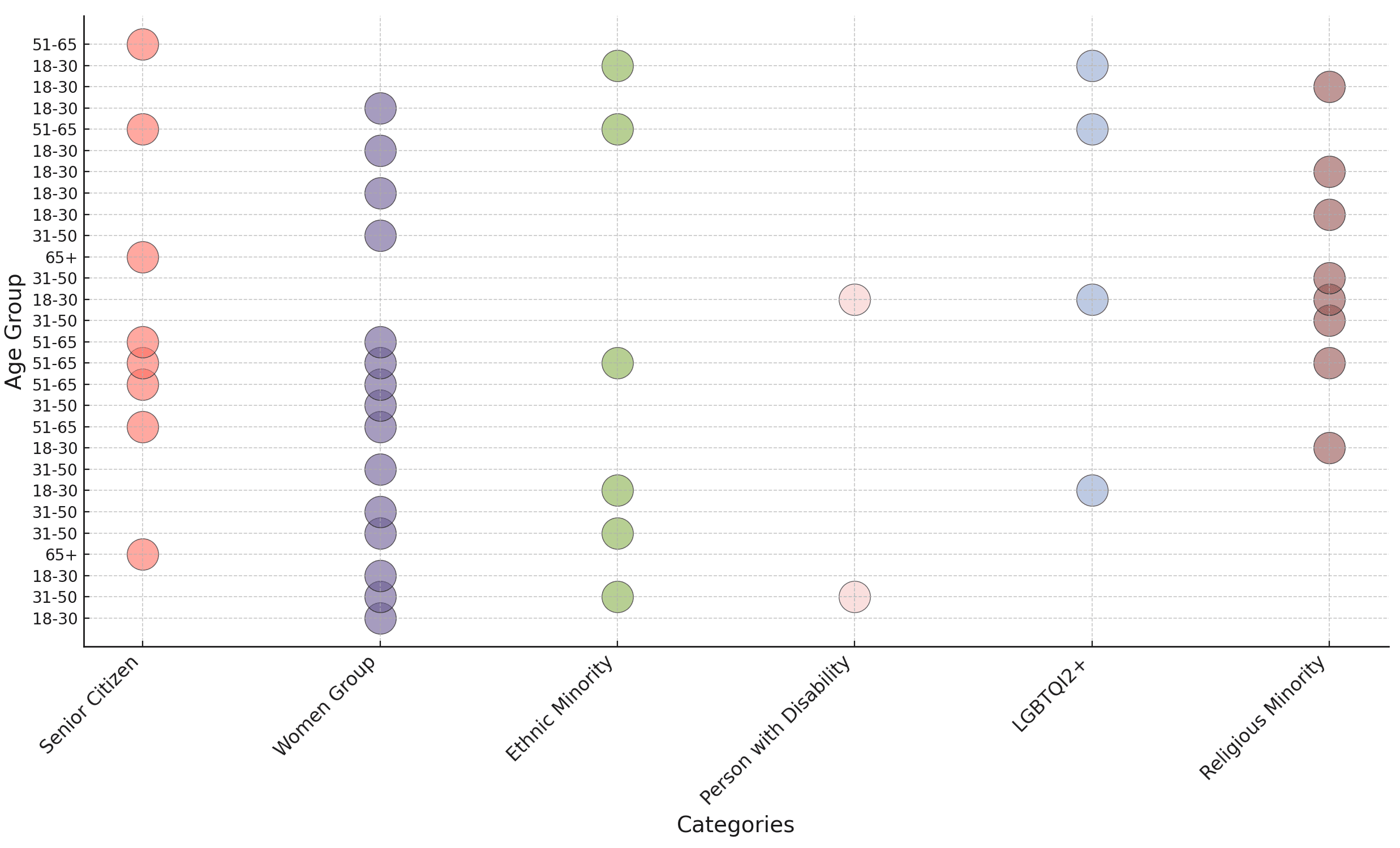}
\caption[Self-declared identity markers in Street Review]{\deleted{Self-declared participant identities across Street Review modules.} \added{Self-declared identity markers in the Street Review participant pool. Markers overlap, and the figure describes the composition of this sample. Source: author data.}}
\label{fig:self_declared_participants}
\label{fig:street-review-identity-markers}
\end{figure}

\begin{figure}[htbp]
\centering
\includegraphics[width=0.95\textwidth]{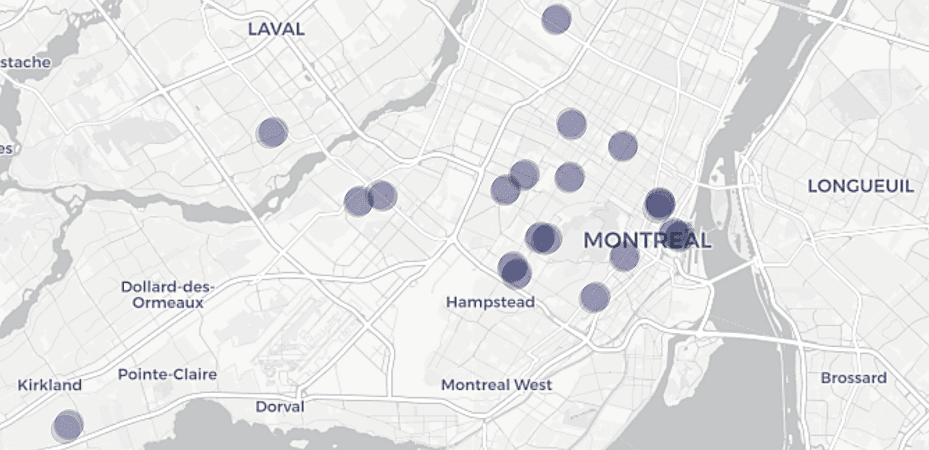}
\caption[Street Review study sites]{Spatial distribution of selected street segments and evaluation sites across Montréal. \added{Source: Author’s map of the Street Review study sites, using OpenStreetMap as the basemap.}}
\label{fig:spatial_distribution_study_sites}
\label{fig:street-review-sites}
\end{figure}

\begin{figure}[htbp]
\centering
\includegraphics[width=1\textwidth]{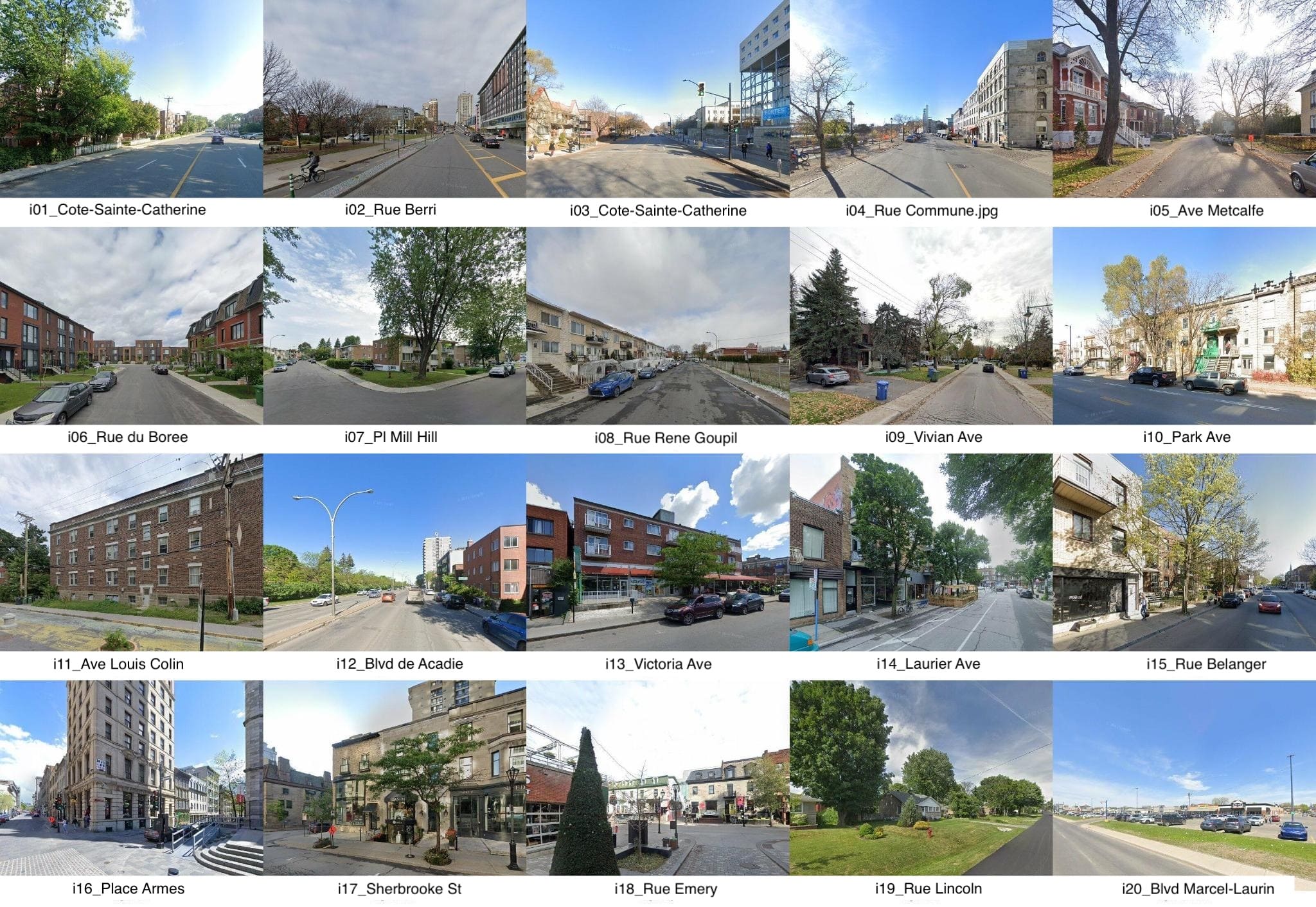}
\caption[Curated Street Review images]{Sample of curated street images used in Street Review evaluation sessions. \added{Source: Author's work.}}
\label{fig:representative_street_images}
\label{fig:street-review-curated-images}
\end{figure}

\begin{figure}[htbp]
\centering
\includegraphics[width=0.95\textwidth]{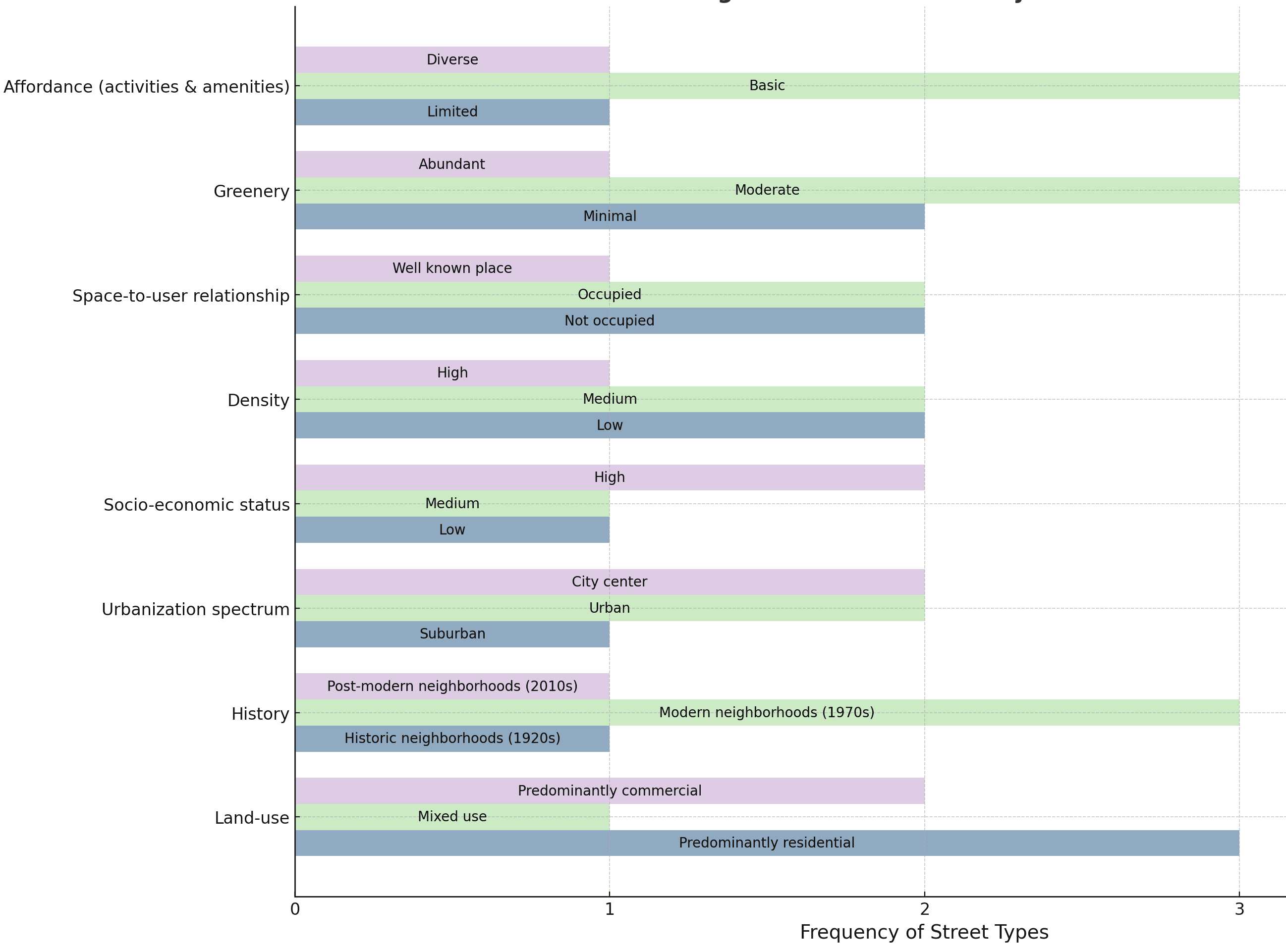}
\caption[Street-context selection matrix]{Distribution of sampled streets across urban-context attributes used to ensure heterogeneity in evaluation stimuli. \added{Source: author analysis of the Street Review sampling matrix.}}
\label{fig:street_diversity_selection_matrix}
\label{fig:street-context-matrix}
\end{figure}

Following this logic, Street Review curated a set of 20 streets, represented through 60 vantage points, spanning commercial corridors, residential blocks, historic districts, and peripheral areas. Figures~\ref{fig:self_declared_participants}--\ref{fig:street_diversity_selection_matrix} present participant diversity and the sampling strategy used to cover multiple street typologies.

\subsection{Image evaluation protocols}
Street Review operationalizes resident descriptors through structured image-based evaluation sessions. Twelve participants formed three heterogeneous focus groups (3--5 participants per group). Participants first scored 20 street images individually, then engaged in moderated discussion, and finally produced a collective group evaluation of the same images. This design separates independent judgments from negotiated collective judgments to make the effects of deliberation empirically measurable.

The same module also included a ranking experiment in which 17 participants ranked seven street images according to twelve criteria, enabling comparison of agreement across a broader set of perceptual and normative attributes. Agreement was assessed using intraclass correlation coefficients (ICC) and Kendall's Tau to distinguish criteria with higher convergence from criteria with persistent  \citep{Bartko1966,PuthEtAl2015}.

In the Street Review scaling module, evaluation stimuli were expanded to 120 images representing 20 streets, with three vantage points per street and two images per vantage point. Two images per data point were used because 360-degree street-view frames can exhibit distortion, and the evaluation protocol aimed to reduce systematic bias from highly distorted views. Participants rated each image on a four-point scale with an accompanying rubric to ground interpretations.

\begin{table}[htbp]
\centering
\caption{Street Review evaluation rubric for image-based scoring on a 1--4 scale.}
\label{tab:evaluation_rubric}
\renewcommand{\arraystretch}{1.2}
\begin{tabular}{p{0.2\textwidth}p{0.72\textwidth}}
\toprule
\textbf{Score} & \textbf{Interpretation (applied across criteria)} \\
\midrule
1 (Poor) & The streetscape fails to meet the criterion and presents significant barriers or negative conditions. \\
2 (Fair) & The streetscape partially meets the criterion but has noticeable limitations that reduce quality or usability. \\
3 (Good) & The streetscape largely meets the criterion, with only minor limitations. \\
4 (Excellent) & The streetscape strongly meets the criterion and offers supportive conditions with minimal barriers. \\
\bottomrule
\end{tabular}
\end{table}

To support communication of results to planning audiences, Street Review also mapped scores to an A--D grading scheme.

\begin{table}[htbp]
\centering
\caption{Grade interpretation used for communicating predicted scores in maps and subgroup summaries.}
\label{tab:grade_interpretation}
\renewcommand{\arraystretch}{1.2}
\begin{tabular}{p{0.18\textwidth}p{0.74\textwidth}}
\toprule
\textbf{Grade} & \textbf{Interpretation} \\
\midrule
A (Excellent) & High-quality streetscape with strong performance on the criterion. \\
B (Good) & Generally supportive streetscape with some limitations. \\
C (Fair) & Mixed conditions; notable deficiencies that reduce perceived quality. \\
D (Poor) & Substantial barriers or negative conditions that undermine perceived quality. \\
\bottomrule
\end{tabular}
\end{table}

\subsection{Theme frequencies and structure}
The Street Review study reports the frequency of key themes across participant groups, providing a qualitative structure that contextualizes subsequent quantitative evaluations. Table~\ref{tab:theme_counts} summarizes these frequencies across four participant groups used in the interview analysis.

\begin{table}[htbp]
\centering
\caption{Frequency of interview themes across participant groups (N=28).}
\label{tab:theme_counts}
\renewcommand{\arraystretch}{1.2}
\small
\begin{tabularx}{\textwidth}{@{}LCCCCC@{}}
\toprule
\textbf{Themes} & \textbf{Elderly} & \textbf{Mobility-impaired} & \textbf{Young adults} & \textbf{\shortstack{LGBTQ2\\+}} & \textbf{Total} \\
\midrule
Safety and security & 20 & 11 & 15 & 20 & 66 \\
Accessibility and mobility & 18 & 15 & 14 & 21 & 68 \\
Functional use and practicality & 15 & 8 & 19 & 22 & 64 \\
Aesthetics and appeal & 15 & 12 & 17 & 10 & 54 \\
Inclusivity and social acceptance & 13 & 6 & 10 & 24 & 53 \\
Maintenance and cleanliness & 11 & 7 & 12 & 11 & 41 \\
Nature and greenery & 8 & 10 & 18 & 12 & 48 \\
\bottomrule
\end{tabularx}
\end{table}

To visualize relationships among themes, a network of co-occurring themes was derived from the interview transcripts, illustrating how participants linked accessibility, safety, aesthetics, and inclusivity-related descriptors. Figure~\ref{fig:network_interview_themes} visualizes this structure. The full network visualization is provided in Appendix~\ref{app:streetreview_supp_figs} (Figure~\ref{fig:network_interview_themes2}).

\begin{figure}[htbp]
\centering
\includegraphics[width=0.95\textwidth]{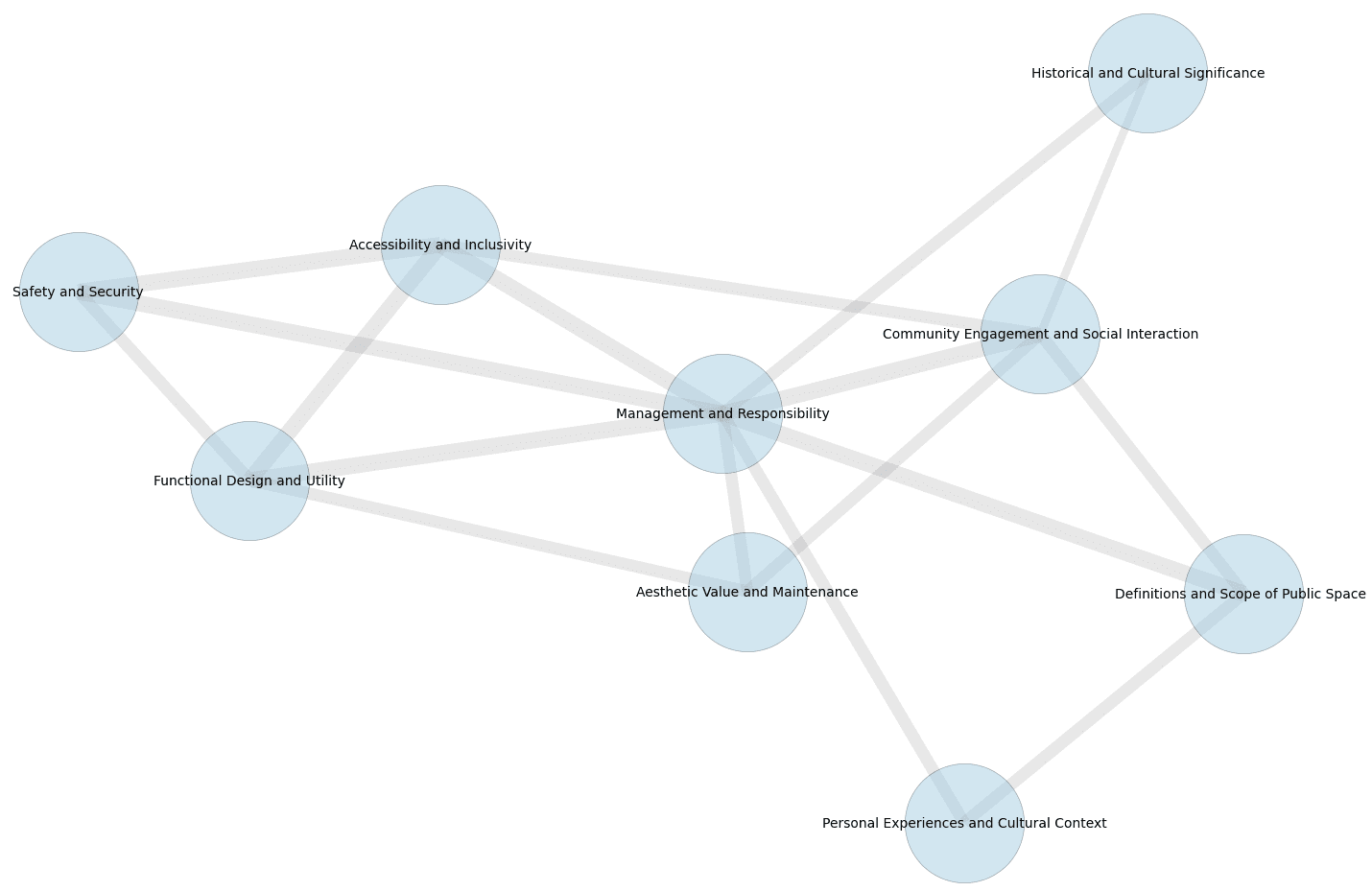}
\caption[Interview-theme network]{Network visualization of interview-derived themes illustrating co-occurrence patterns among accessibility, safety, aesthetics, and inclusivity-related descriptors. \added{Source: author's coding of Street Review interview transcripts.}}
\label{fig:network_interview_themes}
\label{fig:interview-theme-network}
\end{figure}

\subsection{Street Review dataset and citywide imagery}
Street Review links deliberative evaluation to scalable inference through a two-tier imagery strategy. The local dataset comprises 15{,}000 images captured from 60 data points across the 20 sampled streets, with each data point represented by 250 frames in a 360-degree rotation. Participant ratings were collected on 120 images and assigned to the full set of frames for each data point, enabling supervision at scale while keeping the labeling burden feasible.

Because ratings are defined at the \emph{data-point} level and propagated to many near-duplicate frames, the effective number of independent labeled units is $N=60$ data points (not 15{,}000 frames). All model evaluation therefore uses grouped splits that keep frames from the same data point (and, in a stronger test, data points from the same street) in the same fold to avoid leakage across near-duplicate imagery.

For citywide inference, the model was applied to an additional dataset of approximately 45{,}000 street-view images drawn from Mapillary, enabling neighborhood-level mapping of predicted inclusivity and related criteria. This design uses locally collected images as a calibration substrate while leveraging open street-view platforms for spatial coverage.

\begin{tcolorbox}[title=Street Review dataset statistics, colback=white, colframe=black!40]
\small
\begin{itemize}[leftmargin=*]
\item Streets: 20 (3 vantage points per street = 60 data points).
\item Human ratings: 120 rated images (2 images per data point).
\item Local imagery for supervised scaling: $60 \times 250 = 15{,}000$ frames with propagated labels (not independent samples).
\item Citywide inference imagery (unlabeled): $\approx 45{,}000$ Mapillary images.
\end{itemize}
\end{tcolorbox}

\subsection{Machine learning pipeline}
\begin{figure}[htbp]
\centering
\includegraphics[width=0.98\textwidth]{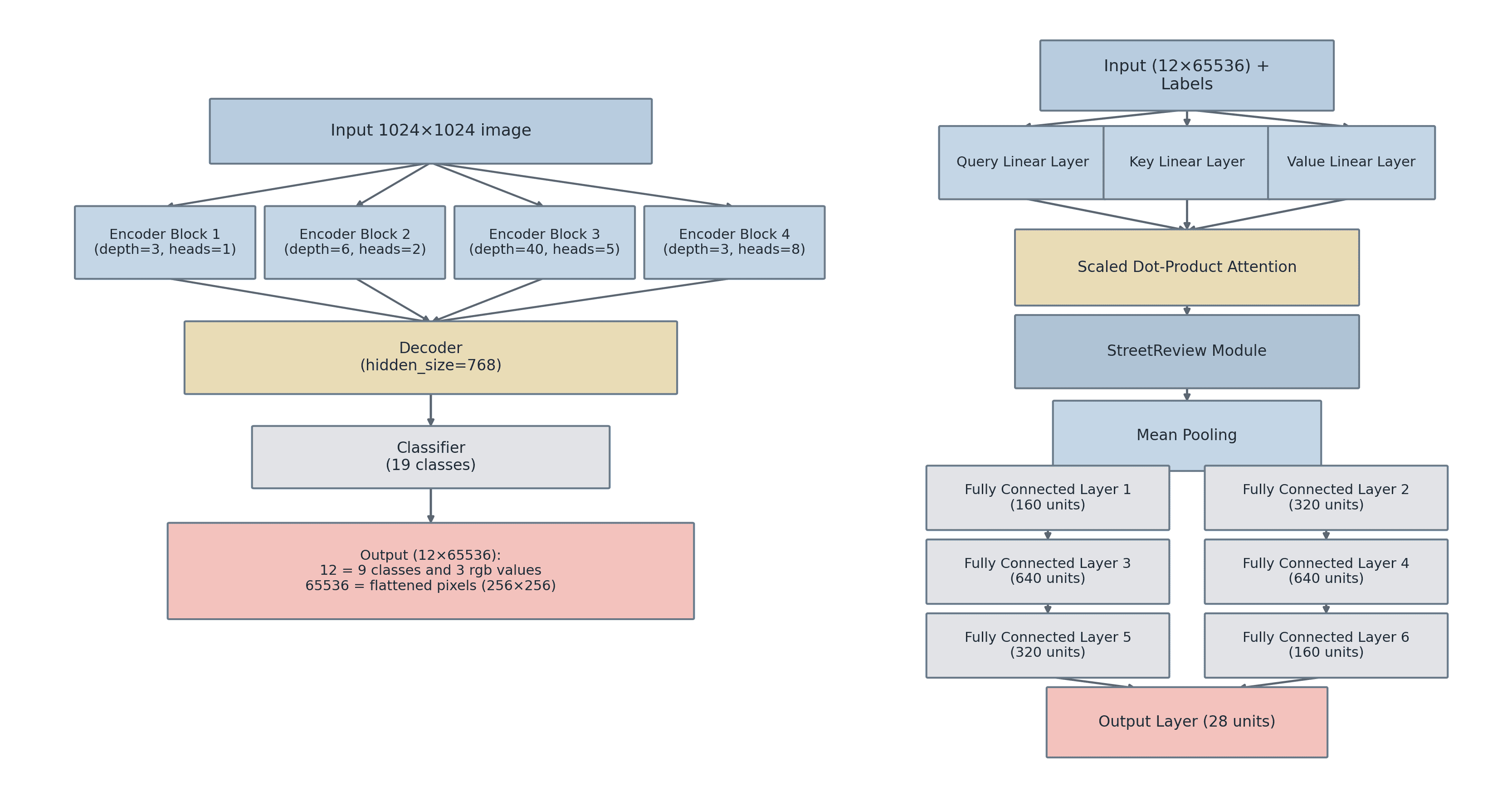}
\caption[Street Review model architecture]{Model architecture used in Street Review. Left: SegFormer segmentation extracts streetscape classes from images. Right: the Street Review attention-based model maps extracted features to predicted scores. \added{Source: author diagram; SegFormer and Cityscapes components follow \citet{Xie2021} and \citet{Cordts2016}.}}
\label{fig:model_architecture}
\label{fig:street-review-model}
\end{figure}

The computational pipeline includes semantic segmentation and multi-output regression (Figure~\ref{fig:model_architecture}).

\medskip

\begin{figure}[htbp]
\centering
\includegraphics[width=0.98\textwidth]{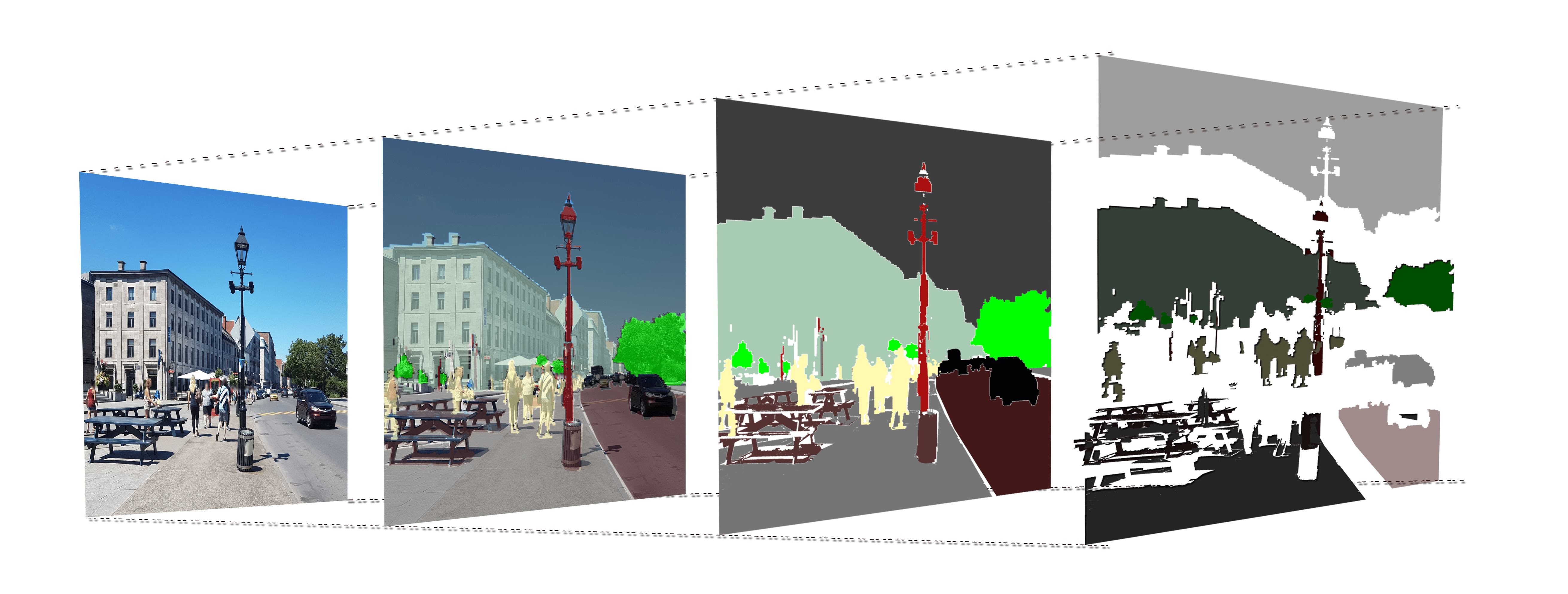}
\caption[Street Review processing pipeline]{Processing pipeline from street-view imagery to segmented classes and filtered features used for prediction. \added{Source: author's work.}}
\label{fig:multistage_processing}
\label{fig:street-review-processing}
\end{figure}

\paragraph{\textbf{Semantic segmentation and feature representation.}}
Street-view images are segmented using SegFormer-B5, fine-tuned on CityScapes at 1024$\times$1024 resolution \citep{Xie2021,Cordts2016}. The pipeline retains a subset of streetscape classes (sidewalk, building, wall, fence, pole, traffic light, traffic sign, vegetation, terrain) and constructs pixel-level feature vectors that include color information and segmentation confidence. Features are aggregated using attention to represent relationships among pixel regions (Figure~\ref{fig:multistage_processing}).

\medskip

\paragraph{\textbf{Multi-output regression.}}
The prediction model is an attention-based multilayer perceptron with six attention heads and 11 fully connected layers \citep{Vaswani2017} (Figure~\ref{fig:model_architecture}). It outputs 28 scores: four criteria for each of six identity-marker groups (24 outputs) plus four group-level scores. Because labels are propagated from 120 rated images to 250 frames per data point, splits are performed with grouping by data point (and, for a stronger test, by street) to avoid leakage across near-duplicate frames \citep{roberts2017crossvalidation}. Train/validation/test splits are created over the 60 data-point groups (70\%, 15\%, 15\%) and then expanded to frames within each split. Training uses mean squared error loss. The Street Review module reports $R^2=0.91$ on validation and $R^2=0.89$ on the test set.

\medskip

\begin{figure}[htbp]
\centering
\includegraphics[width=0.98\textwidth]{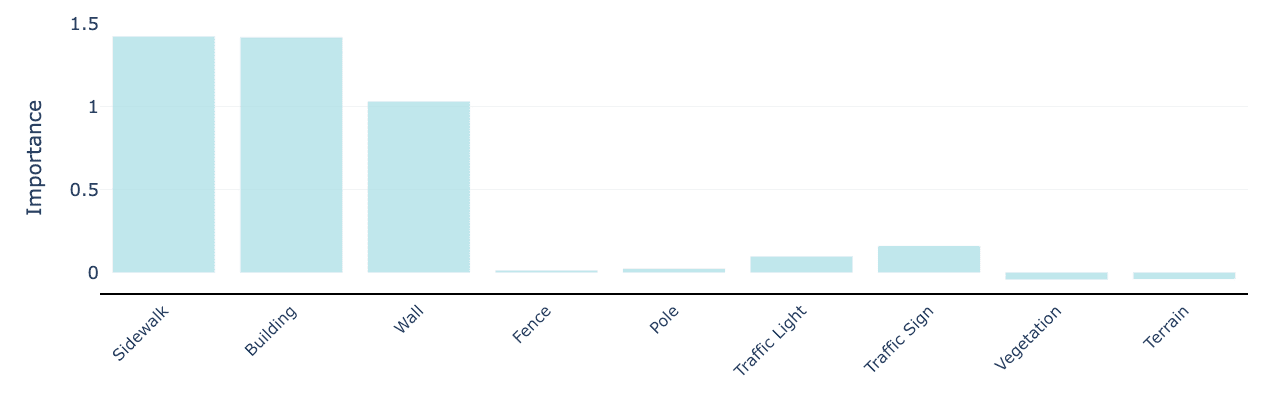}
\caption[Permutation importance by criterion]{Permutation importance of segmented streetscape elements for predicting inclusivity, accessibility, practicality, and aesthetics. \added{Source: author model analysis; see \citet{MushkaniKoseki2026}.}}
\label{fig:permutation_importance}
\label{fig:permutation-importance}
\end{figure}

\paragraph{\textbf{Interpretability.}}
To support interpretability, permutation importance was used to estimate the relative contribution of streetscape elements to predicted outputs. This supports inspection of which visible proxies the model relies on for each criterion and subgroup output (Figure~\ref{fig:permutation_importance}).

\subsection{From predictions to maps}
\begin{figure}[htbp]
\centering
\includegraphics[width=0.98\textwidth]{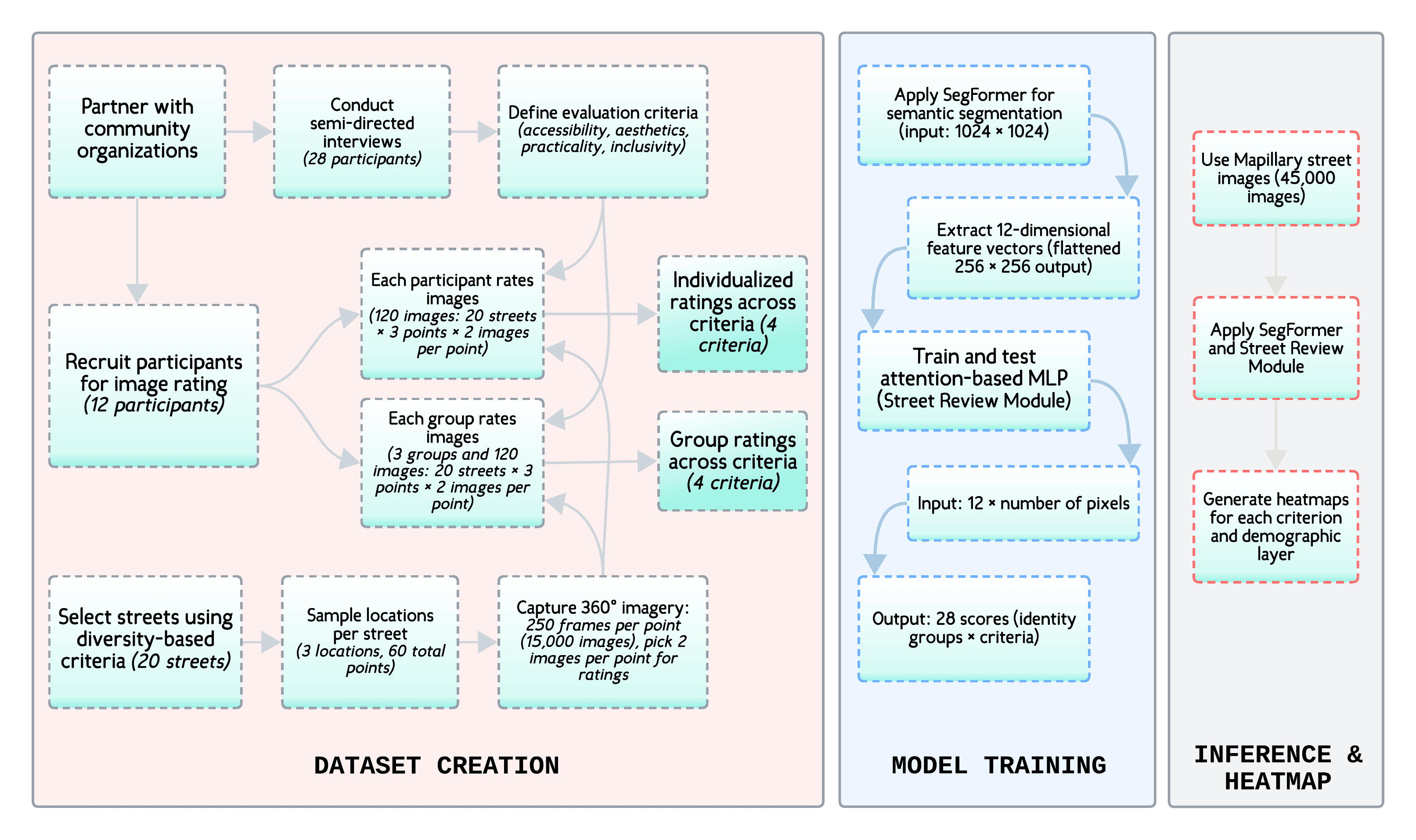}
\caption[Co-production-to-mapping pipeline]{Street Review pipeline from co-production of criteria and descriptors to image evaluation, model training, and citywide mapping. \added{Source: author's synthesis of the Street Review pipeline; see \citet{MushkaniKoseki2026}.}}
\label{fig:streetreview_pipeline}
\label{fig:street-review-pipeline}
\end{figure}

Citywide predictions were aggregated at the street-segment level and visualized as heatmaps using Folium with OpenStreetMap basemaps. Street Review produces group-level maps for citywide auditing and demographic-specific layers to represent differential perceptions. These products are framed as decision-support and accountability artifacts rather than as automated decision rules (Figure~\ref{fig:streetreview_pipeline}).

\section{Findings}
\label{sec:ch5_findings}
This section synthesizes findings from participatory research modules and machine learning analysis, focusing on criteria consolidation, convergence and contestation across criteria and groups, predictive performance and interpretability of the model, and spatial patterns in citywide mapping. Disagreement is treated as an empirical outcome that can inform governance. Persistent divergence on selected criteria is interpreted as a signal of value conflict, ambiguous evidence, or proxy limitations rather than as a measurement defect.

\subsection{Resident descriptors and the meaning of inclusion}
Across interviews, participants articulated inclusion through both material and social registers. Material concerns included sidewalk continuity and width, ramps and curb quality, crossings, and features that shape mobility and navigation. Social and symbolic concerns included lighting, perceived surveillance or policing presence, cultural representation, and the presence of venues or markers associated with particular communities. This descriptor structure spans features that are frequently visible in street-view imagery (sidewalks, building edges, vegetation) and features that are partially visible or not visible (social acceptance, symbolic meaning, temporal dynamics).

Theme frequencies show that some concerns recur across groups, while others are emphasized differently by participant groups. Table~\ref{tab:theme_counts} indicates that safety and accessibility themes appear across groups, while LGBTQ2+ participants refer more frequently to inclusivity and social acceptance. These patterns support the claim that pluralism is structured by social position rather than random variation.

\subsection{Convergence and contestation across criteria}
\begin{figure}[htbp]
\centering
\includegraphics[width=0.98\textwidth]{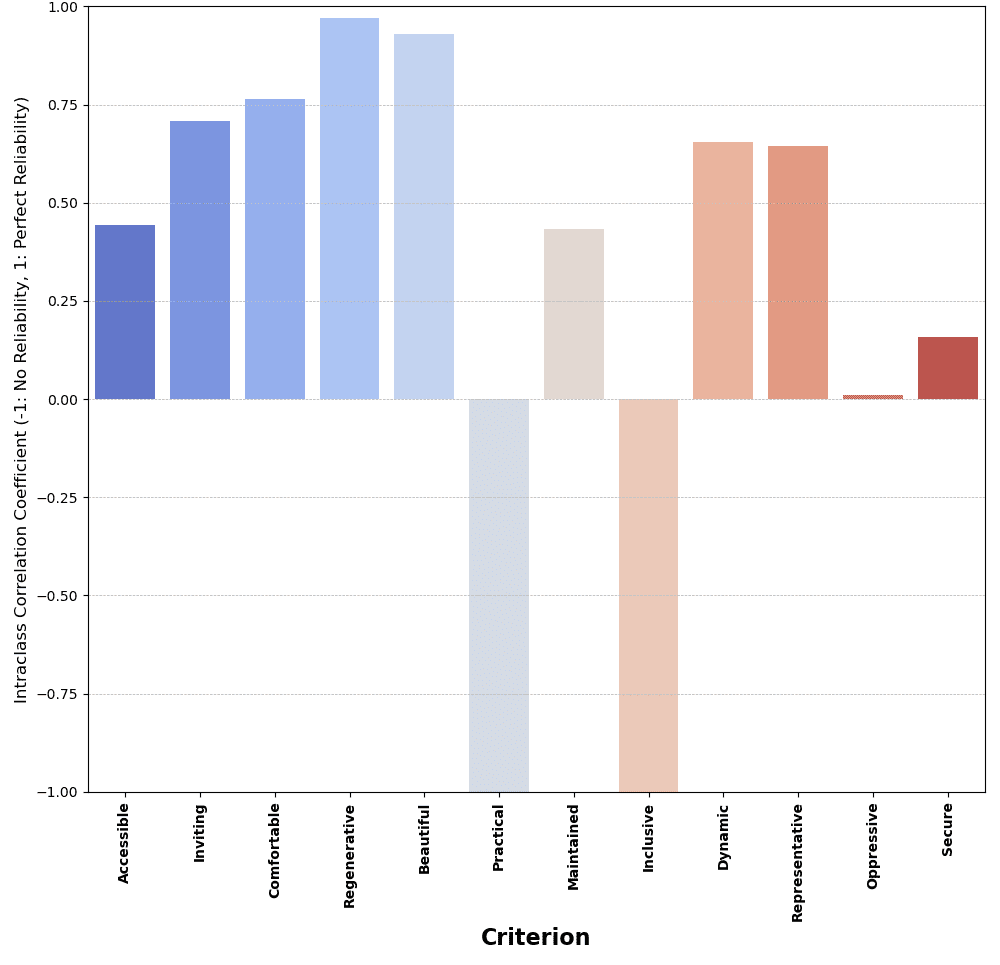}
\caption[Ranking-experiment agreement]{Agreement landscape in the ranking experiment. ICC values across twelve criteria illustrate higher reliability for some perceptual dimensions and lower or negative reliability for criteria such as inclusive and practical. Source: Author’s work \citep{MushkaniJUM2025}.}
\label{fig:jum_icc_ranking}
\label{fig:ranking-icc}
\end{figure}

\begin{figure}[htbp]
\centering
\includegraphics[width=0.99\textwidth]{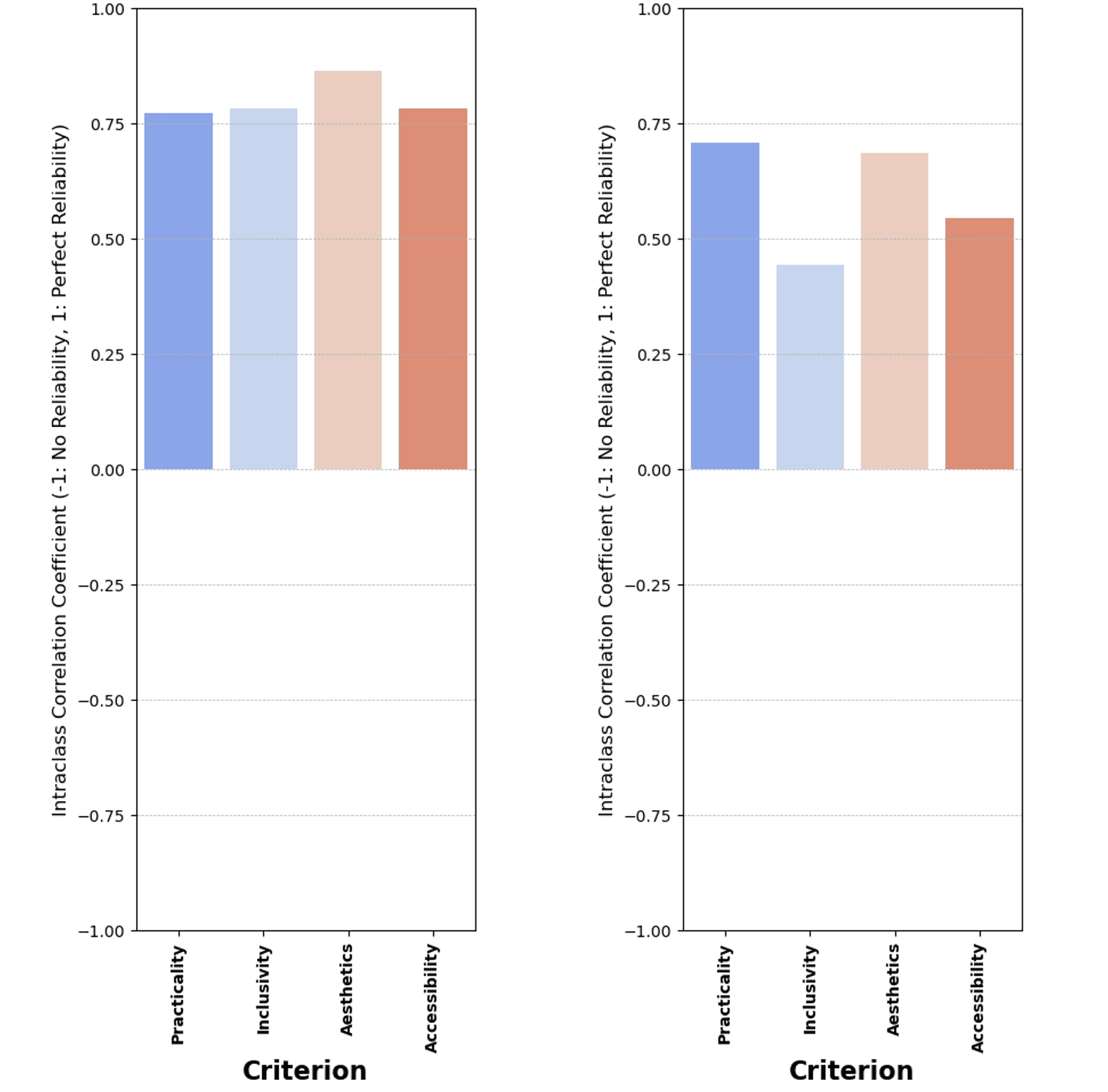}%
\caption[Individual- and group-level reliability]{Individual- versus group-level reliability in the rating experiment. Right: individual ratings; left: group ratings. Intraclass correlation coefficients (ICCs) increase across all four criteria under group deliberation, with consistently higher agreement for aesthetics than for inclusivity. \added{Source: author analysis; see \citet{MushkaniJUM2025}.}}
\label{fig:jum_icc_rating}
\label{fig:rating-icc}
\end{figure}

\begin{figure}[htbp]
\centering

\begin{subfigure}{\textwidth}
\centering
\includegraphics[width=0.73\textwidth]{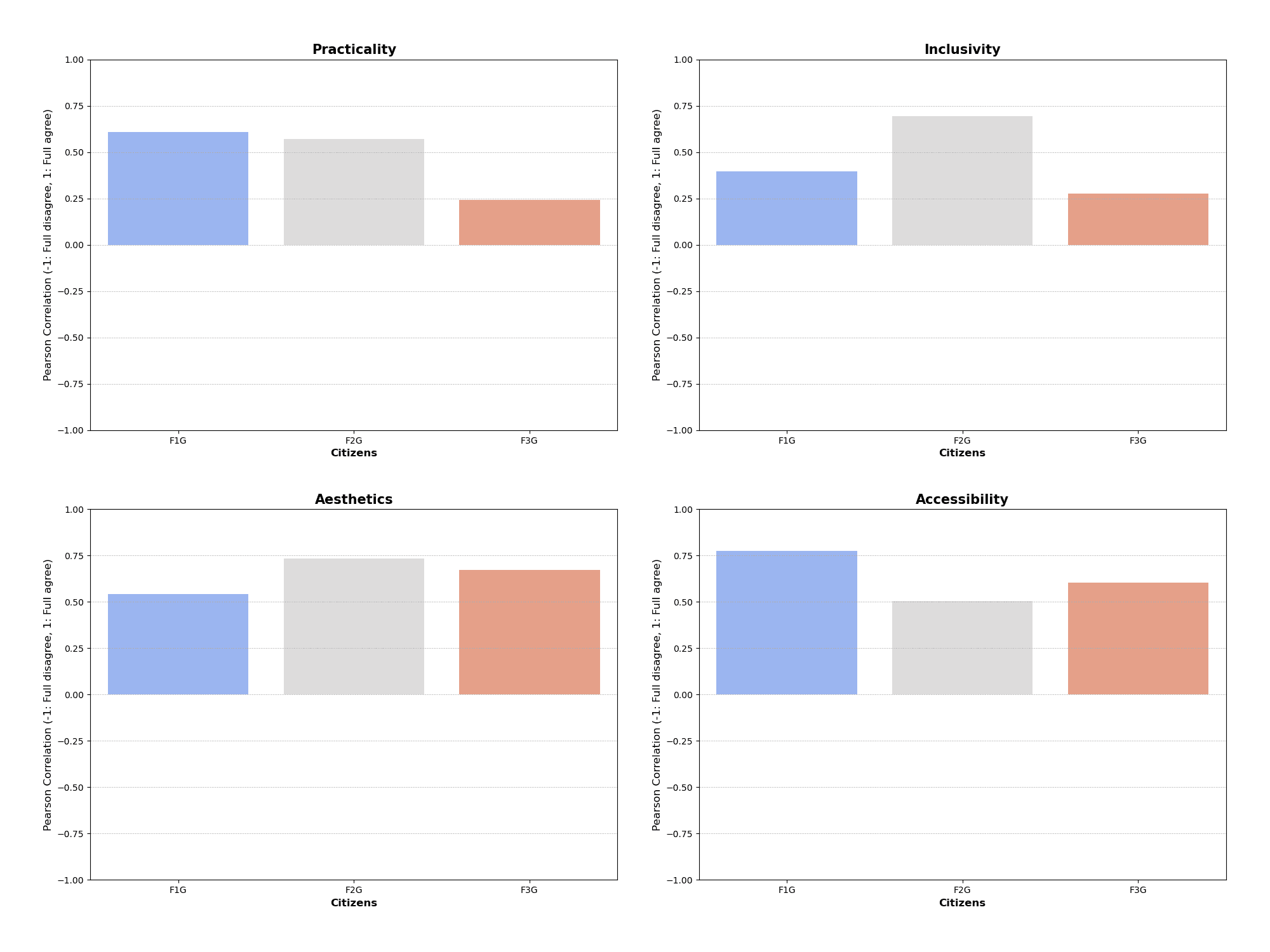}
\caption{Group-level correlations.}
\label{fig:jum_group_correlations}
\end{subfigure}

\begin{subfigure}{\textwidth}
\centering
\includegraphics[width=0.73\textwidth]{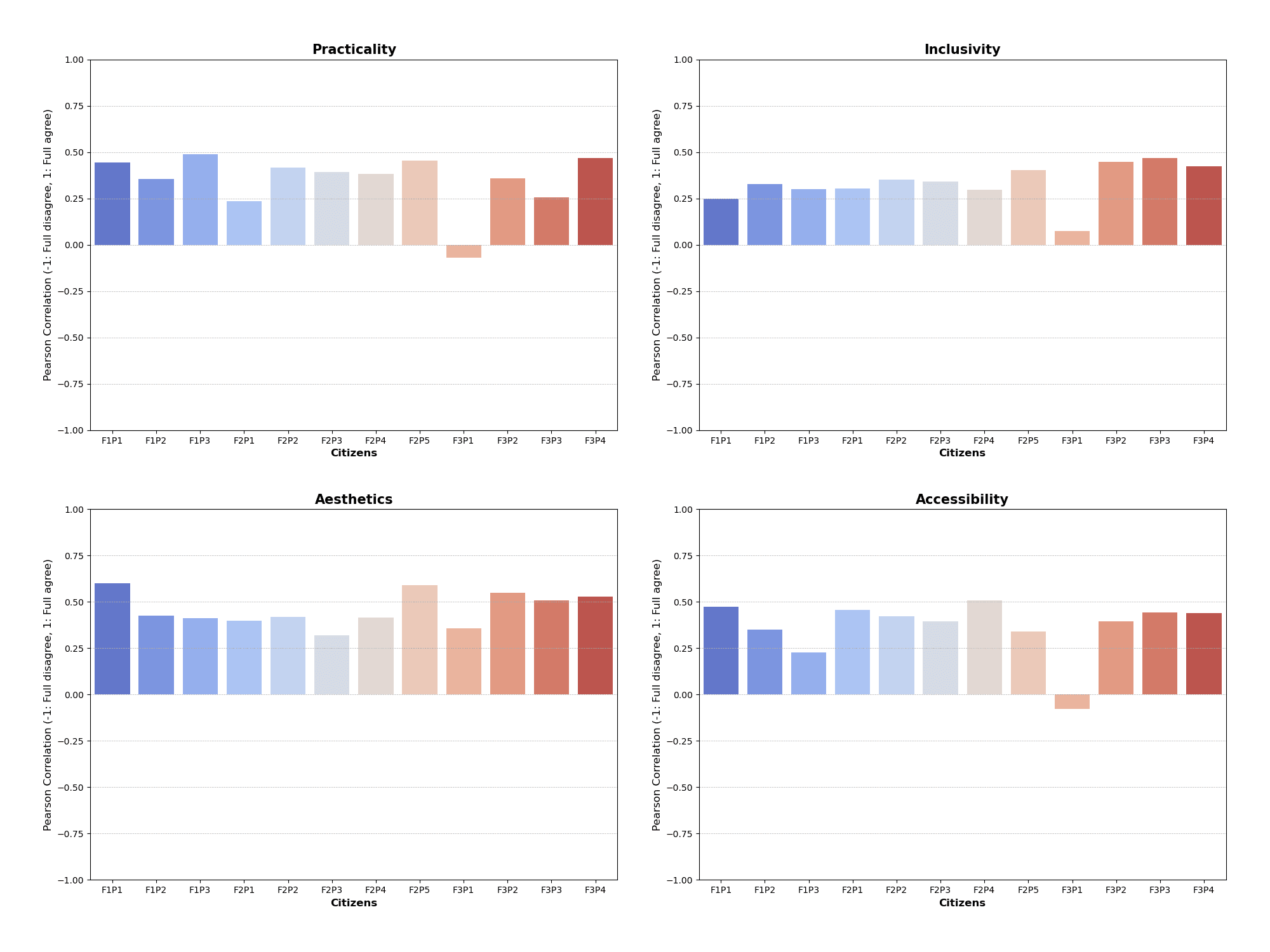}
\caption{Individual-level correlations.}
\label{fig:jum_individual_correlations}
\end{subfigure}

\caption[Individual and group rating correlations]{Pearson correlations between individual and group evaluations across criteria and focus groups, illustrating deliberation-induced convergence patterns. \added{Source: author analysis; see \citet{MushkaniJUM2025}.}}
\label{fig:jum_pearson_correlations}
\label{fig:individual-group-correlations}
\end{figure}

Agreement varies by criterion. In the ranking experiment, intraclass correlations were higher for selected perceptual criteria and lower for criteria such as inclusive and practical, which exhibited near-zero or negative ICC values. This pattern indicates that participants more readily converge on some visually legible attributes than on socially mediated attributes. Figures \ref{fig:jum_icc_ranking} through \ref{fig:jum_pearson_correlations} visualize the results.

In the rating experiment (12 participants, 20 images, four criteria), agreement was higher than in the ranking experiment but remained criterion-dependent. Group-level ICC values were higher than individual-level ICC values across criteria, and group discussion increased convergence. Aesthetics exhibited higher agreement than inclusivity, indicating that socially mediated judgments resist full standardization.

\subsection{Deliberation effects}
This subsection examines how moderated discussion alters the relationship between individual judgments and group outcomes. Pearson correlations between individual- and group-level evaluations show increased alignment following deliberation, with variation across evaluative criteria and groups. Deliberation thus functions as part of the measurement apparatus itself, reshaping how participants justify, contest, and revise their evaluations in relation to others’ experiences.

Deliberation does not eliminate divergence across all criteria. Residual disagreement after discussion indicates that some dimensions remain contested or require context beyond static imagery. For governance, this suggests that the measurement infrastructure should preserve disagreement and represent it explicitly rather than treating it as an error state.

\subsection{Aggregate ratings}
\begin{figure}[htbp]
\centering
\includegraphics[width=0.98\textwidth]{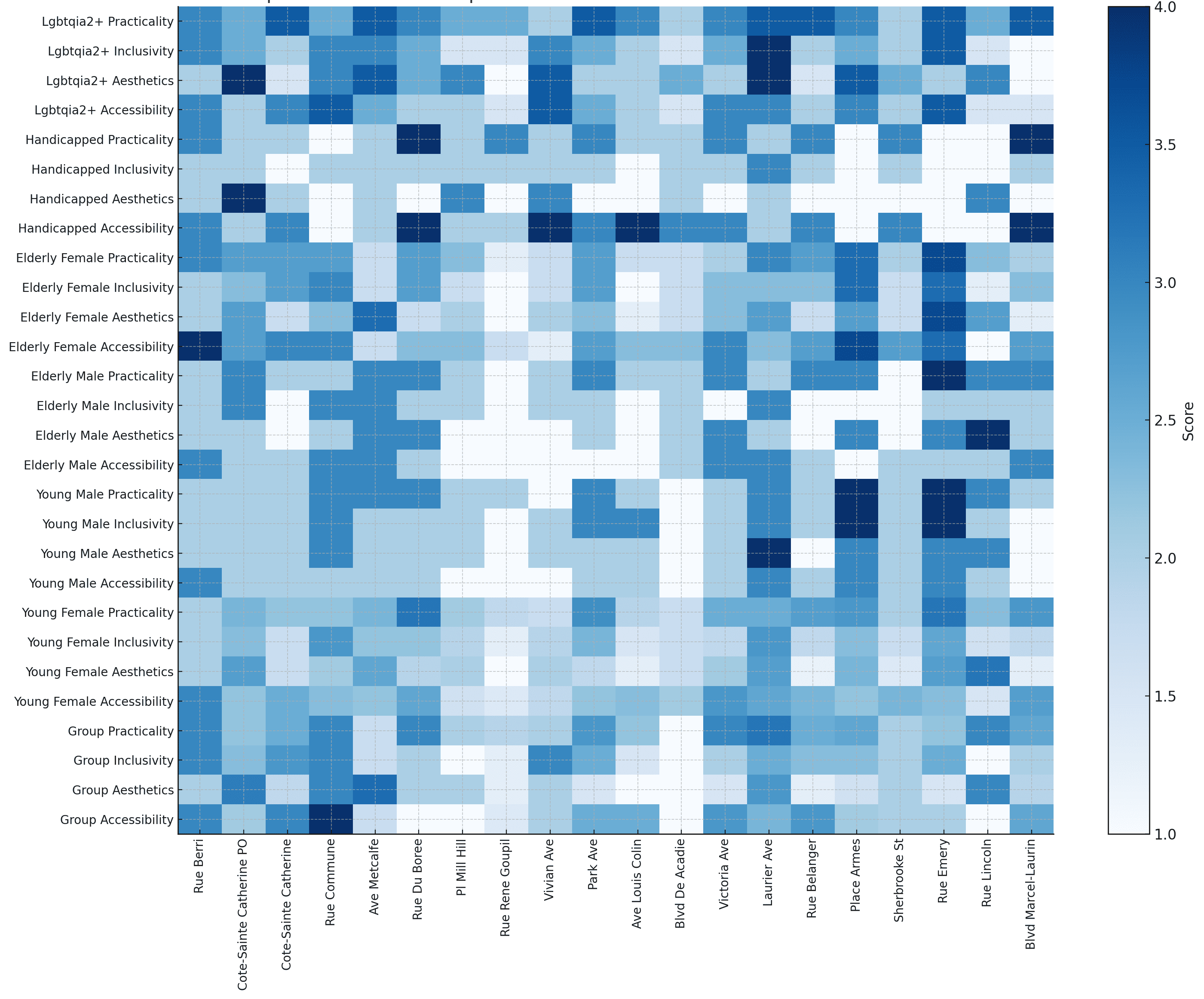}
\caption[Aggregated Street Review rating matrix]{Matrix representation of aggregated ratings for 20 Montréal streets across Street Review criteria. \added{Source: author analysis; see \citet{MushkaniKoseki2026}.}}
\label{fig:aggregated_ratings_matrix}
\label{fig:rating-matrix}
\end{figure}

Figure~\ref{fig:aggregated_ratings_matrix} visualizes aggregated ratings for 20 streets across 60 data points and the four criteria in the Street Review module. Group-level mean scores are in the mid-range of the four-point scale: Group Accessibility averaged 2.12, Group Inclusivity 2.06, Group Practicality 2.39, and Group Aesthetics 1.99. These values indicate that most evaluated streetscapes combine supportive and limiting features.

\begin{figure}[htbp]
\centering
\begin{subfigure}[t]{0.485\textwidth}
\centering
\includegraphics[width=\textwidth]{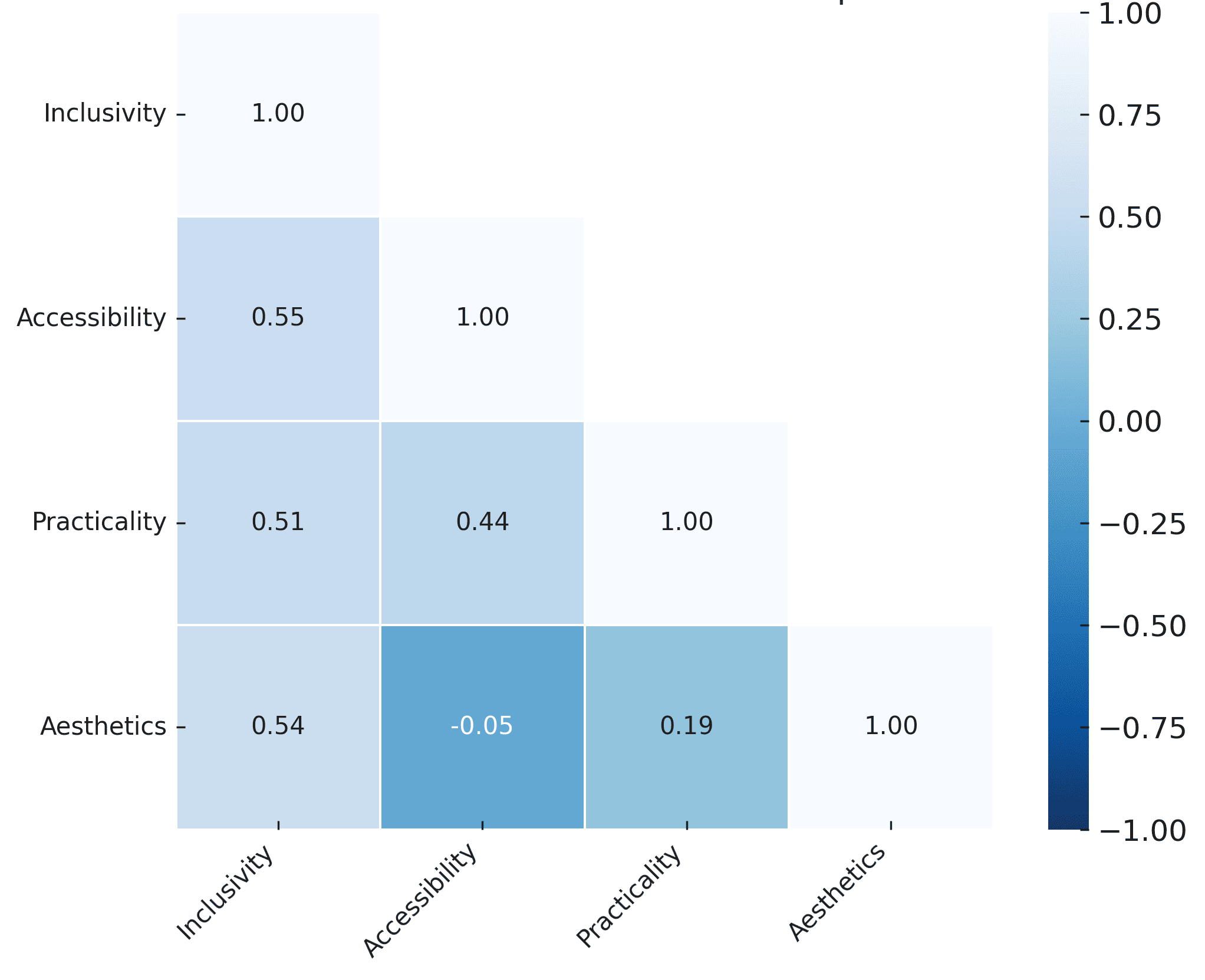}
\caption{Participant evaluation correlations.}
\label{fig:criteria_corr_participant}
\end{subfigure}\hfill
\begin{subfigure}[t]{0.495\textwidth}
\centering
\includegraphics[width=\textwidth]{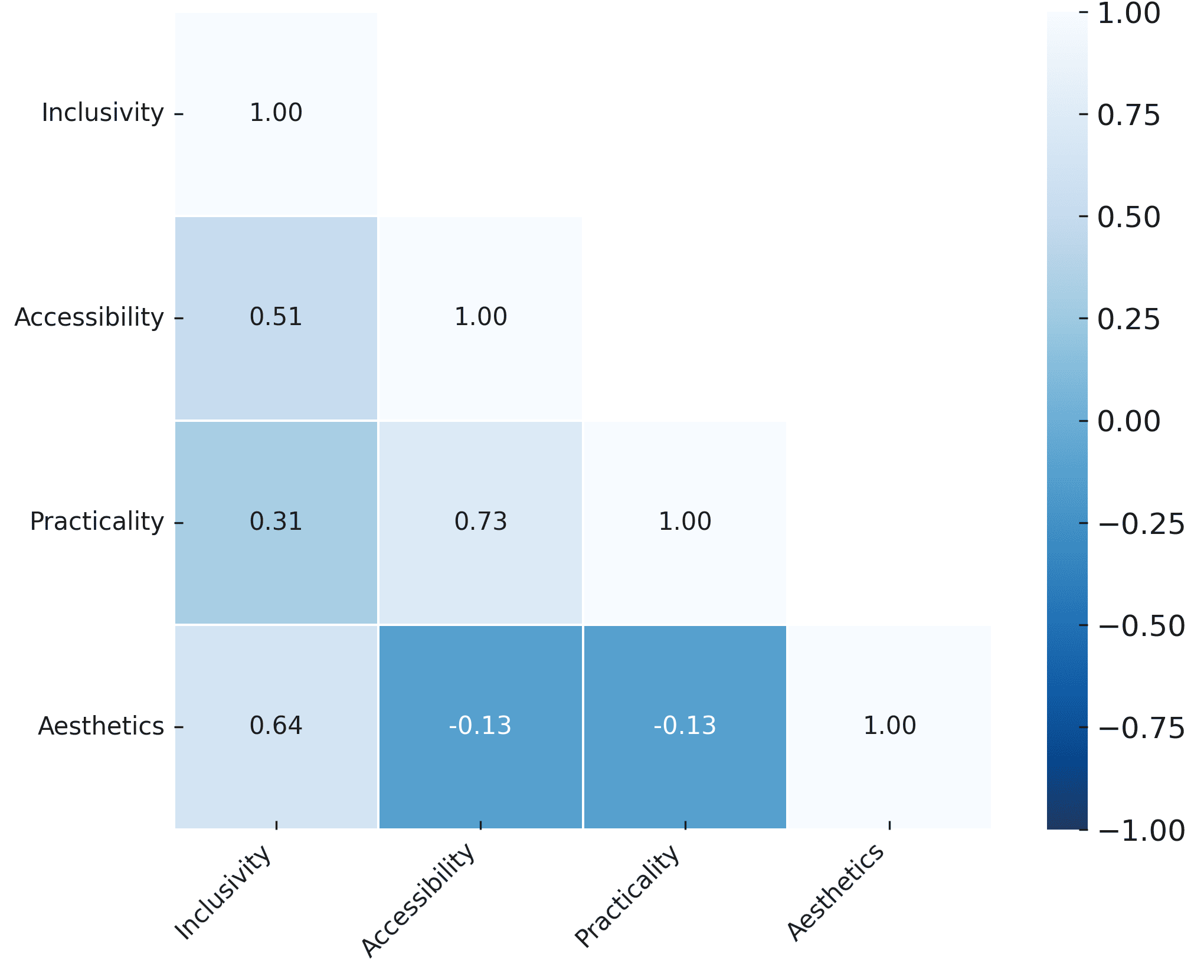}
\caption{Model prediction correlations.}
\label{fig:criteria_corr_model}
\end{subfigure}
\caption[Participant and model correlations]{Correlation between criteria for participant evaluations (left) and for model predictions (right). \added{Source: author analysis; see \citet{MushkaniKoseki2026}.}}
\label{fig:criteria_correlations_human_model}
\label{fig:participant-model-correlations}
\end{figure}

Relationships among criteria indicate why a composite score is analytically and normatively fragile. Participant evaluations show that inclusivity correlates with accessibility (0.55) and aesthetics (0.54), while practicality and aesthetics exhibit a weak negative correlation ($-0.05$). Figure~\ref{fig:criteria_correlations_human_model} compares criterion correlations derived from participant evaluations with those derived from model predictions. The comparison provides a diagnostic representation of where the model reproduces the relational structure of participant judgments and where it may impose its own proxy structure.

\subsection{Subgroup differences}
\begin{figure}[htbp!]
\centering
\includegraphics[width=0.88\textwidth]{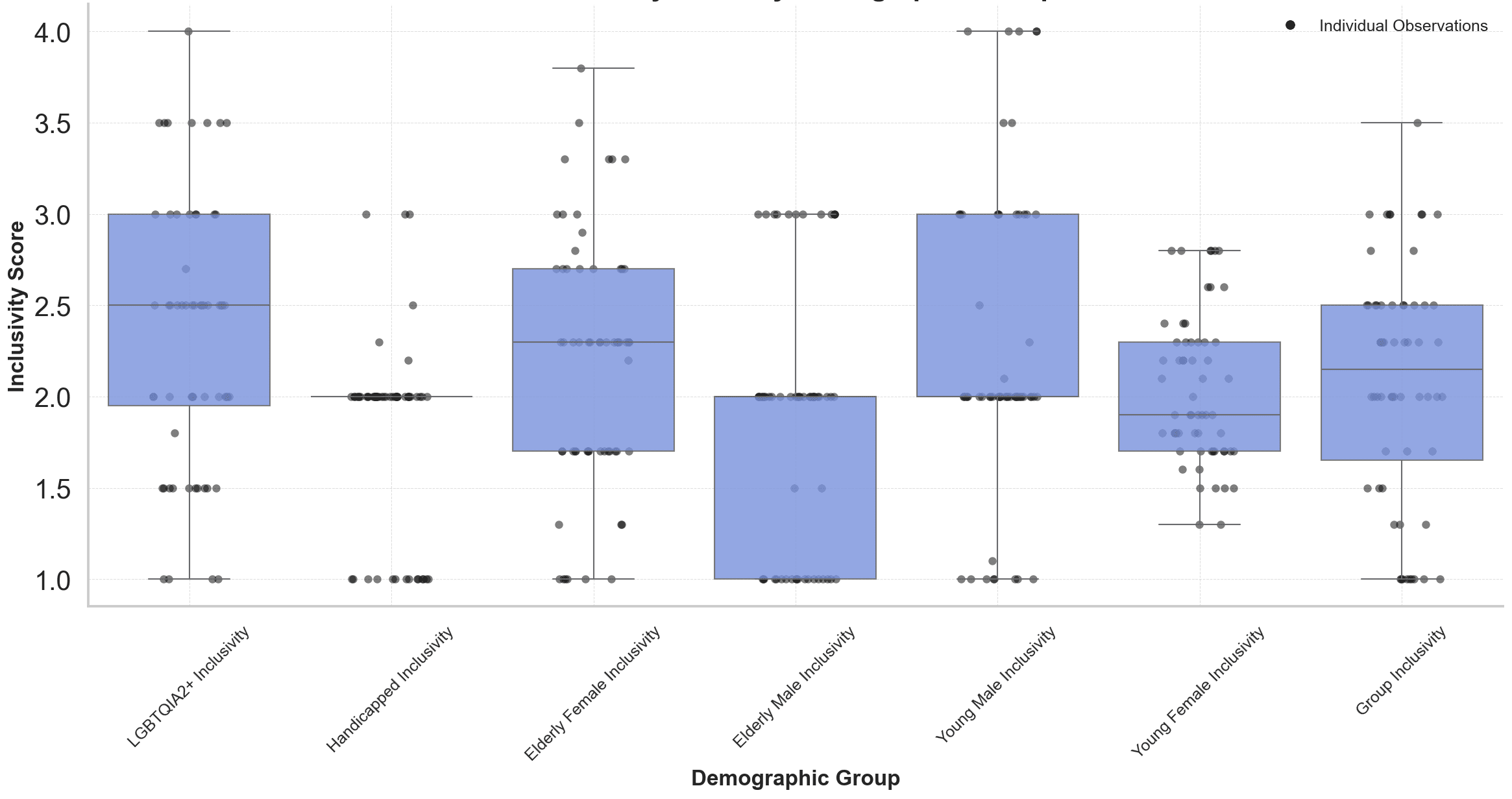}
\caption[Inclusivity ratings by identity marker]{Boxplot analysis of inclusivity ratings across participant groups for 60 data points (20 streets). \added{Source: author analysis; see \citet{MushkaniKoseki2026}.}}
\label{fig:boxplot_inclusivity_by_group}
\label{fig:inclusivity-boxplots}
\end{figure}

Street Review models subgroup differences rather than treating them as residual variance. Disaggregated analysis shows variation in inclusivity perceptions across participant groups evaluating the same data points. Figure~\ref{fig:boxplot_inclusivity_by_group} summarizes inclusivity ratings across groups for the 60 evaluated data points. Participants identifying as elderly males, young females, and individuals with mobility impairments provided lower median inclusivity ratings than other groups.

To provide an intuitive cross-criterion view of subgroup rating profiles, Appendix Figure~\ref{fig:app-streetreview-radar} summarizes mean Street Review scores for accessibility, inclusivity, practicality, and aesthetics by major demographic group. Lines that extend farther outward indicate higher average ratings on that criterion, making visible where groups diverge in both absolute scores and relative emphasis across dimensions.

\begin{figure}[htbp!]
\centering
\includegraphics[width=0.88\textwidth]{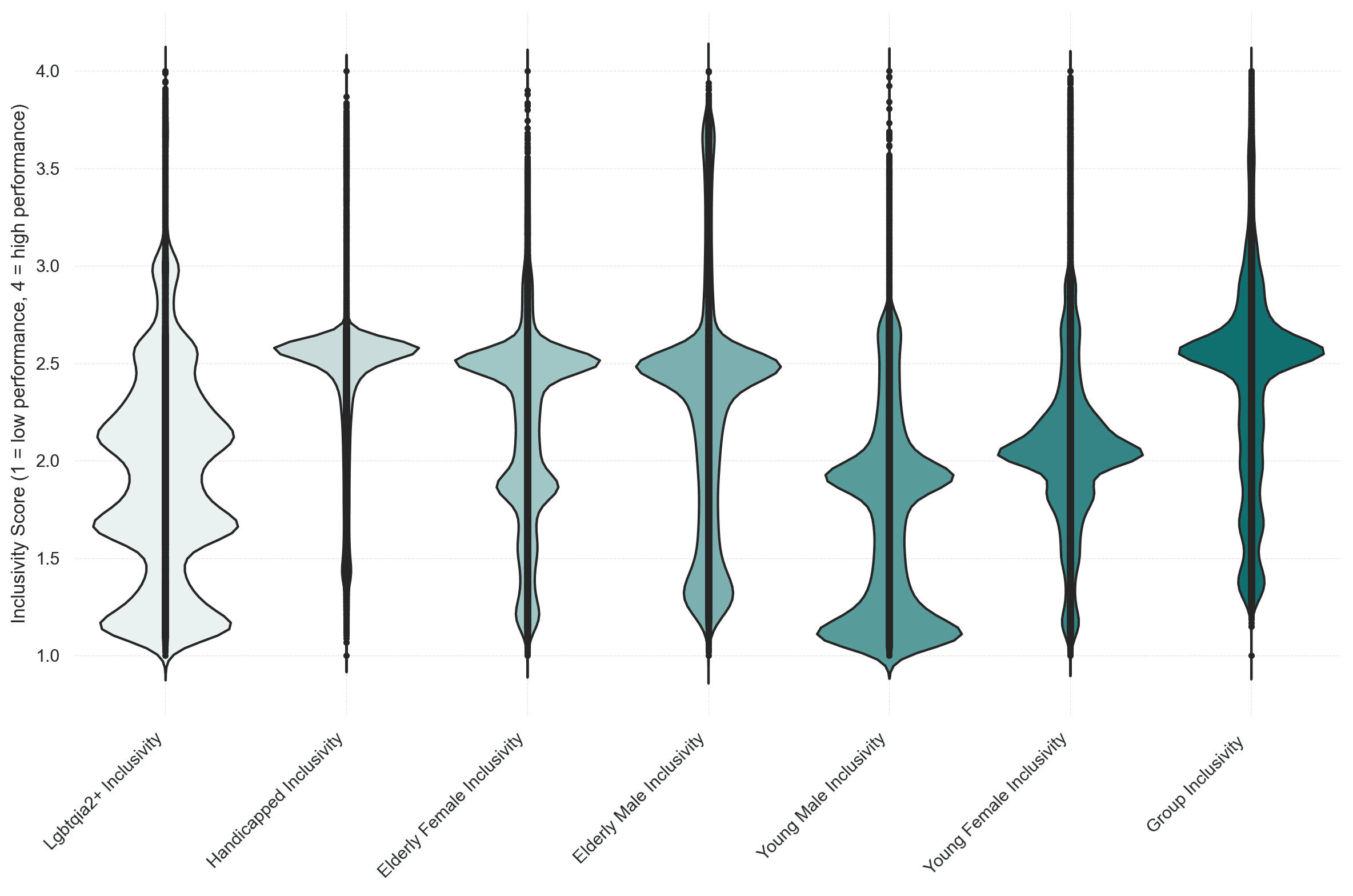}
\caption[Predicted inclusivity distributions]{\deleted{Violin plots of predicted inclusivity ratings across 45{,}000 Mapillary images showing distribution shifts across demographic groups.} \added{Predicted inclusivity distributions for six identity-marker outputs and the aggregate group output across approximately 45{,}000 Mapillary images. Source: author model outputs; see \citet{MushkaniKoseki2026}.}}
\label{fig:violin_citywide_inclusivity}
\label{fig:predicted-inclusivity-distributions}
\end{figure}

\deleted{At city scale, predicted inclusivity across 45{,}000 images clusters in mid-range values, with distribution shifts across groups (Figure~\ref{fig:violin_citywide_inclusivity}).} \added{At city scale, predicted inclusivity across approximately 45{,}000 Mapillary images clusters in mid-range values, with distribution shifts across six identity-marker outputs and the aggregate group output (Figure~\ref{fig:violin_citywide_inclusivity}).} Subgroup differences persist after scaling, and aggregation can obscure variation across groups. The prevalence of mid-range values indicates that many streetscapes are not predicted as uniformly inclusive or exclusionary within the model’s output space.

\begin{figure}[htbp]
\centering
\includegraphics[width=1\textwidth,height=0.8\textheight,keepaspectratio]{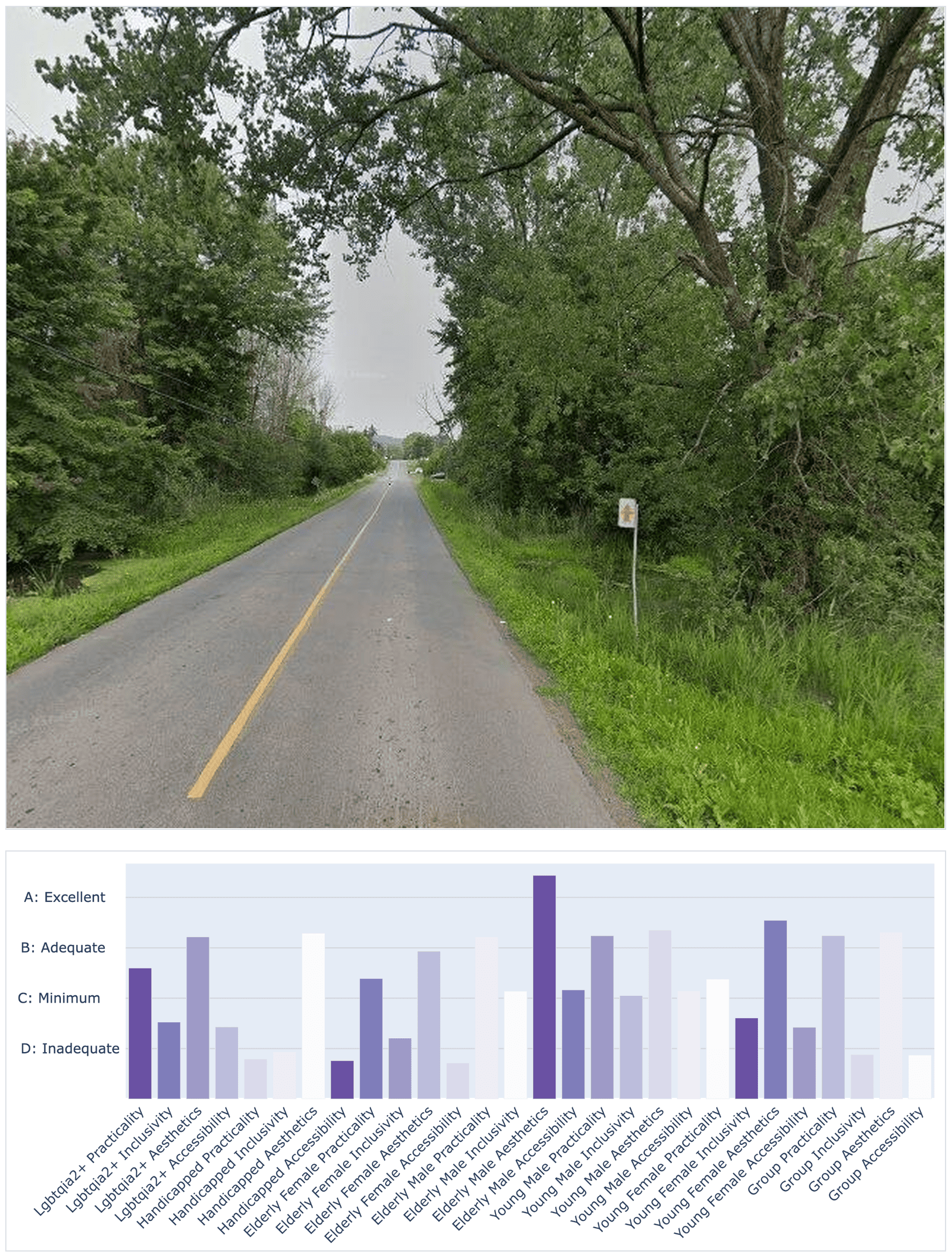}
\caption[Model predictions for one streetscape]{\deleted{Divergent group evaluations for a single streetscape image illustrating variation in predicted grades across demographic-specific models.} \added{Model predictions for one streetscape. Top: source image. Bottom: predicted grades for Inclusivity, Accessibility, Practicality, and Aesthetics for six identity-marker outputs and the aggregate group output. Source: author's work.}}
\label{fig:divergent_group_evaluations}
\label{fig:group-specific-streetscape}
\end{figure}

\deleted{Figure~\ref{fig:divergent_group_evaluations} illustrates how the same streetscape can receive different predicted grades across groups.} \added{Figure~\ref{fig:divergent_group_evaluations} holds the visual stimulus constant and shows variation in model-predicted grades across six identity-marker outputs and the aggregate group output.} Representing divergence as an output is a design choice intended to surface where a street’s perceived conditions vary by social position.

\FloatBarrier

\subsection{Model performance, interpretability, and error modes}
\begin{figure}[htbp]
\centering
\includegraphics[width=0.90\textwidth]{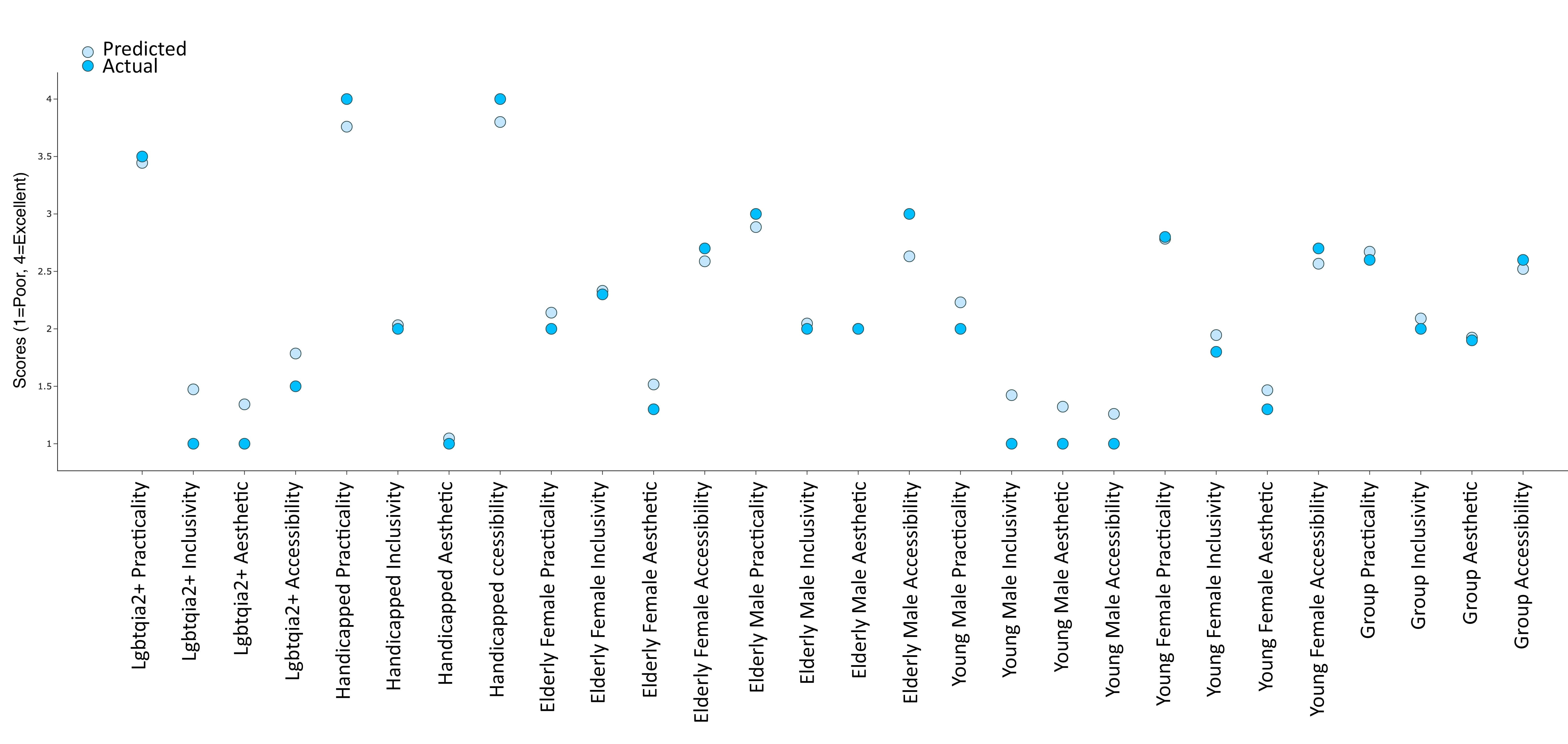}
\caption[Participant ratings and model predictions]{Comparison of participant ratings and model predictions across criteria and participant subgroups. \added{Source: author analysis; see \citet{MushkaniKoseki2026}.}}
\label{fig:actual_vs_predicted}
\label{fig:participant-model-comparison}
\end{figure}

\begin{figure}[htbp]
\centering
\includegraphics[width=0.98\textwidth]{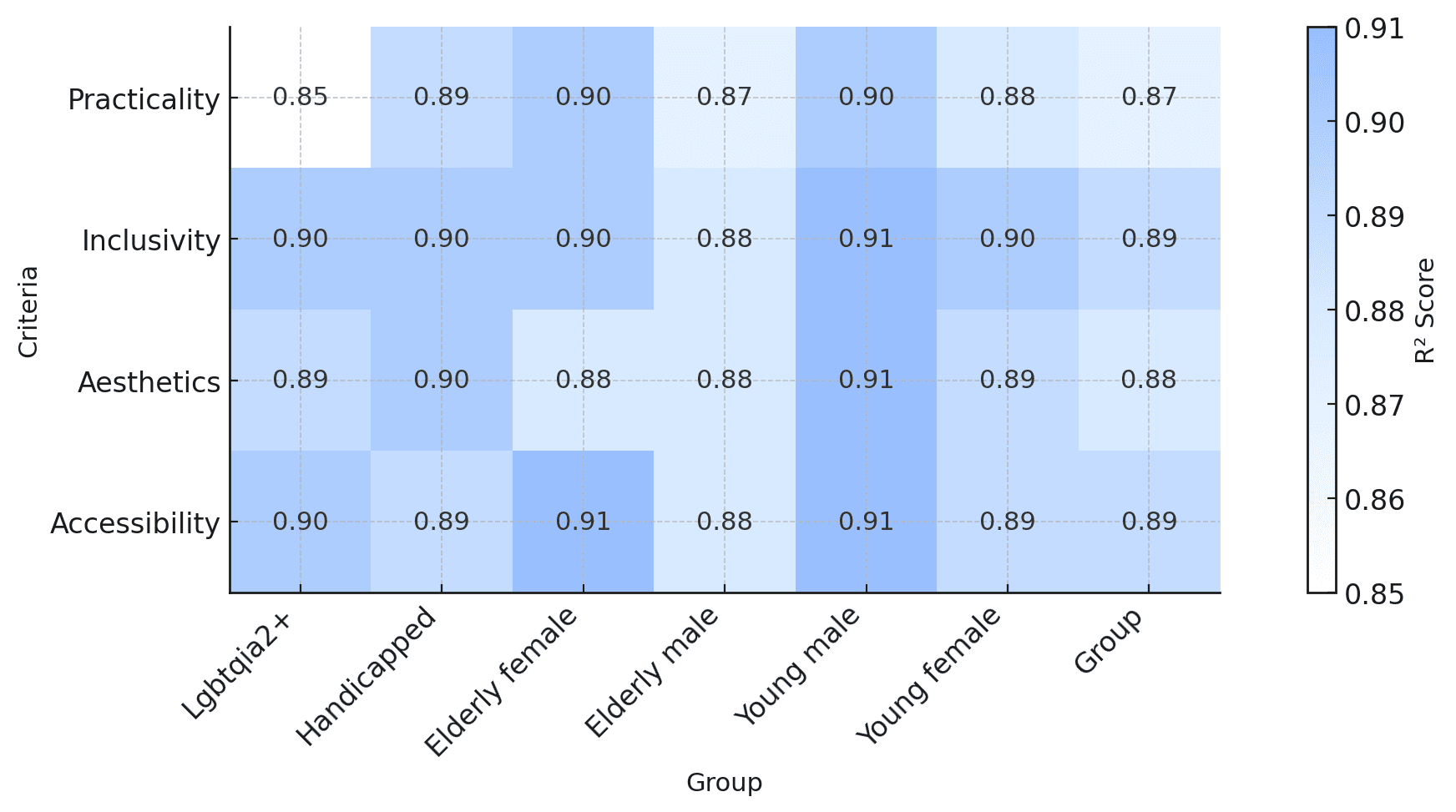}
\caption[Model performance by output and criterion]{$R^2$ scores across demographic subgroups and criteria (28 output combinations). \added{Source: author model analysis; see \citet{MushkaniKoseki2026}.}}
\label{fig:r2_by_group_and_criterion}
\label{fig:r2-by-output}
\end{figure}

The Street Review model achieved $R^2=0.91$ on validation and $R^2=0.89$ on the held-out test set. Figure~\ref{fig:actual_vs_predicted} compares predicted and participant-rated scores. Figure~\ref{fig:r2_by_group_and_criterion} reports $R^2$ by subgroup and criterion, indicating variation across output dimensions.

Interpretability analysis indicates that sidewalks and building-related features are influential predictors, followed by walls and boundary structures. This reliance on visible proxies motivates a bounded interpretation of the model outputs. The Street Review module also documents error modes associated with citywide imagery quality. Mapillary images vary in lighting, blur, distortion, and coverage. Low-quality images were retained to preserve spatial diversity, and this choice can introduce inconsistencies in predicted scores.

\medskip

\paragraph{\textbf{Pluralism and measurement limitations.}}
To distinguish value conflict from measurement artifacts, we tested whether disagreement and residual error co-vary with image quality (blur/lighting) and cue legibility (segmentation confidence and coverage of relevant streetscape classes). Quality degradation explains some variance (e.g., lower legibility increases disagreement), but disagreement persists even under high-quality images for socially contested dimensions such as Inclusivity. We therefore interpret disagreement as pluralism only when quality/legibility diagnostics do not indicate instrumental degradation \citep{Aroyo2015,PavlickKwiatkowski2019}.

\begin{figure}[htbp]
\centering
\includegraphics[width=0.99\textwidth]{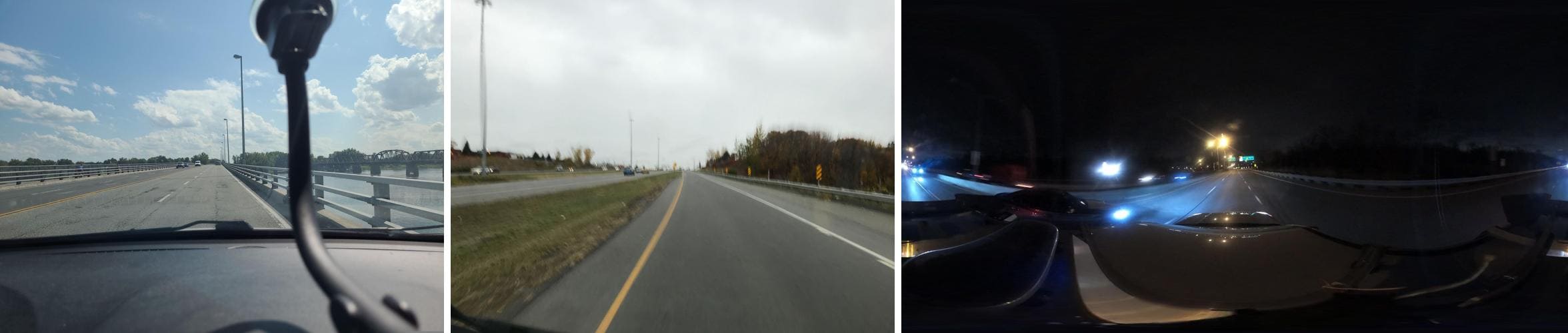}
\caption[Street-view image limitations]{Examples of limitations in street-view imagery (dark, blurry, distorted captures) that constrain feature detection and can introduce inconsistencies in predicted scores. \added{Source: Mapillary street-view imagery, examples selected by the author; see \citet{MushkaniKoseki2026}.}}
\label{fig:mapillary_limitations}
\label{fig:mapillary-image-limits}
\end{figure}

\subsection{Citywide spatial patterns}
\begin{figure}[htbp]
\centering
\includegraphics[width=0.98\textwidth]{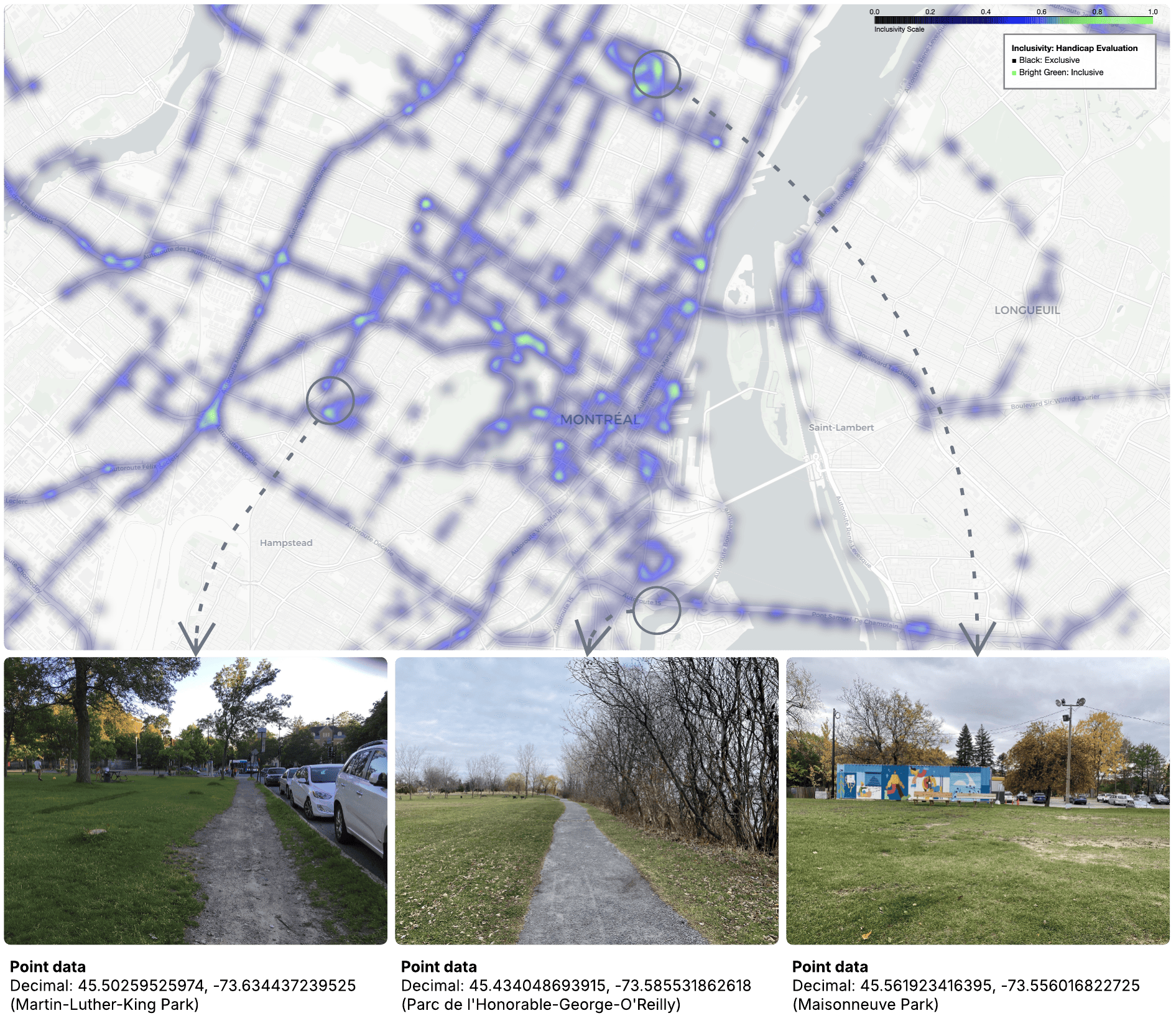}
\caption[Mobility-output inclusivity heatmap]{Inclusivity heatmap based on the mobility-impairment evaluation model predictions. \added{Source: author model outputs from Mapillary imagery; OpenStreetMap basemap; see \citet{MushkaniKoseki2026}.}}
\label{fig:heatmap_handicap_inclusivity}
\label{fig:mobility-inclusivity-heatmap}
\end{figure}

\begin{figure}[htbp]
\centering
\includegraphics[width=0.98\textwidth]{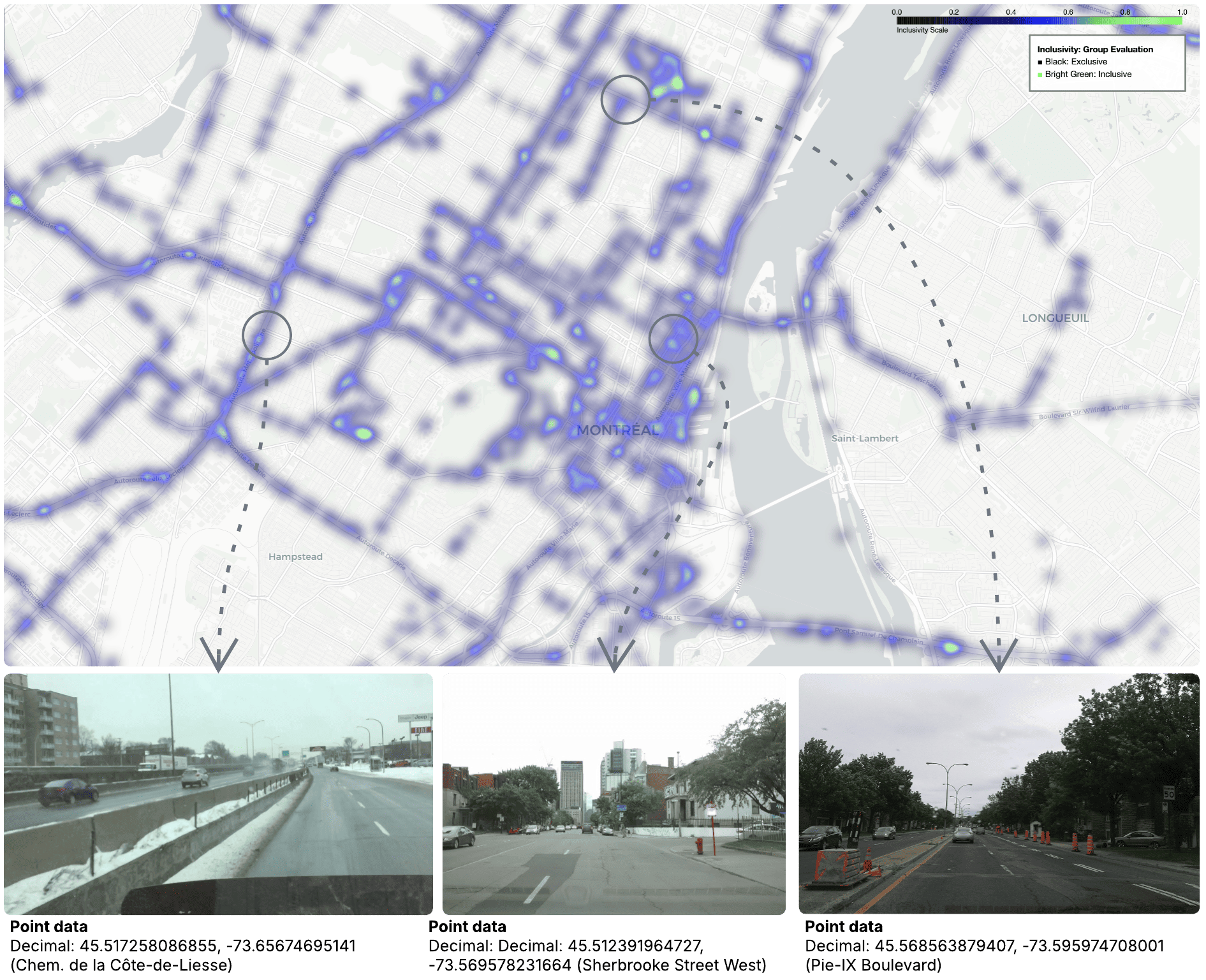}
\caption[Aggregate citywide inclusivity patterns]{Group-level citywide spatial patterns of predicted inclusivity in Montréal. \added{Source: author model outputs from Mapillary imagery; OpenStreetMap basemap; see \citet{MushkaniKoseki2026}.}}
\label{fig:heatmap_group_inclusivity}
\label{fig:aggregate-inclusivity-map}
\end{figure}

After validation, the model was applied to the citywide dataset of approximately 45{,}000 images. Predicted scores were aggregated to street segments and visualized as heatmaps. Figures~\ref{fig:heatmap_handicap_inclusivity} and \ref{fig:heatmap_group_inclusivity} illustrate spatial patterns produced from subgroup and group-level models.

The group-level heatmap indicates higher predicted inclusivity in central districts such as Ville-Marie and Outremont and around major parks, with lower predicted scores in many peripheral areas. These patterns should be interpreted as model outputs derived from image-based proxies. They do not identify causal mechanisms and require complementary field validation when used for planning decisions.

\subsection{A governance-oriented synthesis}
Street Review supports a distinction between dimensions that are amenable to benchmarking and periodic auditing and dimensions where contestation and context dependence warrant deliberation rather than optimization. Table~\ref{tab:convergent_vs_contested} summarizes this distinction grounded in agreement analyses and model limitations.

\vspace{2.5cm}

\begin{table}[htbp]
\centering
\caption{Governance interpretation of convergence and contestation across Street Review criteria, grounded in agreement analyses and scaling and limitation analyses.}
\label{tab:convergent_vs_contested}
\renewcommand{\arraystretch}{1.15}
\small
\setlength{\tabcolsep}{4pt}
\begin{tabular}{
>{\raggedright\arraybackslash}p{0.16\textwidth}
>{\raggedright\arraybackslash}p{0.25\textwidth}
>{\raggedright\arraybackslash}p{0.26\textwidth}
>{\raggedright\arraybackslash}p{0.26\textwidth}
}
\toprule
\textbf{Criterion} & \textbf{Empirical convergence pattern} & \textbf{What the image pipeline captures well} & \textbf{Recommended governance handling} \\
\midrule
Accessibility & Moderate agreement in rating tasks; improves under group deliberation. Often conceptually coupled with inclusivity. & Sidewalk presence and continuity, some barrier proxies, crossings and signage proxies. & Suitable for periodic audits and benchmark targets, while preserving subgroup layers (for example mobility-related models) for accountability. \\
\addlinespace
Aesthetics & Relatively higher agreement and reliability in individual and group contexts. & Visual cues of maintenance, fa\c{c}ade condition, greenery presence (as coarse proxy), streetscape composition. & Suitable for benchmarking with caution: avoid letting aesthetics dominate inclusion policy. \\
\addlinespace
Practicality & Mixed: moderate reliability in rating but low consensus in ranking tasks; sensitive to interpretation of `usefulness.' & Some maintenance and functional cues (signage, obstructions), broad surface conditions. & Treat as context-dependent. Use as a diagnostic layer and deliberation trigger when it conflicts with other criteria or varies strongly across groups. \\
\addlinespace
Inclusivity & Most contested: low agreement in ranking tasks; improves but remains comparatively variable under deliberation. & Some physical proxies (walkability, activity presence) but struggles with intangible cultural markers and social meaning. & Treat as a deliberative criterion. Use subgroup-aware outputs to surface disagreement and to prioritize follow-up engagement. \\
\bottomrule
\end{tabular}
\end{table}

\section{Discussion}
\label{sec:ch5_discussion}
Street Review functions as a methodological contribution and a governance prototype. It demonstrates how the epistemic layer of public-space governance can be treated as civic infrastructure by constructing a documented and revisable measurement system grounded in resident-defined criteria. The empirical results support implications for pluralistic alignment in municipal practice.

First, pluralism alters the role of metrics. Agreement analyses show that convergence varies by criterion and that divergence is patterned by social position. In this setting, measurement cannot be treated as neutral infrastructure. A scalar index of inclusion can obscure contested values by defining what is seen, compared, and acted upon \citep{low2020social,EspelandStevens2008}. Street Review therefore represents evaluation as a criteria-by-group structure rather than as a single target, and it treats disagreement as part of the evidentiary record.

Second, deliberation changes the evidentiary object. The evaluation protocol distinguishes independent individual judgments from negotiated group judgments. Discussion increased convergence and alignment between individual and group evaluations. This indicates that a group score is produced through reason-giving and contestation and therefore changes what counts as collective evidence. Residual disagreement after discussion remains informative. It can indicate value conflict, different experiential baselines, or limits of what can be inferred from imagery.

Scaling extends these measurement properties while introducing proxy constraints. The Street Review module shows that a model trained on co-produced labels can reproduce participant rating structure and generate citywide heatmaps. At the same time, image-based inference cannot capture temporal dynamics and cannot represent many socially salient cues that participants identify as relevant to inclusion \citep{low2020social}. Image quality and uneven platform coverage further constrain reliability. These constraints motivate institutional safeguards that treat outputs as partial evidence rather than as decision rules.

When used within such safeguards, Street Review outputs can support municipal functions that require comparative evidence without presuming consensus. Group-level heatmaps can be used as audit layers to identify corridors where predicted conditions are consistently low across criteria or subgroups and therefore warrant field assessment and engagement. Subgroup layers can support equity diagnostics by revealing cases where aggregated predictions obscure divergence. Outputs can also support iterative design feedback when treated as hypotheses to be evaluated before and after interventions, combined with updated imagery and renewed participatory assessment. Divergence across criteria or across subgroups can function as a deliberation trigger, indicating a need for additional context, negotiation, or complementary data collection rather than for automated optimization.

\medskip

\paragraph{\textbf{Equally important are out-of-scope uses.}} Street Review outputs are not decision rules. They should not be used to justify punitive enforcement or surveillance, nor to rank neighborhoods without contextualization and participatory interpretation. These boundaries follow from the empirical properties of the system. The criteria are co-produced and context-dependent, disagreement persists on contested dimensions, and the predictive pipeline is constrained by visible proxies and data quality.

\medskip

\paragraph{\textbf{Several limitations condition transferability.}} Static street-view images do not capture temporal dynamics and underrepresent tactile and auditory conditions. Demographic groupings are coarse heuristics that do not represent many intersectional combinations and may mask intragroup variation \citep{CostanzaChock2020}. Mapillary coverage and image quality vary across space and can reproduce epistemic inequities when the places that are hardest to map are also those that are under-resourced. The Montréal case does not establish that a model trained on local data generalizes across cities. The transferable claim is a governance pattern: co-produce criteria locally, quantify convergence and contestation, scale with documented proxy limits, and embed outputs in procedures that preserve contestation and enable revision.

\section{Conclusions}
\label{sec:ch5_conclusion}
This chapter presented Street Review as a participatory AI-based framework for assessing streetscape inclusivity. Through interviews and structured evaluation tasks, participants articulated descriptors of street experience that were consolidated into four criteria: Accessibility, Aesthetics, Practicality, and Inclusivity. Agreement analyses demonstrate that convergence varies by criterion and that inclusivity-related judgments remain more contested than dimensions that depend primarily on visible cues, even as structured discussion increases convergence in group evaluations. The Street Review module demonstrates that co-produced values can be scaled through a computer-vision pipeline to produce subgroup-aware predictions and citywide maps, while documenting proxy limits and image-data constraints.

For the dissertation’s argument, Street Review illustrates how a city can commission the epistemic layer of public-space governance as civic infrastructure by building a measurement system that represents heterogeneity, makes disagreement visible, and ties outputs to deliberation and revision. Chapter Six extends this pluralistic alignment stance from measurement to generation by examining how community-derived criteria and preference learning shape inclusive public-space imaginaries in generative models and how neutrality and indeterminacy function as governance signals.


\chapter{LIVS: A Pluralistic Alignment Dataset for Inclusive Public Spaces}
\label{chap:livs}

This chapter introduces the Local Intersectional Visual Spaces (LIVS) dataset and a pluralistic alignment approach for text-to-image (T2I) generative models in the domain of inclusive public spaces. In the dissertation's account of urbanism as a dual infrastructure, generative models operate as part of an epistemic, algorithmic layer that shapes what becomes visualizable, discussable, and actionable in public-space planning. Unlike perception-based auditing and mapping of existing streetscapes, the focus here is the governance of synthetic visual proposals and scenarios that can enter planning consultations, design iteration, and procurement processes \citep{DubeyEtAl2024,guridi2024fake}.

The chapter reports a two-year community-based participatory process conducted in Montréal, in which community organizations collaborated to elicit 634 candidate concepts for inclusive public-space design and iteratively consolidate them into six criteria: Accessibility, Safety, Comfort, Invitingness, Inclusivity, and Diversity. The resulting dataset encodes multi-criteria preference annotations over pairs of T2I-generated images, including neutral judgments that record indifference, ambiguity, or unresolved tradeoffs. Using these data, the chapter fine-tunes a Stable Diffusion XL (SDXL) model \citep{podell2023sdxlimprovinglatentdiffusion} with Direct Preference Optimization (DPO) \citep{rafailov2024} and evaluates whether preference tuning shifts generated outputs toward community preferences, and where alignment remains indeterminate.

Across the evaluation set, preference tuning improves alignment on criteria with higher annotation volume, while a large fraction of comparisons remain neutral. The chapter treats neutrality and residual disagreement as governance-relevant evidence about ambiguity and contested values, consistent with pluralistic alignment approaches that treat heterogeneity as a condition to be represented rather than eliminated \citep{Sorensen2024}. In municipal settings, such evidence motivates procedural responses, including documentation of model limits and, when generated imagery is used in decision-relevant contexts, structured deliberation over criteria tradeoffs rather than further optimization toward a single objective.

The chapter is structured as follows. Section~\ref{sec:livs-introduction} motivates why generative imagery used in planning requires explicit civic authorization and procedural safeguards. Section~\ref{sec:livs-relatedwork} reviews work on inclusive public space, participatory AI, and preference-based alignment. Section~\ref{sec:livs-methods} describes the LIVS co-production process, dataset construction, and the DPO fine-tuning pipeline. Section~\ref{sec:livs-findings} reports empirical findings through four case studies. Section~\ref{sec:livs-discussion} discusses neutrality as a governance signal and outlines documentation obligations and boundary conditions. Section~\ref{sec:livs-conclusion} concludes and links to the governance architecture developed in the next chapter.

\section{Introduction}
\label{sec:livs-introduction}
Visual representations are a planning instrument. Plans, renderings, and scenarios translate spatial proposals into objects that can be debated, revised, and authorized \citep{pallasmaa2011embodied,corner1999}. Diffusion-based T2I models have improved image quality and prompt adherence \citep{podell2023sdxlimprovinglatentdiffusion,zhang2024t2i-survey}, lowering the technical barrier for producing plausible images of streets, plazas, and parks from natural-language prompts. Empirical work indicates that AI-generated imagery can influence attitudes toward urban policy and can support deliberation by enabling participants to examine counterfactual futures \citep{DubeyEtAl2024}. At the same time, synthetic images can be produced and circulated without the provenance cues associated with conventional renderings, creating risks of misrepresentation and manipulation in public processes \citep{guridi2024fake}.

When generative systems are incorporated into planning workflows, they can operate as proposal generators, communication devices, or evidence-like artifacts. In each role, outputs can set the terms of debate by making some design options visually legible while rendering others absent. Alignment in this setting is not limited to preventing explicit harms. It also concerns whether generated scenes reflect locally meaningful criteria of inclusive public space that are intersectional and contested \citep{low2020social,madanipour2010public,Crenshaw1989,MitrasinovicMehta2021}. Mainstream alignment pipelines in generative AI are typically built on large-scale, global preference data and crowdwork \citep{kirk2024prism}, which can under-represent the priorities of specific communities and reproduce dominant visual tropes in depictions of public life \citep{wan2024survey,perrak2024}. Work on representation and aesthetics in T2I systems suggests that model outputs can encode norms embedded in training corpora and evaluation frameworks, with limited capacity to express locally situated criteria without additional intervention \citep{Qadri2023representation,kannen2024aesthetics}.

This chapter responds with a pluralistic alignment approach, in which locally defined criteria and their heterogeneity are treated as first-order alignment inputs \citep{Sorensen2024}. LIVS operationalizes this approach by pairing a co-produced taxonomy of inclusive public-space criteria with a dataset of multi-criteria preference annotations over T2I-generated image pairs. Annotations include neutral judgments to record indifference and indeterminacy rather than forcing binary choices, and consented self-identification markers to support intersectional analysis.

Using this dataset, the chapter reports experiments fine-tuning SDXL with DPO \citep{rafailov2024,wallace2023}. DPO offers a mechanism for aligning a diffusion model with preference data without training an explicit reward model, but it requires a binary preference label for each comparison. As a result, multi-criteria and neutral judgments are partially collapsed during training, creating a gap between pluralistic data collection and single-policy optimization. The empirical results illustrate both the potential and the limit of preference tuning in this domain: shifts are observable on criteria that are more visually legible, while neutrality persists on criteria that are less visually salient or subject to value conflict.

Because generative outputs can shape what becomes thinkable and fundable, the governance question is procedural. Consistent with the dissertation's Right to AI framing, when synthetic imagery is used in decision-relevant settings, cities require disclosure of model use, participatory power over criteria and prompts, and contestability of outputs and associated decisions. LIVS contributes auditable artifacts for such arrangements, including a locally grounded criteria register, a prompt protocol, and preference data that preserve neutrality and disagreement patterns as evidence about where decisions remain political.

\section{Literature Review}
\label{sec:livs-relatedwork}
\subsection{Intersectional public space design}
Public space is a contested civic good. Normative commitments to inclusion extend beyond physical access to encompass recognition, belonging, and the conditions under which diverse users can appropriate space without fear of exclusion \citep{low2020social,madanipour2010public,MitrasinovicMehta2021}. Accounts of street life emphasize the public realm as an everyday infrastructure of encounter and safety through use \citep{Jacobs1961,GehlSvarre2013}. Contemporary planning scholarship argues that inclusion is not a stable checklist but a negotiation over design norms, use patterns, and governance practices \citep{talen2012design,mcandrews2023gender}. Even within a single city, readings of space vary along lines of mobility, gender, age, race and ethnicity, religion, and other intersecting dimensions \citep{Crenshaw1989,mcandrews2023gender}. These variations reflect structured differences in exposure to harm and differential entitlement to occupy space, complicating attempts to operationalize inclusion through single indicators.

\subsection{Text-to-image models in planning}
Diffusion-based T2I models support rapid prototyping and scenario visualization \citep{podell2023sdxlimprovinglatentdiffusion,zhang2024t2i-survey}. In planning contexts, they can be used to communicate design alternatives and to support the exploration of counterfactual futures \citep{DubeyEtAl2024,guridi2024fake,LarteyLaw2025}. Because these models are trained on internet-scale corpora, their outputs reflect global priors and dominant visual tropes. In depictions of public space, this can manifest as systematic misrepresentation of local context, omission of culturally specific markers, and representational bias in who is shown as belonging in a space \citep{wan2024survey,perrak2024,Qadri2023representation,vonBrackelSchmidtEtAl2024}. From a governance perspective, these are legitimacy risks when cities treat generated imagery as evidence or as plausible futures without local authorization.

\subsection{Preference learning}
Preference-based alignment uses comparative judgments to steer generative models. In language models, reinforcement learning from human feedback and related methods rely on preference comparisons to optimize outputs \citep{bai2022rlhf}. For diffusion models, DPO offers a mechanism that directly optimizes a parameterized policy against preference data without training an explicit reward model \citep{rafailov2024,wallace2023}. A limitation of many alignment pipelines is the collapse of diverse judgments into a single objective, which assumes a unitary improvement direction.

Multi-criteria preference learning addresses this limitation by treating judgments as multi-dimensional rather than single-peaked \citep{bhatia2021multi,Chakraborty2024}. In the T2I domain, recent work has proposed datasets and models that incorporate multiple evaluative dimensions, including aesthetic quality, semantic alignment, and detail \citep{xu2023imagereward,zhang2024multidimensionalhp,kirstain2023pickapic,pressmancrowson2022}. These efforts typically remain global in scope and do not foreground locally negotiated criteria in domains where public legitimacy is required.

\subsection{Pluralistic alignment}
Pluralistic alignment argues that, in many domains, there is no single correct objective and alignment must accommodate multiple coexisting, permissible value systems \citep{Sorensen2024}. Participatory approaches address this by involving affected communities in defining objectives, constructing datasets, and evaluating models. HCI scholarship emphasizes that participation does not ensure legitimacy by itself; participation must be tied to power-sharing, resourcing, and institutional commitments to avoid extractive or symbolic engagement \citep{sloane2022,Arnstein1969,CostanzaChock2020,BerditchevskaiaEtAl2021}. Community-based participatory research and participatory action research provide methodological foundations for grounding technical work in local knowledge through iterative, reciprocal learning \citep{Israel1998,Cornish2023,Engestrom2014,Sieber2024PublicsEngaging}.

Within AI systems that may be procured, deployed, and updated over time, documentation functions as a governance interface. It provides an auditable trace through which the public can contest model behavior and through which institutions can constrain use \citep{gebru2021}. For T2I models used in planning, documentation must include not only data and model choices, but also prompt protocols, criteria definitions, and evaluation outcomes, including distributions of neutrality and disagreement.

\section{Methods}
\label{sec:livs-methods}
\subsection{Setting}
LIVS was developed through a two-year collaboration in Montréal. A community-based participatory approach was used to integrate local perspectives that are often absent from top-down datasets \citep{Sieber2024PublicsEngaging,Hosking2024-goldstandard}. The approach positioned community members as collaborators in identifying context-specific priorities for public spaces, including criteria that depend on lived experience of accessibility and safety \citep{Arnstein1969,AnttiroikoDeJong2020}. Engagement was designed as an iterative process of reciprocal learning, in which participants received introductions to T2I systems and researchers learned how criteria were interpreted and contested in local use of public space \citep{Engestrom2014,Israel1998,Cornish2023}.

We initiated outreach by contacting 100 community organizations, including neighborhood councils, disability-focused nonprofits, youth advocacy networks, faith-based groups, and other local stakeholders, with the goal of assembling a demographically and experientially diverse set of participants \citep{StatCan2022,IRCGM2018}. Across the project, we collaborated with 30 organizations through eleven workshops, five annotation batches, and 34 interviews.

All activities were approved by a research ethics board. Participants provided written informed consent and were compensated for their time; personally identifiable information was anonymized. Because generative imagery can surface biased or inaccurate depictions, the protocol was revised during the project in response to participant feedback, and the annotation interface was designed for accessibility to support participation by people with disabilities \citep{Hosking2024-goldstandard}. Figure \ref{fig:self_declared_participants} in the Street Review chapter details the distribution of participants’ self-declared demographics across age groups and identity categories.

\subsection{Engagement sequence}
The LIVS process interleaves knowledge exchange and data production. Over two years, engagement activities included introductory sessions on AI and T2I, criteria elicitation and consolidation workshops, prompt authoring sessions, annotation batches, and an evaluation workshop. The study design reflects a participatory action logic in which criteria and instruments were iteratively refined through interaction with participants rather than fixed in advance \citep{Israel1998,Cornish2023,Engestrom2014}. Table \ref{tab:livs-engagement} provides an overview of the activities and engagement sequence.

\begin{table}[!htbp]
\centering
\small
\caption{Summary of LIVS engagement activities and outputs.}
\label{tab:livs-engagement}
\begin{tabularx}{\textwidth}{P{0.26\textwidth}P{0.30\textwidth}Y}
\toprule
\textbf{Activity} & \textbf{Participation} & \textbf{Primary outputs} \\
\midrule
Introductory sessions & 2 workshops, 25--35 participants each & Orientation to AI and T2I; open-ended reflections on public space and AI \\
Criteria brainstorming & 6 workshops, 28 participants total & 634 initial concepts describing inclusive public-space attributes \\
Tutorial and platform feedback & 1 workshop, 20 participants & Annotation interface and protocol refinement \\
Criteria validation & 1 workshop, 18 participants; 34 interviews & Consolidation to 35 intermediate criteria; ranking to six final criteria \\
Prompt authoring & 1 workshop, 24 participants & 440 human-authored prompts \\
Annotations & 5 batches, 18 participants & 42,235 raw comparisons; cleaned to 35,510 high-quality annotations \\
Evaluation & 1 workshop & Post-fine-tuning evaluation and discussion of model outputs \\
\bottomrule
\end{tabularx}
\end{table}
\FloatBarrier

\subsection{Criteria elicitation and consolidation}
\begin{figure}[!htbp]
\centering
\includegraphics[width=1\textwidth]{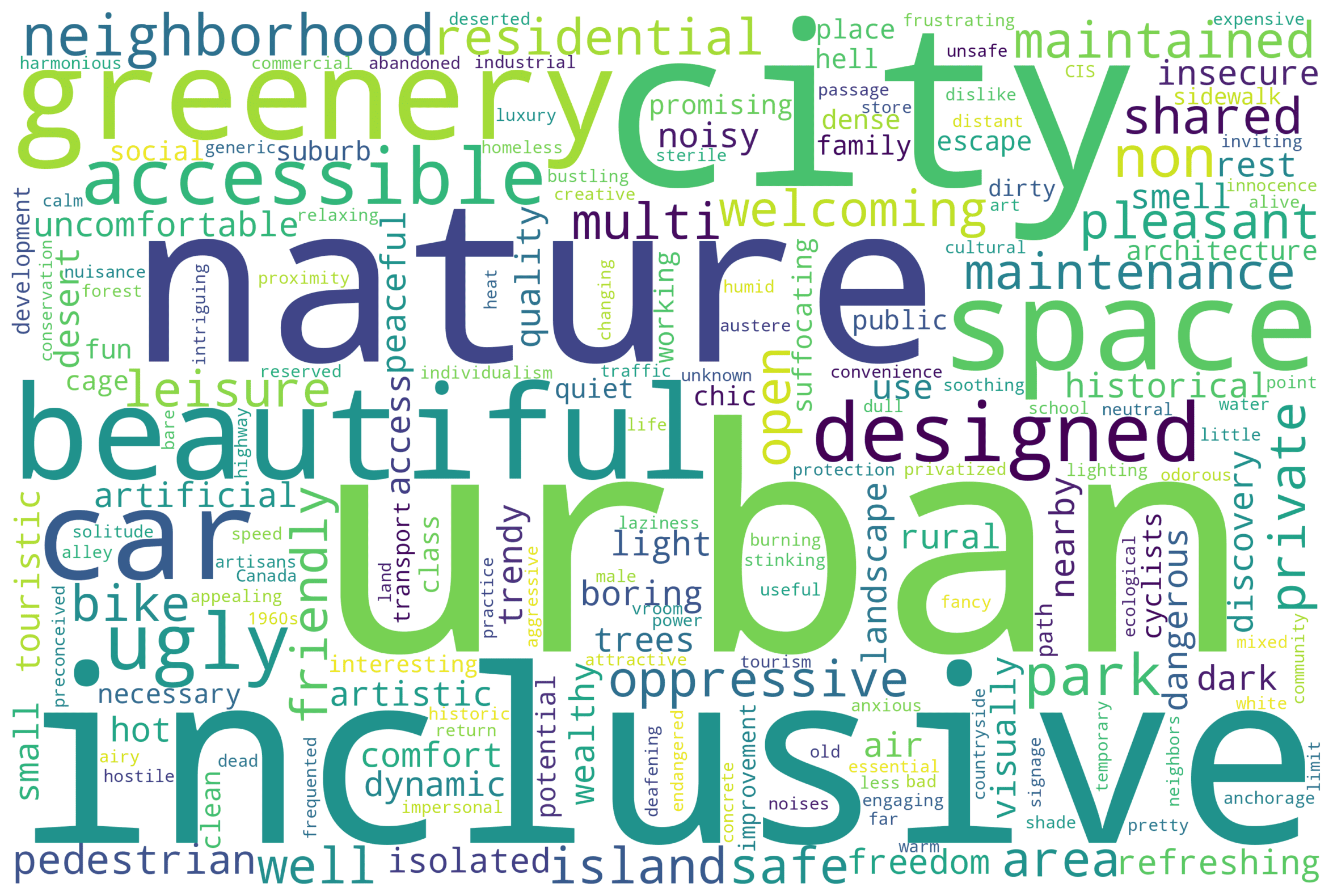}
\caption[Criteria-brainstorming word cloud]{Word cloud of concepts surfaced during criteria brainstorming workshops. \added{Source data: LIVS criteria workshops; see \citet{pmlr-v267-mushkani25a}.}}
\label{fig:livs-criteria-wordcloud}
\label{fig:criteria-word-cloud}
\end{figure}

The project began with multi-stakeholder workshops focused on defining inclusive public space in local terms. In criteria brainstorming workshops, participants reviewed pairs of images of existing public spaces and described reactions and concerns, producing 634 initial concepts spanning physical, social, and psychological attributes. Figure~\ref{fig:livs-criteria-wordcloud} illustrates the diversity of terms surfaced during this stage.

\FloatBarrier

\begin{figure}[!htbp]
\centering
\includegraphics[width=0.95\textwidth]{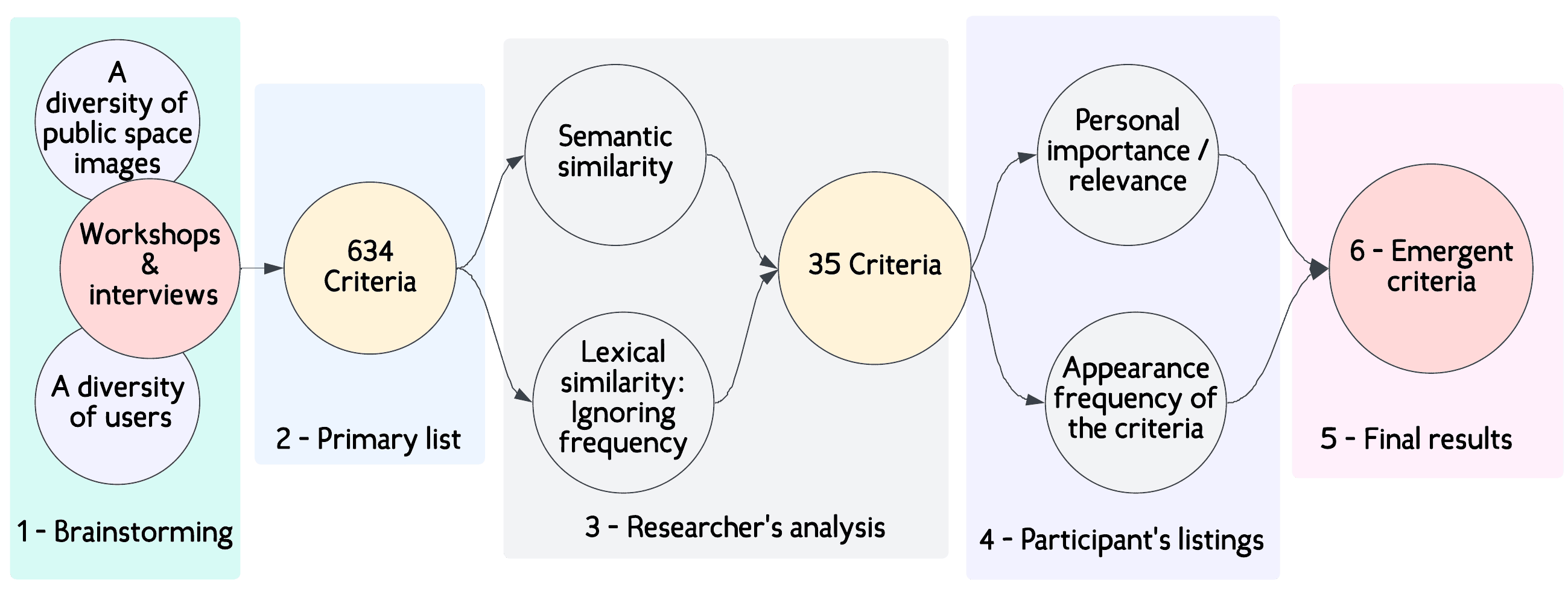}
\caption[Criteria consolidation]{Iterative criteria consolidation process, from brainstorming through researcher analysis and participant ranking to six criteria. \added{Source: Author’s work; see \citet{pmlr-v267-mushkani25a}.}}
\label{fig:livs-method}
\label{fig:criteria-consolidation}
\end{figure}

These concepts were consolidated through iterative merging and semantic grouping into 35 intermediate criteria, which were then ranked and refined during a validation workshop and follow-up interviews, yielding six criteria. The consolidation process treated frequency as insufficient for legitimacy: criteria were selected based on judgments of importance and local relevance, rather than frequency alone. Figure~\ref{fig:livs-method} illustrates the iterative criteria consolidation process.

\FloatBarrier

\noindent\textbf{Accessibility.} Physical and cognitive usability for people of all abilities, including ramps, tactile indicators, and clear signage.

\noindent\textbf{Safety.} Freedom from crime, hazards, or harassment, often reflected in lighting, visibility, active presence, or protective barriers.

\noindent\textbf{Comfort.} Amenities and mitigation of environmental conditions, including seating, shade, cleanliness, noise, and shelter.

\noindent\textbf{Invitingness.} Features that encourage people to enter and remain, such as greenery, open layouts, and welcoming signage.

\noindent\textbf{Inclusivity.} Avoidance of exclusionary design, support for cultural or religious needs, and cues such as multilingual signage.

\noindent\textbf{Diversity.} Representation of different demographic groups and a range of potential uses.

\subsection{Prompt collection and expansion}
\begin{figure}[!htbp]
\centering
\includegraphics[width=0.95\textwidth]{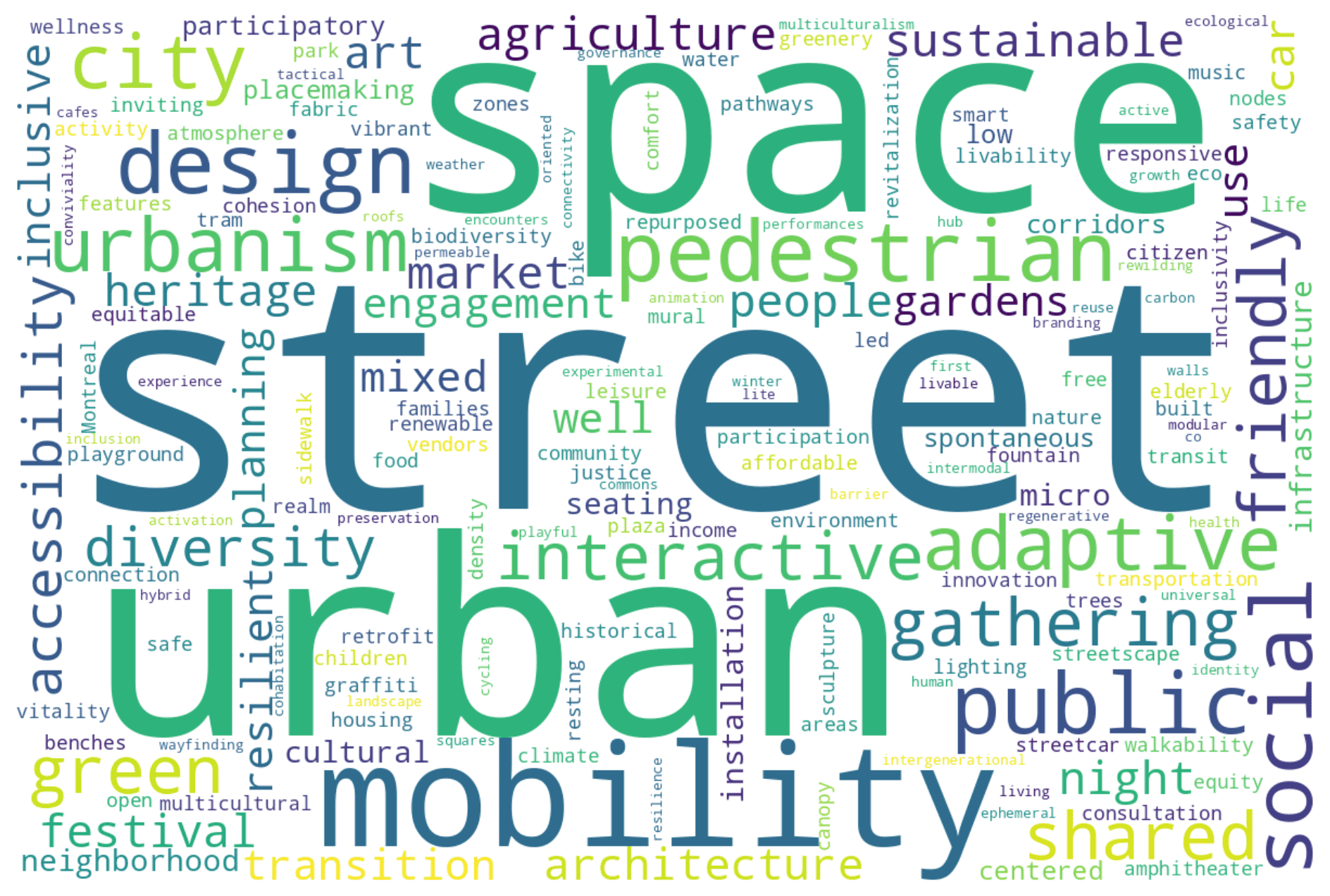}
\caption[Human-authored prompt word cloud]{Word cloud of the 440 human-authored prompts. Word size reflects frequency. \added{Source data: LIVS human-authored prompts; author’s work, see \citet{pmlr-v267-mushkani25a}.}}
\label{fig:livs-wordcloud}
\label{fig:prompt-word-cloud}
\end{figure}

To generate T2I scenes reflecting local public-space typologies, participants co-authored 440 prompts during a prompting workshop. Groups composed of residents, a computer scientist, and an urban architect collaboratively wrote prompts describing spatial scenarios and desired features. To expand the dataset and increase prompt variety, workshop transcripts were used with a language model (GPT-4) and three prompting strategies to generate an additional 2,910 synthetic prompts \citep{openai2024gpt4}. Differences between human-written and model-generated prompt sets were assessed using Jensen--Shannon Divergence \citep{plank2011jsd}. Figure \ref{fig:livs-wordcloud} visualizes key concepts in the 440 human-authored prompts.

\FloatBarrier

\subsection{Image generation}
For each prompt, up to 20 images were generated using SDXL \citep{podell2023sdxlimprovinglatentdiffusion}. To reduce redundancy and avoid comparisons between near-identical images, we selected the four most distinct images per prompt using a greedy algorithm based on CLIP similarity scores \citep{radford2021clip}. In total, 16,693 images were generated, with a subset reserved for quality checks, leaving 13,462 images for annotation.

\subsection{Annotation protocol and dataset structure}
Preference data were collected through an accessible web-based platform designed to accommodate participants from diverse backgrounds, including those with disabilities \citep{Hosking2024-goldstandard}. Annotation tasks were divided into five batches, each lasting two weeks. Eighteen participants completed approximately 750 pairwise comparisons per batch, totaling 42,235 raw comparisons.

In each task, two images were presented side by side with three randomly selected criteria from the six. Annotators rated each criterion using a slider from $-1$ (strong preference for the left image) to $+1$ (strong preference for the right image), with $0$ indicating neutrality. Presenting all six criteria simultaneously was found to increase cognitive fatigue and reduce usability; restricting each comparison to three criteria improved engagement and preserved image visibility. Annotators were required to provide at least one criterion-level annotation per pair.

\begin{figure}[htbp]
\centering
\includegraphics[width=0.95\textwidth]{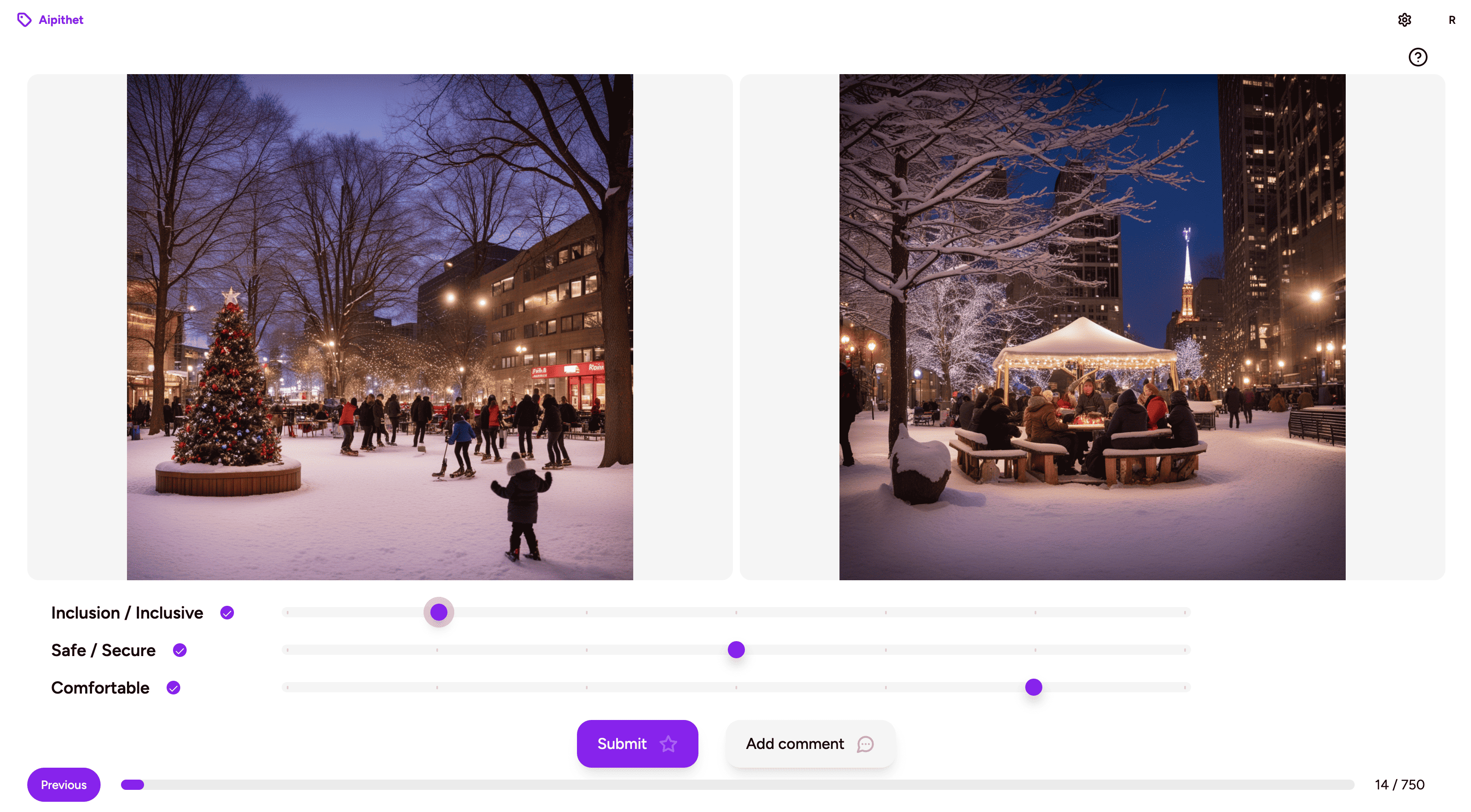}
\caption[Preference-annotation interface]{Annotation interface used to collect multi-criteria preference judgments, with three criteria shown per pairwise comparison. \added{Source: LIVS annotation interface; see \citet{pmlr-v267-mushkani25a}.}}
\label{fig:livs-interface}
\label{fig:annotation-interface}
\end{figure}
\FloatBarrier

\begin{tcolorbox}[title=LIVS dataset statistics, colback=white, colframe=black!40]
\small
\begin{itemize}[leftmargin=*]
\item Image pairs annotated (raw; including Quality Control (QC) and incomplete submissions): 42{,}235.
\item Filtering removed QC items and incomplete submissions.
\item Usable image pairs after filtering: 37{,}710 (three criteria shown per pair; $\approx 113{,}130$ criterion-level judgments).
\item Image pairs used for DPO fine-tuning: 35{,}510.
\item Unique images retained for annotation: 13{,}462.
\item Evaluation workshop: 2{,}100 new comparisons (700 DPO-preferred; 300 baseline-preferred; 1{,}100 neutral).
\end{itemize}
\end{tcolorbox}

A multi-stage data cleaning process removed quality-control items and incomplete submissions, yielding 35,510 high-quality multi-criteria preference annotations for model fine-tuning. Across the broader collection process, the pipeline produced 37,710 image-level comparisons, corresponding to approximately 113,130 criterion-level annotations. Participants reported that some criteria, particularly Inclusivity, were harder to infer from generated imagery, contributing to higher rates of neutral judgments.

\begin{figure}[!htbp]
\centering
\includegraphics[width=0.95\textwidth]{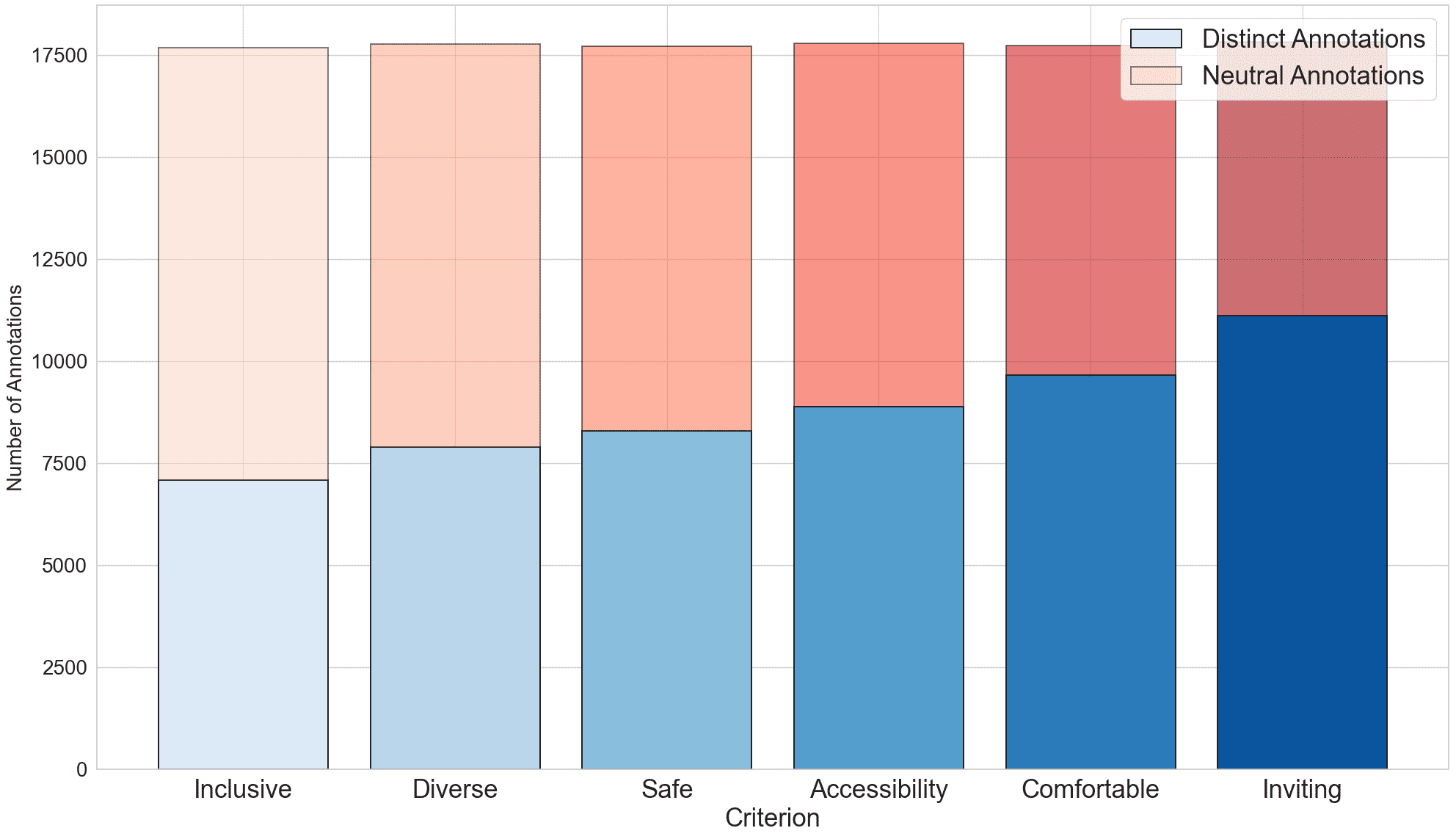}
\caption[Annotation frequency by criterion]{Distribution of annotation frequencies by criterion after data cleaning. Neutral preferences are shown alongside distinct preferences. \added{Source: Author’s work, reported in \citet{pmlr-v267-mushkani25a}.}}
\label{fig:livs-criteria-distribution}
\label{fig:annotation-frequency}
\end{figure}
\FloatBarrier

\subsection{Fine-tuning with Direct Preference Optimization}
We fine-tuned SDXL using DPO on the LIVS training annotations \citep{podell2023sdxlimprovinglatentdiffusion,rafailov2024,wallace2023}. Because DPO requires binary preference labels, multi-criteria feedback for each image pair was collapsed into a single preferred versus not-preferred label via majority voting across the criteria assigned to that pair. When annotators expressed conflicting preferences across the three criteria, the final label followed the majority. Neutral comparisons were not used as positive training signals. This constraint is analytically important because it makes explicit how plural criteria can be collapsed during optimization, and why neutrality can persist even when aggregate metrics improve.

\section{Findings}
\label{sec:livs-findings}
\subsection{Case Study I}
\begin{figure}[htbp]
\centering
\includegraphics[width=0.95\textwidth]{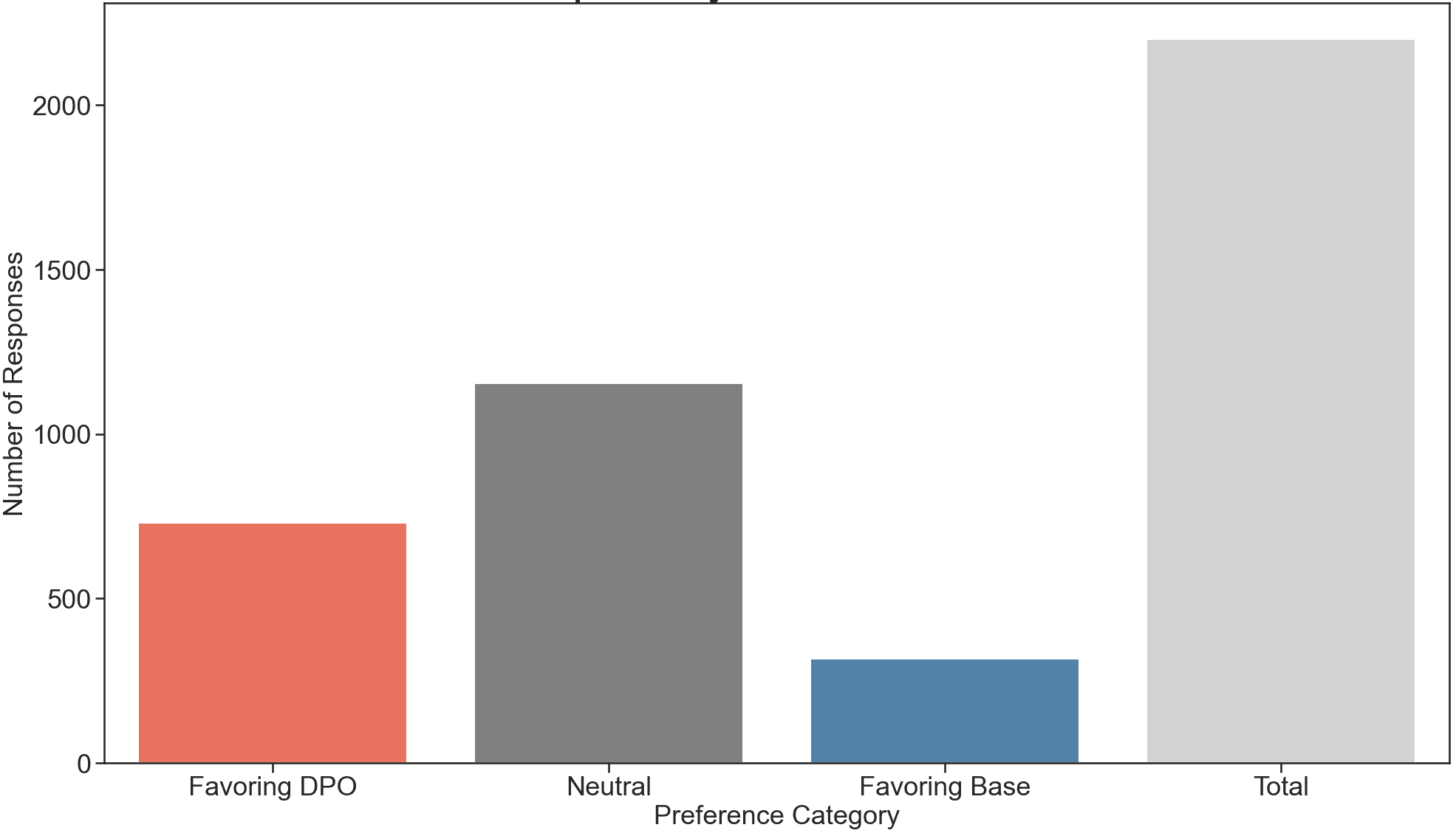}
\caption[Aggregate evaluation outcomes]{Aggregate outcomes of the 2,100 evaluation comparisons: preferences favoring the DPO-fine-tuned model, preferences favoring the baseline, and neutral outcomes. \added{Source: Author’s work.}}
\label{fig:livs-total-responses}
\label{fig:evaluation-outcomes}
\end{figure}

To evaluate whether a multi-criteria dataset used with DPO improves alignment with community preferences, we compared outputs from the DPO fine-tuned SDXL model against the baseline SDXL model in an evaluation workshop. Across 2,100 new comparisons, annotators favored the DPO fine-tuned model in 700 instances, favored the baseline model in 300 instances, and marked 1,100 comparisons as neutral. Figure~\ref{fig:livs-total-responses} summarizes the aggregate outcomes of the 2,100 evaluation comparisons.

\FloatBarrier

\begin{figure}[htbp]
\centering
\includegraphics[width=0.8\textwidth]{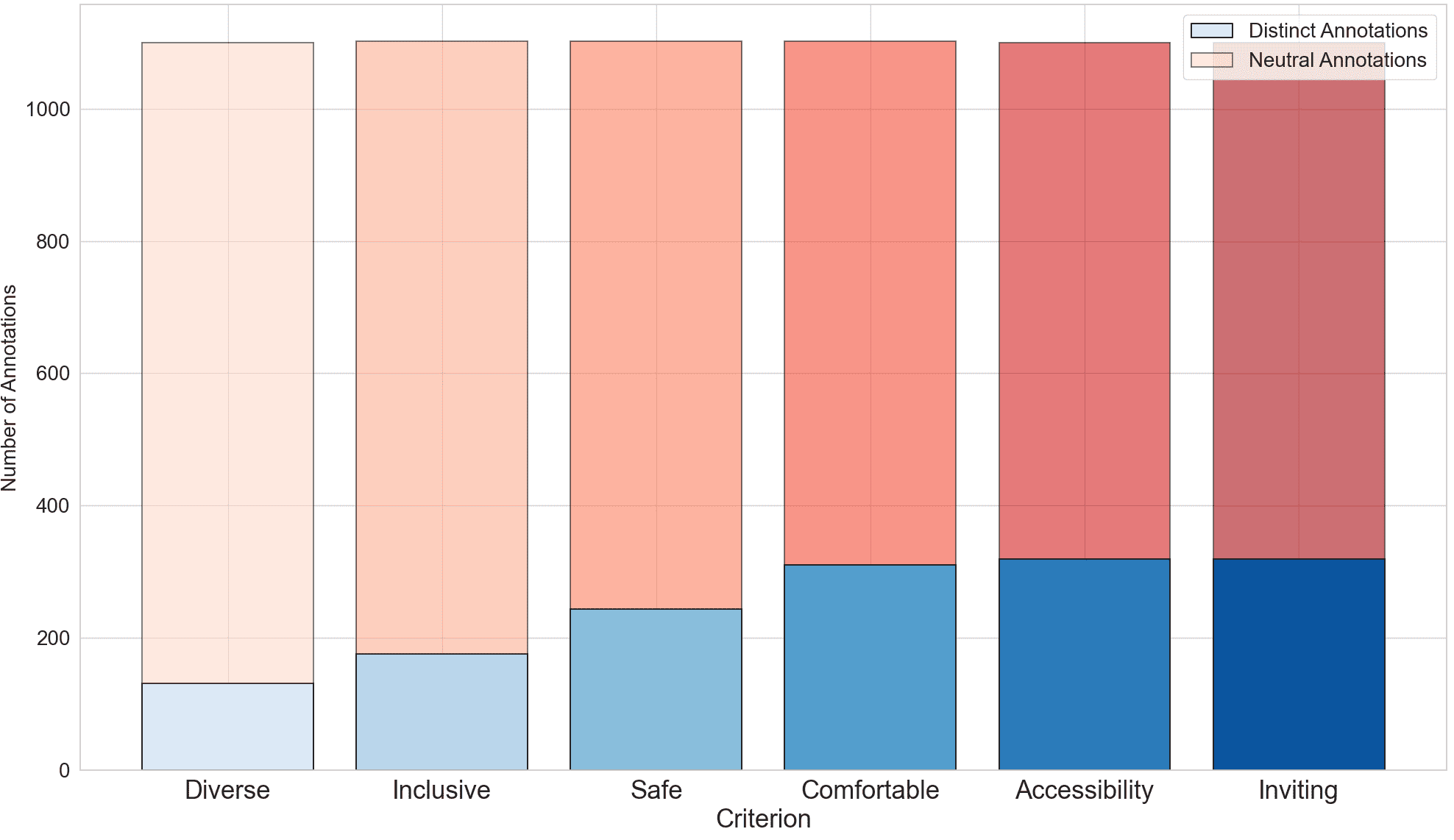}
\caption[Evaluation outcomes by criterion]{Criterion-level outcomes on the evaluation dataset. Neutral outcomes are more common for Inclusivity and Diversity. \added{Source: Author’s work.}}
\label{fig:livs-eval-criterion}
\label{fig:criterion-evaluation-outcomes}
\end{figure}

\begin{figure}[htbp]
\centering
\includegraphics[width=0.8\textwidth]{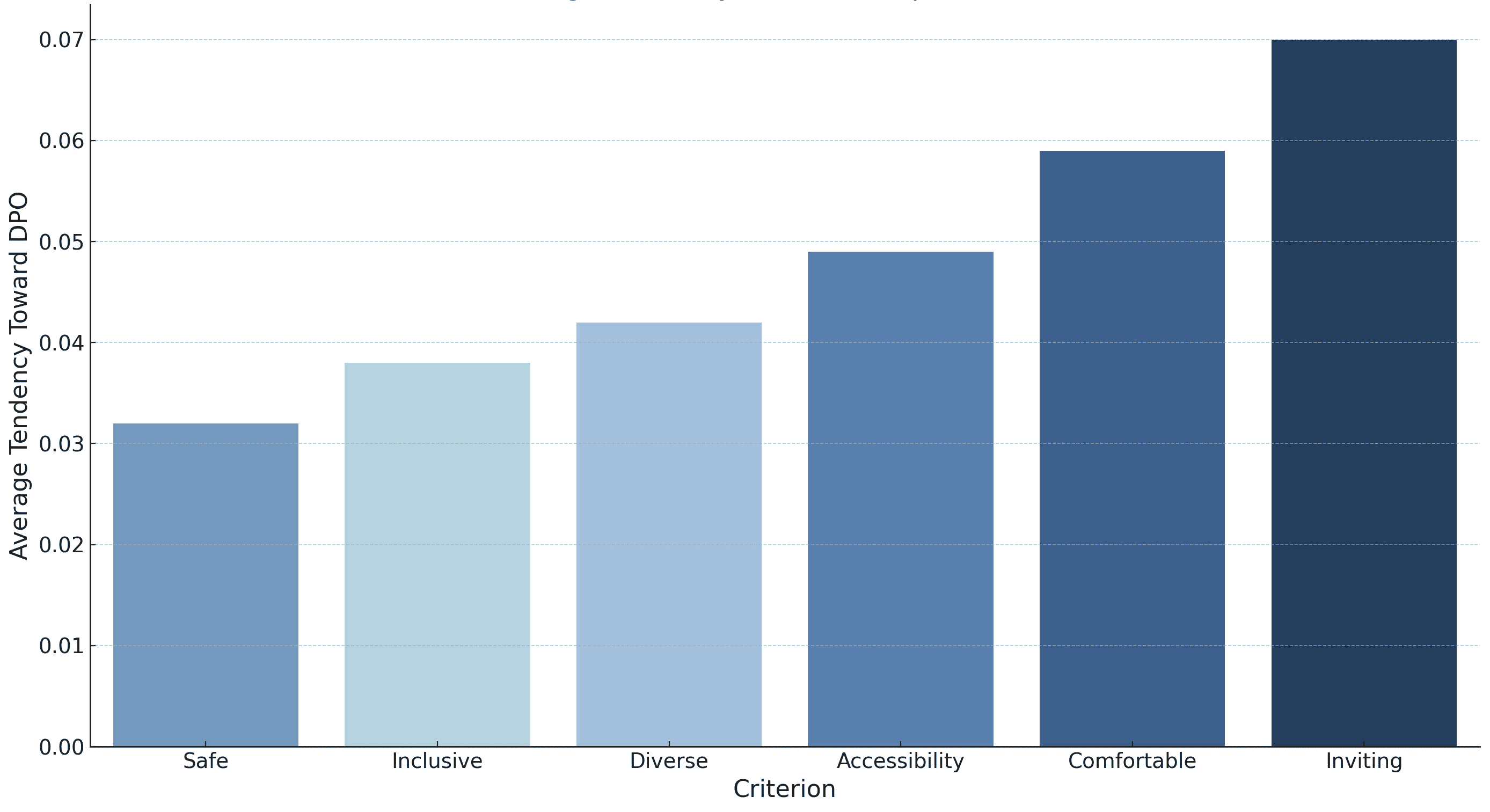}
\caption[DPO preference tendency by criterion]{Average preference tendency toward DPO outputs by criterion. Performance tends to improve in criteria with higher annotation counts. \added{Source: Author’s work.}}
\label{fig:livs-tendency}
\label{fig:dpo-preference-tendency}
\end{figure}

At the criterion level, shifts were not uniform. Criteria with more training annotations, such as Comfort and Invitingness, exhibited stronger preference shifts toward DPO outputs, whereas Inclusivity and Diversity exhibited higher neutral rates. Figure~\ref{fig:livs-eval-criterion} summarizes the neutral and distinct outcomes per criterion on the evaluation dataset, and Figure~\ref{fig:livs-tendency} reports the mean tendency toward DPO outputs by criterion.

\begin{figure}[!htbp]
\centering
\includegraphics[width=0.99\textwidth]{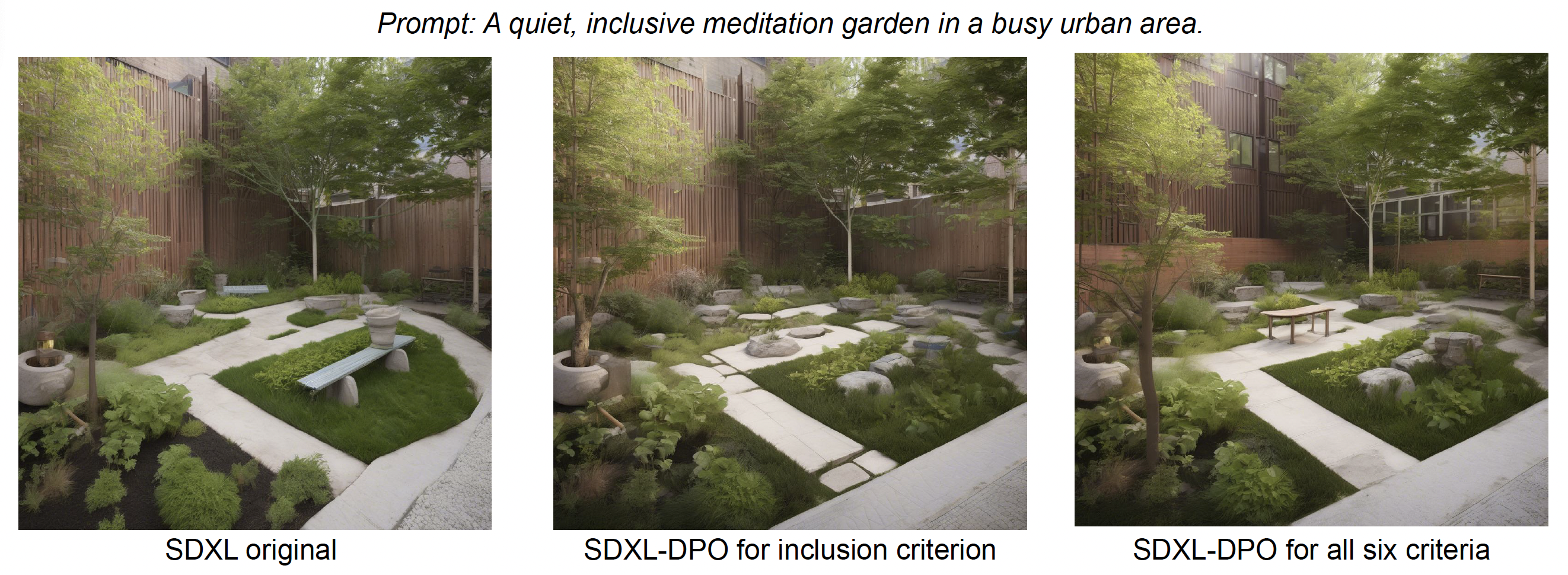}
\caption[Baseline and tuned SDXL outputs]{SDXL outputs generated with identical prompt, seed, and hyperparameters. Left: baseline SDXL. Middle: single-criterion fine-tuning (Inclusivity). Right: multi-criteria fine-tuning. \added{Source: LIVS model outputs generated with SDXL; see \citet{pmlr-v267-mushkani25a}.}}
\label{fig:livs-example}
\label{fig:sdxl-output-comparison}
\end{figure}

Qualitative inspection indicated that DPO outputs often produced clearer walkways and seating configurations but did not consistently generate detailed features associated with accessibility and inclusion, such as ramps, tactile paving, or multilingual signage. Figure~\ref{fig:livs-example} presents a comparison under identical prompt, seed, and hyperparameters.

The prevalence of neutral outcomes indicates that, even when a model improves on average, many comparisons remain ambiguous or represent balanced tradeoffs that are not resolvable through a single fine-tuned model.

\FloatBarrier

\FloatBarrier

\subsection{Case Study II}
\begin{figure}[htbp!]
\centering
\includegraphics[width=0.85\textwidth]{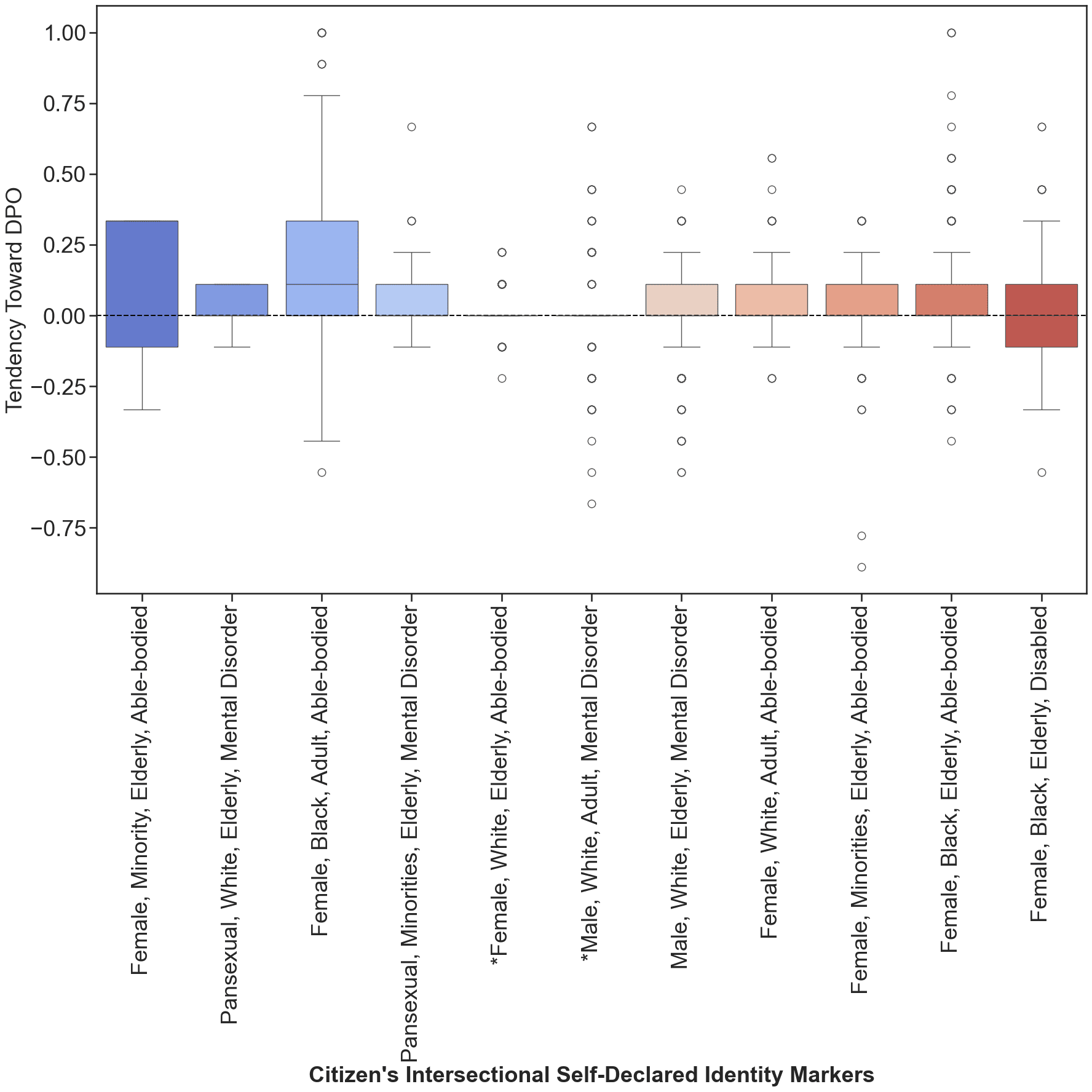}
\caption[Participant-level preference variation]{Participant-level variation in preference tendency toward DPO outputs, grouped by intersectional self-declared identity markers. A value of 1 indicates consistent preference for DPO, while -1 indicates consistent preference for the baseline. \added{Source: Author’s work.}}
\label{fig:livs-user-variation}
\label{fig:participant-preference-variation}
\end{figure}

We examined whether variation in self-declared demographics influenced preferences between DPO-fine-tuned and baseline outputs by aggregating each participant's total tendency to favor DPO versus baseline across the six criteria. Most participants showed a modest tendency toward DPO outputs, while two late-joining participants exhibited no clear tendency. These individuals joined after the workshops that established the criteria taxonomy, indicating that involvement in the knowledge-exchange process can shape how alignment improvements are perceived. Figure~\ref{fig:livs-user-variation} visualizes the participant-level variation in preference tendency toward DPO outputs.

While some annotators who reported mobility challenges tended to favor DPO outputs more often, no single demographic factor dominated preferences, indicating that improvement cannot be treated as a single global shift across users.

\FloatBarrier

\subsection{Case Study III}
\begin{figure}[!htbp]
\centering
\includegraphics[width=0.95\textwidth]{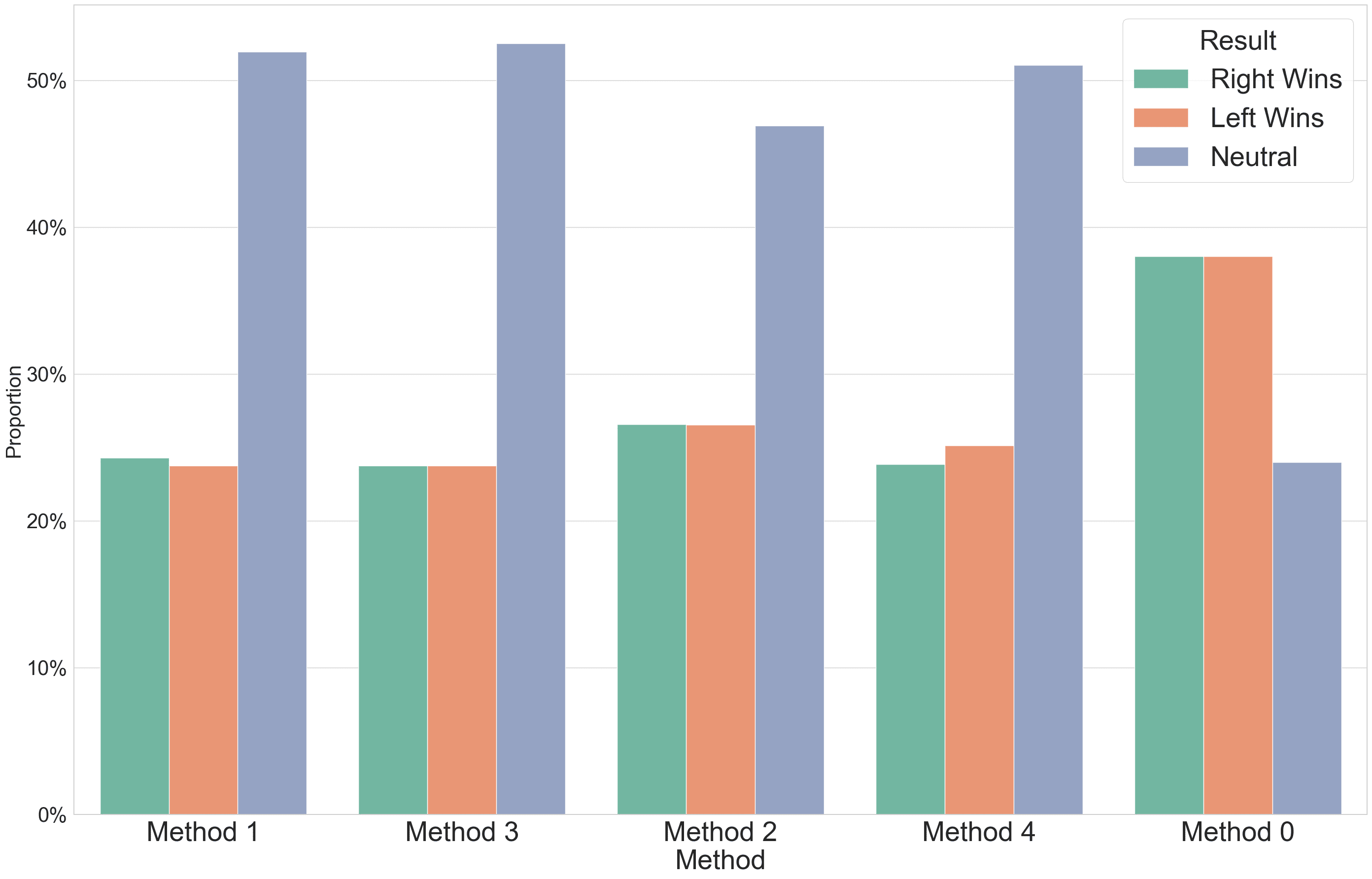}
\caption[Neutrality by prompt method]{Proportion of neutral annotations for human-authored prompts (Method 0) and GPT-4-generated prompt sets (Methods 1--4). Human-authored prompts elicited lower neutral rates. \added{Source: Author’s work.}}
\label{fig:livs-prompt-performance}
\label{fig:neutrality-by-prompt-method}
\end{figure}

To assess whether prompt composition influences rating consistency, we compared evaluations of images generated from the 440 human-authored prompts against four prompt sets generated using GPT-4 \citep{openai2024gpt4}. Human-authored prompts were associated with lower neutral rates, indicating higher decisiveness. Figure~\ref{fig:livs-prompt-performance} summarizes prompt performance based on their creation method.

\FloatBarrier

This pattern indicates that prompt protocols are governance-relevant. If prompts are generic or detached from local context, generated images can collapse toward similar global priors, reducing the visibility of value tradeoffs and increasing neutrality.

\subsection{Case Study IV}
\begin{figure}[!htbp]
\centering
\includegraphics[width=0.95\textwidth]{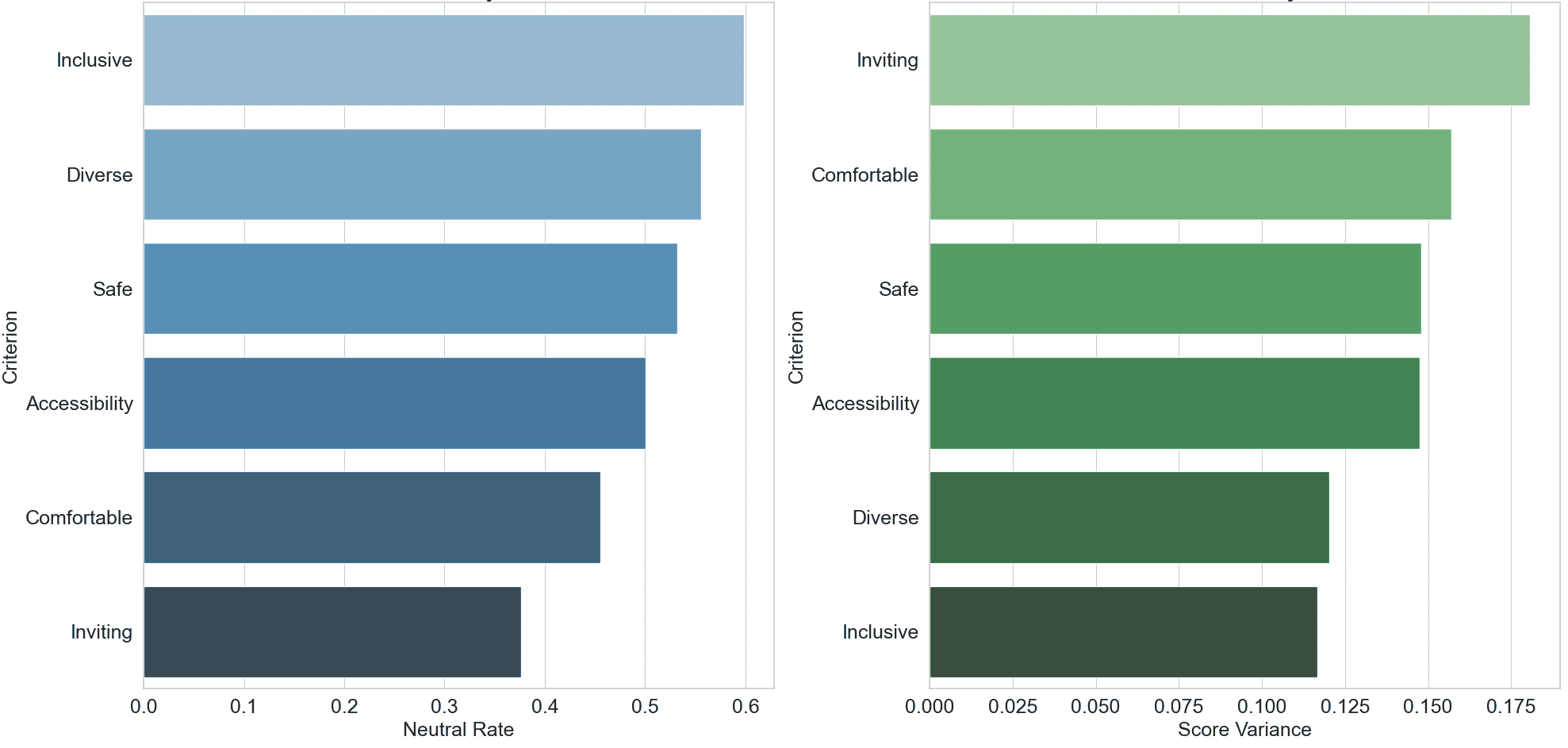}
\caption[Neutrality and score variance]{Neutral rates and score variance by criterion in the LIVS dataset. Neutrality is higher for Inclusivity and Diversity. \added{Source: Author’s work.}}
\label{fig:livs-variance}
\label{fig:neutrality-and-score-variance}
\end{figure}

Criteria differ systematically in how often participants issue neutral judgments and in the variance of criterion scores. Figure~\ref{fig:livs-variance} summarizes these patterns for the dataset and indicates that Inclusivity and Diversity are associated with higher neutrality, consistent with participant reports that these criteria rely on cues that are subtle, symbolic, or context-dependent. These patterns distinguish criteria that are more visually legible and therefore more amenable to image-based alignment from criteria for which image modality and consensus are constrained.

\FloatBarrier

\section{Discussion}
\label{sec:livs-discussion}
\subsection{Neutrality as governance-relevant evidence}
A central empirical outcome of the LIVS evaluation is the prevalence of neutral judgments: 1,100 of 2,100 evaluation comparisons were marked as neutral. In standard alignment settings, neutrality is often treated as indecision to be reduced through prompt refinement, stronger training, or more data. The LIVS results support a differentiated interpretation.

Neutrality can arise because criteria are not visually salient. Participants reported that Inclusivity and Diversity often depend on cues that are symbolic or contextual and may not be expressed reliably in generated imagery. Neutrality can also reflect value balance, in which image pairs represent different but acceptable tradeoffs among criteria. Neutrality can further arise from within-pair conflict across criteria: an image can be preferred for one criterion and dispreferred for another, and collapsing such feedback into a single binary label can obscure the structure of disagreement. Finally, neutrality can be induced by prompt underspecification, consistent with the prompt-composition experiment.

In a governance frame, these interpretations matter. Neutrality is not only a measurement artifact; it is documentation of contested or ambiguous evaluative space. In municipal practice, such space should trigger deliberation, additional context gathering, or explicit articulation of decision rules rather than being forced into optimization by a single objective \citep{Sorensen2024}. Table~\ref{tab:neutral-governance} summarizes a governance interpretation grounded in the LIVS results.

\begin{table}[!htbp]
\centering
\small
\caption{Governance interpretation of neutral judgments in pluralistic alignment, grounded in LIVS findings that neutrality is substantial and criterion-dependent.}
\label{tab:neutral-governance}
\begin{tabularx}{\textwidth}{P{0.20\textwidth}P{0.28\textwidth}Y}
\toprule
\textbf{Neutrality pattern} & \textbf{Plausible meaning} & \textbf{Governance response} \\
\midrule
High neutrality on a criterion (e.g., Inclusivity, Diversity) & Criterion not reliably visible in image modality; symbolic or contextual cues dominate & Treat the criterion as contested or under-represented; require deliberation and documentation of limits; restrict use of outputs as decision evidence for this criterion \\
Neutrality concentrated in specific prompt sets & Prompt underspecification yields similar global priors; low contrast between alternatives & Revise participatory prompt templates; re-run generation with context-specific prompts; document prompt protocols as part of procurement or audit artifacts \\
Neutrality with stable, mixed preferences across participants & Balanced tradeoffs or legitimate pluralism & Do not force a single optimum; represent distributions; use deliberative cells or structured discussion to negotiate context-specific standards \\
Neutrality arising from multi-criteria conflict within pairs & Competing criteria create internal disagreement hidden by majority vote & Avoid collapsing conflicts in governance decisions; preserve criterion-level records; consider multi-objective decision rules in subsequent commissioning \\
\bottomrule
\end{tabularx}
\end{table}
\FloatBarrier

\paragraph{\textbf{Pluralism and measurement limitations.}}
To separate value indeterminacy from measurement artifacts, we examined whether neutrality and disagreement are concentrated in (i) high within-pair visual similarity (low contrast; measured via CLIP cosine similarity), (ii) low-quality or artifact-laden images, and (iii) prompt origin (human-authored vs.\ GPT-4-generated) and criterion legibility. Neutrality increases for high-similarity pairs and lower-quality images, but remains substantial even among low-similarity, high-quality pairs for Inclusivity and Diversity, suggesting genuine contested evaluative space beyond instrumentation limits \citep{Aroyo2015,PavlickKwiatkowski2019}.

\subsection{What pluralistic alignment changes in a planning workflow}
LIVS operationalizes pluralistic alignment by pairing locally co-produced criteria with multi-criteria preference data. This changes planning workflows in two ways.

First, it changes the object of accountability. A generative model used in planning becomes a governed artifact whose outputs can be evaluated against a locally negotiated criteria taxonomy rather than treated as a neutral visualization device. Second, it produces civic artifacts that can be versioned, contested, and revisited as local priorities shift, including criteria definitions, prompt protocols, and preference records. These artifacts function as governance interfaces in procurement and oversight, enabling contestation and re-commissioning.

\subsection{Limits}
The LIVS approach is bounded by dataset size and scope and by modality limits. The dataset is specific to one city and to a set of collaborating organizations. DPO requires binary labels; majority voting collapses multi-criteria conflicts and neutral comparisons are not used as training signals \citep{rafailov2024}. Image-only alignment is limited in its capacity to capture cultural and symbolic markers that are not visually expressed or that require local context. Claims of alignment therefore must be scoped to what the modality and criteria can represent.

These constraints translate into documentation obligations. If a city uses generative tools in planning processes, it should document intended uses and disallowed uses, the criteria taxonomy and its provenance, prompt protocols, and the neutrality and variance patterns that indicate where outputs are not decision-relevant \citep{gebru2021}. The central implication is that preference tuning does not eliminate pluralism and should not be treated as a substitute for legitimate negotiation.

\section{Conclusions}
\label{sec:livs-conclusion}
This chapter introduced LIVS as a pluralistic alignment dataset for inclusive public spaces, developed through a two-year participatory process with community organizations in Montréal. The chapter showed that multi-criteria preference data can be used to fine-tune a T2I model with DPO and can increase alignment with local preferences on selected criteria. At the same time, the prevalence of neutral outcomes indicates persistent heterogeneity and ambiguity, particularly for criteria that rely on cues that are not consistently represented in image outputs.

In the dissertation's argument about the algorithmic layer of urbanism, LIVS addresses the generative side of that infrastructure. Governing this layer requires methods that represent pluralism rather than collapsing it into a single objective. The LIVS dataset, including consented self-identification markers, is available for research purposes at \url{https://mid-space.one}. The next chapter builds on this result to specify institutional mechanisms through which cities can treat both measurement models and generative models as public infrastructure governed by enforceable rights, lifecycle procedures, and deliberative oversight.


\chapter{Governance and Implementation Architecture}
\label{chap:governance-implementation}

This chapter specifies a municipal architecture for governing AI systems that evaluate, classify, predict, or generate representations of public space. It translates the normative commitments developed in Chapters~\ref{chap:right-to-ai} and~\ref{chap:coproducing_ai_lifecycle}, and the empirical constraints documented in Chapters~\ref{chap:streetreview} and~\ref{chap:livs}, into procedures that can be embedded in municipal practice. The central premise is that public-space AI systems are not neutral decision aids. They operate as epistemic infrastructure that shapes what becomes visible, measurable, and comparable, and they redistribute authority by stabilizing some categories and evaluation rules while rendering others less legible \citep{star1999ethnography,BowkerStar1999,Jasanoff2004,Kitchin2016,Beer2016}. This infrastructural role is consistent with research on quantification and commensuration, which shows that indicators and rankings transform governance by defining what counts as evidence and by reorganizing accountability around what can be audited \citep{EspelandStevens2008,Porter1995,Strathern2000}.

In cities, this dynamic is amplified by platform-mediated sensing and data infrastructures that mediate how streets and neighborhoods are represented for planning, maintenance, and design \citep{Kitchin2014a,plantin2018infrastructure,LarteyLaw2025,Mattern2017a}. When AI systems are introduced into contested civic domains, the aggregation of heterogeneous judgments is not a technical nuisance to be averaged away. It is a governance choice that can either preserve plural claims about inclusion, safety, recognition, and access, or suppress them through commensuration \citep{Desrosieres1998,Merry2016}.

The empirical chapters motivate four constraints that this architecture treats as design requirements rather than as residual uncertainty. First, evaluative judgments about streetscapes are structured and position-dependent, with higher divergence for socially contested dimensions than for visually legible dimensions. Second, deliberation changes the evidentiary object: structured discussion can produce shared standards for some criteria while preserving residual disagreement on contested civic dimensions. Third, co-produced scaling can generate citywide representations from locally grounded standards, yet the representational adequacy of these representations is bounded by modality limits, uneven coverage, and data quality. Fourth, preference-based evaluation of inclusive public-space imagery yields substantial indeterminacy, indicating that attempts to force singular alignment should trigger deliberation and constraint rather than optimization toward one normative target.

The chapter's goal is operational. It specifies how a municipality can implement the civic \RightToAI{} as a governance layer that constrains procurement, authorizes deployment, and supports recourse across the lifecycle of both streets and models. The architecture is designed to address a recurring limitation in applied AI ethics practice: principles without institutionalized procedures, documentation, and accountability mechanisms are difficult to enforce and tend to be selectively applied \citep{Schiff2021a,larsson_governance_2020,lazar2024lectureigoverningalgorithmic}.

The chapter proceeds as follows. Section~\ref{sec:ch7-requirements} derives governance requirements from propositions P1 to P4 and formalizes disagreement-preserving governance. Section~\ref{sec:ch7-right-to-ai-layer} specifies the \RightToAI{} as a municipal governance layer, including minimum entitlements, bright-line prohibitions, and mandated procedures. Section~\ref{sec:ch7-workflow} maps the five co-produced lifecycle phases into a municipal workflow with checkpoints, artifacts, and pause authority. Section~\ref{sec:ch7-procurement} specifies procurement and commissioning rules for both models and AI-mediated street projects, emphasizing commissioning symmetry and re-commissioning triggers. Section~\ref{sec:ch7-oversight} designs oversight and recourse centered on recurring \DelibCells{}. Section~\ref{sec:ch7-feasibility} evaluates feasibility, capacity, and legitimacy risks, and proposes a tiered implementation strategy. Section~\ref{sec:ch7-conclusions} concludes with a synthesis and leads into the dissertation conclusion.

\section{When Aggregation Is Permitted}
\label{sec:ch7-requirements}

The governance architecture in this chapter treats propositions P1 to P4 as constraints on institutional design. The aim is not to convert plural judgments into a single metric, but to specify when aggregation is permitted, when it is prohibited, and which procedures are required for decisions to be legitimate under persistent disagreement.

\subsection{Pluralism as procedure}

In public space contexts, pluralism is not a distributional inconvenience to be reduced through aggregation. It is a condition of governance that should shape what is measured, how evidence is combined, and how decisions are authorized. Urban theory and planning scholarship have long treated public space as a site of contested meanings and negotiated access rather than as a domain of stable, universally shared evaluation criteria \citep{Lefebvre1968,Jacobs1961,Gehl2011,low2020social}. When AI systems translate contested judgments into scores, maps, or design rankings, the choice of measurement and aggregation rules becomes a political and administrative decision that can erase minority harms through commensuration \citep{EspelandStevens2008,Porter1995,Strathern2000}.

Municipal settings amplify this problem because quantified outputs are administratively portable. Once produced, they can circulate across departments, contract cycles, and public consultations as if they were neutral facts rather than constructed indicators with limits \citep{Desrosieres1998,Merry2016}. AI-mediated representations can therefore shift decision authority from deliberative settings to technical pipelines and vendor-defined defaults, particularly when documentation is incomplete or inaccessible \citep{Batty2024,LarteyLaw2025,Pasquale2015,Eubanks2018}. The empirical chapters show that disagreement and indeterminacy are not measurement failures to be eliminated, but governance-relevant states that should determine when deliberation is required and when the system's decision boundary should be narrowed.

\subsection{Disagreement-preserving governance}

This chapter uses \emph{disagreement-preserving governance} to denote a procedural stance in which persistent divergence in judgments about contested civic dimensions is treated as actionable evidence. Disagreement-preserving governance does not reject quantification. It constrains how quantification is produced and used by requiring disaggregation, documentation of contestation, and explicit decision gates that determine when deliberation is required before a measure can authorize action. This stance draws on infrastructure and classification scholarship, which emphasizes that standards embed political choices and become difficult to contest once routinized, and on the sociology of quantification, which documents how measurement practices reorganize accountability and authority \citep{BowkerStar1999,star1999ethnography,Porter1995,Strathern2000}.

Two implications follow for municipal practice. First, the evidentiary object should include both convergence and disagreement. Municipal artifacts should preserve heterogeneity over time and across versions, rather than collapsing divergence into a single scalar that can travel without context. Second, deliberation should be treated as part of the measurement-to-governance pipeline, because legitimacy in plural settings depends on institutionalized opportunities for reasons-giving, contestation, and revision that connect public input to decision authority \citep{Dewey1927,Habermas1996,Healey1997,Ercan02015}.

\subsection{Core governance requirements}

The dissertation's evidence chain supports seven requirements for governing public-space AI systems as civic infrastructure. These requirements are stated here as design requirements for municipal procedure.

\begin{enumerate}
\item \textbf{Plural representation without forced aggregation.} Governance should represent heterogeneous values in a form that does not erase minority harms through default majority averaging. This implies subgroup-aware outputs where appropriate, explicit handling of heterogeneity in evaluation, and decision rules that recognize distributional and recognitional conflict rather than treating it as noise \citep{Crenshaw1989,BarocasSelbst2016,BarocasHardtNarayanan2023}.
\item \textbf{Disagreement-preserving documentation.} Residual disagreement should be recorded as part of the evidentiary object, including where and why views diverge and what decision consequences follow. This requirement aligns with the view that metrics are institutional artifacts whose legitimacy depends on documented construction, interpretation, and contestability \citep{EspelandStevens2008,Merry2016,Power1997}.
\item \textbf{Deliberation as an institutional component of measurement.} Structured deliberation should be built into the governance pipeline because it transforms individual judgments into revisable collective standards and clarifies which dimensions remain contested. Participation should connect to decision authority rather than serving only as consultation \citep{Arnstein1969,Habermas1996,Healey1997,Fischer2000}.
\item \textbf{Lifecycle governance and versioned accountability.} Models that mediate public-space decisions should be treated as evolving infrastructure that requires monitoring, review, and re-commissioning. Accountability should attach across phases, not only at the time of initial consultation or deployment \citep{Sculley2015}.
\item \textbf{Auditability proportional to impact.} Municipal governance should require traceable artifacts, including data documentation, model cards, evaluation logs, and decision memos, with independent auditing pathways commensurate with the system's impact on residents and public resources \citep{Mitchell2019,gebru2021,NIST2023,Leslie2019}.
\item \textbf{Recourse and contestability with standing.} Impacted residents and groups should have standing to contest outputs and uses, with pathways to request review, trigger pause authority, and compel revision or decommissioning where harms are systematic \citep{CohenSuzor2024,Pasquale2015}.
\item \textbf{Indeterminacy as a governance signal.} High rates of neutral or indeterminate judgments in evaluation signal that values are contested or that the evidentiary basis is insufficient for closure. Governance should respond through deliberation, additional evidence collection, narrowed decision boundaries, or revised standards rather than forcing optimization toward a single target \citep{Sorensen2024}.
\end{enumerate}

\subsection{Key artifacts: the \ValueRegister{} and disagreement logs}

To implement the requirements above, the governance architecture relies on two persistent artifacts that stabilize negotiated standards without suppressing contestation and preserve institutional memory across project cycles and system updates.

\medskip

\begin{quote}
\noindent\textbf{Definition (\ValueRegister{}).} A \ValueRegister{} is a municipally maintained, publicly accessible record of the evaluative dimensions used to govern a specific class of public-space interventions or AI systems. It distinguishes (i) dimensions that have reached sufficient convergence to function as benchmarks, (ii) dimensions that remain contested and therefore require deliberation triggers and stronger procedural safeguards, and (iii) operational definitions, measurement limits, and disallowed uses associated with each dimension. The \ValueRegister{} is versioned and updated through re-commissioning.
\end{quote}

\medskip

\begin{quote}
\noindent\textbf{Definition (Disagreement log).} A disagreement log is a structured record of where evaluators, subgroups, or \DelibCells{} differ in their assessments or preferences, including the reasons offered, the criteria at stake, and the decision consequences. It preserves residual contestation as legitimate evidence rather than treating it as measurement error.
\end{quote}

\medskip

These artifacts are consistent with the premise that indicators gain authority through institutional embedding and travel across contexts as administrative facts. They also respond to the audit literature's warning that metrics can substitute for judgment when their construction and limits are not documented in durable records \citep{Porter1995,EspelandStevens2008,Power1997}.

\medskip

\subsection{Requirements crosswalk}
\label{sec:ch7-requirements-table}

Table~\ref{tab:requirements-p1-p4} provides a crosswalk from propositions P1–P4 to corresponding governance requirements and failure modes. It makes explicit how empirical conditions translate into procedural constraints on measurement, deliberation, and use in municipal contexts.

\begin{table}[!htbp]
\centering
\small
\caption{Governance requirements derived from propositions P1 to P4. The table links empirically grounded constraints to procedural implications for municipal governance.}
\label{tab:requirements-p1-p4}

\renewcommand{\arraystretch}{1.15}

\begin{tabularx}{\textwidth}{
  >{\raggedright\arraybackslash}p{0.11\textwidth}
  >{\raggedright\arraybackslash}p{0.24\textwidth}
  >{\raggedright\arraybackslash}X
  >{\raggedright\arraybackslash}p{0.24\textwidth}
}
\toprule
\textbf{Prop.} & \textbf{Empirical condition} & \textbf{Governance requirement} & \textbf{Primary failure mode if ignored} \\
\midrule
P1 &
Pluralism is structured; socially contested dimensions exhibit lower agreement than visually legible dimensions &
Preserve subgroup-aware evidence; avoid default majority averaging; maintain a \ValueRegister{} that separates benchmarks from contested dimensions \citep{Crenshaw1989,BarocasSelbst2016}. &
Illegitimate universal metric; minority harms erased through commensuration; governance capture by dominant norms \citep{EspelandStevens2008,Strathern2000}. \\
\midrule
P2 &
Deliberation can increase convergence for some criteria while preserving residual contestation &
Institutionalize \DelibCells{}; maintain disagreement logs; treat residual disagreement as evidence that constrains permissible uses \citep{Dewey1927,Habermas1996,Healey1997}. &
Technocratic closure through forced consensus or administrative paralysis without negotiation mechanisms. \\
\midrule
P3 &
Co-produced scaling is feasible but bounded by modality limits and uneven data quality &
Constrain permissible uses; require auditability and documentation; incorporate data quality checks and drift monitoring; ensure recourse for systematic misrepresentation \citep{Sculley2015,NIST2023,Ibrahim2020}. &
Overtrust in maps and scores; mission creep; uneven visibility leads to uneven governance capacity \citep{star1999ethnography,Kitchin2016}. \\
\midrule
P4 &
Indeterminacy persists in preference-based evaluation of contested values &
Treat neutral and indeterminate outcomes as triggers for deliberation or scope narrowing; avoid optimization toward a single normative target \citep{Sorensen2024,ConitzerEtAl2024}. &
False precision; procedural illegitimacy; alignment claims used to justify contested decisions without deliberation \citep{Pasquale2015,Merry2016}. \\
\bottomrule
\end{tabularx}
\end{table}

\FloatBarrier

\section{The \RightToAI{} as a Municipal Governance Layer}
\label{sec:ch7-right-to-ai-layer}

Urban governance traditions emphasize that public space is a political object through which inclusion, recognition, and participation in urban life are organized \citep{Lefebvre1968,Harvey2003,low2020social,MitrasinovicMehta2021,Gehl2011}. These traditions also treat publics as constituted through problems and their institutional handling, making the organization of inquiry and decision-making a core site of democratic legitimacy \citep{Dewey1927}. The \RightToAI{} extends this framing to settings where AI systems classify, score, predict, or generate representations that inform public-space decisions. In municipal contexts, reliance on AI-mediated representations for planning, maintenance prioritization, design optioning, or resource allocation turns the system into a governance object that requires democratic authorization, procedural legitimacy, and contestability.

This chapter operationalizes the \RightToAI{} as a municipal charter layer that sits above individual projects. It establishes a governance floor consisting of minimum entitlements, prohibited practices, and mandated procedures that every public-space AI system should satisfy before it is commissioned, deployed, or renewed. The aim is to prevent municipal practice from defaulting to vendor-defined standards, opaque technical constraints, or discretionary use that shifts public decision authority without explicit authorization \citep{Kitchin2014a,plantin2018infrastructure,burrell_how_2016,Crawford2021}.

\subsection{Minimum entitlements in public-space contexts}
\label{sec:ch7-minimum-entitlements}

The \RightToAI{} can be operationalized as a set of minimum entitlements that apply whenever an AI system materially shapes public-space decisions, representations, or resource allocation. These entitlements translate rights language into municipal duties and required artifacts.

\subsubsection{Right to disclosure and notice}

Residents should receive notice when AI systems materially affect public-space decisions or representations, including systems that operate as representational infrastructure within planning workflows. Disclosure should be intelligible, updated as systems change, and specific about decision boundaries, data sources, and known limitations \citep{burrell_how_2016,NIST2023,Leslie2019}.

\subsubsection{Right to participate as power}

The \RightToAI{} treats participation as a distribution of decision power rather than as consultation \citep{Arnstein1969,Fung2004,Jacobs1961}. Municipal operationalization requires that participatory authority be tied to impact and risk tiering. Where systems can materially shape access, stigma, or distributive outcomes, participation should include co-framing authority and oversight with pause capacity. This requirement aligns with deliberative accounts that connect inclusive public reasoning to binding decision authority \citep{Habermas1996,Ercan02015}.

\subsubsection{Right to documentation and intelligible artifacts}

Meaningful participation requires intelligible artifacts that connect technical choices to negotiated values. At a minimum, this includes model cards, data documentation, evaluation summaries, and decision memos that record how objectives, thresholds, and use constraints were selected and revised \citep{Mitchell2019,gebru2021}. In municipal settings, these artifacts function as administrative records that enable oversight, audit, and challenge rather than as optional transparency supplements \citep{NIST2023,Leslie2019,kaminski2018right}.

\subsubsection{Right to contestation and recourse}

Impacted residents must have standing to contest outputs and uses, request review, and trigger reconsideration, suspension, or decommissioning where harms are systematic. In public-space contexts, recourse should cover representational harms, such as stigmatizing labels or decontextualized maps, and procedural harms, such as use beyond approved scope \citep{airight2025,CohenSuzor2024,kaminski2021right}.

\subsubsection{Right to non-extractive data practices}

Participatory datasets and community knowledge are governance assets that require stewardship. Municipal practice should limit extractive reuse, require consent and benefit-sharing, and maintain data stewardship agreements specifying retention, access, and secondary-use restrictions. This entitlement also requires attention to contextual integrity, aligning data flows with the institutional purposes and social expectations under which data were collected \citep{Datafeminism,CostanzaChock2020,zhang2024right,Nissenbaum2010}.

\subsubsection{Right to pluralistic representation}

Where evaluation shows structured heterogeneity, municipal governance should protect against aggregation rules that erase minority harms. This entitlement does not require essentializing identity categories. It requires that disaggregated evidence be available as a safeguard against false universality and that decisions document how tradeoffs were handled \citep{Crenshaw1989,BarocasSelbst2016,BarocasHardtNarayanan2023,Sorensen2024}.

\subsection{Bright-line prohibitions}
\label{sec:ch7-prohibitions}

The \RightToAI{} also implies bright-line prohibitions that protect against predictable municipal failure modes in AI governance.

\begin{enumerate}
\item \textbf{No high-impact deployment without a co-framing charter.} A city should not deploy a high-impact public-space AI system without a documented co-framing charter specifying objectives, non-objectives, decision boundaries, disallowed uses, and who has standing for recourse.
\item \textbf{No black-box procurement.} Procurement should not result in systems the city cannot inspect, audit, or re-commission. Contractual inability to access documentation, evaluate performance, or constrain secondary use undermines democratic accountability \citep{Pasquale2015,NIST2023,burrell_how_2016}.
\item \textbf{No punitive repurposing of evaluative outputs without explicit authorization and due process safeguards.} Outputs designed for audit and deliberation should be treated as out of scope for punitive enforcement by default. Any enforcement-adjacent repurposing requires separate authorization, documented necessity, and procedural safeguards \citep{Eubanks2018,Leslie2019}.
\item \textbf{No mission creep without re-authorization.} Using a system beyond its approved scope, including new decision contexts, new populations, or new objectives, requires re-commissioning and renewed participatory authorization \citep{NIST2023}.
\end{enumerate}

\subsection{Mandated procedures}
\label{sec:ch7-mandates}

The \RightToAI{} becomes implementable when it mandates municipal procedures.

\begin{enumerate}
\item \textbf{Lifecycle governance with checkpoints.} Every public-space AI system should pass explicit checkpoints across co-framing, co-design, co-implementation, co-deployment, and co-maintenance, with required artifacts and sign-offs \citep{Koseki2022,NIST2023}.
\item \textbf{Standing oversight via \DelibCells{}.} A city should maintain recurring \DelibCells{} with authority to review contested dimensions, update the \ValueRegister{}, and recommend pause, revision, or decommissioning when triggers occur \citep{Dewey1927,Habermas1996,negotiativealignment}.
\item \textbf{Re-commissioning schedules and triggers.} Both streets and models should be governed as maintainable infrastructure requiring periodic re-commissioning under drift, demographic change, scope change, or sustained contestation \citep{Sculley2015,NIST2023}.
\end{enumerate}

These obligations are consistent with risk-based governance approaches that operationalize accountability through documentation, monitoring, and constraints on high-impact uses \citep{NIST2023,EuropeanUnion2024}.

\subsection{A municipal governance schematic}
\label{sec:ch7-governance-schematic}

\begin{figure}[!htbp]
\centering
\begin{adjustbox}{width=\linewidth}
\begin{tikzpicture}[
  box/.style={draw, rounded corners, align=left, inner sep=10pt, text width=0.28\linewidth},
  arrow/.style={-Latex, thick},
  font=\small,
  node distance=14pt and 16pt
]
\node[box] (rights) {\textbf{\RightToAI{}}\\[2pt]
  Minimum entitlements\\
  Bright-line prohibitions\\
  Mandated procedures};

\node[box, right=of rights] (duties) {\textbf{Municipal duties}\\[2pt]
  Disclosure and notice\\
  Participation as power\\
  Auditability and documentation\\
  Recourse and standing\\
  Data stewardship};

\node[box, right=of duties] (artifacts) {\textbf{Governance artifacts}\\[2pt]
  Public Space AI Charter\\
  \ValueRegister{} (versioned)\\
  Model card and data documentation\\
  Evaluation logs and audits\\
  Disagreement log};

\node[box, right=of artifacts] (recourse) {\textbf{Recourse and re-commissioning}\\[2pt]
  Complaint intake\\
  Review and appeal\\
  Pause and rollback triggers\\
  Revision of standards\\
  Decommissioning};

\draw[arrow] (rights) -- (duties);
\draw[arrow] (duties) -- (artifacts);
\draw[arrow] (artifacts) -- (recourse);

\node[
  anchor=north west,
  text width=\textwidth,
  align=center,
  font=\footnotesize
] at ([yshift=-14pt]current bounding box.south west) {Entitlements imply duties. Duties require artifacts. Artifacts enable contestation, revision, and accountability over time.
};

\end{tikzpicture}
\end{adjustbox}
\caption[Right to AI municipal governance layer]{The \RightToAI{} as a municipal governance layer connecting entitlements, duties, artifacts, and recourse. \added{Source: author's synthesis.}}
\label{fig:right-to-ai-layer}
\label{fig:right-to-ai-municipal-layer}
\end{figure}
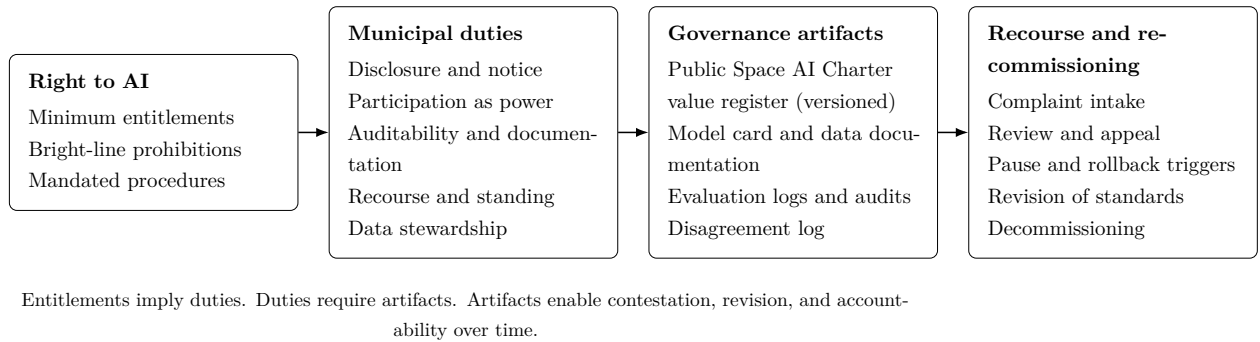

Figure~\ref{fig:right-to-ai-layer} illustrates how rights give rise to duties, duties are realized through governance artifacts, and those artifacts enable ongoing recourse, revision, and accountability.

\FloatBarrier

\section{Operational Procedure}
\label{sec:ch7-workflow}

\subsection{Why a lifecycle workflow is necessary}

Participation that ends at consultation is structurally mismatched to AI systems that update, drift, and change meaning over time. The augmented participatory lifecycle developed in Chapter~\ref{chap:coproducing_ai_lifecycle} (Figure~\ref{fig:augmented_ai_lifecycle_ch4}) responds to this mismatch by specifying five co-produced phases and by mapping participation to risk mitigation tasks. In public-space contexts, lifecycle governance is reinforced by three empirical considerations. Representational infrastructures derived from platform imagery or digital traces change as streets change, as coverage shifts, and as institutional practices evolve. Model behavior can drift with changes in input distributions and operational coupling \citep{Sculley2015}. Deliberation changes what can be treated as a legitimate collective standard under plural judgments \citep{Habermas1996,Dewey1927}. Finally, the persistent indeterminacy observed in preference evaluation indicates that technical alignment cannot substitute for procedures that authorize and constrain value-laden decisions \citep{Sorensen2024,ConitzerEtAl2024}.

This section translates the lifecycle into a municipal workflow. The workflow is designed to align with municipal realities including departmental roles, procurement constraints, pilot cycles, and maintenance budgeting.

\subsection{Municipal workflow mapping}
\label{sec:ch7-workflow-checkpoints}

The workflow implements the five phases through decision gates that authorize continuation. Each gate requires a minimum set of artifacts and a determination of whether contested dimensions require deliberation before outputs can be used to justify action.

\subsubsection{Phase 1: Co-framing}

Co-framing establishes a democratically authorized problem definition, specifies decision boundaries, and produces an initial \ValueRegister{} that distinguishes benchmark dimensions from contested dimensions.

The phase produces a Public Space AI Charter that specifies scope, objectives, non-objectives, intended uses, and disallowed uses; a \ValueRegister{} (v0) that records provisional operational definitions and classifies dimensions as benchmark or contested; and a risk register that identifies representational, distributional, privacy, and mission creep risks.

A \DelibCell{} should be able to block continuation when the framing collapses contested values into a single score without justification, when disallowed uses are not constrained in the charter, or when governance artifacts do not specify standing for recourse.

\subsubsection{Phase 2: Co-design}

Co-design translates the charter and \ValueRegister{} into data practices, measurement protocols, and interface commitments that preserve pluralism and communicate limits.

The phase produces a data governance plan covering consent, retention, access control, and restrictions on secondary use; an evaluation protocol that specifies how disagreement and indeterminate cases are recorded and interpreted; an evaluation plan that determines where disaggregated outputs are required and how they will be interpreted; and an interface and communication plan that specifies how uncertainty, limitations, and use constraints are presented to municipal staff and the public to reduce overinterpretation of quantified outputs \citep{Nissenbaum2010,Porter1995,EspelandStevens2008,Mitchell2019,gebru2021}.

Procurement and implementation should not proceed when data stewardship is unspecified, when disaggregation requirements are absent despite known heterogeneity, or when recourse pathways and decision boundaries are not operationalized in the interface.

\subsubsection{Phase 3: Co-implementation}

Co-implementation builds or configures the system and produces traceable artifacts that link technical choices to the charter and \ValueRegister{}.

The phase produces a model card specifying intended uses, disallowed uses, limitations, and evaluation results; data documentation that records dataset composition and representational limits; evaluation logs that document error patterns and uncertainty; and a traceability mapping from \ValueRegister{} items to implementation decisions such as features, prompts, thresholds, and outputs \citep{Mitchell2019,gebru2021,Sculley2015}.

The gate to pilot should fail when critical limitations are undocumented, when evaluation indicates unacceptable errors for identified populations or geographies, or when the pipeline cannot be independently audited.

\subsubsection{Phase 4: Co-deployment}

Co-deployment introduces the system through a pilot that is constrained to approved uses and accompanied by public notice, logging, and an assessment gate.

The phase produces a pilot authorization memo defining scope, affected geographies, decision uses, and limits; public communication materials describing the system's role, limitations, and recourse pathways; and logging infrastructure that records outputs, uses, and downstream decisions.

Continued deployment requires an assessment report that covers performance and error hotspots, compliance with disallowed uses and decision boundaries, patterns of disagreement and indeterminacy that signal contested values, and the outcome of deliberation on contested dimensions. A \DelibCell{} should be able to recommend suspension, scope narrowing, or rollback when the pilot reveals systematic harms, mission creep, or persistent indeterminacy that exceeds the decision boundary specified in the charter.

\subsubsection{Phase 5: Co-maintenance}

Co-maintenance treats the system as living infrastructure subject to drift, changing norms, and evolving municipal uses.

The phase produces versioned documentation updates, drift monitoring reports, re-commissioning reports that update the \ValueRegister{}, and a sunset policy defining conditions for discontinuation and data retirement \citep{Sculley2015,Koseki2022}.

\subsection{Checkpoint table}
\label{sec:ch7-checkpoint-table}

To operationalize pause authority within the augmented participatory lifecycle, Table~\ref{tab:workflow-checkpoints} defines mandatory decision gates, required artifacts, and explicit pause triggers across each lifecycle phase.

\begin{table}[!htbp]
\centering
\small
\caption{Municipal workflow checkpoints and pause authority mapped to the augmented participatory lifecycle.}
\label{tab:workflow-checkpoints}
\begin{tabularx}{\textwidth}{
>{\raggedright\arraybackslash}P{0.18\textwidth}
>{\raggedright\arraybackslash}P{0.20\textwidth}
>{\raggedright\arraybackslash}X
>{\raggedright\arraybackslash}P{0.28\textwidth}
}
\toprule
\textbf{Phase} &
\textbf{Decision gate} &
\textbf{Required artifacts (minimum)} &
\textbf{Pause trigger} \\
\midrule
Co-framing &
Authorization to proceed to design &
Project charter; \ValueRegister{} v0; risk register; documented decision boundaries &
Contested values collapsed into a single metric; disallowed uses not explicitly constrained \\
\midrule
Co-design &
Authorization to implement and procure &
Data governance plan; evaluation protocol including indeterminacy; communication plan; oversight and recourse plan &
Missing data stewardship provisions; absence of required disaggregation plans; no recourse pathway defined \\
\midrule
Co-implementation &
Authorization to pilot &
Model card; data documentation; baseline evaluation and error analysis; traceability mapping &
Unacceptable error rates for identified populations or geographies; undocumented limitations; non-auditable pipeline \\
\midrule
Co-deployment &
Authorization to continue or scale &
Pilot memo; public notice; logging infrastructure; assessment report &
Use beyond approved scope; complaint clusters indicating systematic harm; failure to meet assessment gate conditions \\
\midrule
Co-maintenance &
Authorization to renew, revise, or sunset &
Versioned documentation; drift report; re-commissioning report; updated \ValueRegister{} &
Detected drift or data coverage changes; scope expansion; sustained indeterminacy on contested dimensions; governance non-compliance \\
\bottomrule
\end{tabularx}
\end{table}
\FloatBarrier

\section{Procurement and Commissioning Rules}
\label{sec:ch7-procurement}

\subsection{Commissioning symmetry}

Infrastructure scholarship emphasizes that standards and classifications embed political choices and become difficult to contest once embedded in routine practice \citep{star1999ethnography,BowkerStar1999}. Commissioning symmetry matters because both the city and the evaluative standards used to govern it change over time, and because systems can drift in meaning as well as in technical performance.

\deleted{Commissioning symmetry implies that procurement must secure not only system functionality but also the city's ability to govern the system over time. In practice, this requires contractual audit rights, version control, access to documentation, and a budgeted pathway for updates, pauses, and decommissioning. It also requires attention to brittleness and hidden coupling in ML systems, where minor pipeline changes can produce consequential shifts in behavior and accountability \citep{Sculley2015,burrell_how_2016,Crawford2021,LarteyLaw2025,yigitcanlar2025editorial}.}

\added{Commissioning symmetry places the technical system and the public-space intervention it informs within the same cycle of authorization, maintenance, and review. Studies of municipal adoption provide the administrative context. \citet{LarteyLaw2025} review practical and sociopolitical conditions affecting AI adoption in urban planning, including limited integration with public administration and unresolved concerns about equity, transparency, and public trust. \citet{yigitcanlar2025editorial} situate these systems within the wider modernization of urban planning and management.}

\added{Engineering and opacity create one set of lifecycle liabilities. \citet{Sculley2015} show how hidden technical debt accumulates through data dependencies, feedback loops, configuration, and changing system boundaries. \citet{burrell_how_2016} distinguishes opacity arising from institutional secrecy, gaps in technical literacy, and model complexity. These conditions justify versioned documentation, drift testing, dependency review, and renewed authorization after material changes to data, models, interfaces, or purpose.}

\added{Material extraction creates a further set of liabilities. \citet{Crawford2021} places AI systems within chains of energy, minerals, labor, and data extraction. Recommissioning therefore includes an accounting of energy use, hardware replacement, data labor, vendor supply chains, and burdens shifted to affected communities. Software review addresses the integrity and maintainability of the system. Material review addresses the conditions under which the system is produced and sustained. The commissioning record assigns a municipal authority empowered to pause, revise, or terminate the arrangement when either review fails its stated conditions.}
\label{rev:commissioning-symmetry}

\subsection{Procurement rules for public-space AI systems}

This section specifies procurement clauses that translate the \RightToAI{} and lifecycle checkpoints into contract requirements. The clauses apply both to vendor-provided systems and to internally developed tools when they are deployed operationally.

\subsubsection{Disclosure and decision boundary clauses}

Contracts should require an explicit statement of intended uses and disallowed uses, a decision boundary description specifying what the tool may inform and what it must not determine, and a public notice template describing when and where the tool is used. These requirements operationalize disclosure and contestability and reduce the risk that systems are adopted as general-purpose decision authorities \citep{Pasquale2015,NIST2023}.

\subsubsection{Auditability and access clauses}

\deleted{Contracts should require access to technical artifacts sufficient for audit, including model cards, data documentation, evaluation results, and update logs. They should grant independent audit rights, specify minimum evaluation reporting, and require logging of outputs and downstream uses. These clauses are necessary because performance claims do not substitute for traceability and because many harms arise from secondary use and institutional coupling rather than from isolated prediction error \citep{Mitchell2019,gebru2021,Power1997,Pasquale2015,Eubanks2018,NIST2023}.}

\added{Documentation instruments serve distinct administrative functions. Model cards record intended uses, evaluation conditions, performance, and limitations at model level \citep{Mitchell2019}. Datasheets document dataset composition, provenance, collection, preprocessing, maintenance, and recommended uses \citep{gebru2021}. The NIST AI Risk Management Framework organizes institutional work across the govern, map, measure, and manage functions \citep{NIST2023}. Together, these instruments establish a versioned record for review.}

\added{Algorithmic impact assessments provide a separate administrative mechanism. \citet{Reisman2018} propose agency self-assessment before acquisition, public notice and comment, and disclosure of the resulting assessment. \citet{selbst2021institutional} shows that an assessment's effect depends on its legal and institutional setting and on whether collaborative processes carry decision authority. Assessment findings must therefore connect directly to procurement, deployment, and review decisions.}

\added{Structural critiques define the conditions under which that record may fail to protect affected people. \citet{Pasquale2015} traces the power asymmetries created by opaque institutions. \citet{Eubanks2018} documents how automated public administration can intensify exclusion through eligibility rules, classification, and surveillance. \citet{Power1997} shows how verification can become ritualized, allowing checklist completion to displace substantive judgment. These accounts establish the need for authority and remedy alongside documentation.}

\added{Every public-space AI contract therefore specifies municipal and independent access to current and historical documentation; the data, models, logs, interfaces, and designated personnel available for inspection or interview; review rights extending to suppliers whose work is material to the system; procedures for investigating complaints and systematic error; and the office authorized to pause use, require correction, order recommissioning, or terminate the system. Each artifact is linked to a decision gate, a responsible officer, a review trigger, and an available remedy. Missing, obsolete, or inaccessible documentation activates the corresponding contractual response.}
\label{rev:auditability}

\subsubsection{Data governance and non-extractive participation clauses}

Contracts should require a data governance plan specifying consent, retention, access control, and secondary-use restrictions. Where participatory data are collected, contracts should specify stewardship terms, conditions for reuse, and compensation and accessibility supports. These clauses respond to design justice critiques of extractive participation and to contextual integrity constraints on cross-context data flows \citep{CostanzaChock2020,Nissenbaum2010,Datafeminism}.

\subsubsection{Re-commissioning, update, and sunset clauses}

Contracts should include an explicit re-commissioning schedule and pricing structure for updates, audits, and retraining; conditions under which the city may pause or roll back the system; and a sunset clause specifying conditions for decommissioning and data retirement. These clauses operationalize the lifecycle requirement that systems be governed as evolving infrastructure rather than as one-time procurements \citep{Koseki2022,Sculley2015,NIST2023}.

\subsection{Procurement rules for commissioned streets}

Commissioning symmetry also implies that street procurement should incorporate algorithmic governance artifacts when AI-mediated evaluation or generation informs street design or maintenance prioritization. In such cases, the street project should include a project-specific excerpt of the \ValueRegister{} that identifies benchmark and contested dimensions; an evaluation protocol that specifies how disagreement and indeterminacy will be handled; and a re-evaluation cadence aligned with physical maintenance cycles. This requirement treats indicator selection and aggregation rules as part of commissioning, rather than as technical defaults, and it locates deliberation and documentation as prerequisites for using AI-mediated representations in high-impact decisions \citep{EspelandStevens2008,Porter1995}.

\subsection{Re-commissioning triggers}

Re-commissioning should be calendar-based and signal-based. Trigger categories include demographic and use change that alter the meaning of inclusion; drift or shifts in data coverage and quality; sustained indeterminacy on contested dimensions; complaint clusters indicating recurring harms or misunderstandings; and policy shifts that change objectives or decision boundaries \citep{Sculley2015,NIST2023}.

\subsection{Procurement summary}

Table~\ref{tab:procurement-clauses} summarizes baseline procurement and commissioning clauses that translate participatory governance commitments into enforceable contractual requirements for public-space AI systems.

\begin{table}[!htbp]
\centering
\small
\caption{Procurement and commissioning clauses that operationalize the \RightToAI{} and lifecycle governance for public-space AI systems and AI-mediated street projects.}
\label{tab:procurement-clauses}
\begin{tabularx}{\textwidth}{
>{\raggedright\arraybackslash}P{0.26\textwidth}
>{\raggedright\arraybackslash}X
}
\toprule
\textbf{Clause category} & \textbf{Minimum contract requirements} \\
\midrule
Disclosure and boundaries &
Intended uses and disallowed uses; decision boundary statement; public notice; prohibition on enforcement-adjacent repurposing without explicit authorization and safeguards. \\
\midrule
Auditability and access &
Model card and data documentation; evaluation results and update logs; independent audit rights; logging of outputs and downstream uses; access to versioned documentation. \\
\midrule
Data stewardship &
Consent and retention terms; restrictions on secondary use; governance of participatory data; privacy safeguards where applicable; compensation and accessibility supports for participation. \\
\midrule
Plural representation &
Evaluation plan specifying where disaggregation is required; explicit handling of disagreement and indeterminacy; documentation of contested dimensions and associated safeguards. \\
\midrule
Re-commissioning and sunset &
Update schedule and costs; triggers for pause and rollback; re-authorization requirements for scope changes; decommissioning and data retirement plan. \\
\bottomrule
\end{tabularx}
\end{table}
\FloatBarrier

\subsection{Example: PAU in Montréal}
\label{subsec:ch7-worked-example}

To reduce the gap between the governance architecture and municipal operations (RQ5), this subsection provides a compact worked example showing where the lifecycle checkpoints and artifacts attach to a typical public-works cycle in Montréal. The example is illustrative (roles and gates vary across departments and boroughs), but it makes explicit what must exist for the architecture to be implementable: decision authority, contractual access, resourcing for participation, and a scheduled pathway for updates and sunset.

\begin{table}[!htbp]
\centering
\small
\caption{Worked example mapping of PAU artifacts to municipal decision points in a Montréal public-works procurement and maintenance cycle.}
\label{tab:ch7-worked-example}
\begin{tabularx}{\textwidth}{P{0.16\textwidth}P{0.37\textwidth}Y}
\toprule
\textbf{Phase} & \textbf{Decision point (municipal hook)} & \textbf{Artifacts and outputs (minimum)} \\
\midrule
Co-framing & Department authorizes a pilot to audit a target issue (e.g., sidewalk accessibility) and sets a decision boundary for use & Charter (scope, disallowed uses, decision boundary); initial \ValueRegister{}; participation plan and budget; risk register \\
\midrule
Co-design & Procurement team drafts requirements / Request for Proposal and evaluation plan & Evaluation protocol including disagreement handling; data governance plan; documentation plan; procurement clauses from Table~\ref{tab:procurement-clauses} \\
\midrule
Co-implementation & Build and acceptance testing prior to pilot & Model card and data documentation; baseline evaluation (incl. subgroup reporting); traceability mapping; audit access configured \\
\midrule
Co-deployment & Pilot launch and operational use in prioritization workflow & Public notice; logging and complaint intake; pilot assessment memo; disagreement log; pause triggers and escalation routing \\
\midrule
Co-maintenance & Scheduled review aligned with annual budgeting and maintenance programming & Drift/coverage report; re-commissioning report; updated \ValueRegister{}; renew/rollback/sunset decision \\
\bottomrule
\end{tabularx}
\end{table}
\FloatBarrier

\section{Oversight Design}
\label{sec:ch7-oversight}

\subsection{Why \DelibCells{} are the institutional unit}

The governance architecture centers recurring \DelibCells{} because the empirical chapters show that deliberation can produce partially shared standards while preserving residual contestation on socially contested dimensions. A \DelibCell{} is not a one-off consultation session. It is a standing institutional unit that updates the \ValueRegister{}, reviews disagreement logs and monitoring reports, and exercises oversight with the capacity to recommend pause, revision, or decommissioning when governance triggers occur.

This design addresses two municipal failure modes. The first is technocratic closure, in which decisions are justified by a single metric produced by a tool that appears neutral but embeds contested assumptions \citep{EspelandStevens2008,Pasquale2015}. The second is consultation without power, in which participation exists but is weakly connected to decision authority \citep{Arnstein1969}. By tying \DelibCells{} to lifecycle checkpoints and pause authority, deliberation becomes part of how municipal institutions authorize and constrain the use of quantified representations under conditions of pluralism \citep{Habermas1996,Dewey1927}.

\subsection{Composition and selection}

For public-space contexts, composition should balance representativeness, lived experience expertise, and feasibility. A feasible baseline is 12 to 18 members selected through stratified sampling for demographic diversity, complemented by a limited number of stakeholder seats with transparent criteria. Members participate as voting participants; municipal staff provide technical and operational input as non-voting advisors; facilitation and audit expertise should be independent. Resourcing should include compensation, childcare, translation, accessibility supports, and plain-language materials to reduce participation bias \citep{CostanzaChock2020,BerditchevskaiaEtAl2021,Cornish2023,Sieber2024CivicParticipation}.

Because identity categories do not function as fixed proxies for perception, selection should be paired with disagreement-preserving practices that record divergence rather than treating it as failure. The governance purpose is to reduce systematic exclusion of those most likely to experience harms, while avoiding claims that any individual speaks as a stable representative for a group \citep{Crenshaw1989}.

\subsection{Cadence and municipal cycles}

A feasible cadence is quarterly sessions, with additional sessions triggered by high-impact projects, drift events, complaint clusters, or policy shifts. The cadence should align with municipal planning and maintenance cycles, including annual budgeting, capital planning, seasonal maintenance scheduling, and periodic updates to design guidelines and accessibility standards. Alignment supports commissioning symmetry by making the revision of evaluative standards part of routine institutional practice.

\subsection{Escalation and decision authority}

A \DelibCell{} can contribute to governance only if it is connected to decision authority. The architecture proposes a layered topology.

\begin{enumerate}
\item \textbf{Project team.} Implements the system, maintains logs, and responds to issues within the approved scope.
\item \textbf{Departmental AI and ethics office.} Ensures compliance with procurement clauses, documentation, monitoring, and audit readiness \citep{Leslie2019,NIST2023}.
\item \textbf{Deliberative cell.} Reviews contested dimensions, updates the \ValueRegister{}, and recommends pause, revision, or decommissioning under defined triggers.
\item \textbf{Council committee.} Authorizes high-impact deployments, approves major scope changes, and adjudicates policy shifts that alter decision boundaries.
\item \textbf{Independent office.} Provides an appeal pathway, monitors complaints, and can commission independent audits.
\end{enumerate}

\subsection{Recourse mechanisms}

Recourse should be actionable and time-bounded. The architecture specifies four tiers.

\begin{enumerate}
\item \textbf{Informal correction.} Residents can request correction of inaccurate documentation, public materials, or miscommunication about scope and limitations.
\item \textbf{Output contestation.} Residents can contest specific outputs and request review of underlying data, documentation, and interpretation.
\item \textbf{Use contestation.} Residents can contest use beyond scope, including mission creep, and request a pause pending review.
\item \textbf{System-level challenge.} Residents and groups can trigger an audit, re-commissioning, or decommissioning review when harms are systematic.
\end{enumerate}

Because representational outputs can be interpreted as objective truth, recourse should include communicative remedies such as public corrections, revisions to how uncertainty is presented, and changes to default aggregation and display rules when these create predictable misinterpretation risks \citep{Porter1995,EspelandStevens2008,burrell_how_2016}.

\subsection{Oversight topology}
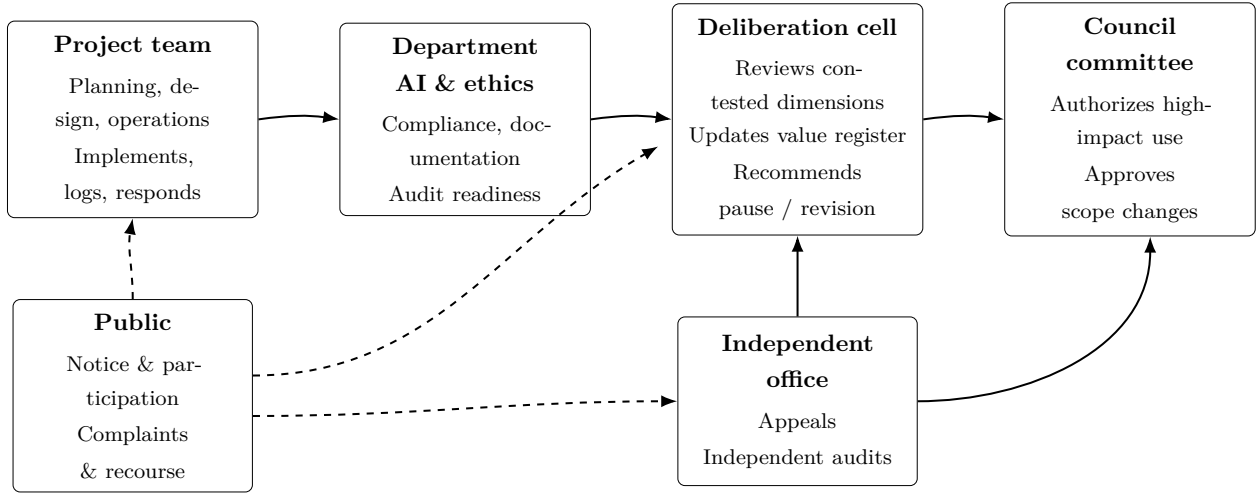
\begin{figure}[htbp]
\centering
\begin{adjustbox}{max width=\linewidth}
\begin{tikzpicture}[
  font=\footnotesize,
  >={Latex},
  box/.style={
    draw, rounded corners=3pt, align=center,
    inner xsep=7pt, inner ysep=6pt,
    text width=0.195\linewidth,
    minimum height=27mm
  },
  sbox/.style={
    box,
    text width=0.185\linewidth,
    minimum height=25mm
  },
  arrow/.style={->, line width=0.85pt},
  dashedarrow/.style={->, line width=0.85pt, dashed}
]

\node[box] (proj) {%
  {\bfseries Project team}\\[3pt]%
  {\scriptsize Planning, design, operations\\
  Implements, logs, responds}%
};

\node[box, right=12mm of proj] (ai) {%
  {\bfseries Department AI \& ethics}\\[3pt]%
  {\scriptsize Compliance, documentation\\
  Audit readiness}%
};

\node[box, right=12mm of ai] (cell) {%
  {\bfseries Deliberation cell}\\[3pt]%
  {\scriptsize Reviews contested dimensions\\
  Updates \ValueRegister{}\\
  Recommends pause / revision}%
};

\node[box, right=12mm of cell] (council) {%
  {\bfseries Council committee}\\[3pt]%
  {\scriptsize Authorizes high-impact use\\
  Approves scope changes}%
};

\node[sbox, below=12mm of proj] (public) {%
  {\bfseries Public}\\[3pt]%
  {\scriptsize Notice \& participation\\
  Complaints \& recourse}%
};

\node[sbox, below=12mm of cell] (omb) {%
  {\bfseries Independent office}\\[3pt]%
  {\scriptsize Appeals\\
  Independent audits}%
};

\draw[arrow] (proj.east) to[out=8, in=172] (ai.west);
\draw[arrow] (ai.east)   to[out=8, in=172] (cell.west);
\draw[arrow] (cell.east) to[out=8, in=172] (council.west);

\draw[dashedarrow] (public.north) to[out=90, in=260] (proj.south);

\draw[dashedarrow]
  ([yshift=3mm]public.east)
  to[out=0, in=210] ([xshift=-2mm,yshift=-4mm]cell.west);

\draw[dashedarrow]
  ([yshift=-3mm]public.east)
  to[out=0, in=180] (omb.west);

\draw[arrow] (omb.north) to[out=90, in=270] (cell.south);

\draw[arrow]
  (omb.east)
  to[out=0, in=270] ([xshift=3mm]council.south);

\end{tikzpicture}
\end{adjustbox}
\caption[Municipal oversight topology]{Oversight topology for municipal public-space AI governance. Deliberation cell provide recurring participatory oversight connected to departmental compliance functions and democratic authorization, with an independent recourse pathway. \added{Source: author's synthesis.}}
\label{fig:oversight-topology}
\end{figure}

Figure~\ref{fig:oversight-topology} shows how delivery, compliance, participatory review, and democratic authorization connect through an oversight and recourse topology for municipal public-space AI systems.

\FloatBarrier

\section{Feasibility and Risks}
\label{sec:ch7-feasibility}

\subsection{Capacity constraints}
Municipalities vary in technical capacity, legal resources, and participatory infrastructure. A governance architecture that assumes extensive in-house expertise or sustained consultation for every system risks tokenism or non-compliance. Risk-based governance frameworks emphasize proportionality, requiring stronger safeguards where systems have greater capacity to affect rights, access, or distributive outcomes \citep{NIST2023,EuropeanUnion2024}.

\medskip

A tiered implementation strategy scales governance requirements to impact. Tier~1 systems are low-impact internal analytics and require documentation, internal disclosure, and periodic review. Tier~2 systems support planning and maintenance decisions and require public notice, a project-specific charter and \ValueRegister{}, and \DelibCell{} review at checkpoints. Tier~3 systems produce public-facing representations or materially influence resource allocation and require full lifecycle checkpoints, independent audit readiness, and enforceable recourse. Tier~4 systems are enforcement-adjacent or otherwise rights-sensitive and require special authorization, due process safeguards, and, depending on context and democratic decision, prohibition or moratoria. This dissertation focuses on planning and design settings; governance for enforcement systems requires additional domain-specific constraints beyond the scope of the empirical demonstrations.

\medskip

\subsection{Legitimacy risks}

Deliberative mechanisms can be captured by organized interests, reproduce exclusions, or become symbolic exercises. The architecture mitigates these risks through transparent selection, rotation and term limits, conflict-of-interest rules, independent facilitation, and disagreement-preserving documentation that avoids treating consensus as a procedural requirement \citep{Arnstein1969,CostanzaChock2020}.
A related legitimacy risk is the tendency to treat quantified outputs as neutral truth. The sociology of quantification shows that the authority of metrics depends on institutional embedding and can displace contestation when outputs are treated as objective facts rather than as constructed indicators with limits \citep{Porter1995,EspelandStevens2008,Merry2016}. In municipal contexts, this risk can be reduced through decision boundary enforcement, documentation requirements, and recourse pathways that make the contestability of representations an administrative routine rather than an exceptional intervention.

\subsection{Data and modality limits}
The empirical chapters rely on visual and platform-mediated representations of streetscapes and design alternatives. Such representations have predictable limits. Static imagery omits on-site experience and dynamic conditions; coverage and image quality vary; and some markers of inclusion are context-specific and not reliably captured by a single modality. These are governance risks because they can produce uneven visibility across neighborhoods and systematic blind spots in what becomes measurable \citep{Ibrahim2020,Zhu2025,Kitchin2014a}.

Municipal procedures should treat data gaps, low-confidence regions, and persistent disagreement as triggers for additional evidence collection, such as participatory walks, on-site audits, or multimodal inputs, rather than treating model outputs as definitive. These triggers should be recorded in disagreement logs and reflected in the \ValueRegister{} as measurement limits and use constraints.

\subsection{Indeterminacy}
The LIVS results reported in Chapter~\ref{chap:livs} show that preference-based evaluation of inclusive public-space imagery yields substantial indeterminacy, particularly for contested dimensions (Figure~\ref{fig:livs-eval-criterion}). Indeterminacy is not reducible to annotation error. It can reflect value conflict, insufficient contextual information, or balanced tradeoffs that are not resolvable through a single aligned model \citep{pmlr-v267-mushkani25a,Sorensen2024}.

The governance response to sustained indeterminacy is procedural rather than purely technical. Appropriate responses include convening a \DelibCell{} to negotiate constraints and acceptable ranges; collecting additional context or multimodal evidence; narrowing the decision boundary so that outputs cannot authorize high-impact actions; and updating the \ValueRegister{} to classify the dimension as contested and to require stronger safeguards.

\subsection{Implementation}
A minimal viable implementation of the \RightToAI{} in municipal practice consists of public disclosure and decision boundaries for each system; a project-specific charter and \ValueRegister{}; model card and data documentation; an accessible complaint and review pathway; and scheduled re-commissioning. A full implementation adds recurring \DelibCells{} with pause authority, independent audit capacity and contractual audit rights, formal escalation to council committees for high-impact deployments, and continuous monitoring with drift reporting and versioned updates.

\added{The framework has governance validity only when its procedures can change an institutional decision. Its current evidence concerns framework design, participatory studies, bounded modeling, and preference evaluation. It has yet to be evaluated through a longitudinal municipal procurement, deployment, complaint, update, and sunset cycle. Procurement law, vendor intellectual-property claims, staffing, budgets, community compensation, records management, and the independence of review bodies can each narrow the authority described here.}

\added{Procedural compliance also creates a risk of hollow implementation. A completed model card, impact assessment, workshop, or audit can coexist with unchanged scope and authority. The framework addresses this risk by attaching each artifact to a named decision gate, an accountable office, a trigger, and a remedy. Empirical evaluation must test whether these links alter procurement terms, deployment boundaries, resource allocation, or system operation when disagreement, drift, or harm is documented.}

\added{Deliberative cells introduce risks of selection, fatigue, capture, and unequal capacity to participate. Their legitimacy depends on transparent selection, accessible formats, compensation, renewal, public reasons, conflict-of-interest rules, and an appeals route beyond the project team. The city must also record whose experience remains absent. A disagreement log gains institutional value through the standing of affected residents to initiate review and through an authority capable of changing the decision.}

\added{Cross-city use requires recommissioning. Montréal's criteria, subgroup distributions, partner relationships, imagery, and administrative arrangements supply the empirical basis for this thesis. Another city must identify its affected publics, elicit local criteria, sample its own spatial and temporal conditions, revalidate technical components, and negotiate the authority of oversight institutions. The framework supplies a procedure and a set of accountable artifacts; the legitimacy of their content is produced locally.}
\label{rev:ch7-feasibility}

\section{Conclusions}
\label{sec:ch7-conclusions}
This chapter specified a governance and implementation architecture that treats public-space AI systems as civic infrastructure governed by a municipal \RightToAI{}. The architecture is constrained by empirical conditions documented in the dissertation: evaluations are plural and dimension-dependent; deliberation can produce partially shared standards while preserving residual contestation; scaling through locally co-produced models is feasible but bounded by representational and modality limits; and preference evaluation yields substantial indeterminacy that should be handled procedurally rather than optimized away.
Operationally, the chapter specified governance requirements derived from propositions P1 to P4, including disagreement-preserving documentation and deliberation triggers; the \RightToAI{} as a municipal governance layer with minimum entitlements, prohibitions, and mandated procedures; a municipal workflow mapped to the augmented participatory lifecycle with checkpoints and pause authority; procurement clauses that secure auditability, data stewardship, and commissioning symmetry for both streets and models; oversight design centered on recurring \DelibCells{} linked to administrative and democratic decision authority; and a tiered implementation strategy for feasibility across municipal capacities.
The practical implication is that legitimacy in AI-mediated public space requires institutions that can represent and negotiate plural values, encode negotiated standards into both physical and algorithmic infrastructures, and maintain those standards through re-commissioning, audit, and recourse. The next chapter synthesizes the dissertation's contributions, answers the research questions, and specifies boundary conditions and future research directions.


\chapter{Conclusion}
\label{chap:conclusion}

Public-space governance increasingly depends on computational representations that render streets and their users legible as scores, rankings, forecasts, and generated alternatives. When these representations become embedded in municipal routines of problem definition, justification, and intervention, they operate as epistemic infrastructure: they shape what institutions treat as evidence, what forms of contestation are actionable, and which distributions of costs and benefits are normalized as technical output rather than political choice \citep{star1999ethnography,BowkerStar1999,Jasanoff2004,scott1998seeing,Kitchin2014a,Mattern2017a}. This shift is not reducible to a change in tools. It reflects a governance transformation in which commensuration, auditability, and procedural authority are increasingly delegated to metrics and models \citep{EspelandStevens1998,Porter1995,Desrosieres1998,Power1997,Strathern2000,Merry2016,LarteyLaw2025,Batty2024,Kitchin2023}. Under these conditions, legitimacy depends on whether the epistemic layer is authorized, maintainable, and contestable as civic infrastructure rather than treated as a neutral technical supplement.

\medskip

The dissertation advances an alternative: an urban Right to AI and an approach to pluralistic alignment in which value conflict is represented and negotiated as part of legitimate governance rather than averaged away. The normative premise is continuous with traditions of urban rights and democratic planning that treat public space as a distributive and recognitional good whose terms of access and belonging require justification to those affected \citep{Lefebvre1968,lefebvre1996writings,Harvey2003,Arnstein1969,Healey1997,Jacobs1961}. The methodological premise is that pluralism is not an exception state that disappears with more data. It is a structural condition of public-space evaluation that must be surfaced in forms that institutions can act on without converting disagreement into a single objective function \citep{Berlin1969,rawls1993political,Mouffe2000,selbst2019fairness,Binns2020,BarocasHardtNarayanan2023}. The empirical chapters demonstrate that disagreement and indeterminacy persist even under improved measurement and preference learning. The dissertation interprets this persistence as a governance signal that requires procedural handling rather than additional optimization pressure.

\section{Summary of chapters}
\label{sec:ch8_summary}

The dissertation links a planning claim, an HCI claim, and an AI-governance claim into a single evidence chain. The planning claim is that computational representations of public space should be governed as civic infrastructure when they become part of the institutional conditions under which public-space action is defined and justified \citep{scott1998seeing,Kitchin2014a,shelton2015actually,Mattern2017b,Batty2024}. The HCI claim is that, in civic contexts, system evaluation cannot be reduced to a single aggregate measure without importing unexamined assumptions about whose experience counts and which kinds of reasons are admissible \citep{CostanzaChock2020,selbst2019fairness,Sorensen2024}. The AI-governance claim is that municipal legitimacy requires more than technical performance: it requires enforceable entitlements to documentation, contestation, and recourse, alongside limits on high-impact uses and secondary uses that convert evaluative systems into instruments of punishment or exclusion \citep{Pasquale2015,Noble2018,Eubanks2018,Benjamin2019,EuropeanUnion2024,wachter2019reasonable,mushkani2025urbanaigovernanceembed}.

\medskip

Across the dissertation, pluralistic alignment is treated as a representation-and-negotiation problem. This framing is responsive to established results in sociotechnical systems research: abstraction and aggregation can hide materially consequential differences, fairness notions can conflict across levels of analysis, and measurement choices can relocate political disagreement into technical artifacts \citep{EspelandStevens1998,Porter1995,selbst2019fairness,Binns2020,BarocasHardtNarayanan2023}. In public space, this implies a governance requirement: institutions need procedures that can preserve and interpret heterogeneity, identify when standards are underdetermined, and specify how decisions will proceed when value conflict remains unresolved.

\medskip

Within this chain, the dissertation answers the research questions by establishing that (i) the algorithmic layer can function as epistemic infrastructure and therefore requires commissioning, maintenance, and oversight obligations analogous to other municipal infrastructures \citep{star1999ethnography,BowkerStar1999,scott1998seeing}; (ii) a rights-based framing is necessary to specify enforceable floors for documentation, contestability, and recourse under conditions of asymmetric expertise and vendor control \citep{Mitchell2019,gebru2021,EuropeanUnion2024,wachter2019reasonable}; and (iii) pluralism and indeterminacy must be treated as first-order features of the governance problem, with procedures that constrain inappropriate uses of model outputs and that route contested cases into deliberative and administrative processes rather than forced aggregation \citep{Habermas1996,Fung2004,Fishkin2009,NIST2023}.

\section{Contributions}
\label{sec:ch8_contributions}

The dissertation makes five contributions that are intended to be portable across municipal contexts while remaining sensitive to local political economies and histories.

First, it articulates an urban Right to AI tailored to public-space governance. The contribution is to specify rights not as aspirational principles but as design constraints on commissioning, deployment, updating, and decommissioning of systems that influence access and belonging in the city. This extends rights-based urban theory into the domain of computational representation, in which the capacity to contest and revise evidence artifacts becomes part of the conditions of democratic membership \citep{Lefebvre1968,lefebvre1996writings,Harvey2003,Jasanoff2004}.

\medskip

Second, it proposes pluralistic-alignment urbanism as a theoretical synthesis for governing coupled material and epistemic infrastructures. The contribution is to define alignment as a legitimacy problem under value pluralism, rather than as convergence to a single target. This clarifies why disagreements about public-space values cannot be treated as annotation error and why ``improving the model'' is often insufficient when the object of governance is contested \citep{Berlin1969,rawls1993political,Mouffe2000,selbst2019fairness,Sorensen2024,ConitzerEtAl2024}.

\medskip

Third, it contributes a values-to-infrastructure methodology that links participatory value articulation to scalable technical systems without requiring premature commensuration. This contribution aligns with established documentation and accountability practices for datasets and models by treating data, evaluation traces, and update decisions as civic artifacts that must be interpretable and auditable in institutional settings \citep{Cornish2023,sloane2022,Mitchell2019,Desrosieres1998,Porter1995}. It also situates intersectional and group-differentiated evaluation as a pragmatic response to known limitations of one-size-fits-all measures \citep{Crenshaw1989,BuolamwiniGebru2018,BarocasSelbst2016}.

\medskip

Fourth, it provides empirical evidence that disagreement and indeterminacy persist even under improved measurement and preference learning, and that deliberation can change patterns of convergence without eliminating residual conflict. The contribution is not the claim that deliberation produces consensus, but the demonstration that deliberation can transform the evidentiary status of judgments and reasons, making some standards more justifiable while clarifying where standards remain underdetermined \citep{Habermas1996,Fishkin2009,Fung2004}.

\medskip

Fifth, it offers an implementation architecture that connects rights to lifecycle governance and procurement. The contribution is to treat contracting, documentation, audit access, and update governance as central municipal levers, consistent with risk-based approaches that distinguish between low-impact and high-impact deployments and require proportionate controls \citep{EuropeanUnion2024,NIST2023,Kitchin2016}. This addresses a recurrent weakness in civic technology programs: participation is frequently decoupled from the procurement and maintenance mechanisms that determine whether public input has durable institutional effect.

\section{Implications}
\label{sec:ch8_implications}

\subsection{Implications for urban planning}

For planning, the core implication is that representational systems that shape problem definition and justification should be treated as part of planning's object of governance. This extends familiar concerns about legibility, expertise, and standard-setting into contexts where representations are computed, updated, and integrated into administrative routines \citep{scott1998seeing,Porter1995,Kitchin2014a,Batty2024,LarteyLaw2025}. Two practical consequences follow. First, the choice of evaluative criteria and their operationalization is a planning decision with distributive effects, not an implementation detail of analytics \citep{EspelandStevens1998,lazar2024lectureigoverningalgorithmic,Mehta2014}. Second, commissioning must incorporate maintenance and re-authorization. If streets require periodic re-commissioning through repair and redesign, then models that mediate how streets are evaluated require analogous cycles of review, documentation, and decommissioning, with clear boundaries on permissible uses \citep{star1999ethnography,BowkerStar1999,Kitchin2016,AnttiroikoDeJong2020,MitrasinovicMehta2021}.

\subsection{Implications for HCI}

For HCI, the implication is that civic systems should be designed and evaluated under conditions of pluralism, where disagreement is informative and where forced aggregation can misrepresent lived experience. This shifts evaluation from maximizing a single reliability or satisfaction measure to developing interfaces and methods that support contestability, reason-giving, and documented uncertainty \citep{CostanzaChock2020,selbst2019fairness}. It also reframes explanation and recourse as interaction design problems that are inseparable from institutional procedure. When a system produces outputs that are uncertain, indeterminate, or contested, interfaces should be able to represent these states without converting them into false precision, and should connect users to meaningful pathways for challenge and review \citep{Mitchell2019,gebru2021,wachter2019reasonable,sloane2022}.

\subsection{Implications for AI governance}

For AI governance, the dissertation emphasizes that municipal settings require governance of aggregation and secondary use, not only technical bias mitigation. Established critiques of algorithmic systems show how opacity, automation, and institutional incentives can entrench disadvantage when tools are used to allocate risk, resources, or enforcement attention \citep{Pasquale2015,Eubanks2018,Noble2018,Benjamin2019,Crawford2021}. In municipalities, a further risk is that an evaluative tool built for planning support can be repurposed for surveillance or punitive management. Risk-based regulatory and standards frameworks reinforce the need for proportional controls, documentation, and monitoring, and they provide actionable templates for operationalizing such requirements in procurement and oversight \citep{EuropeanUnion2024,NIST2023}. The dissertation adds that legitimacy also depends on how governance handles cases where values do not admit a determinate ordering. Under pluralistic alignment, residual disagreement and neutrality are conditions that trigger procedure, including deliberation, evidentiary supplementation, and explicit constraint of model use, rather than being treated as noise to be removed \citep{Habermas1996,Fung2004,BarocasHardtNarayanan2023}.

\section{Limitations}
\label{sec:ch8_limitations}

\added{The thesis supports claims at three levels: findings within the participating samples, bounded demonstrations of technical scaling and preference tuning, and a proposed architecture for municipal governance. The distinctions among these levels define the limits of the contribution.}

\added{At the empirical and procedural levels, purposive recruitment through community partners tied the samples to those networks, and small, overlapping participant categories restrict subgroup findings to within-sample comparisons. Agreement formed after small-group discussion also depends on facilitation, composition, time, fatigue, compromise, and willingness to dissent. The methods chapter documents these limits in detail. Audio records, reasons, and disagreement logs are needed to distinguish reasoned convergence from accommodation or silence and to retain unresolved claims in later decisions.}

\added{At the measurement level, street images omit non-visual and temporal conditions, while the trained models remain bounded by their sampled streets, labels, imagery, and validation design. LIVS adds a generative boundary: neutral judgments can reflect indeterminacy, ambiguity, fatigue, prompt sensitivity, or limited visual difference. The resulting maps and generated alternatives are auditable estimates under stated conditions. Field observation, renewed public inquiry, and dedicated review of harmful symbolism, stereotyping, and representational absence remain necessary before consequential use.}

\added{At the institutional level, PAU is an implementation architecture awaiting longitudinal municipal evaluation. Its operation presumes funds for accessible and compensated participation, access to vendor systems and records, staff with technical and legal capacity, independent review, public complaint channels, and officials willing to use pause or termination powers. Compliance rituals and institutional capture remain possible. Evaluation must therefore examine decisions changed, uses constrained, resources redistributed, and remedies delivered over time.}

\added{These limits leave two bounded empirical findings and one conceptual proposition. Plural judgments about public space can be elicited and represented with greater traceability than a single aggregate score. Image-based scaling and preference tuning can produce maps and generated alternatives within explicit sampling, modality, and evaluation bounds. The governance framework proposes institutions that preserve disagreement, assign standing, and connect documented review to the authority to change or halt a use.}
\label{rev:ch8-limitations}

\section{Future research}
\label{sec:ch8_future}

Future research should prioritize comparative replication across cities, with explicit attention to how local histories of segregation, mobility regimes, and governance structures shape both the content of public-space values and the feasibility of lifecycle oversight \citep{shelton2015actually,yigitcanlar2025editorial,LarteyLaw2025}. Such work can clarify which patterns of disagreement are context-dependent and which governance mechanisms are robust under varied institutional arrangements.

\medskip

A second line of work concerns multi-modal and situated measurement under privacy constraints. Integrating complementary modalities can reduce reliance on any single proxy, but it introduces new governance problems around data minimization, consent, and stewardship. Research is needed on how to combine heterogeneous evidence while preserving pluralism, documenting uncertainty, and preventing new forms of surveillance \citep{Gabrys2016,Crawford2021,NIST2023}. This agenda also motivates work on community data stewardship models that treat civic datasets as governed resources, with benefit-sharing and enforceable limits on secondary use \citep{Ostrom1990,Ostrom2009,marmor2015privacy,Datafeminism}.

\medskip

A third line concerns institutional experiments that evaluate deliberative and administrative procedures for handling contested cases, including how deliberation interacts with documentation design, facilitation practices, and accountability mechanisms over repeated commissioning cycles \citep{Habermas1996,Fishkin2009,Fung2004}. In parallel, applied legal and operational research on procurement templates and audit rights can test which contractual mechanisms reliably support maintainability, contestability, and decommissioning in municipal contexts \citep{EuropeanUnion2024,Mitchell2019,gebru2021,zhang2024right,kaminski2021right}.

\clearpage

\section{Concluding remarks}
\label{sec:ch8_concluding}

\deleted{The dissertation argues that contemporary public-space governance is increasingly enacted through a dual infrastructure: streets and the computational representations through which streets are evaluated and governed. Under conditions of value pluralism, legitimacy cannot be achieved by treating models and metrics as neutral arbiters. It requires institutions that can authorize, document, contest, and maintain the epistemic layer as civic infrastructure, and that can route disagreement and indeterminacy into procedures capable of producing justified standards and accountable action. A Right to AI specifies minimum entitlements and limits for such systems. Pluralistic alignment specifies how heterogeneous values can be represented without premature commensuration, so that the city can make explicit, contestable choices about what is measured, what is optimized, and what must remain subject to democratic judgment.}

\added{Public-space AI participates in defining, comparing, and acting on urban conditions. Once scores, maps, and synthetic images enter budgeting, prioritization, design, or public justification, choices about criteria, data, aggregation, and evaluation become municipal responsibilities.}

\added{The empirical studies establish bounded findings. Evaluations of shared streetscape stimuli vary by criterion and social position. Structured discussion changes the evidentiary record and leaves some judgments contested. Co-produced labels can support subgroup-aware prediction and mapping within stated sampling and modality limits. Neutral preferences remain common in generative evaluation and constrain what preference tuning can justify. Municipal review must therefore retain recorded disagreement, disclose the procedure and imagery coverage, and treat neutral judgments as limits on use.}

\added{The thesis's principal conceptual contribution is to make the evidence-production chain an object of municipal governance. Civic experience becomes categories, labels, models, maps, visualizations, and administrative action through that chain, which operates as an epistemic infrastructure alongside streets themselves. The \RightToAI{} states the corresponding procedural floor: notice, authority at defined gates, access to documentation, standing to contest, and recourse capable of changing a use.}

\added{Pluralistic-Alignment Urbanism supplies an implementation architecture for those duties. Co-produced standards move from framing to procurement, deployment, maintenance, and recourse through named decision gates, versioned artifacts, disagreement logs, review triggers, and assigned remedies. Three procedural conditions make claims of municipal legitimacy assessable: standing for affected communities, visible disagreement, and review authority capable of changing a decision. Each city must identify affected publics, elicit local criteria, document known gaps in participation and evidence, and establish authority through which review can change a decision.}
\label{rev:conclusion}

\clearpage
\bibliographystyle{apalike2-doi}
\bibliography{0_01_references}

@article{Aizenberg2020,
  author  = {Aizenberg, Evgeni and van den Hoven, Jeroen},
  title   = {Designing for Human Rights in {AI}},
  journal = {Big Data \& Society},
  year    = {2020},
  volume  = {7},
  number  = {2},
  pages   = {2053951720949566},
  doi     = {10.1177/2053951720949566},
  url     = {https://doi.org/10.1177/2053951720949566}
}

@article{AmeriDutta2005,
  author  = {Ameri, Farhad and Dutta, Deba},
  title   = {Product Lifecycle Management: Closing the Knowledge Loops},
  journal = {Computer-Aided Design and Applications},
  year    = {2005},
  volume  = {2},
  number  = {5},
  pages   = {577--590},
  doi     = {10.1080/16864360.2005.10738322},
  url     = {https://www.cad-journal.net/files/vol_2/Vol2No5.html}
}

@misc{AngwinEtAl2016,
  author       = {Angwin, Julia and Larson, Jeff and Mattu, Surya and Kirchner, Lauren},
  title        = {Machine Bias: There's Software Used Across the Country to Predict Future Criminals. And It's Biased Against Blacks},
  howpublished = {ProPublica},
  year         = {2016},
  month        = {May},
  note         = {Published May 23, 2016},
  url          = {https://www.propublica.org/article/machine-bias-risk-assessments-in-criminal-sentencing}
}

@book{AnttiroikoDeJong2020,
  author    = {Anttiroiko, Ari-Veikko and de Jong, Martin},
  title     = {{The Inclusive City: The Theory and Practice of Creating Shared Urban Prosperity}},
  year      = {2020},
  publisher = {Springer International Publishing},
  address   = {Cham},
  isbn      = {9783030613648},
  doi       = {10.1007/978-3-030-61365-5},
  url       = {https://doi.org/10.1007/978-3-030-61365-5}
}

@article{Arnstein1969,
  author  = {Arnstein, Sherry R.},
  title   = {{A Ladder Of Citizen Participation}},
  journal = {Journal of the American Institute of Planners},
  year    = {1969},
  volume  = {35},
  number  = {4},
  pages   = {216--224},
  doi     = {10.1080/01944366908977225},
  url     = {https://doi.org/10.1080/01944366908977225}
}

@article{Aroyo2015,
  author  = {Aroyo, Lora and Welty, Chris},
  title   = {{Truth Is a Lie: Crowd Truth and the Seven Myths of Human Annotation}},
  journal = {AI Magazine},
  year    = {2015},
  volume  = {36},
  number  = {1},
  pages   = {15--24},
  doi     = {10.1609/aimag.v36i1.2564},
  url     = {https://doi.org/10.1609/aimag.v36i1.2564}
}

@article{asaro_transforming_2000,
  author  = {Asaro, Peter M.},
  title   = {{Transforming society by transforming technology: the science and politics of participatory design}},
  journal = {Accounting, Management and Information Technologies},
  year    = {2000},
  volume  = {10},
  number  = {4},
  pages   = {257--290},
  doi     = {10.1016/s0959-8022(00)00004-7},
  url     = {https://doi.org/10.1016/s0959-8022(00)00004-7}
}

@inproceedings{AssalChiasson2018,
  author    = {Assal, Hala and Chiasson, Sonia},
  title     = {Security in the Software Development Lifecycle},
  booktitle = {Fourteenth Symposium on Usable Privacy and Security (SOUPS 2018)},
  year      = {2018},
  pages     = {281--296},
  publisher = {USENIX Association},
  address   = {Baltimore, MD},
  isbn      = {9781939133106},
  url       = {https://www.usenix.org/conference/soups2018/presentation/assal}
}

@misc{bai2022rlhf,
  author        = {Bai, Yuntao and Jones, Andy and Ndousse, Kamal and Askell, Amanda and Chen, Anna and DasSarma, Nova and Drain, Dawn and Fort, Stanislav and Ganguli, Deep and Henighan, Tom and Joseph, Nicholas and Kadavath, Saurav and Kernion, Jackson and Conerly, Tom and El-Showk, Sheer and Elhage, Nelson and Hatfield-Dodds, Zac and Hernandez, Danny and Hume, Tristan and Johnston, Scott and Kravec, Shauna and Lovitt, Liane and Nanda, Neel and Olsson, Catherine and Amodei, Dario and Brown, Tom and Clark, Jack and McCandlish, Sam and Olah, Chris and Mann, Ben and Kaplan, Jared},
  title         = {Training a Helpful and Harmless Assistant with Reinforcement Learning from Human Feedback},
  year          = {2022},
  eprint        = {2204.05862},
  archivePrefix = {arXiv},
  primaryClass  = {cs.CL},
  doi           = {10.48550/arXiv.2204.05862},
  url           = {https://arxiv.org/abs/2204.05862},
  note          = {arXiv preprint}
}

@article{BarocasSelbst2016,
  author  = {Barocas, Solon and Selbst, Andrew D.},
  title   = {Big Data's Disparate Impact},
  journal = {California Law Review},
  year    = {2016},
  volume  = {104},
  number  = {3},
  pages   = {671--732},
  doi     = {10.15779/Z38BG31},
  url     = {https://lawcat.berkeley.edu/record/1127463}
}

@book{BarocasHardtNarayanan2023,
  author    = {Barocas, Solon and Hardt, Moritz and Narayanan, Arvind},
  title     = {Fairness and Machine Learning: Limitations and Opportunities},
  year      = {2023},
  publisher = {The MIT Press},
  address   = {Cambridge, MA},
  isbn      = {9780262048613},
  url       = {https://mitpress.mit.edu/9780262048613/fairness-and-machine-learning/}
}

@article{BarredoArrieta2020,
  author  = {Barredo Arrieta, Alejandro and Díaz-Rodríguez, Natalia and Del Ser, Javier and Bennetot, Adrien and Tabik, Siham and Barbado, Alberto and Garcia, Salvador and Gil-Lopez, Sergio and Molina, Daniel and Benjamins, Richard and Chatila, Raja and Herrera, Francisco},
  title   = {{Explainable Artificial Intelligence (XAI): Concepts, taxonomies, opportunities and challenges toward responsible AI}},
  journal = {Information Fusion},
  year    = {2020},
  volume  = {58},
  pages   = {82--115},
  doi     = {10.1016/j.inffus.2019.12.012},
  url     = {https://doi.org/10.1016/j.inffus.2019.12.012}
}

@article{Bartko1966,
  author  = {Bartko, John J.},
  title   = {The Intraclass Correlation Coefficient as a Measure of Reliability},
  journal = {Psychological Reports},
  volume  = {19},
  number  = {1},
  pages   = {3--11},
  year    = {1966},
  doi     = {10.2466/pr0.1966.19.1.3},
  url     = {https://journals.sagepub.com/doi/10.2466/pr0.1966.19.1.3}
}

@book{Batty2018,
  author    = {Batty, Michael},
  title     = {Inventing Future Cities},
  year      = {2018},
  publisher = {The MIT Press},
  address   = {Cambridge, MA},
  isbn      = {9780262038959},
  url       = {https://mitpress.mit.edu/9780262038959/inventing-future-cities/}
}

@book{Batty2024,
  author    = {Batty, Michael},
  title     = {The Computable City: Histories, Technologies, Stories, Predictions},
  publisher = {The MIT Press},
  address   = {Cambridge, MA},
  year      = {2024},
  isbn      = {9780262547574},
  url       = {https://mitpress.mit.edu/9780262547574/the-computable-city/}
}

@book{Baudrillard1970,
  author    = {Baudrillard, Jean},
  title     = {La société de consommation: ses mythes, ses structures},
  year      = {1970},
  publisher = {Denoël},
  address   = {Paris},
  url       = {https://archive.org/details/lasocietedeconso0000baud}
}

@misc{BeckEtAl2001,
  author = {Beck, Kent and Beedle, Mike and van Bennekum, Arie and Cockburn, Alistair and Cunningham, Ward and Fowler, Martin and Grenning, James and Highsmith, Jim and Hunt, Andrew and Jeffries, Ron and Kern, Jon and Marick, Brian and Martin, Robert C. and Mellor, Steve and Schwaber, Ken and Sutherland, Jeff and Thomas, Dave},
  title  = {Manifesto for Agile Software Development},
  year   = {2001},
  url    = {https://agilemanifesto.org/}
}

@article{Beer2016,
  author  = {Beer, David},
  title   = {The Social Power of Algorithms},
  journal = {Information, Communication \& Society},
  year    = {2017},
  volume  = {20},
  number  = {1},
  pages   = {1--13},
  doi     = {10.1080/1369118X.2016.1216147},
  url     = {https://doi.org/10.1080/1369118X.2016.1216147}
}

@book{Benjamin2019,
  author    = {Benjamin, Ruha},
  title     = {Race After Technology: Abolitionist Tools for the New Jim Code},
  year      = {2019},
  publisher = {Polity},
  address   = {Medford, MA},
  isbn      = {9781509526406},
  url       = {https://www.wiley.com/en-in/Race\%2BAfter\%2BTechnology\%3A\%2BAbolitionist\%2BTools\%2Bfor\%2Bthe\%2BNew\%2BJim\%2BCode-p-9781509526406}
}

@book{Benkler2006,
  author    = {Benkler, Yochai},
  title     = {The Wealth of Networks: How Social Production Transforms Markets and Freedom},
  year      = {2006},
  publisher = {Yale University Press},
  address   = {New Haven, CT},
  isbn      = {9780300110562},
  url       = {https://www.jstor.org/stable/j.ctt1njknw}
}

@techreport{BerditchevskaiaEtAl2021,
  author      = {Berditchevskaia, Aleks and Peach, Kathy and Malliaraki, Eirini},
  title       = {Participatory {AI} for Humanitarian Innovation: A Briefing Paper},
  institution = {Nesta},
  address     = {London},
  type        = {Briefing paper},
  month       = sep,
  year        = {2021},
  url         = {https://www.nesta.org.uk/report/participatory-ai-humanitarian-innovation-briefing-paper/}
}

@book{Berlin1969,
  author    = {Berlin, Isaiah},
  title     = {Four Essays on Liberty},
  publisher = {Oxford University Press},
  address   = {London},
  year      = {1969},
  isbn      = {9780192810342},
  url       = {https://search.worldcat.org/title/Four-essays-on-liberty/oclc/34866}
}

@inproceedings{bhatia2021multi,
  author    = {Bhatia, Kush and Pananjady, Ashwin and Bartlett, Peter L. and Dragan, Anca and Wainwright, Martin J.},
  title     = {Preference learning along multiple criteria: A game-theoretic perspective},
  booktitle = {Advances in Neural Information Processing Systems},
  editor    = {Larochelle, Hugo and Ranzato, Marc'Aurelio and Hadsell, Raia and Balcan, Maria-Florina and Lin, Hsuan-Tien},
  volume    = {33},
  pages     = {7413--7424},
  publisher = {Curran Associates, Inc.},
  year      = {2020},
  url       = {https://proceedings.neurips.cc/paper/2020/hash/52f4691a4de70b3c441bca6c546979d9-Abstract.html}
}

@inproceedings{Binns2020,
  author    = {Binns, Reuben},
  title     = {{On the apparent conflict between individual and group fairness}},
  booktitle = {Proceedings of the 2020 Conference on Fairness, Accountability, and Transparency},
  year      = {2020},
  pages     = {514--524},
  publisher = {ACM},
  doi       = {10.1145/3351095.3372864},
  url       = {https://doi.org/10.1145/3351095.3372864}
}

@inproceedings{Birhane2022-power,
  author    = {Birhane, Abeba and Isaac, William and Prabhakaran, Vinodkumar and D{\'i}az, Mark and Elish, Madeleine Clare and Gabriel, Iason and Mohamed, Shakir},
  title     = {Power to the People? Opportunities and Challenges for Participatory {AI}},
  booktitle = {Equity and Access in Algorithms, Mechanisms, and Optimization ({EAAMO} '22)},
  year      = {2022},
  month     = oct,
  pages     = {1--8},
  publisher = {Association for Computing Machinery},
  isbn      = {9781450394772},
  doi       = {10.1145/3551624.3555290},
  url       = {https://doi.org/10.1145/3551624.3555290}
}

@article{BiscontiEtAl2023,
  author  = {Bisconti, Piercosma and Orsitto, Davide and Fedorczyk, Federica and Brau, Fabio and Capasso, Marianna and De Marinis, Lorenzo and Eken, Hüseyin and Merenda, Federica and Forti, Mirko and Pacini, Marco and Schettini, Claudia},
  title   = {Maximizing Team Synergy in {AI}-Related Interdisciplinary Groups: An Interdisciplinary-by-Design Iterative Methodology},
  journal = {AI \& Society},
  year    = {2023},
  volume  = {38},
  number  = {4},
  pages   = {1443--1452},
  doi     = {10.1007/s00146-022-01518-8},
  url     = {https://link.springer.com/article/10.1007/s00146-022-01518-8}
}

@book{Bishop2006,
  author    = {Bishop, Christopher M.},
  title     = {Pattern Recognition and Machine Learning},
  series    = {Information Science and Statistics},
  year      = {2006},
  edition   = {1},
  publisher = {Springer},
  address   = {New York, NY},
  isbn      = {9780387310732},
  url       = {https://link.springer.com/book/9780387310732}
}

@book{Bohman2000,
  author    = {Bohman, James},
  title     = {Public Deliberation: Pluralism, Complexity, and Democracy},
  year      = {2000},
  publisher = {The MIT Press},
  address   = {Cambridge, MA},
  isbn      = {9780262522786},
  url       = {https://mitpress.mit.edu/9780262522786/public-deliberation/}
}

@misc{BommasaniEtAl2022,
  author        = {Bommasani, Rishi and Hudson, Drew A. and Adeli, Ehsan and Altman, Russ and Arora, Simran and von Arx, Sydney and Bernstein, Michael S. and Bohg, Jeannette and Bosselut, Antoine and Brunskill, Emma and Brynjolfsson, Erik and Buch, Shyamal and Card, Dallas and Castellon, Rodrigo and Chatterji, Niladri and Chen, Annie and Creel, Kathleen and Davis, Jared Quincy and Demszky, Dora and Donahue, Chris and Doumbouya, Moussa and Durmus, Esin and Ermon, Stefano and Etchemendy, John and Ethayarajh, Kawin and Fei-Fei, Li and Finn, Chelsea and Gale, Trevor and Gillespie, Lauren and Goel, Karan and Goodman, Noah and Grossman, Shelby and Guha, Neel and Hashimoto, Tatsunori and Henderson, Peter and Hewitt, John and Ho, Daniel E. and Hong, Jenny and Hsu, Kyle and Huang, Jing and Icard, Thomas and Jain, Saahil and Jurafsky, Dan and Kalluri, Pratyusha and Karamcheti, Siddharth and Keeling, Geoff and Khani, Fereshte and Khattab, Omar and Koh, Pang Wei and Krass, Mark and Krishna, Ranjay and Kuditipudi, Rohith and Kumar, Ananya and Ladhak, Faisal and Lee, Mina and Lee, Tony and Leskovec, Jure and Levent, Isabelle and Li, Xiang Lisa and Li, Xuechen and Ma, Tengyu and Malik, Ali and Manning, Christopher D. and Mirchandani, Suvir and Mitchell, Eric and Munyikwa, Zanele and Nair, Suraj and Narayan, Avanika and Narayanan, Deepak and Newman, Ben and Nie, Allen and Niebles, Juan Carlos and Nilforoshan, Hamed and Nyarko, Julian and Ogut, Giray and Orr, Laurel and Papadimitriou, Isabel and Park, Joon Sung and Piech, Chris and Portelance, Eva and Potts, Christopher and Raghunathan, Aditi and Reich, Rob and Ren, Hongyu and Rong, Frieda and Roohani, Yusuf and Ruiz, Camilo and Ryan, Jack and Ré, Christopher and Sadigh, Dorsa and Sagawa, Shiori and Santhanam, Keshav and Shih, Andy and Srinivasan, Krishnan and Tamkin, Alex and Taori, Rohan and Thomas, Armin W. and Tramèr, Florian and Wang, Rose E. and Wang, William and Wu, Bohan and Wu, Jiajun and Wu, Yuhuai and Xie, Sang Michael and Yasunaga, Michihiro and You, Jiaxuan and Zaharia, Matei and Zhang, Michael and Zhang, Tianyi and Zhang, Xikun and Zhang, Yuhui and Zheng, Lucia and Zhou, Kaitlyn and Liang, Percy},
  title         = {On the Opportunities and Risks of Foundation Models},
  year          = {2022},
  eprint        = {2108.07258},
  archiveprefix = {arXiv},
  primaryclass  = {cs.LG},
  note          = {arXiv:2108.07258, version 3, revised July 12, 2022},
  doi           = {10.48550/arXiv.2108.07258},
  url           = {https://arxiv.org/abs/2108.07258}
}

@book{BowkerStar1999,
  author    = {Bowker, Geoffrey C. and Star, Susan Leigh},
  title     = {Sorting Things Out: Classification and Its Consequences},
  publisher = {The MIT Press},
  address   = {Cambridge, MA},
  year      = {1999},
  doi       = {10.7551/mitpress/6352.001.0001},
  isbn      = {9780262024617},
  url       = {https://mitpress.mit.edu/9780262522953/sorting-things-out/}
}

@article{Brabham2009,
  author   = {Brabham, Daren C.},
  title    = {Crowdsourcing the Public Participation Process for Planning Projects},
  journal  = {Planning Theory},
  year     = {2009},
  volume   = {8},
  number   = {3},
  pages    = {242--262},
  doi      = {10.1177/1473095209104824},
  url      = {https://doi.org/10.1177/1473095209104824}
}

@article{BraunClarke2006,
  author  = {Braun, Virginia and Clarke, Victoria},
  title   = {Using Thematic Analysis in Psychology},
  journal = {Qualitative Research in Psychology},
  volume  = {3},
  number  = {2},
  pages   = {77--101},
  year    = {2006},
  doi     = {10.1191/1478088706qp063oa},
  url     = {https://www.tandfonline.com/doi/abs/10.1191/1478088706qp063oa}
}

@article{Brayne2017,
  author  = {Brayne, Sarah},
  title   = {{Big Data Surveillance: The Case of Policing}},
  journal = {American Sociological Review},
  year    = {2017},
  volume  = {82},
  number  = {5},
  pages   = {977--1008},
  doi     = {10.1177/0003122417725865},
  url     = {https://doi.org/10.1177/0003122417725865}
}

@article{Broderick2022,
  author  = {Broderick, Lliam Anthony},
  title   = {Homeless Encampments, Hotels, and Equitable Access to Public Space: Socially Sustainable Post-Pandemic Public Spaces in British Columbia, Canada},
  journal = {Tourism Cases},
  year    = {2022},
  volume  = {2022},
  number  = {1},
  pages   = {tourism20220010},
  doi     = {10.1079/tourism.2022.0010},
  url     = {https://doi.org/10.1079/tourism.2022.0010}
}

@book{Bryman2012,
  author    = {Bryman, Alan},
  title     = {Social Research Methods},
  edition   = {4},
  publisher = {Oxford University Press},
  address   = {Oxford},
  year      = {2012},
  isbn      = {9780199588053},
  url       = {https://books.google.com/books?id=vCq5m2hPkOMC}
}

@inproceedings{BuolamwiniGebru2018,
  author    = {Buolamwini, Joy and Gebru, Timnit},
  title     = {Gender Shades: Intersectional Accuracy Disparities in Commercial Gender Classification},
  booktitle = {Proceedings of the 1st Conference on Fairness, Accountability and Transparency},
  editor    = {Friedler, Sorelle A. and Wilson, Christo},
  year      = {2018},
  series    = {Proceedings of Machine Learning Research},
  volume    = {81},
  pages     = {77--91},
  publisher = {PMLR},
  url       = {https://proceedings.mlr.press/v81/buolamwini18a.html}
}

@article{burrell_how_2016,
  author  = {Burrell, Jenna},
  title   = {How the Machine `Thinks': Understanding Opacity in Machine Learning Algorithms},
  journal = {Big Data \& Society},
  year    = {2016},
  volume  = {3},
  number  = {1},
  pages   = {2053951715622512},
  doi     = {10.1177/2053951715622512},
  url     = {https://doi.org/10.1177/2053951715622512}
}

@article{cachat-rosset_diversity_2023,
  author  = {Cachat-Rosset, Gaelle and Klarsfeld, Alain},
  title   = {Diversity, Equity, and Inclusion in Artificial Intelligence: An Evaluation of Guidelines},
  journal = {Applied Artificial Intelligence},
  year    = {2023},
  volume  = {37},
  number  = {1},
  pages   = {2176618},
  doi     = {10.1080/08839514.2023.2176618},
  url     = {https://doi.org/10.1080/08839514.2023.2176618}
}

@article{calabrese_barton_beyond_2020,
  author  = {Calabrese Barton, Angela and Tan, Edna},
  title   = {{Beyond Equity as Inclusion: A Framework of “Rightful Presence” for Guiding Justice-Oriented Studies in Teaching and Learning}},
  journal = {Educational Researcher},
  year    = {2020},
  volume  = {49},
  number  = {6},
  pages   = {433--440},
  doi     = {10.3102/0013189x20927363},
  url     = {https://doi.org/10.3102/0013189x20927363}
}

@article{CarifioPerla2008,
  author  = {Carifio, James and Perla, Rocco},
  title   = {Resolving the 50-Year Debate Around Using and Misusing {Likert} Scales},
  journal = {Medical Education},
  volume  = {42},
  number  = {12},
  pages   = {1150--1152},
  year    = {2008},
  doi     = {10.1111/j.1365-2923.2008.03172.x},
  url     = {https://asmepublications.onlinelibrary.wiley.com/doi/10.1111/j.1365-2923.2008.03172.x}
}

@article{CarnemollaEtAl2021,
  author  = {Carnemolla, Phillippa and Robinson, Sally and Lay, Kiri},
  title   = {Towards Inclusive Cities and Social Sustainability: A Scoping Review of Initiatives to Support the Inclusion of People with Intellectual Disability in Civic and Social Activities},
  journal = {City, Culture and Society},
  volume  = {25},
  pages   = {100398},
  year    = {2021},
  doi     = {10.1016/j.ccs.2021.100398},
  url     = {https://www.sciencedirect.com/science/article/pii/S187791662100028X}
}

@inproceedings{Chakraborty2024,
  author    = {Chakraborty, Souradip and Qiu, Jiahao and Yuan, Hui and Koppel, Alec and Manocha, Dinesh and Huang, Furong and Bedi, Amrit and Wang, Mengdi},
  title     = {{M}ax{M}in-{RLHF}: Alignment with Diverse Human Preferences},
  booktitle = {Proceedings of the 41st International Conference on Machine Learning},
  editor    = {Salakhutdinov, Ruslan and Kolter, Zico and Heller, Katherine and Weller, Adrian and Oliver, Nuria and Scarlett, Jonathan and Berkenkamp, Felix},
  series    = {Proceedings of Machine Learning Research},
  volume    = {235},
  pages     = {6116--6135},
  publisher = {PMLR},
  month     = jul,
  year      = {2024},
  url       = {https://proceedings.mlr.press/v235/chakraborty24b.html}
}

@article{Chouldechova2017,
  author  = {Chouldechova, Alexandra},
  title   = {{Fair Prediction with Disparate Impact: A Study of Bias in Recidivism Prediction Instruments}},
  journal = {Big Data},
  year    = {2017},
  volume  = {5},
  number  = {2},
  pages   = {153--163},
  doi     = {10.1089/big.2016.0047},
  url     = {https://doi.org/10.1089/big.2016.0047}
}

@inproceedings{ChristianoEtAl2017,
  author    = {Christiano, Paul F. and Leike, Jan and Brown, Tom and Martic, Miljan and Legg, Shane and Amodei, Dario},
  title     = {Deep Reinforcement Learning from Human Preferences},
  booktitle = {Advances in Neural Information Processing Systems},
  editor    = {Guyon, Isabelle and von Luxburg, Ulrike and Bengio, Samy and Wallach, Hanna and Fergus, Rob and Vishwanathan, S. V. N. and Garnett, Roman},
  volume    = {30},
  pages     = {4299--4307},
  publisher = {Curran Associates, Inc.},
  year      = {2017},
  url       = {https://proceedings.neurips.cc/paper/2017/hash/d5e2c0adad503c91f91df240d0cd4e49-Abstract.html}
}

@article{CohenSuzor2024,
  author  = {Cohen, Tegan and Suzor, Nicolas P.},
  title   = {{Contesting the public interest in AI governance}},
  journal = {Internet Policy Review},
  year    = {2024},
  volume  = {13},
  number  = {3},
  doi     = {10.14763/2024.3.1794},
  url     = {https://doi.org/10.14763/2024.3.1794}
}

@inproceedings{ConitzerEtAl2024,
  author    = {Conitzer, Vincent and Freedman, Rachel and Heitzig, Jobst and Holliday, Wesley H. and Jacobs, Bob M. and Lambert, Nathan and Mosse, Milan and Pacuit, Eric and Russell, Stuart and Schoelkopf, Hailey and Tewolde, Emanuel and Zwicker, William S.},
  title     = {Position: Social Choice Should Guide {AI} Alignment in Dealing with Diverse Human Feedback},
  booktitle = {Proceedings of the 41st International Conference on Machine Learning},
  editor    = {Salakhutdinov, Ruslan and Kolter, Zico and Heller, Katherine and Weller, Adrian and Oliver, Nuria and Scarlett, Jonathan and Berkenkamp, Felix},
  year      = {2024},
  series    = {Proceedings of Machine Learning Research},
  volume    = {235},
  pages     = {9346--9360},
  publisher = {PMLR},
  url       = {https://proceedings.mlr.press/v235/conitzer24a.html}
}

@inproceedings{Cordts2016,
  author    = {Cordts, Marius and Omran, Mohamed and Ramos, Sebastian and Rehfeld, Timo and Enzweiler, Markus and Benenson, Rodrigo and Franke, Uwe and Roth, Stefan and Schiele, Bernt},
  title     = {{The Cityscapes Dataset for Semantic Urban Scene Understanding}},
  booktitle = {2016 IEEE Conference on Computer Vision and Pattern Recognition (CVPR)},
  year      = {2016},
  pages     = {3213--3223},
  publisher = {IEEE},
  doi       = {10.1109/cvpr.2016.350},
  url       = {https://doi.org/10.1109/cvpr.2016.350}
}

@incollection{corner1999,
  author    = {Corner, James},
  title     = {The Agency of Mapping: Speculation, Critique, and Invention},
  booktitle = {Mappings},
  editor    = {Cosgrove, Denis},
  publisher = {Reaktion Books},
  address   = {London},
  year      = {1999},
  pages     = {213--252},
  isbn      = {9781861890214},
  url       = {https://reaktionbooks.co.uk/work/mappings}
}

@article{Cornish2023,
  author  = {Cornish, Flora and Breton, Nancy and Moreno-Tabarez, Ulises and Delgado, Jenna and Rua, Mohi and {de-Graft Aikins}, Ama and Hodgetts, Darrin},
  title   = {Participatory action research},
  journal = {Nature Reviews Methods Primers},
  volume  = {3},
  number  = {1},
  pages   = {34},
  year    = {2023},
  doi     = {10.1038/s43586-023-00214-1},
  url     = {https://www.nature.com/articles/s43586-023-00214-1}
}

@book{CostanzaChock2020,
  author    = {Costanza-Chock, Sasha},
  title     = {Design Justice: Community-Led Practices to Build the Worlds We Need},
  year      = {2020},
  publisher = {The MIT Press},
  address   = {Cambridge, MA},
  isbn      = {9780262043458},
  doi       = {10.7551/mitpress/12255.001.0001},
  url       = {https://direct.mit.edu/books/oa-monograph/4605/Design-JusticeCommunity-Led-Practices-to-Build-the}
}

@book{Crawford2021,
  author    = {Crawford, Kate},
  title     = {Atlas of {AI}: Power, Politics, and the Planetary Costs of Artificial Intelligence},
  year      = {2021},
  publisher = {Yale University Press},
  address   = {New Haven, CT},
  isbn      = {9780300209570},
  url       = {https://yalebooks.yale.edu/book/9780300264630/atlas-of-ai/}
}

@article{Crenshaw1989,
  author  = {Crenshaw, Kimberle},
  title   = {Demarginalizing the Intersection of Race and Sex: A Black Feminist Critique of Antidiscrimination Doctrine, Feminist Theory and Antiracist Politics},
  journal = {University of Chicago Legal Forum},
  year    = {1989},
  volume  = {1989},
  number  = {1},
  pages   = {139--167},
  issn    = {0892-5593},
  note    = {Article 8},
  url     = {https://chicagounbound.uchicago.edu/uclf/vol1989/iss1/8/}
}

@book{CreswellCreswell2022,
  author    = {Creswell, John W. and Creswell, J. David},
  title     = {Research Design: Qualitative, Quantitative, and Mixed Methods Approaches},
  year      = {2022},
  edition   = {6},
  publisher = {SAGE Publications},
  address   = {Thousand Oaks, CA},
  isbn      = {9781071817940},
  url       = {https://www.sagepub.com/shop/buy-a-book/research-design-6-270550}
}

@book{Dahl1971,
  author    = {Dahl, Robert A.},
  title     = {Polyarchy: Participation and Opposition},
  year      = {1971},
  publisher = {Yale University Press},
  address   = {New Haven, CT},
  isbn      = {9780300013917},
  url       = {https://archive.org/details/polyarchypartici0000dahl}
}

@article{Danish2025,
  author  = {Danish, Matthew and Labib, S.M. and Ricker, Britta and Helbich, Marco},
  title   = {{A citizen science toolkit to collect human perceptions of urban environments using open street view images}},
  journal = {Computers, Environment and Urban Systems},
  year    = {2025},
  volume  = {116},
  pages   = {102207},
  doi     = {10.1016/j.compenvurbsys.2024.102207},
  url     = {https://doi.org/10.1016/j.compenvurbsys.2024.102207}
}

@incollection{Dastin2022,
  author    = {Dastin, Jeffrey},
  title     = {Amazon Scraps Secret {AI} Recruiting Tool That Showed Bias against Women},
  booktitle = {Ethics of Data and Analytics: Concepts and Cases},
  editor    = {Martin, Kirsten},
  year      = {2022},
  edition   = {1},
  pages     = {296--299},
  publisher = {Auerbach Publications},
  isbn      = {9781003278290},
  doi       = {10.1201/9781003278290-44},
  url       = {https://www.taylorfrancis.com/chapters/edit/10.1201/9781003278290-44/amazon-scraps-secret-ai-recruiting-tool-showed-bias-women-jeffrey-dastin}
}

@article{Davidoff1965,
  author  = {Davidoff, Paul},
  title   = {Advocacy and Pluralism in Planning},
  journal = {Journal of the American Institute of Planners},
  volume  = {31},
  number  = {4},
  pages   = {331--338},
  year    = {1965},
  doi     = {10.1080/01944366508978187},
  url     = {https://www.tandfonline.com/doi/abs/10.1080/01944366508978187}
}

@article{DeHond2022,
  author  = {de Hond, Anne A H and van Buchem, Marieke M and Hernandez-Boussard, Tina},
  title   = {{Picture a data scientist: a call to action for increasing diversity, equity, and inclusion in the age of AI}},
  journal = {Journal of the American Medical Informatics Association},
  year    = {2022},
  volume  = {29},
  number  = {12},
  pages   = {2178--2181},
  doi     = {10.1093/jamia/ocac156},
  url     = {https://doi.org/10.1093/jamia/ocac156}
}

@article{demarcellis-warin2022ai,
  author  = {de Marcellis-Warin, Nathalie and Marty, Frédéric and Thelisson, Eva and Warin, Thierry},
  title   = {{Artificial intelligence and consumer manipulations: from consumer's counter algorithms to firm's self-regulation tools}},
  journal = {AI and Ethics},
  year    = {2022},
  volume  = {2},
  number  = {2},
  pages   = {259--268},
  doi     = {10.1007/s43681-022-00149-5},
  url     = {https://doi.org/10.1007/s43681-022-00149-5}
}

@book{Deming1986,
  author    = {Deming, W. Edwards},
  title     = {Out of the Crisis},
  year      = {1986},
  publisher = {Massachusetts Institute of Technology, Center for Advanced Engineering Study},
  address   = {Cambridge, MA},
  isbn      = {9780911379013},
  url       = {https://archive.org/details/outofcrisis000demi}
}

@article{DeSilvaAlahakoon2022,
  author  = {De Silva, Daswin and Alahakoon, Damminda},
  title   = {An Artificial Intelligence Life Cycle: From Conception to Production},
  journal = {Patterns},
  year    = {2022},
  volume  = {3},
  number  = {6},
  pages   = {100489},
  note    = {Article 100489},
  doi     = {10.1016/j.patter.2022.100489},
  url     = {https://www.sciencedirect.com/science/article/pii/S2666389922000745}
}

@book{Desrosieres1998,
  author     = {Desrosi{\`e}res, Alain},
  title      = {The Politics of Large Numbers: A History of Statistical Reasoning},
  translator = {Naish, Camille},
  publisher  = {Harvard University Press},
  address    = {Cambridge, MA},
  year       = {1998},
  isbn       = {9780674689329},
  url        = {https://www.hup.harvard.edu/books/9780674009691}
}

@book{Dewey1927,
  author    = {Dewey, John},
  title     = {The Public and Its Problems},
  year      = {1927},
  publisher = {Henry Holt and Company},
  address   = {New York, NY},
  url       = {https://catalog.hathitrust.org/Record/001432743}
}

@article{Dignam2020,
  author  = {Dignam, Alan},
  title   = {{Artificial intelligence, tech corporate governance and the public interest regulatory response}},
  journal = {Cambridge Journal of Regions, Economy and Society},
  year    = {2020},
  volume  = {13},
  number  = {1},
  pages   = {37--54},
  doi     = {10.1093/cjres/rsaa002},
  url     = {https://doi.org/10.1093/cjres/rsaa002}
}

@book{Datafeminism,
  author    = {D'Ignazio, Catherine and Klein, Lauren F.},
  title     = {Data Feminism},
  series    = {Strong Ideas},
  publisher = {The MIT Press},
  address   = {Cambridge, MA},
  year      = {2020},
  isbn      = {9780262358521},
  doi       = {10.7551/mitpress/11805.001.0001},
  url       = {https://direct.mit.edu/books/book/4660/Data-Feminism}
}

@article{Donia2021,
  author  = {Donia, Joseph and Shaw, James A.},
  title   = {Co-Design and Ethical Artificial Intelligence for Health: An Agenda for Critical Research and Practice},
  journal = {Big Data \& Society},
  year    = {2021},
  volume  = {8},
  number  = {2},
  pages   = {20539517211065248},
  doi     = {10.1177/20539517211065248},
  url     = {https://doi.org/10.1177/20539517211065248}
}

@inproceedings{Dubey2016,
  author    = {Dubey, Abhimanyu and Naik, Nikhil and Parikh, Devi and Raskar, Ramesh and Hidalgo, C{\'e}sar A.},
  title     = {{Deep Learning the City: Quantifying Urban Perception at a Global Scale}},
  booktitle = {Computer Vision--ECCV 2016},
  editor    = {Leibe, Bastian and Matas, Jiri and Sebe, Nicu and Welling, Max},
  series    = {Lecture Notes in Computer Science},
  volume    = {9905},
  pages     = {196--212},
  publisher = {Springer International Publishing},
  address   = {Cham},
  year      = {2016},
  isbn      = {9783319464473},
  doi       = {10.1007/978-3-319-46448-0_12},
  url       = {https://doi.org/10.1007/978-3-319-46448-0_12}
}

@article{DubeyEtAl2024,
  author  = {Dubey, Rachit and Hardy, Mathew D. and Griffiths, Thomas L. and Bhui, Rahul},
  title   = {{AI}-Generated Visuals of Car-Free {US} Cities Help Improve Support for Sustainable Policies},
  journal = {Nature Sustainability},
  year    = {2024},
  volume  = {7},
  number  = {4},
  pages   = {399--403},
  doi     = {10.1038/s41893-024-01299-6},
  url     = {https://www.nature.com/articles/s41893-024-01299-6}
}

@article{Dwivedi2021,
  author  = {Dwivedi, Yogesh K. and Hughes, Laurie and Ismagilova, Elvira and Aarts, Gert and Coombs, Crispin and Crick, Tom and Duan, Yanqing and Dwivedi, Rohita and Edwards, John and Eirug, Aled and Galanos, Vassilis and Ilavarasan, P. Vigneswara and Janssen, Marijn and Jones, Paul and Kar, Arpan Kumar and Kizgin, Hatice and Kronemann, Bianca and Lal, Banita and Lucini, Biagio and Medaglia, Rony and Le Meunier-FitzHugh, Kenneth and Le Meunier-FitzHugh, Leslie Caroline and Misra, Santosh and Mogaji, Emmanuel and Sharma, Sujeet Kumar and Singh, Jang Bahadur and Raghavan, Vishnupriya and Raman, Ramakrishnan and Rana, Nripendra P. and Samothrakis, Spyridon and Spencer, Jak and Tamilmani, Kuttimani and Tubadji, Annie and Walton, Paul and Williams, Michael D.},
  title   = {{Artificial Intelligence (AI): Multidisciplinary perspectives on emerging challenges, opportunities, and agenda for research, practice and policy}},
  journal = {International Journal of Information Management},
  year    = {2021},
  volume  = {57},
  pages   = {101994},
  doi     = {10.1016/j.ijinfomgt.2019.08.002},
  url     = {https://doi.org/10.1016/j.ijinfomgt.2019.08.002}
}

@inproceedings{Dwork2012,
  author    = {Dwork, Cynthia and Hardt, Moritz and Pitassi, Toniann and Reingold, Omer and Zemel, Richard},
  title     = {{Fairness through awareness}},
  booktitle = {Proceedings of the 3rd Innovations in Theoretical Computer Science Conference},
  year      = {2012},
  pages     = {214--226},
  publisher = {ACM},
  doi       = {10.1145/2090236.2090255},
  url       = {https://doi.org/10.1145/2090236.2090255}
}

@article{Eitzel2021,
  author  = {Eitzel, M.V. and Solera, Jon and Mhike Hove, Emmanuel and Wilson, K.B. and Mawere Ndlovu, Abraham and Ndlovu, Daniel and Changarara, Abraham and Ndlovu, Alice and Neves, Kleber and Chirindira, Adnomore and Omoju, Oluwasola and Fisher, Aaron C. and Veski, André},
  title   = {{Assessing the Potential of Participatory Modeling for Decolonial Restoration of an Agro-Pastoral System in Rural Zimbabwe}},
  journal = {Citizen Science: Theory and Practice},
  year    = {2021},
  volume  = {6},
  number  = {1},
  pages   = {2},
  doi     = {10.5334/cstp.339},
  url     = {https://doi.org/10.5334/cstp.339}
}

@article{Engestrm2010,
  author  = {Engeström, Yrjö and Sannino, Annalisa},
  title   = {{Studies of expansive learning: Foundations, findings and future challenges}},
  journal = {Educational Research Review},
  year    = {2010},
  volume  = {5},
  number  = {1},
  pages   = {1--24},
  doi     = {10.1016/j.edurev.2009.12.002},
  url     = {https://doi.org/10.1016/j.edurev.2009.12.002}
}

@book{Engestrom2014,
  author    = {Engeström, Yrjö},
  title     = {{Learning by Expanding: An Activity-Theoretical Approach to Developmental Research}},
  year      = {2014},
  publisher = {Cambridge University Press},
  isbn      = {9781107074422},
  doi       = {10.1017/cbo9781139814744},
  url       = {https://doi.org/10.1017/cbo9781139814744}
}

@article{Engin2020,
  author  = {Engin, Zeynep and van Dijk, Justin and Lan, Tian and Longley, Paul A. and Treleaven, Philip and Batty, Michael and Penn, Alan},
  title   = {{Data-driven urban management: Mapping the landscape}},
  journal = {Journal of Urban Management},
  year    = {2020},
  volume  = {9},
  number  = {2},
  pages   = {140--150},
  doi     = {10.1016/j.jum.2019.12.001},
  url     = {https://doi.org/10.1016/j.jum.2019.12.001}
}

@article{EspelandStevens1998,
  author  = {Espeland, Wendy Nelson and Stevens, Mitchell L.},
  title   = {Commensuration as a Social Process},
  journal = {Annual Review of Sociology},
  volume  = {24},
  number  = {1},
  pages   = {313--343},
  year    = {1998},
  doi     = {10.1146/annurev.soc.24.1.313},
  url     = {https://www.annualreviews.org/content/journals/10.1146/annurev.soc.24.1.313}
}

@article{EspelandStevens2008,
  author  = {Espeland, Wendy Nelson and Stevens, Mitchell L.},
  title   = {A Sociology of Quantification},
  journal = {European Journal of Sociology},
  volume  = {49},
  number  = {3},
  pages   = {401--436},
  year    = {2008},
  doi     = {10.1017/S0003975609000150},
  url     = {https://www.cambridge.org/core/journals/european-journal-of-sociology-archives-europeennes-de-sociologie/article/sociology-of-quantification/53D563E0E4A75A05E877B27C06E957F9}
}

@book{Eubanks2018,
  author    = {Eubanks, Virginia},
  title     = {Automating Inequality: How High-Tech Tools Profile, Police, and Punish the Poor},
  publisher = {St. Martin's Press},
  address   = {New York},
  year      = {2018},
  month     = {January},
  isbn      = {9781250074317},
  url       = {https://panmacmillan.co.in/authors/virginia-eubanks/automating-inequality/9781250074317}
}

@misc{EuropeanUnion2024,
  author       = {{European Parliament} and {Council of the European Union}},
  title        = {{Regulation (EU) 2024/1689 of the European Parliament and of the Council of 13 June 2024 Laying Down Harmonised Rules on Artificial Intelligence and Amending Regulations (EC) No 300/2008, (EU) No 167/2013, (EU) No 168/2013, (EU) 2018/858, (EU) 2018/1139 and (EU) 2019/2144 and Directives 2014/90/EU, (EU) 2016/797 and (EU) 2020/1828 (Artificial Intelligence Act) (Text with EEA Relevance)}},
  howpublished = {Official Journal of the European Union, L 2024/1689},
  year         = {2024},
  month        = {July},
  note         = {Published 12 July 2024; CELEX 32024R1689; ELI http://data.europa.eu/eli/reg/2024/1689/oj},
  url          = {https://eur-lex.europa.eu/eli/reg/2024/1689/oj}
}

@book{Fainstein2010,
  author    = {Fainstein, Susan S.},
  title     = {The Just City},
  publisher = {Cornell University Press},
  address   = {Ithaca, NY},
  year      = {2010},
  doi       = {10.7591/9780801460487},
  isbn      = {9780801476907},
  url       = {https://cornellpress.cornell.edu/book/9780801476907/the-just-city/}
}

@book{Fischer2000,
  author    = {Fischer, Frank},
  title     = {Citizens, Experts, and the Environment: The Politics of Local Knowledge},
  publisher = {Duke University Press},
  address   = {Durham, NC},
  year      = {2000},
  month     = {December},
  doi       = {10.1215/9780822380283},
  isbn      = {9780822326229},
  url       = {https://www.dukeupress.edu/citizens-experts-and-the-environment}
}

@book{Fishkin2009,
  author    = {Fishkin, James S.},
  title     = {When the People Speak: Deliberative Democracy and Public Consultation},
  publisher = {Oxford University Press},
  address   = {Oxford},
  year      = {2009},
  isbn      = {9780199572106},
  url       = {https://cddrl.fsi.stanford.edu/publications/when_the_people_speak_deliberative_democracy_and_public_consultation}
}

@book{Flyvbjerg1998,
  author     = {Flyvbjerg, Bent},
  title      = {Rationality and Power: Democracy in Practice},
  translator = {Sampson, Steven},
  publisher  = {University of Chicago Press},
  address    = {Chicago},
  year       = {1998},
  isbn       = {9780226254517},
  url        = {https://press.uchicago.edu/ucp/books/book/chicago/R/bo3640330.html}
}

@book{Forester1988,
  author    = {Forester, John},
  title     = {Planning in the Face of Power},
  publisher = {University of California Press},
  address   = {Berkeley},
  year      = {1988},
  isbn      = {9780520064133},
  note      = {Paperback edition},
  url       = {https://www.ucpress.edu/book/9780520064133/planning-in-the-face-of-power}
}

@book{Forester1999,
  author    = {Forester, John F.},
  title     = {The Deliberative Practitioner: Encouraging Participatory Planning Processes},
  publisher = {The MIT Press},
  address   = {Cambridge, MA},
  year      = {1999},
  isbn      = {9780262561228},
  url       = {https://mitpress.mit.edu/9780262561228/the-deliberative-practitioner/}
}

@article{Fraser1995,
  author  = {Fraser, Nancy},
  title   = {From Redistribution to Recognition? Dilemmas of Justice in a `Post-Socialist' Age},
  journal = {New Left Review},
  volume  = {I},
  number  = {212},
  pages   = {68--93},
  year    = {1995},
  month   = {July--August},
  doi     = {10.64590/4rl},
  url     = {https://newleftreview.org/issues/i212/articles/nancy-fraser-from-redistribution-to-recognition-dilemmas-of-justice-in-a-post-socialist-age}
}

@book{Fricker2007,
  author    = {Fricker, Miranda},
  title     = {Epistemic Injustice: Power and the Ethics of Knowing},
  publisher = {Oxford University Press},
  address   = {Oxford},
  year      = {2007},
  month     = {June},
  doi       = {10.1093/acprof:oso/9780198237907.001.0001},
  isbn      = {9780198237907},
  url       = {https://global.oup.com/academic/product/epistemic-injustice-9780198237907}
}

@book{Fung2004,
  author    = {Fung, Archon},
  title     = {Empowered Participation: Reinventing Urban Democracy},
  publisher = {Princeton University Press},
  address   = {Princeton, NJ},
  year      = {2004},
  doi       = {10.1515/9781400835638},
  isbn      = {9780691115351},
  url       = {https://www.degruyter.com/document/doi/10.1515/9781400835638/html}
}

@book{Gabrys2016,
  author    = {Gabrys, Jennifer},
  title     = {Program Earth: Environmental Sensing Technology and the Making of a Computational Planet},
  publisher = {University of Minnesota Press},
  address   = {Minneapolis},
  year      = {2016},
  doi       = {10.5749/minnesota/9780816693122.001.0001},
  isbn      = {9780816693122},
  url       = {https://doi.org/10.5749/minnesota/9780816693122.001.0001}
}

@article{GalazEtAl2021,
  author  = {Galaz, Victor and Centeno, Miguel A. and Callahan, Peter W. and Causevic, Amar and Patterson, Thayer and Brass, Irina and Baum, Seth and Farber, Darryl and Fischer, Joern and Garcia, David and McPhearson, Timon and Jimenez, Daniel and King, Brian and Larcey, Paul and Levy, Karen},
  title   = {Artificial Intelligence, Systemic Risks, and Sustainability},
  journal = {Technology in Society},
  volume  = {67},
  pages   = {101741},
  year    = {2021},
  month   = {November},
  doi     = {10.1016/j.techsoc.2021.101741},
  url     = {https://www.sciencedirect.com/science/article/pii/S0160791X21002165}
}

@book{Galston2002,
  author    = {Galston, William A.},
  title     = {Liberal Pluralism: The Implications of Value Pluralism for Political Theory and Practice},
  publisher = {Cambridge University Press},
  address   = {Cambridge},
  year      = {2002},
  doi       = {10.1017/CBO9780511613579},
  isbn      = {9780521813044},
  url       = {https://www.cambridge.org/core/books/liberal-pluralism/B7B1CC377F1E093457A525CDC14EA008}
}

@article{gebru2021,
  author  = {Gebru, Timnit and Morgenstern, Jamie and Vecchione, Briana and Vaughan, Jennifer Wortman and Wallach, Hanna and {Daum{\'e} III}, Hal and Crawford, Kate},
  title   = {Datasheets for Datasets},
  journal = {Communications of the ACM},
  year    = {2021},
  volume  = {64},
  number  = {12},
  pages   = {86--92},
  doi     = {10.1145/3458723},
  url     = {https://doi.org/10.1145/3458723}
}

@book{Gehl2011,
  author     = {Gehl, Jan},
  title      = {Life Between Buildings: Using Public Space},
  translator = {Koch, Jo},
  edition    = {6},
  publisher  = {Island Press},
  address    = {Washington, DC},
  year       = {2011},
  month      = {January},
  isbn       = {9781597268271},
  url        = {https://islandpress.org/books/life-between-buildings}
}

@book{GehlSvarre2013,
  author     = {Gehl, Jan and Svarre, Birgitte},
  translator = {Steenhard, Karen Ann},
  title      = {How to Study Public Life},
  publisher  = {Island Press},
  address    = {Washington, DC},
  year       = {2013},
  isbn       = {9781610915250},
  doi        = {10.5822/978-1-61091-525-0},
  url        = {https://doi.org/10.5822/978-1-61091-525-0}
}

@article{Gerdes2022,
  author  = {Gerdes, Anne},
  title   = {A Participatory Data-Centric Approach to {AI} Ethics by Design},
  journal = {Applied Artificial Intelligence},
  year    = {2022},
  volume  = {36},
  number  = {1},
  pages   = {2009222},
  doi     = {10.1080/08839514.2021.2009222},
  url     = {https://doi.org/10.1080/08839514.2021.2009222}
}

@article{Goodchild2007,
  author   = {Goodchild, Michael F.},
  title    = {Citizens as Sensors: The World of Volunteered Geography},
  journal  = {GeoJournal},
  year     = {2007},
  volume   = {69},
  number   = {4},
  pages    = {211--221},
  doi      = {10.1007/s10708-007-9111-y},
  url      = {https://doi.org/10.1007/s10708-007-9111-y}
}

@article{GowaikarEtAl2024,
  author  = {Gowaikar, Shreeyash and Berard, Hugo and Mushkani, Rashid and Koseki, Shin},
  title   = {From Efficiency to Equity: Measuring Fairness in Preference Learning},
  journal = {Proceedings of the AAAI/ACM Conference on AI, Ethics, and Society},
  year    = {2025},
  volume  = {8},
  number  = {2},
  pages   = {1134--1143},
  doi     = {10.1609/aies.v8i2.36616},
  url     = {https://doi.org/10.1609/aies.v8i2.36616}
}

@book{Graham2001,
  author    = {Graham, Steve and Marvin, Simon},
  title     = {{Splintering Urbanism: Networked Infrastructures, Technological Mobilities and the Urban Condition}},
  year      = {2001},
  publisher = {Routledge},
  address   = {London},
  isbn      = {9780203452202},
  doi       = {10.4324/9780203452202},
  url       = {https://doi.org/10.4324/9780203452202}
}

@article{guridi2024fake,
  author    = {Guridi, Jose A. and Hwang, Angel Hsing-Chi and Santo, Duarte and Goula, Maria and Cheyre, Cristobal and Humphreys, Lee and Rangel, Marco},
  title     = {From Fake Perfects to Conversational Imperfects: Exploring Image-Generative {AI} as a Boundary Object for Participatory Design of Public Spaces},
  journal   = {Proceedings of the ACM on Human-Computer Interaction},
  volume    = {9},
  number    = {N2},
  articleno = {CSCW014},
  pages     = {1--33},
  numpages  = {33},
  year      = {2025},
  issn      = {2573-0142},
  doi       = {10.1145/3710912},
  url       = {https://dl.acm.org/doi/10.1145/3710912}
}

@article{Haakman2021,
  author  = {Haakman, Mark and Cruz, Lu{\'i}s and Huijgens, Hennie and van Deursen, Arie},
  title   = {{AI} Lifecycle Models Need to Be Revised: An Exploratory Study in Fintech},
  journal = {Empirical Software Engineering},
  year    = {2021},
  volume  = {26},
  number  = {5},
  pages   = {95},
  doi     = {10.1007/s10664-021-09993-1},
  url     = {https://doi.org/10.1007/s10664-021-09993-1}
}

@book{Habermas1996,
  author     = {Habermas, J{\"u}rgen},
  title      = {Between Facts and Norms: Contributions to a Discourse Theory of Law and Democracy},
  translator = {Rehg, William},
  series     = {Studies in Contemporary German Social Thought},
  publisher  = {The MIT Press},
  address    = {Cambridge, MA},
  year       = {1996},
  month      = {May},
  doi        = {10.7551/mitpress/1564.001.0001},
  isbn       = {9780262082433},
  url        = {https://mitpress.mit.edu/9780262581622/between-facts-and-norms/}
}

@article{HadfieldClark2023,
  author  = {Hadfield, Gillian K. and Clark, Jack},
  title   = {Regulatory Markets: The Future of {AI} Governance},
  journal = {Jurimetrics},
  year    = {2026},
  volume  = {65},
  pages   = {195--240},
  url     = {https://www.americanbar.org/groups/science_technology/resources/jurimetrics/2026-winter/regulatory-markets-future-ai-governance/}
}

@incollection{Haklay2013,
  author    = {Haklay, Muki},
  title     = {Citizen Science and Volunteered Geographic Information: Overview and Typology of Participation},
  booktitle = {Crowdsourcing Geographic Knowledge},
  editor    = {Sui, Daniel and Elwood, Sarah and Goodchild, Michael},
  year      = {2013},
  pages     = {105--122},
  publisher = {Springer Netherlands},
  address   = {Dordrecht},
  isbn      = {9789400745865},
  doi       = {10.1007/978-94-007-4587-2_7},
  url       = {https://doi.org/10.1007/978-94-007-4587-2_7}
}

@inproceedings{HardtEtAl2016,
  author    = {Hardt, Moritz and Price, Eric and Srebro, Nati},
  title     = {Equality of Opportunity in Supervised Learning},
  booktitle = {Advances in Neural Information Processing Systems},
  editor    = {Lee, Daniel D. and Sugiyama, Masashi and von Luxburg, Ulrike and Guyon, Isabelle and Garnett, Roman},
  volume    = {29},
  pages     = {3315--3323},
  publisher = {Curran Associates, Inc.},
  year      = {2016},
  url       = {https://proceedings.neurips.cc/paper_files/paper/2016/hash/6a9659feb1216f14f7384ba499518b38-Abstract.html}
}

@article{Harvey2003,
  author  = {Harvey, David},
  title   = {The Right to the City},
  journal = {International Journal of Urban and Regional Research},
  volume  = {27},
  number  = {4},
  pages   = {939--941},
  year    = {2003},
  doi     = {10.1111/j.0309-1317.2003.00492.x},
  url     = {https://onlinelibrary.wiley.com/doi/10.1111/j.0309-1317.2003.00492.x}
}

@book{Harvey2012,
  author    = {Harvey, David},
  title     = {Rebel Cities: From the Right to the City to the Urban Revolution},
  publisher = {Verso},
  address   = {London and New York},
  year      = {2012},
  isbn      = {9781844678822},
  url       = {https://books.google.com/books?id=s4s5f7NnaZAC}
}

@book{Healey1997,
  author    = {Healey, Patsy},
  title     = {Collaborative Planning: Shaping Places in Fragmented Societies},
  series    = {Planning, Environment, Cities},
  publisher = {Macmillan Education UK},
  address   = {London},
  year      = {1997},
  doi       = {10.1007/978-1-349-25538-2},
  isbn      = {9780333495742},
  url       = {https://link.springer.com/book/10.1007/978-1-349-25538-2}
}

@article{HelbingEtAl2023,
  author  = {Helbing, Dirk and Mahajan, Sachit and Fricker, Regula H{\"a}nggli and Musso, Andrea and Hausladen, Carina I. and Carissimo, Cesare and Carpentras, Dino and Stockinger, Elisabeth and Argota Sanchez-Vaquerizo, Javier and Yang, Joshua C. and Ballandies, Mark C. and Korecki, Marcin and Dubey, Rohit K. and Pournaras, Evangelos},
  title   = {Democracy by Design: Perspectives for Digitally Assisted, Participatory Upgrades of Society},
  journal = {Journal of Computational Science},
  volume  = {71},
  pages   = {102061},
  year    = {2023},
  month   = {July},
  doi     = {10.1016/j.jocs.2023.102061},
  url     = {https://www.sciencedirect.com/science/article/pii/S1877750323001217}
}

@techreport{AIHLEG2018,
  author      = {{High-Level Expert Group on Artificial Intelligence}},
  title       = {Draft Ethics Guidelines for Trustworthy {AI}},
  institution = {European Commission},
  address     = {Brussels},
  type        = {Working document for stakeholders' consultation},
  year        = {2018},
  month       = {December},
  url         = {https://digital-strategy.ec.europa.eu/en/library/draft-ethics-guidelines-trustworthy-ai}
}

@inproceedings{Hosking2024-goldstandard,
  author        = {Hosking, Tom and Blunsom, Phil and Bartolo, Max},
  title         = {Human Feedback is not Gold Standard},
  booktitle     = {The Twelfth International Conference on Learning Representations},
  year          = {2024},
  eprint        = {2309.16349},
  archivePrefix = {arXiv},
  primaryClass  = {cs.CL},
  url           = {https://openreview.net/forum?id=7W3GLNImfS}
}

@inproceedings{Huang2024,
  author    = {Huang, Saffron and Siddarth, Divya and Lovitt, Liane and Liao, Thomas I. and Durmus, Esin and Tamkin, Alex and Ganguli, Deep},
  title     = {Collective Constitutional {AI}: Aligning a Language Model with Public Input},
  booktitle = {The 2024 ACM Conference on Fairness, Accountability, and Transparency ({FAccT} '24)},
  year      = {2024},
  month     = jun,
  pages     = {1395--1417},
  publisher = {Association for Computing Machinery},
  isbn      = {9798400704505},
  doi       = {10.1145/3630106.3658979},
  url       = {https://doi.org/10.1145/3630106.3658979}
}

@article{Huang2023,
  author  = {Huang, Yingjing and Zhang, Fan and Gao, Yong and Tu, Wei and Duarte, Fabio and Ratti, Carlo and Guo, Diansheng and Liu, Yu},
  title   = {{Comprehensive urban space representation with varying numbers of street-level images}},
  journal = {Computers, Environment and Urban Systems},
  year    = {2023},
  volume  = {106},
  pages   = {102043},
  doi     = {10.1016/j.compenvurbsys.2023.102043},
  url     = {https://doi.org/10.1016/j.compenvurbsys.2023.102043}
}

@article{Ibrahim2020,
  author  = {Ibrahim, Mohamed R. and Haworth, James and Cheng, Tao},
  title   = {{Understanding cities with machine eyes: A review of deep computer vision in urban analytics}},
  journal = {Cities},
  year    = {2020},
  volume  = {96},
  pages   = {102481},
  doi     = {10.1016/j.cities.2019.102481},
  url     = {https://doi.org/10.1016/j.cities.2019.102481}
}

@misc{IRCGM2018,
  author       = {{Information and Referral Centre of Greater Montr{\'e}al}},
  title        = {Directory of Community and Social Organizations},
  howpublished = {211 Qu{\'e}bec},
  year         = {n.d.},
  url          = {https://www.211qc.ca/en/directory},
  urldate      = {2026-07-21},
  note         = {Online community-resource directory; the cited page supplies no publication date}
}

@book{InnesBooher2010,
  author    = {Innes, Judith E. and Booher, David E.},
  title     = {Planning with Complexity: An Introduction to Collaborative Rationality for Public Policy},
  edition   = {1},
  publisher = {Routledge},
  address   = {Abingdon},
  year      = {2010},
  doi       = {10.4324/9780203864302},
  url       = {https://www.taylorfrancis.com/books/mono/10.4324/9780203864302/planning-complexity-judith-innes-david-booher}
}

@article{Israel1998,
  author  = {Israel, Barbara A. and Schulz, Amy J. and Parker, Edith A. and Becker, Adam B.},
  title   = {Review of Community-Based Research: Assessing Partnership Approaches to Improve Public Health},
  journal = {Annual Review of Public Health},
  volume  = {19},
  number  = {1},
  pages   = {173--202},
  month   = may,
  year    = {1998},
  doi     = {10.1146/annurev.publhealth.19.1.173},
  url     = {https://www.annualreviews.org/content/journals/10.1146/annurev.publhealth.19.1.173}
}

@book{Jacobs1961,
  author    = {Jacobs, Jane},
  title     = {The Death and Life of Great American Cities},
  publisher = {Random House},
  address   = {New York},
  year      = {1961},
  url       = {https://catalogue.nla.gov.au/catalog/2173507}
}

@inproceedings{Jain_2024,
  author    = {Jain, Shomik and Suriyakumar, Vinith and Creel, Kathleen and Wilson, Ashia},
  title     = {Algorithmic Pluralism: A Structural Approach to Equal Opportunity},
  booktitle = {The 2024 ACM Conference on Fairness, Accountability, and Transparency ({FAccT} '24)},
  year      = {2024},
  month     = jun,
  pages     = {197--206},
  publisher = {Association for Computing Machinery},
  isbn      = {9798400704505},
  doi       = {10.1145/3630106.3658899},
  url       = {https://doi.org/10.1145/3630106.3658899}
}

@book{Jasanoff2004,
  editor    = {Jasanoff, Sheila},
  title     = {States of Knowledge: The Co-Production of Science and the Social Order},
  edition   = {1},
  publisher = {Routledge},
  address   = {London},
  year      = {2004},
  doi       = {10.4324/9780203413845},
  isbn      = {9780203413845},
  url       = {https://www.taylorfrancis.com/books/edit/10.4324/9780203413845/states-knowledge-sheila-jasanoff}
}

@article{Jobin2019,
  author  = {Jobin, Anna and Ienca, Marcello and Vayena, Effy},
  title   = {{The global landscape of AI ethics guidelines}},
  journal = {Nature Machine Intelligence},
  year    = {2019},
  volume  = {1},
  number  = {9},
  pages   = {389--399},
  doi     = {10.1038/s42256-019-0088-2},
  url     = {https://doi.org/10.1038/s42256-019-0088-2}
}

@article{kaminski2018right,
  author  = {Kaminski, Margot E.},
  title   = {The Right to Explanation, Explained},
  journal = {Berkeley Technology Law Journal},
  year    = {2019},
  volume  = {34},
  number  = {1},
  pages   = {189--218},
  doi     = {10.15779/Z38TD9N83H},
  url     = {https://doi.org/10.15779/Z38TD9N83H}
}

@article{kaminski2021right,
  author  = {Kaminski, Margot E. and Urban, Jennifer M.},
  title   = {The Right to Contest {AI}},
  journal = {Columbia Law Review},
  volume  = {121},
  number  = {7},
  pages   = {1957--2048},
  year    = {2021},
  url     = {https://www.columbialawreview.org/content/the-right-to-contest-ai/},
  note    = {JSTOR stable URL: https://www.jstor.org/stable/27083420}
}

@inproceedings{kannen2024aesthetics,
  author    = {Kannen, Nithish and Ahmad, Arif and Andreetto, Marco and Prabhakaran, Vinodkumar and Prabhu, Utsav and Dieng, Adji Bousso and Bhattacharyya, Pushpak and Dave, Shachi},
  title     = {Beyond Aesthetics: Cultural Competence in Text-to-Image Models},
  booktitle = {Advances in Neural Information Processing Systems},
  editor    = {A. Globerson and L. Mackey and D. Belgrave and A. Fan and U. Paquet and J. Tomczak and C. Zhang},
  volume    = {37},
  pages     = {13716--13747},
  publisher = {Curran Associates, Inc.},
  year      = {2024},
  doi       = {10.52202/079017-0439},
  url       = {https://proceedings.neurips.cc/paper_files/paper/2024/hash/18c669b80d1a8f589713b768bc8fe9a4-Abstract-Datasets_and_Benchmarks_Track.html},
  note      = {Datasets and Benchmarks Track}
}

@article{KiritsisEtAl2003,
  author  = {Kiritsis, Dimitris and Bufardi, Ahmed and Xirouchakis, Paul},
  title   = {Research issues on product lifecycle management and information tracking using smart embedded systems},
  journal = {Advanced Engineering Informatics},
  volume  = {17},
  number  = {3--4},
  pages   = {189--202},
  year    = {2003},
  doi     = {10.1016/j.aei.2004.09.005},
  url     = {https://www.sciencedirect.com/science/article/pii/S1474034604000187}
}

@inproceedings{kirk2024prism,
  author    = {Kirk, Hannah Rose and Whitefield, Alexander and R{\"o}ttger, Paul and Bean, Andrew and Margatina, Katerina and Ciro, Juan and Mosquera, Rafael and Bartolo, Max and Williams, Adina and He, He and Vidgen, Bertie and Hale, Scott A.},
  title     = {The {PRISM} Alignment Dataset: What Participatory, Representative and Individualised Human Feedback Reveals About the Subjective and Multicultural Alignment of Large Language Models},
  booktitle = {Advances in Neural Information Processing Systems},
  volume    = {37},
  pages     = {105236--105344},
  publisher = {Curran Associates, Inc.},
  year      = {2024},
  doi       = {10.52202/079017-3342},
  url       = {https://proceedings.neurips.cc/paper_files/paper/2024/hash/be2e1b68b44f2419e19f6c35a1b8cf35-Abstract-Datasets_and_Benchmarks_Track.html},
  note      = {Datasets and Benchmarks Track}
}

@inproceedings{kirstain2023pickapic,
  author    = {Kirstain, Yuval and Polyak, Adam and Singer, Uriel and Matiana, Shahbuland and Penna, Joe and Levy, Omer},
  title     = {Pick-a-Pic: An Open Dataset of User Preferences for Text-to-Image Generation},
  booktitle = {Advances in Neural Information Processing Systems},
  editor    = {Oh, Alice and Naumann, Tristan and Globerson, Amir and Saenko, Kate and Hardt, Moritz and Levine, Sergey},
  volume    = {36},
  pages     = {36652--36663},
  publisher = {Curran Associates, Inc.},
  year      = {2023},
  doi       = {10.52202/075280-1594},
  url       = {https://proceedings.neurips.cc/paper_files/paper/2023/hash/73aacd8b3b05b4b503d58310b523553c-Abstract-Conference.html}
}

@book{Kitchin2014b,
  author    = {Kitchin, Rob},
  title     = {The Data Revolution: Big Data, Open Data, Data Infrastructures \& Their Consequences},
  publisher = {SAGE Publications Ltd},
  address   = {London},
  year      = {2014},
  isbn      = {9781446287484},
  doi       = {10.4135/9781473909472},
  url       = {https://doi.org/10.4135/9781473909472}
}

@article{Kitchin2014a,
  author  = {Kitchin, Rob},
  title   = {The Real-Time City? Big Data and Smart Urbanism},
  journal = {GeoJournal},
  year    = {2014},
  volume  = {79},
  number  = {1},
  pages   = {1--14},
  doi     = {10.1007/s10708-013-9516-8},
  url     = {https://doi.org/10.1007/s10708-013-9516-8}
}

@article{Kitchin2016,
  author  = {Kitchin, Rob},
  title   = {Thinking Critically about and Researching Algorithms},
  journal = {Information, Communication \& Society},
  year    = {2017},
  volume  = {20},
  number  = {1},
  pages   = {14--29},
  doi     = {10.1080/1369118X.2016.1154087},
  url     = {https://doi.org/10.1080/1369118X.2016.1154087}
}

@incollection{Kitchin2023,
  author    = {Kitchin, Rob},
  title     = {Urban Data Power: Capitalism, Governance, Ethics, and Justice},
  booktitle = {Data Power in Action: Urban Data Politics in Times of Crisis},
  editor    = {S{\"o}derstr{\"o}m, Ola and Datta, Ayona},
  publisher = {Bristol University Press},
  address   = {Bristol},
  year      = {2023},
  pages     = {21--41},
  doi       = {10.56687/9781529233551-005},
  url       = {https://doi.org/10.56687/9781529233551-005}
}

@article{KooLi2016,
  author  = {Koo, Terry K. and Li, Mae Y.},
  title   = {A Guideline of Selecting and Reporting Intraclass Correlation Coefficients for Reliability Research},
  journal = {Journal of Chiropractic Medicine},
  volume  = {15},
  number  = {2},
  pages   = {155--163},
  year    = {2016},
  doi     = {10.1016/j.jcm.2016.02.012},
  url     = {https://pubmed.ncbi.nlm.nih.gov/27330520/}
}

@techreport{Koseki2022,
  author      = {Koseki, Shin and Jameson, Shazade and Farnadi, Golnoosh and Rolnick, David and R{\'e}gis, Catherine and Denis, Jean-Louis and Leal, Amanda and de Bezenac, Cecile and Occhini, Giulia and Lefebvre, Hugo and Gallego-Posada, Jose and Chehbouni, Khaoula and Molamohammadi, Maryam and Sefala, Raesetje and Salganik, Rebecca and Yahaya, Safiya and T{\'e}hinian, Shoghig},
  title       = {{AI} and Cities: Risks, Applications and Governance},
  institution = {United Nations Human Settlements Programme ({UN-Habitat}) and Mila -- Quebec {AI} Institute},
  year        = {2022},
  url         = {https://unhabitat.org/ai-cities-risks-applications-and-governance}
}

@misc{Krueger2002,
  author       = {Krueger, Richard A.},
  title        = {Designing and Conducting Focus Group Interviews},
  howpublished = {Presentation notes, University of Minnesota},
  month        = oct,
  year         = {2002},
  url          = {https://www.eiu.edu/ihec/Krueger-FocusGroupInterviews.pdf}
}

@book{KrumholzForester1990,
  author    = {Krumholz, Norman and Forester, John},
  title     = {Making Equity Planning Work: Leadership in the Public Sector},
  publisher = {Temple University Press},
  address   = {Philadelphia},
  year      = {1990},
  isbn      = {9780877227014},
  url       = {https://tupress.temple.edu/books/making-equity-planning-work}
}

@book{KukutaiTaylor2016,
  editor    = {Kukutai, Tahu and Taylor, John},
  title     = {Indigenous Data Sovereignty: Toward an Agenda},
  series    = {{CAEPR} Research Monograph},
  number    = {38},
  publisher = {{ANU Press}},
  address   = {Canberra},
  month     = nov,
  year      = {2016},
  doi       = {10.22459/CAEPR38.11.2016},
  isbn      = {9781760460303},
  url       = {https://press.anu.edu.au/publications/series/caepr/indigenous-data-sovereignty},
  note      = {Online ISBN 9781760460310; edited volume}
}

@article{LarteyLaw2025,
  author  = {Lartey, Desmond and Law, Kris M. Y.},
  title   = {Artificial intelligence adoption in urban planning governance: A systematic review of advancements in decision-making, and policy making},
  journal = {Landscape and Urban Planning},
  volume  = {258},
  pages   = {105337},
  year    = {2025},
  doi     = {10.1016/j.landurbplan.2025.105337},
  url     = {https://www.sciencedirect.com/science/article/pii/S0169204625000441}
}

@article{lazar2024lectureigoverningalgorithmic,
  author  = {Lazar, Seth},
  title   = {Governing the Algorithmic City},
  journal = {Philosophy \& Public Affairs},
  volume  = {53},
  number  = {2},
  pages   = {102--168},
  month   = apr,
  year    = {2025},
  doi     = {10.1111/papa.12279},
  url     = {https://onlinelibrary.wiley.com/doi/10.1111/papa.12279},
  note    = {Published online 27 January 2025; supersedes the 2024 arXiv lecture}
}

@article{Lee2019,
  author  = {Lee, Min Kyung and Kusbit, Daniel and Kahng, Anson and Kim, Ji Tae and Yuan, Xinran and Chan, Allissa and See, Daniel and Noothigattu, Ritesh and Lee, Siheon and Psomas, Alexandros and Procaccia, Ariel D.},
  title   = {{WeBuildAI: Participatory Framework for Algorithmic Governance}},
  journal = {Proceedings of the ACM on Human-Computer Interaction},
  year    = {2019},
  volume  = {3},
  number  = {CSCW},
  pages   = {1--35},
  doi     = {10.1145/3359283},
  url     = {https://doi.org/10.1145/3359283}
}

@book{Lefebvre1968,
  author    = {Lefebvre, Henri},
  title     = {Le droit {\`a} la ville},
  series    = {Soci{\'e}t{\'e} et urbanisme},
  publisher = {Anthropos},
  address   = {Paris},
  year      = {1968},
  url       = {https://books.google.com/books?id=4DFXwAEACAAJ}
}

@article{Lepri2018,
  author  = {Lepri, Bruno and Oliver, Nuria and Letouzé, Emmanuel and Pentland, Alex and Vinck, Patrick},
  title   = {{Fair, Transparent, and Accountable Algorithmic Decision-making Processes: The Premise, the Proposed Solutions, and the Open Challenges}},
  journal = {Philosophy \& Technology},
  year    = {2018},
  volume  = {31},
  number  = {4},
  pages   = {611--627},
  doi     = {10.1007/s13347-017-0279-x},
  url     = {https://doi.org/10.1007/s13347-017-0279-x}
}

@book{Lewis2020,
  editor    = {Lewis, Jason Edward},
  title     = {Indigenous Protocol and Artificial Intelligence Position Paper},
  publisher = {The Initiative for Indigenous Futures and the Canadian Institute for Advanced Research},
  address   = {Honolulu, Hawai'i},
  month     = jan,
  year      = {2020},
  doi       = {10.11573/spectrum.library.concordia.ca.00986506},
  note      = {Position paper; report author: Indigenous Protocol and Artificial Intelligence Working Group},
  url       = {https://spectrum.library.concordia.ca/id/eprint/986506/}
}

@techreport{Litman2024,
  author      = {Litman, Todd},
  title       = {Learning from Montreal: An Affordable and Inclusive City / Le{\c{c}}ons de Montr{\'e}al: Une Ville Abordable et Inclusive},
  institution = {Victoria Transport Policy Institute},
  address     = {Victoria, BC},
  month       = dec,
  year        = {2025},
  url         = {https://www.vtpi.org/montreal.pdf},
  note        = {Report dated 21 December 2025; copyright 2021--2025}
}

@book{LowSmith2006,
  editor    = {Low, Setha and Smith, Neil},
  title     = {The Politics of Public Space},
  edition   = {1},
  publisher = {Routledge},
  address   = {New York},
  year      = {2006},
  doi       = {10.4324/9780203390306},
  isbn      = {9780415951395},
  url       = {https://www.taylorfrancis.com/books/edit/10.4324/9780203390306/politics-public-space-setha-low-neil-smith}
}

@incollection{low2020social,
  author    = {Low, Setha},
  title     = {Social Justice as a Framework for Evaluating Public Space},
  booktitle = {Companion to Public Space},
  editor    = {Mehta, Vikas and Palazzo, Danilo},
  publisher = {Routledge},
  year      = {2020},
  pages     = {59--69},
  doi       = {10.4324/9781351002189-6},
  url       = {https://doi.org/10.4324/9781351002189-6}
}

@book{Low2000,
  author    = {Low, Setha M.},
  title     = {On the Plaza: The Politics of Public Space and Culture},
  publisher = {University of Texas Press},
  address   = {Austin},
  year      = {2000},
  doi       = {10.7560/747135},
  isbn      = {9780292747142},
  url       = {https://utpress.utexas.edu/9780292747142/}
}

@book{madanipour2010public,
  editor    = {Madanipour, Ali},
  title     = {Whose Public Space? International Case Studies in Urban Design and Development},
  edition   = {1},
  publisher = {Routledge},
  year      = {2010},
  isbn      = {978-0-415-55386-5},
  doi       = {10.4324/9780203860946},
  url       = {https://www.routledge.com/Whose-Public-Space-International-Case-Studies-in-Urban-Design-and-Development/Madanipour/p/book/9780415553865}
}

@book{MaddenMarcuse2017,
  author    = {Madden, David and Marcuse, Peter},
  title     = {In Defense of Housing: The Politics of Crisis},
  publisher = {Verso},
  address   = {London and New York},
  year      = {2016},
  isbn      = {9781784783549},
  url       = {https://www.versobooks.com/products/191-in-defense-of-housing}
}

@article{Mattern2017a,
  author  = {Mattern, Shannon},
  title   = {A City Is Not a Computer},
  journal = {Places Journal},
  month   = feb,
  year    = {2017},
  doi     = {10.22269/170207},
  url     = {https://placesjournal.org/article/a-city-is-not-a-computer/}
}

@book{Mattern2017b,
  author    = {Mattern, Shannon},
  title     = {Code and Clay, Data and Dirt: Five Thousand Years of Urban Media},
  publisher = {University of Minnesota Press},
  address   = {Minneapolis},
  year      = {2017},
  doi       = {10.5749/minnesota/9781517902438.001.0001},
  isbn      = {9781517902438},
  url       = {https://academic.oup.com/minnesota-scholarship-online/book/17223}
}

@article{mcandrews2023gender,
  author  = {McAndrews, Carolyn and Schneider, Robert J. and Yang, Yicong and Kohn, Genevieve and Schmitz, Andrew and Elliott, Forrest and Pittner, Jessica and Purisch, Hans},
  title   = {Toward a Gender-Inclusive Complete Streets Movement},
  journal = {Journal of Planning Literature},
  volume  = {38},
  number  = {1},
  pages   = {3--18},
  month   = feb,
  year    = {2023},
  doi     = {10.1177/08854122221087472},
  url     = {https://journals.sagepub.com/doi/10.1177/08854122221087472},
  note    = {First published online March 21, 2022}
}

@book{McKercher2020,
  author    = {McKercher, KA},
  title     = {Beyond Sticky Notes: Doing Co-design for Real: Mindsets, Methods and Movements},
  edition   = {1},
  publisher = {Beyond Sticky Notes},
  address   = {Sydney, NSW},
  year      = {2020},
  isbn      = {9780648787501},
  url       = {https://www.beyondstickynotes.com/tellmemore}
}

@article{Mehrabi2021,
  author  = {Mehrabi, Ninareh and Morstatter, Fred and Saxena, Nripsuta and Lerman, Kristina and Galstyan, Aram},
  title   = {{A Survey on Bias and Fairness in Machine Learning}},
  journal = {ACM Computing Surveys},
  year    = {2021},
  volume  = {54},
  number  = {6},
  pages   = {1--35},
  doi     = {10.1145/3457607},
  url     = {https://doi.org/10.1145/3457607}
}

@article{Mehta2014,
  author  = {Mehta, Vikas},
  title   = {Evaluating Public Space},
  journal = {Journal of Urban Design},
  volume  = {19},
  number  = {1},
  pages   = {53--88},
  year    = {2014},
  doi     = {10.1080/13574809.2013.854698},
  url     = {https://www.tandfonline.com/doi/full/10.1080/13574809.2013.854698}
}

@article{Mehta2019,
  author  = {Mehta, Vikas},
  title   = {The Continued Quest to Assess Public Space},
  journal = {Journal of Urban Design},
  volume  = {24},
  number  = {3},
  pages   = {365--367},
  year    = {2019},
  doi     = {10.1080/13574809.2019.1594075},
  url     = {https://www.tandfonline.com/doi/full/10.1080/13574809.2019.1594075}
}

@book{Merry2016,
  author    = {Merry, Sally Engle},
  title     = {The Seductions of Quantification: Measuring Human Rights, Gender Violence, and Sex Trafficking},
  publisher = {University of Chicago Press},
  address   = {Chicago},
  year      = {2016},
  doi       = {10.7208/chicago/9780226261317.001.0001},
  isbn      = {9780226261287},
  url       = {https://press.uchicago.edu/ucp/books/book/chicago/S/bo23044232.html}
}

@article{MikalefEtAl2022,
  author  = {Mikalef, Patrick and Conboy, Kieran and Lundström, Jenny Eriksson and Popovič, Aleš},
  title   = {{Thinking responsibly about responsible AI and ‘the dark side’ of AI}},
  journal = {European Journal of Information Systems},
  year    = {2022},
  volume  = {31},
  number  = {3},
  pages   = {257--268},
  doi     = {10.1080/0960085x.2022.2026621},
  url     = {https://doi.org/10.1080/0960085x.2022.2026621}
}

@article{miller_explanation_2019,
  author  = {Miller, Tim},
  title   = {{Explanation in artificial intelligence: Insights from the social sciences}},
  journal = {Artificial Intelligence},
  year    = {2019},
  volume  = {267},
  pages   = {1--38},
  doi     = {10.1016/j.artint.2018.07.007},
  url     = {https://doi.org/10.1016/j.artint.2018.07.007}
}

@article{Mitchell1995,
  author  = {Mitchell, Don},
  title   = {The End of Public Space? People's Park, Definitions of the Public, and Democracy},
  journal = {Annals of the Association of American Geographers},
  volume  = {85},
  number  = {1},
  pages   = {108--133},
  year    = {1995},
  doi     = {10.1111/j.1467-8306.1995.tb01797.x},
  url     = {https://onlinelibrary.wiley.com/doi/10.1111/j.1467-8306.1995.tb01797.x}
}

@book{Mitchell2003,
  author    = {Mitchell, Don},
  title     = {The Right to the City: Social Justice and the Fight for Public Space},
  publisher = {Guilford Press},
  address   = {New York},
  year      = {2003},
  isbn      = {9781572308473},
  url       = {https://www.guilford.com/books/The-Right-to-the-City/Don-Mitchell/9781572308473}
}

@inproceedings{Mitchell2019,
  author    = {Mitchell, Margaret and Wu, Simone and Zaldivar, Andrew and Barnes, Parker and Vasserman, Lucy and Hutchinson, Ben and Spitzer, Elena and Raji, Inioluwa Deborah and Gebru, Timnit},
  title     = {{Model Cards for Model Reporting}},
  booktitle = {Proceedings of the Conference on Fairness, Accountability, and Transparency},
  year      = {2019},
  pages     = {220--229},
  publisher = {ACM},
  doi       = {10.1145/3287560.3287596},
  url       = {https://doi.org/10.1145/3287560.3287596}
}

@book{MitrasinovicMehta2021,
  title     = {{Public Space Reader}},
  editor    = {Mitrašinović, Miodrag and Mehta, Vikas},
  year      = {2021},
  publisher = {Routledge},
  isbn      = {9781351202558},
  doi       = {10.4324/9781351202558},
  url       = {https://doi.org/10.4324/9781351202558}
}

@article{Mohammed2017,
  author  = {Mohammed, Nabil M. and Niazi, Mahmood and Alshayeb, Mohammad and Mahmood, Sajjad},
  title   = {{Exploring software security approaches in software development lifecycle: A systematic mapping study}},
  journal = {Computer Standards \& Interfaces},
  year    = {2017},
  volume  = {50},
  pages   = {107--115},
  doi     = {10.1016/j.csi.2016.10.001},
  url     = {https://doi.org/10.1016/j.csi.2016.10.001}
}

@book{Molnar2025,
  author    = {Molnar, Christoph},
  title     = {Interpretable Machine Learning: A Guide for Making Black Box Models Explainable},
  edition   = {3},
  publisher = {Self-published},
  year      = {2025},
  isbn      = {978-3-911578-03-5},
  url       = {https://christophm.github.io/interpretable-ml-book/}
}

@incollection{Morison2020,
  author    = {Morison, John},
  title     = {Towards a Democratic Singularity? Algorithmic Governmentality, the Eradication of Politics---and the Possibility of Resistance},
  booktitle = {Is Law Computable? Critical Perspectives on Law and Artificial Intelligence},
  editor    = {Deakin, Simon and Markou, Christopher},
  chapter   = {4},
  pages     = {85--106},
  edition   = {1},
  publisher = {Hart Publishing},
  address   = {Oxford},
  year      = {2020},
  doi       = {10.5040/9781509937097.ch-004},
  isbn      = {9781509937066},
  url       = {https://doi.org/10.5040/9781509937097.ch-004}
}

@book{Morozov2013,
  author    = {Morozov, Evgeny},
  title     = {To Save Everything, Click Here: The Folly of Technological Solutionism},
  edition   = {1},
  publisher = {PublicAffairs},
  address   = {New York},
  year      = {2013},
  isbn      = {9781610391382},
  url       = {https://www.publicaffairsbooks.com/titles/evgeny-morozov/to-save-everything-click-here/9781610391399/}
}

@book{Mouffe2000,
  author    = {Mouffe, Chantal},
  title     = {The Democratic Paradox},
  publisher = {Verso},
  address   = {London},
  year      = {2000},
  isbn      = {9781859842799},
  url       = {https://books.google.com/books?id=5rwu0FA9aO4C}
}

@book{Murray2017,
  author    = {Murray, Cameron Keith and Frijters, Paul},
  title     = {Game of Mates: How Favours Bleed the Nation},
  publisher = {Publicious Pty Ltd},
  address   = {Brisbane, Queensland},
  year      = {2017},
  pagetotal = {204},
  isbn      = {9780648061106},
  url       = {https://catalogue.nla.gov.au/catalog/7343776}
}

@article{mushkani2025position,
  author  = {Mushkani, Rashid and Berard, Hugo and Ammar, Toumadher and Chatonnier, Cassandre and Koseki, Shin},
  title   = {Co-Producing {AI}: Toward an Augmented, Participatory Lifecycle},
  journal = {Proceedings of the AAAI/ACM Conference on AI, Ethics, and Society},
  year    = {2025},
  volume  = {8},
  number  = {2},
  pages   = {1785--1799},
  doi     = {10.1609/aies.v8i2.36674},
  url     = {https://doi.org/10.1609/aies.v8i2.36674}
}

@article{MushkaniHabitat2025,
  author  = {Mushkani, Rashid and Koseki, Shin},
  title   = {Intersecting Perspectives: A Participatory Street Review Framework for Urban Inclusivity},
  journal = {Habitat International},
  year    = {2025},
  volume  = {164},
  pages   = {103536},
  doi     = {10.1016/j.habitatint.2025.103536},
  url     = {https://doi.org/10.1016/j.habitatint.2025.103536}
}

@inproceedings{pmlr-v267-mushkani25a,
  author    = {Mushkani, Rashid and Nayak, Shravan and Berard, Hugo and Cohen, Allison and Koseki, Shin and Bertrand, Hadrien},
  title     = {{LIVS}: A Pluralistic Alignment Dataset for Inclusive Public Spaces},
  booktitle = {Proceedings of the 42nd International Conference on Machine Learning},
  series    = {Proceedings of Machine Learning Research},
  volume    = {267},
  pages     = {45311--45341},
  publisher = {PMLR},
  year      = {2025},
  url       = {https://proceedings.mlr.press/v267/mushkani25a.html}
}

@misc{negotiativealignment,
  author        = {Mushkani, Rashid and Berard, Hugo and Koseki, Shin},
  title         = {Negotiative Alignment: Embracing Disagreement to Achieve Fairer Outcomes---Insights from Urban Studies},
  year          = {2025},
  eprint        = {2503.12613},
  archivePrefix = {arXiv},
  primaryClass  = {cs.HC},
  doi           = {10.48550/arXiv.2503.12613},
  url           = {https://arxiv.org/abs/2503.12613}
}

@inproceedings{airight2025,
  author    = {Mushkani, Rashid and Berard, Hugo and Cohen, Allison and Koseki, Shin},
  title     = {Position: The Right to {AI}},
  booktitle = {Proceedings of the 42nd International Conference on Machine Learning},
  series    = {Proceedings of Machine Learning Research},
  volume    = {267},
  pages     = {81876--81896},
  publisher = {PMLR},
  year      = {2025},
  url       = {https://proceedings.mlr.press/v267/mushkani25b.html}
}

@article{MushkaniJUM2025,
  author  = {Mushkani, Rashid and Berard, Hugo and Ammar, Toumadher and Koseki, Shin},
  title   = {Public Perceptions of Montr{\'e}al's Streets: Implications for Inclusive Public Space Making and Management},
  journal = {Journal of Urban Management},
  year    = {2026},
  volume  = {15},
  number  = {1},
  pages   = {9--25},
  doi     = {10.1016/j.jum.2025.07.004},
  url     = {https://doi.org/10.1016/j.jum.2025.07.004}
}

@article{MushkaniKoseki2026,
  author  = {Mushkani, Rashid and Koseki, Shin},
  title   = {Street Review: A Participatory {AI}-Based Framework for Assessing Streetscape Inclusivity},
  journal = {Cities},
  year    = {2026},
  volume  = {170},
  pages   = {106602},
  doi     = {10.1016/j.cities.2025.106602},
  url     = {https://doi.org/10.1016/j.cities.2025.106602}
}

@inproceedings{Naik2014,
  author    = {Naik, Nikhil and Philipoom, Jade and Raskar, Ramesh and Hidalgo, C{\'e}sar A.},
  title     = {Streetscore---Predicting the Perceived Safety of One Million Streetscapes},
  booktitle = {2014 IEEE Conference on Computer Vision and Pattern Recognition Workshops},
  publisher = {IEEE},
  address   = {Columbus, OH},
  month     = jun,
  year      = {2014},
  pages     = {793--799},
  isbn      = {9781479943081},
  doi       = {10.1109/CVPRW.2014.121},
  url       = {https://doi.org/10.1109/CVPRW.2014.121}
}

@inproceedings{NajafizadehFroehlich2018,
  author    = {Najafizadeh, Ladan and Froehlich, Jon E.},
  title     = {A Feasibility Study of Using {Google Street View} and Computer Vision to Track the Evolution of Urban Accessibility},
  booktitle = {Proceedings of the 20th International {ACM SIGACCESS} Conference on Computers and Accessibility},
  series    = {{ASSETS} '18},
  pages     = {340--342},
  publisher = {ACM},
  year      = {2018},
  doi       = {10.1145/3234695.3240999},
  url       = {https://dl.acm.org/doi/10.1145/3234695.3240999}
}

@inproceedings{Nekoto2020,
  author    = {Nekoto, Wilhelmina and Marivate, Vukosi and Matsila, Tshinondiwa and Fasubaa, Timi and Fagbohungbe, Taiwo and Akinola, Solomon Oluwole and Muhammad, Shamsuddeen and Kabongo Kabenamualu, Salomon and Osei, Salomey and Sackey, Freshia and Niyongabo, Rubungo Andre and Macharm, Ricky and Ogayo, Perez and Ahia, Orevaoghene and Berhe, Musie Meressa and Adeyemi, Mofetoluwa and Mokgesi-Selinga, Masabata and Okegbemi, Lawrence and Martinus, Laura and Tajudeen, Kolawole and Degila, Kevin and Ogueji, Kelechi and Siminyu, Kathleen and Kreutzer, Julia and Webster, Jason and Ali, Jamiil Toure and Abbott, Jade and Orife, Iroro and Ezeani, Ignatius and Dangana, Idris Abdulkadir and Kamper, Herman and Elsahar, Hady and Duru, Goodness and Kioko, Ghollah and Espoir, Murhabazi and van Biljon, Elan and Whitenack, Daniel and Onyefuluchi, Christopher and Emezue, Chris Chinenye and Dossou, Bonaventure F. P. and Sibanda, Blessing and Bassey, Blessing and Olabiyi, Ayodele and Ramkilowan, Arshath and Öktem, Alp and Akinfaderin, Adewale and Bashir, Abdallah},
  title     = {{Participatory Research for Low-resourced Machine Translation: A Case Study in African Languages}},
  booktitle = {Findings of the Association for Computational Linguistics: EMNLP 2020},
  year      = {2020},
  pages     = {2144--2160},
  publisher = {Association for Computational Linguistics},
  doi       = {10.18653/v1/2020.findings-emnlp.195},
  url       = {https://doi.org/10.18653/v1/2020.findings-emnlp.195}
}

@book{Nissenbaum2010,
  author    = {Nissenbaum, Helen},
  title     = {Privacy in Context: Technology, Policy, and the Integrity of Social Life},
  series    = {Stanford Law Books},
  publisher = {Stanford University Press},
  address   = {Stanford, CA},
  year      = {2010},
  doi       = {10.1515/9780804772891},
  isbn      = {9780804752374},
  url       = {https://www.sup.org/books/law/privacy-context},
  note      = {Catalogued and copyrighted as 2010; publisher release date 24 November 2009}
}

@book{Noble2018,
  author    = {Noble, Safiya Umoja},
  title     = {Algorithms of Oppression: How Search Engines Reinforce Racism},
  publisher = {NYU Press},
  address   = {New York},
  year      = {2018},
  doi       = {10.2307/j.ctt1pwt9w5},
  isbn      = {9781479837243},
  url       = {https://nyupress.org/9781479837243/algorithms-of-oppression/}
}

@article{Norman2010,
  author  = {Norman, Geoff},
  title   = {Likert scales, levels of measurement and the ``laws'' of statistics},
  journal = {Advances in Health Sciences Education},
  volume  = {15},
  number  = {5},
  pages   = {625--632},
  year    = {2010},
  doi     = {10.1007/s10459-010-9222-y},
  url     = {https://link.springer.com/article/10.1007/s10459-010-9222-y}
}

@book{NuceraOnuoha2018,
  author    = {Onuoha, Mimi and Nucera, Diana},
  title     = {A People's Guide to {AI}},
  publisher = {Allied Media Projects},
  address   = {Detroit},
  month     = aug,
  year      = {2018},
  url       = {https://alliedmedia.org/wp-content/uploads/2020/09/peoples-guide-ai.pdf},
  note      = {Diana Nucera is also known as Mother Cyborg; 78 pages}
}

@misc{openai2024gpt4,
  author        = {{OpenAI}},
  title         = {{GPT-4o} System Card},
  month         = oct,
  year          = {2024},
  eprint        = {2410.21276},
  archivePrefix = {arXiv},
  primaryClass  = {cs.CL},
  doi           = {10.48550/arXiv.2410.21276},
  url           = {https://arxiv.org/abs/2410.21276},
  note          = {System card; first released by OpenAI on August 8, 2024}
}

@article{Ostrom1996,
  author  = {Ostrom, Elinor},
  title   = {{Crossing the great divide: Coproduction, synergy, and development}},
  journal = {World Development},
  year    = {1996},
  volume  = {24},
  number  = {6},
  pages   = {1073--1087},
  doi     = {10.1016/0305-750x(96)00023-x},
  url     = {https://doi.org/10.1016/0305-750x(96)00023-x}
}

@book{Ostrom2009,
  author    = {Ostrom, Elinor},
  title     = {Understanding Institutional Diversity},
  publisher = {Princeton University Press},
  address   = {Princeton, NJ},
  year      = {2009},
  isbn      = {9781400831739},
  note      = {Electronic edition; original edition published 2005},
  doi       = {10.2307/j.ctt7s7wm},
  url       = {https://doi.org/10.2307/j.ctt7s7wm}
}

@inproceedings{ouyang2022,
  author    = {Ouyang, Long and Wu, Jeffrey and Jiang, Xu and Almeida, Diogo and Wainwright, Carroll and Mishkin, Pamela and Zhang, Chong and Agarwal, Sandhini and Slama, Katarina and Ray, Alex and Schulman, John and Hilton, Jacob and Kelton, Fraser and Miller, Luke and Simens, Maddie and Askell, Amanda and Welinder, Peter and Christiano, Paul F. and Leike, Jan and Lowe, Ryan},
  title     = {Training Language Models to Follow Instructions with Human Feedback},
  booktitle = {Advances in Neural Information Processing Systems},
  volume    = {35},
  pages     = {27730--27744},
  publisher = {Curran Associates, Inc.},
  year      = {2022},
  doi       = {10.52202/068431-2011},
  url       = {https://proceedings.neurips.cc/paper_files/paper/2022/hash/b1efde53be364a73914f58805a001731-Abstract-Conference.html},
  note      = {Main Conference Track}
}

@book{pallasmaa2011embodied,
  author    = {Pallasmaa, Juhani},
  title     = {The Embodied Image: Imagination and Imagery in Architecture},
  series    = {{AD} Primers},
  edition   = {1},
  publisher = {John Wiley \& Sons},
  address   = {Chichester},
  year      = {2011},
  isbn      = {978-0-470-71190-3},
  url       = {https://www.cca.qc.ca/en/search/details/library/publication/670479708}
}

@book{Pasquale2015,
  author    = {Pasquale, Frank},
  title     = {The Black Box Society: The Secret Algorithms That Control Money and Information},
  publisher = {Harvard University Press},
  address   = {Cambridge, MA},
  year      = {2015},
  doi       = {10.4159/harvard.9780674736061},
  isbn      = {9780674368279},
  url       = {https://doi.org/10.4159/harvard.9780674736061}
}

@article{PavlickKwiatkowski2019,
  author  = {Pavlick, Ellie and Kwiatkowski, Tom},
  title   = {Inherent Disagreements in Human Textual Inferences},
  journal = {Transactions of the Association for Computational Linguistics},
  volume  = {7},
  pages   = {677--694},
  year    = {2019},
  doi     = {10.1162/tacl_a_00293},
  url     = {https://direct.mit.edu/tacl/article/doi/10.1162/tacl_a_00293/43531/Inherent-Disagreements-in-Human-Textual-Inferences}
}

@inproceedings{PawarEtAl2018,
  author    = {Pawar, Ankita and Ahirrao, Snehal and Churi, Prathamesh P.},
  title     = {Anonymization Techniques for Protecting Privacy: A Survey},
  booktitle = {2018 {IEEE} PuneCon},
  address   = {Pune, India},
  pages     = {1--6},
  publisher = {IEEE},
  year      = {2018},
  doi       = {10.1109/PUNECON.2018.8745425},
  url       = {https://ieeexplore.ieee.org/document/8745425/}
}

@inproceedings{plank2011jsd,
  author    = {Plank, Barbara and {van Noord}, Gertjan},
  editor    = {Lin, Dekang and Matsumoto, Yuji and Mihalcea, Rada},
  title     = {Effective Measures of Domain Similarity for Parsing},
  booktitle = {Proceedings of the 49th Annual Meeting of the Association for Computational Linguistics: Human Language Technologies},
  address   = {Portland, Oregon, USA},
  pages     = {1566--1576},
  publisher = {Association for Computational Linguistics},
  month     = jun,
  year      = {2011},
  doi       = {10.5555/2002472.2002661},
  url       = {https://aclanthology.org/P11-1157/}
}

@article{plantin2018infrastructure,
  author  = {Plantin, Jean-Christophe and Lagoze, Carl and Edwards, Paul N and Sandvig, Christian},
  title   = {{Infrastructure studies meet platform studies in the age of Google and Facebook}},
  journal = {New Media \& Society},
  year    = {2018},
  volume  = {20},
  number  = {1},
  pages   = {293--310},
  doi     = {10.1177/1461444816661553},
  url     = {https://doi.org/10.1177/1461444816661553}
}

@inproceedings{podell2023sdxlimprovinglatentdiffusion,
  author    = {Podell, Dustin and English, Zion and Lacey, Kyle and Blattmann, Andreas and Dockhorn, Tim and M{\"u}ller, Jonas and Penna, Joe and Rombach, Robin},
  title     = {{SDXL}: Improving Latent Diffusion Models for High-Resolution Image Synthesis},
  booktitle = {International Conference on Learning Representations},
  year      = {2024},
  url       = {https://proceedings.iclr.cc/paper_files/paper/2024/hash/081b08068e4733ae3e7ad019fe8d172f-Abstract-Conference.html},
  note      = {ICLR 2024 conference paper; no publisher DOI assigned}
}

@book{Porter1995,
  author    = {Porter, Theodore M.},
  title     = {Trust in Numbers: The Pursuit of Objectivity in Science and Public Life},
  publisher = {Princeton University Press},
  address   = {Princeton, NJ},
  year      = {1995},
  doi       = {10.1515/9781400821617},
  isbn      = {9780691037769},
  url       = {https://doi.org/10.1515/9781400821617}
}

@book{Power1997,
  author    = {Power, Michael},
  title     = {The Audit Society: Rituals of Verification},
  publisher = {Oxford University Press},
  address   = {Oxford},
  year      = {1997},
  doi       = {10.1093/acprof:oso/9780198296034.001.0001},
  isbn      = {9780198289470},
  url       = {https://global.oup.com/academic/product/the-audit-society-9780198289470}
}

@techreport{pressmancrowson2022,
  author      = {Pressman, John David and Crowson, Katherine and {Simulacra Captions Contributors}},
  title       = {Simulacra Aesthetic Captions},
  institution = {Stability AI},
  type        = {Technical report},
  number      = {Version 1.0},
  year        = {2022},
  url         = {https://github.com/JD-P/simulacra-aesthetic-captions},
  urldate     = {2026-07-21}
}

@article{Purcell2014,
  author  = {Purcell, Mark},
  title   = {Possible Worlds: Henri Lefebvre and the Right to the City},
  journal = {Journal of Urban Affairs},
  volume  = {36},
  number  = {1},
  pages   = {141--154},
  month   = feb,
  year    = {2014},
  doi     = {10.1111/juaf.12034},
  url     = {https://www.tandfonline.com/doi/full/10.1111/juaf.12034}
}

@article{PuthEtAl2015,
  author  = {Puth, Marie-Therese and Neuhäuser, Markus and Ruxton, Graeme D.},
  title   = {Effective use of Spearman's and Kendall's correlation coefficients for association between two measured traits},
  journal = {Animal Behaviour},
  volume  = {102},
  pages   = {77--84},
  year    = {2015},
  doi     = {10.1016/j.anbehav.2015.01.010},
  url     = {https://www.sciencedirect.com/science/article/pii/S0003347215000196}
}

@inproceedings{Qadri2023representation,
  author    = {Qadri, Rida and Shelby, Renee and Bennett, Cynthia L. and Denton, Emily},
  title     = {{AI}'s Regimes of Representation: A Community-Centered Study of Text-to-Image Models in South Asia},
  booktitle = {Proceedings of the 2023 ACM Conference on Fairness, Accountability, and Transparency},
  series    = {FAccT '23},
  address   = {New York, NY, USA},
  publisher = {Association for Computing Machinery},
  pages     = {506--517},
  year      = {2023},
  isbn      = {9798400701924},
  doi       = {10.1145/3593013.3594016},
  url       = {https://dl.acm.org/doi/10.1145/3593013.3594016}
}

@inproceedings{rafailov2024,
  author    = {Rafailov, Rafael and Sharma, Archit and Mitchell, Eric and Manning, Christopher D. and Ermon, Stefano and Finn, Chelsea},
  title     = {Direct Preference Optimization: Your Language Model Is Secretly a Reward Model},
  booktitle = {Advances in Neural Information Processing Systems},
  volume    = {36},
  pages     = {53728--53741},
  publisher = {Curran Associates, Inc.},
  year      = {2023},
  doi       = {10.5555/3666122.3668460},
  url       = {https://proceedings.neurips.cc/paper_files/paper/2023/hash/a85b405ed65c6477a4fe8302b5e06ce7-Abstract-Conference.html}
}

@book{rawls1993political,
  author         = {Rawls, John},
  title          = {Political Liberalism},
  year           = {1993},
  publisher      = {Columbia University Press},
  address        = {New York, NY},
  isbn           = {9780231052481},
  url            = {https://search.worldcat.org/title/Political-liberalism/oclc/782135666}
}

@techreport{Reisman2018,
  author      = {Reisman, Dillon and Schultz, Jason and Crawford, Kate and Whittaker, Meredith},
  title       = {Algorithmic Impact Assessments Report: A Practical Framework for Public Agency Accountability},
  institution = {AI Now Institute},
  address     = {New York},
  day         = {9},
  month       = apr,
  year        = {2018},
  url         = {https://ainowinstitute.org/publications/algorithmic-impact-assessments-report-2}
}

@article{roberts2017crossvalidation,
  author         = {Roberts, David R. and Bahn, Volker and Ciuti, Simone and Boyce, Mark S. and Elith, Jane and Guillera-Arroita, Gurutzeta and Hauenstein, Severin and Lahoz-Monfort, Jos{\'e} J. and Schr{\"o}der, Boris and Thuiller, Wilfried and Warton, David I. and Wintle, Brendan A. and Hartig, Florian and Dormann, Carsten F.},
  title          = {Cross-validation Strategies for Data with Temporal, Spatial, Hierarchical, or Phylogenetic Structure},
  journal        = {Ecography},
  year           = {2017},
  volume         = {40},
  number         = {8},
  pages          = {913--929},
  doi            = {10.1111/ecog.02881},
  issn           = {0906-7590},
  url            = {https://doi.org/10.1111/ecog.02881}
}

@inproceedings{SahaEtAl2019,
  author    = {Saha, Manaswi and Saugstad, Michael and Maddali, Hanuma Teja and Zeng, Aileen and Holland, Ryan and Bower, Steven and Dash, Aditya and Chen, Sage and Li, Anthony and Hara, Kotaro and Froehlich, Jon},
  title     = {{Project Sidewalk}: A Web-Based Crowdsourcing Tool for Collecting Sidewalk Accessibility Data at Scale},
  booktitle = {Proceedings of the 2019 {CHI} Conference on Human Factors in Computing Systems},
  series    = {{CHI} '19},
  pages     = {1--14},
  publisher = {ACM},
  year      = {2019},
  doi       = {10.1145/3290605.3300292},
  url       = {https://dl.acm.org/doi/10.1145/3290605.3300292}
}

@article{Saheb2024,
  author  = {Saheb, Tahereh and Saheb, Tayebeh},
  title   = {Mapping Ethical Artificial Intelligence Policy Landscape: A Mixed Method Analysis},
  journal = {Science and Engineering Ethics},
  year    = {2024},
  volume  = {30},
  number  = {2},
  pages   = {9},
  doi     = {10.1007/s11948-024-00472-6},
  url     = {https://doi.org/10.1007/s11948-024-00472-6}
}

@article{Schiff2021a,
  author  = {Schiff, Daniel and Borenstein, Jason and Biddle, Justin and Laas, Kelly},
  title   = {{AI} Ethics in the Public, Private, and {NGO} Sectors: A Review of a Global Document Collection},
  journal = {IEEE Transactions on Technology and Society},
  year    = {2021},
  volume  = {2},
  number  = {1},
  pages   = {31--42},
  doi     = {10.1109/TTS.2021.3052127},
  url     = {https://doi.org/10.1109/TTS.2021.3052127}
}

@article{Schiff2021b,
  author  = {Schiff, Daniel and Rakova, Bogdana and Ayesh, Aladdin and Fanti, Anat and Lennon, Michael},
  title   = {Explaining the Principles to Practices Gap in {AI}},
  journal = {IEEE Technology and Society Magazine},
  year    = {2021},
  volume  = {40},
  number  = {2},
  pages   = {81--94},
  doi     = {10.1109/MTS.2021.3056286},
  url     = {https://doi.org/10.1109/MTS.2021.3056286}
}

@book{scott1998seeing,
  author         = {Scott, James C.},
  title          = {Seeing Like a State: How Certain Schemes to Improve the Human Condition Have Failed},
  year           = {1998},
  publisher      = {Yale University Press},
  address        = {New Haven, CT},
  series         = {Yale Agrarian Studies},
  isbn           = {9780300070163},
  url            = {https://www.jstor.org/stable/j.ctt1nq3vk}
}

@inproceedings{Sculley2015,
  author    = {Sculley, D. and Holt, Gary and Golovin, Daniel and Davydov, Eugene and Phillips, Todd and Ebner, Dietmar and Chaudhary, Vinay and Young, Michael and Crespo, Jean-Fran{\c c}ois and Dennison, Dan},
  title     = {Hidden Technical Debt in Machine Learning Systems},
  booktitle = {Advances in Neural Information Processing Systems},
  volume    = {28},
  pages     = {2503--2511},
  publisher = {Curran Associates, Inc.},
  year      = {2015},
  doi       = {10.5555/2969442.2969519},
  url       = {https://proceedings.neurips.cc/paper_files/paper/2015/hash/86df7dcfd896fcaf2674f757a2463eba-Abstract.html}
}

@inproceedings{selbst2019fairness,
  author         = {Selbst, Andrew D. and Boyd, Danah and Friedler, Sorelle A. and Venkatasubramanian, Suresh and Vertesi, Janet},
  title          = {Fairness and Abstraction in Sociotechnical Systems},
  booktitle      = {Proceedings of the Conference on Fairness, Accountability, and Transparency},
  series         = {FAT* '19},
  year           = {2019},
  pages          = {59--68},
  publisher      = {Association for Computing Machinery},
  address        = {New York, NY},
  location       = {Atlanta, GA, USA},
  isbn           = {9781450361255},
  doi            = {10.1145/3287560.3287598},
  url            = {https://doi.org/10.1145/3287560.3287598}
}

@article{selbst2021institutional,
  author         = {Selbst, Andrew D.},
  title          = {An Institutional View of Algorithmic Impact Assessments},
  journal        = {Harvard Journal of Law \& Technology},
  year           = {2021},
  volume         = {35},
  number         = {1},
  pages          = {117--191},
  issn           = {0897-3393},
  url            = {https://jolt.law.harvard.edu/assets/articlePDFs/v35/Selbst-An-Institutional-View-of-Algorithmic-Impact-Assessments.pdf}
}

@inproceedings{perrak2024,
  author    = {Shah, Prerak},
  title     = {Addressing Bias in Text-to-Image Generation: A Review of Mitigation Methods},
  booktitle = {2024 Third International Conference on Smart Technologies and Systems for Next Generation Computing (ICSTSN)},
  address   = {Villupuram, India},
  pages     = {1--6},
  publisher = {IEEE},
  month     = jul,
  year      = {2024},
  doi       = {10.1109/ICSTSN61422.2024.10671230},
  url       = {https://ieeexplore.ieee.org/document/10671230/}
}

@article{shelton2015actually,
  author         = {Shelton, Taylor and Zook, Matthew and Wiig, Alan},
  title          = {The ``Actually Existing Smart City''},
  journal        = {Cambridge Journal of Regions, Economy and Society},
  year           = {2015},
  volume         = {8},
  number         = {1},
  pages          = {13--25},
  doi            = {10.1093/cjres/rsu026},
  issn           = {1752-1378},
  url            = {https://doi.org/10.1093/cjres/rsu026}
}

@article{Sieber2006,
  author   = {Sieber, Ren{\'e}e},
  title    = {Public Participation Geographic Information Systems: A Literature Review and Framework},
  journal  = {Annals of the Association of American Geographers},
  year     = {2006},
  volume   = {96},
  number   = {3},
  pages    = {491--507},
  doi      = {10.1111/j.1467-8306.2006.00702.x},
  url      = {https://doi.org/10.1111/j.1467-8306.2006.00702.x}
}

@article{Sieber2024PublicsEngaging,
  author  = {Sieber, Ren{\'e}e and Brandusescu, Ana and Adu-Daako, Abigail and Sangiambut, Suthee},
  title   = {Who Are the Publics Engaging in {AI}?},
  journal = {Public Understanding of Science},
  year    = {2024},
  volume  = {33},
  number  = {5},
  pages   = {634--653},
  doi     = {10.1177/09636625231219853},
  url     = {https://doi.org/10.1177/09636625231219853}
}

@article{Sieber2024CivicParticipation,
  author  = {Sieber, Ren{\'e}e and Brandusescu, Ana and Sangiambut, Suthee and Adu-Daako, Abigail},
  title   = {What Is Civic Participation in Artificial Intelligence?},
  journal = {Environment and Planning B: Urban Analytics and City Science},
  year    = {2025},
  volume  = {52},
  number  = {6},
  pages   = {1388--1406},
  doi     = {10.1177/23998083241296200},
  url     = {https://doi.org/10.1177/23998083241296200}
}

@inproceedings{sloane2022,
  author    = {Sloane, Mona and Moss, Emanuel and Awomolo, Olaitan and Forlano, Laura},
  title     = {Participation Is Not a Design Fix for Machine Learning},
  booktitle = {Equity and Access in Algorithms, Mechanisms, and Optimization ({EAAMO} '22)},
  year      = {2022},
  month     = oct,
  pages     = {1--6},
  publisher = {Association for Computing Machinery},
  isbn      = {9781450394772},
  doi       = {10.1145/3551624.3555285},
  url       = {https://doi.org/10.1145/3551624.3555285}
}

@article{Sloane2024,
  author  = {Sloane, Mona},
  title   = {Controversies, Contradiction, and ``Participation'' in {AI}},
  journal = {Big Data \& Society},
  year    = {2024},
  volume  = {11},
  number  = {1},
  pages   = {20539517241235862},
  doi     = {10.1177/20539517241235862},
  url     = {https://doi.org/10.1177/20539517241235862}
}

@inproceedings{Sorensen2024,
  author    = {Sorensen, Taylor and Moore, Jared and Fisher, Jillian and Gordon, Mitchell L. and Mireshghallah, Niloofar and Rytting, Christopher Michael and Ye, Andre and Jiang, Liwei and Lu, Ximing and Dziri, Nouha and Althoff, Tim and Choi, Yejin},
  title     = {Position: A Roadmap to Pluralistic Alignment},
  booktitle = {Proceedings of the 41st International Conference on Machine Learning},
  editor    = {Salakhutdinov, Ruslan and Kolter, Zico and Heller, Katherine and Weller, Adrian and Oliver, Nuria and Scarlett, Jonathan and Berkenkamp, Felix},
  series    = {Proceedings of Machine Learning Research},
  volume    = {235},
  pages     = {46280--46302},
  publisher = {PMLR},
  month     = jul,
  year      = {2024},
  url       = {https://proceedings.mlr.press/v235/sorensen24a.html}
}

@article{star1999ethnography,
  author         = {Star, Susan Leigh},
  title          = {The Ethnography of Infrastructure},
  journal        = {American Behavioral Scientist},
  year           = {1999},
  volume         = {43},
  number         = {3},
  pages          = {377--391},
  doi            = {10.1177/00027649921955326},
  issn           = {0002-7642},
  url            = {https://doi.org/10.1177/00027649921955326}
}

@misc{StatCan2022,
  author       = {{Statistics Canada}},
  title        = {Census Profile, 2021 Census of Population},
  howpublished = {Government of Canada},
  year         = {2022},
  month        = feb,
  note         = {Released February 9, 2022; updated November 15, 2023},
  url          = {https://www12.statcan.gc.ca/census-recensement/2021/dp-pd/prof/index.cfm?Lang=E},
  urldate      = {2026-07-21}
}

@article{Sun2020,
  author  = {Sun, Haochen},
  title   = {Reinvigorating the Human Right to Technology},
  journal = {Michigan Journal of International Law},
  year    = {2020},
  volume  = {41},
  number  = {2},
  pages   = {279--326},
  doi     = {10.36642/mjil.41.2.reinvigorating},
  url     = {https://doi.org/10.36642/mjil.41.2.reinvigorating}
}

@techreport{NIST2023,
  author      = {Tabassi, Elham},
  title       = {Artificial Intelligence Risk Management Framework ({AI RMF} 1.0)},
  institution = {National Institute of Standards and Technology},
  address     = {Gaithersburg, MD},
  type        = {NIST AI},
  number      = {100-1},
  year        = {2023},
  doi         = {10.6028/NIST.AI.100-1},
  url         = {https://doi.org/10.6028/NIST.AI.100-1}
}

@article{Taeih2021,
  author  = {Taeihagh, Araz},
  title   = {{Governance of artificial intelligence}},
  journal = {Policy and Society},
  year    = {2021},
  volume  = {40},
  number  = {2},
  pages   = {137--157},
  doi     = {10.1080/14494035.2021.1928377},
  url     = {https://doi.org/10.1080/14494035.2021.1928377}
}

@book{talen2012design,
  author    = {Talen, Emily},
  title     = {Design for Diversity: Exploring Socially Mixed Neighborhoods},
  edition   = {1},
  publisher = {Taylor \& Francis},
  year      = {2012},
  isbn      = {9780080557601},
  doi       = {10.4324/9780080557601},
  url       = {https://www.taylorfrancis.com/books/mono/10.4324/9780080557601/design-diversity-emily-talen}
}

@article{Terzi2010,
  author  = {Terzi, Sergio and Bouras, Abdelaziz and Dutta, Debashi and Garetti, Marco and Kiritsis, Dimitris},
  title   = {Product Lifecycle Management -- from Its History to Its New Role},
  journal = {International Journal of Product Lifecycle Management},
  year    = {2010},
  volume  = {4},
  number  = {4},
  pages   = {360--389},
  doi     = {10.1504/IJPLM.2010.036489},
  url     = {https://doi.org/10.1504/IJPLM.2010.036489}
}

@techreport{ieee2019ethically,
  author         = {{The IEEE Global Initiative on Ethics of Autonomous and Intelligent Systems}},
  title          = {Ethically Aligned Design: A Vision for Prioritizing Human Well-being with Autonomous and Intelligent Systems},
  year           = {2019},
  institution    = {IEEE},
  type           = {First Edition},
  pagetotal      = {294},
  url            = {https://standards.ieee.org/wp-content/uploads/import/documents/other/ead1e.pdf}
}

@article{tong2007coreq,
  author         = {Tong, Allison and Sainsbury, Peter and Craig, Jonathan},
  title          = {Consolidated Criteria for Reporting Qualitative Research ({COREQ}): A 32-item Checklist for Interviews and Focus Groups},
  journal        = {International Journal for Quality in Health Care},
  year           = {2007},
  volume         = {19},
  number         = {6},
  pages          = {349--357},
  doi            = {10.1093/intqhc/mzm042},
  issn           = {1353-4505},
  url            = {https://doi.org/10.1093/intqhc/mzm042}
}

@book{townsend2013smart,
  author         = {Townsend, Anthony M.},
  title          = {Smart Cities: Big Data, Civic Hackers, and the Quest for a New Utopia},
  year           = {2013},
  publisher      = {W. W. Norton \& Company},
  address        = {New York, NY},
  isbn           = {9780393082876},
  url            = {https://wwnorton.com/books/smart-cities/}
}

@misc{Turchin2019-valueDoNOtExist,
  author       = {Turchin, Alexey},
  title        = {{AI} Alignment Problem: ``Human Values'' Don't Actually Exist},
  howpublished = {PhilArchive},
  day          = {22},
  month        = apr,
  year         = {2019},
  url          = {https://philarchive.org/rec/TURAAP},
  note         = {Repository manuscript}
}

@article{turnhout2010participation,
  author  = {Turnhout, Esther and Van Bommel, Severine and Aarts, Noelle},
  title   = {How Participation Creates Citizens: Participatory Governance as Performative Practice},
  journal = {Ecology and Society},
  year    = {2010},
  volume  = {15},
  number  = {4},
  pages   = {26},
  doi     = {10.5751/ES-03701-150426},
  issn    = {1708-3087},
  url     = {https://doi.org/10.5751/ES-03701-150426}
}

@article{Ulnicane2024,
  author  = {Ulnicane, Inga},
  title   = {Governance Fix? Power and Politics in Controversies about Governing Generative {AI}},
  journal = {Policy and Society},
  year    = {2025},
  volume  = {44},
  number  = {1},
  pages   = {70--84},
  doi     = {10.1093/polsoc/puae022},
  url     = {https://doi.org/10.1093/polsoc/puae022}
}

@techreport{montreal2018declaration,
  author      = {{Universit{\'e} de Montr{\'e}al}},
  title       = {Montr{\'e}al Declaration for a Responsible Development of Artificial Intelligence},
  institution = {Universit{\'e} de Montr{\'e}al},
  address     = {Montr{\'e}al, Qu{\'e}bec},
  month       = dec,
  year        = {2018},
  url         = {https://montrealdeclaration-responsibleai.com/the-declaration/}
}

@article{valenca2025visualizations,
  author         = {Valen{\c{c}}a, Gabriel and Azevedo, Carlos M. Lima and Moura, Filipe and de S{\'a}, Ana Morais},
  title          = {Creating Visualizations Using Generative {AI} to Guide Decision-making in Street Designs: A Viewpoint},
  journal        = {Journal of Urban Mobility},
  year           = {2025},
  volume         = {7},
  pages          = {100104},
  articleno      = {100104},
  doi            = {10.1016/j.urbmob.2025.100104},
  issn           = {2667-0917},
  url            = {https://doi.org/10.1016/j.urbmob.2025.100104}
}

@inproceedings{Varanasi2023,
  author    = {Varanasi, Rama Adithya and Goyal, Nitesh},
  title     = {{“It is currently hodgepodge”: Examining AI/ML Practitioners’ Challenges during Co-production of Responsible AI Values}},
  booktitle = {Proceedings of the 2023 CHI Conference on Human Factors in Computing Systems},
  year      = {2023},
  pages     = {1--17},
  publisher = {ACM},
  doi       = {10.1145/3544548.3580903},
  url       = {https://doi.org/10.1145/3544548.3580903}
}

@article{varna2010publicness,
  author         = {Varna, George and Tiesdell, Steve},
  title          = {Assessing the Publicness of Public Space: The Star Model of Publicness},
  journal        = {Journal of Urban Design},
  year           = {2010},
  volume         = {15},
  number         = {4},
  pages          = {575--598},
  doi            = {10.1080/13574809.2010.502350},
  issn           = {1357-4809},
  url            = {https://doi.org/10.1080/13574809.2010.502350}
}

@inproceedings{Vaswani2017,
  author    = {Vaswani, Ashish and Shazeer, Noam and Parmar, Niki and Uszkoreit, Jakob and Jones, Llion and Gomez, Aidan N. and Kaiser, {\L}ukasz and Polosukhin, Illia},
  title     = {Attention Is All You Need},
  booktitle = {Advances in Neural Information Processing Systems},
  editor    = {Guyon, Isabelle and von Luxburg, Ulrike and Bengio, Samy and Wallach, Hanna M. and Fergus, Rob and Vishwanathan, S. V. N. and Garnett, Roman},
  volume    = {30},
  pages     = {5998--6008},
  publisher = {Curran Associates, Inc.},
  year      = {2017},
  isbn      = {9781510860964},
  url       = {https://proceedings.neurips.cc/paper_files/paper/2017/hash/3f5ee243547dee91fbd053c1c4a845aa-Abstract.html}
}

@book{Vliz2021,
  author    = {V{\'e}liz, Carissa},
  title     = {Privacy Is Power: Why and How You Should Take Back Control of Your Data},
  publisher = {Corgi},
  address   = {London},
  month     = jul,
  year      = {2021},
  isbn      = {9780552177719},
  url       = {https://www.penguin.co.uk/books/442343/privacy-is-power-by-carissa-veliz/9780552177719}
}

@incollection{vonBrackelSchmidtEtAl2024,
  author    = {von Brackel-Schmidt, Constantin and Ku{\v{c}}evi{\'c}, Emir and Leible, Stephan and Simic, Dejan and G{\"u}c{\"u}k, Gian-Luca and Schmidt, Felix N.},
  title     = {Equipping Participation Formats with Generative {AI}: A Case Study Predicting the Future of a Metropolitan City in the Year 2040},
  booktitle = {{HCI} in Business, Government and Organizations},
  editor    = {Nah, Fiona Fui-Hoon and Siau, Keng Leng},
  series    = {Lecture Notes in Computer Science},
  volume    = {14720},
  pages     = {270--285},
  publisher = {Springer Nature Switzerland},
  address   = {Cham},
  year      = {2024},
  doi       = {10.1007/978-3-031-61315-9_19},
  isbn      = {9783031613142},
  url       = {https://link.springer.com/chapter/10.1007/978-3-031-61315-9_19}
}

@inproceedings{wallace2023,
  author    = {Wallace, Bram and Dang, Meihua and Rafailov, Rafael and Zhou, Linqi and Lou, Aaron and Purushwalkam, Senthil and Ermon, Stefano and Xiong, Caiming and Joty, Shafiq and Naik, Nikhil},
  title     = {Diffusion Model Alignment Using Direct Preference Optimization},
  booktitle = {Proceedings of the IEEE/CVF Conference on Computer Vision and Pattern Recognition ({CVPR})},
  year      = {2024},
  month     = jun,
  pages     = {8228--8238},
  publisher = {IEEE},
  doi       = {10.1109/CVPR52733.2024.00786},
  url       = {https://doi.org/10.1109/CVPR52733.2024.00786}
}

@misc{wan2024survey,
  author        = {Wan, Yixin and Subramonian, Arjun and Ovalle, Anaelia and Lin, Zongyu and Suvarna, Ashima and Chance, Christina and Bansal, Hritik and Pattichis, Rebecca and Chang, Kai-Wei},
  title         = {Survey of Bias in Text-to-Image Generation: Definition, Evaluation, and Mitigation},
  month         = apr,
  year          = {2024},
  eprint        = {2404.01030},
  archivePrefix = {arXiv},
  primaryClass  = {cs.CV},
  doi           = {10.48550/arXiv.2404.01030},
  url           = {https://arxiv.org/abs/2404.01030},
  note          = {arXiv preprint}
}

@incollection{Wenar2023,
  author    = {Wenar, Leif},
  title     = {Rights},
  booktitle = {The Stanford Encyclopedia of Philosophy},
  editor    = {Zalta, Edward N. and Nodelman, Uri},
  edition   = {Spring 2023},
  publisher = {Metaphysics Research Lab, Stanford University},
  year      = {2023},
  issn      = {1095-5054},
  url       = {https://plato.stanford.edu/archives/spr2023/entries/rights/}
}

@book{Whyte2001,
  author    = {Whyte, William H.},
  title     = {The Social Life of Small Urban Spaces},
  edition   = {8},
  publisher = {Project for Public Spaces},
  address   = {New York, NY},
  year      = {2001},
  isbn      = {9780970632418},
  note      = {Reprint; original edition published in 1980},
  url       = {https://www.pps.org/product/the-social-life-of-small-urban-spaces}
}

@article{wong2023toolkit,
  author         = {Wong, Richmond Y. and Madaio, Michael A. and Merrill, Nick},
  title          = {Seeing Like a Toolkit: How Toolkits Envision the Work of {AI} Ethics},
  journal        = {Proceedings of the ACM on Human-Computer Interaction},
  year           = {2023},
  volume         = {7},
  number         = {CSCW1},
  articleno      = {145},
  pages          = {1--27},
  doi            = {10.1145/3579621},
  issn           = {2573-0142},
  url            = {https://doi.org/10.1145/3579621}
}

@inproceedings{Xie2021,
  author    = {Xie, Enze and Wang, Wenhai and Yu, Zhiding and Anandkumar, Anima and Alvarez, Jose M. and Luo, Ping},
  title     = {{SegFormer}: Simple and Efficient Design for Semantic Segmentation with Transformers},
  booktitle = {Advances in Neural Information Processing Systems},
  volume    = {34},
  pages     = {12077--12090},
  publisher = {Curran Associates, Inc.},
  year      = {2021},
  doi       = {10.5555/3540261.3541185},
  url       = {https://proceedings.neurips.cc/paper/2021/hash/64f1f27bf1b4ec22924fd0acb550c235-Abstract.html}
}

@inproceedings{xu2023imagereward,
  author    = {Xu, Jiazheng and Liu, Xiao and Wu, Yuchen and Tong, Yuxuan and Li, Qinkai and Ding, Ming and Tang, Jie and Dong, Yuxiao},
  title     = {{ImageReward}: Learning and Evaluating Human Preferences for Text-to-Image Generation},
  booktitle = {Advances in Neural Information Processing Systems},
  volume    = {36},
  pages     = {15903--15935},
  publisher = {Curran Associates, Inc.},
  year      = {2023},
  doi       = {10.5555/3666122.3666822},
  url       = {https://openreview.net/forum?id=JVzeOYEx6d}
}

@article{Ye2019,
  author  = {Ye, Junjia},
  title   = {{Re-orienting geographies of urban diversity and coexistence: Analyzing inclusion and difference in public space}},
  journal = {Progress in Human Geography},
  year    = {2019},
  volume  = {43},
  number  = {3},
  pages   = {478--495},
  doi     = {10.1177/0309132518768405},
  url     = {https://doi.org/10.1177/0309132518768405}
}

@article{yigitcanlar2025editorial,
  author         = {Yigitcanlar, Tan and Desouza, Kevin C. and Mossberger, Karen and Cheong, Pauline H. and Li, Rita Yi Man and Mehmood, Rashid and Corchado, Juan M.},
  title          = {Artificial Intelligence and the City: An Editorial Perspective},
  journal        = {Journal of Urban Technology},
  year           = {2025},
  volume         = {32},
  number         = {3},
  pages          = {1--7},
  doi            = {10.1080/10630732.2025.2500822},
  issn           = {1063-0732},
  url            = {https://doi.org/10.1080/10630732.2025.2500822}
}

@article{Youngbloom2023,
  author  = {Youngbloom, Amy J. and Thierry, Benoit and Fuller, Daniel and Kestens, Yan and Winters, Meghan and Hirsch, Jana A. and Michael, Yvonne L. and Firth, Caislin},
  title   = {{Gentrification, perceptions of neighborhood change, and mental health in Montréal, Québec}},
  journal = {SSM - Population Health},
  year    = {2023},
  volume  = {22},
  pages   = {101406},
  doi     = {10.1016/j.ssmph.2023.101406},
  url     = {https://doi.org/10.1016/j.ssmph.2023.101406}
}

@book{young2000inclusion,
  author         = {Young, Iris Marion},
  title          = {Inclusion and Democracy},
  year           = {2000},
  series         = {Oxford Political Theory},
  publisher      = {Oxford University Press},
  address        = {Oxford},
  isbn           = {9780198297543, 9780198297550},
  doi            = {10.1093/0198297556.001.0001},
  url            = {https://doi.org/10.1093/0198297556.001.0001}
}

@article{ZaidanIbrahim2024,
  author  = {Zaidan, Esmat and Ibrahim, Imad Antoine},
  title   = {{AI} Governance in a Complex and Rapidly Changing Regulatory Landscape: A Global Perspective},
  journal = {Humanities and Social Sciences Communications},
  year    = {2024},
  volume  = {11},
  number  = {1},
  pages   = {1121},
  doi     = {10.1057/s41599-024-03560-x},
  url     = {https://doi.org/10.1057/s41599-024-03560-x}
}

@article{Zamanifard2019,
  author  = {Zamanifard, Hadi and Alizadeh, Tooran and Bosman, Caryl and Coiacetto, Eddo},
  title   = {Measuring Experiential Qualities of Urban Public Spaces: Users' Perspective},
  journal = {Journal of Urban Design},
  year    = {2019},
  volume  = {24},
  number  = {3},
  pages   = {340--364},
  doi     = {10.1080/13574809.2018.1484664},
  url     = {https://doi.org/10.1080/13574809.2018.1484664}
}

@misc{zhang2024t2i-survey,
  author        = {Zhang, Chenshuang and Zhang, Chaoning and Zhang, Mengchun and Kweon, In So and Kim, Junmo},
  title         = {Text-to-Image Diffusion Models in Generative {AI}: A Survey},
  month         = mar,
  year          = {2023},
  eprint        = {2303.07909},
  archivePrefix = {arXiv},
  primaryClass  = {cs.CV},
  doi           = {10.48550/arXiv.2303.07909},
  url           = {https://arxiv.org/abs/2303.07909},
  note          = {arXiv preprint; version 3 revised 8 November 2024}
}

@article{zhang2024right,
  author  = {Zhang, Dawen and Finckenberg-Broman, Pamela and Hoang, Thong and Pan, Shidong and Xing, Zhenchang and Staples, Mark and Xu, Xiwei},
  title   = {Right to Be Forgotten in the Era of Large Language Models: Implications, Challenges, and Solutions},
  journal = {AI and Ethics},
  year    = {2025},
  volume  = {5},
  number  = {3},
  pages   = {2445--2454},
  doi     = {10.1007/s43681-024-00573-9},
  url     = {https://doi.org/10.1007/s43681-024-00573-9}
}

@inproceedings{zhang2024multidimensionalhp,
  author    = {Zhang, Sixian and Wang, Bohan and Wu, Junqiang and Li, Yan and Gao, Tingting and Zhang, Di and Wang, Zhongyuan},
  title     = {Learning Multi-Dimensional Human Preference for Text-to-Image Generation},
  booktitle = {Proceedings of the IEEE/CVF Conference on Computer Vision and Pattern Recognition},
  pages     = {8018--8027},
  publisher = {IEEE},
  month     = jun,
  year      = {2024},
  doi       = {10.1109/CVPR52733.2024.00766},
  url       = {https://openaccess.thecvf.com/content/CVPR2024/html/Zhang_Learning_Multi-Dimensional_Human_Preference_for_Text-to-Image_Generation_CVPR_2024_paper.html}
}

@article{Zhu2025,
  author  = {Zhu, Yihan and Zhang, Ye and Biljecki, Filip},
  title   = {{Understanding the user perspective on urban public spaces: A systematic review and opportunities for machine learning}},
  journal = {Cities},
  year    = {2025},
  volume  = {156},
  pages   = {105535},
  doi     = {10.1016/j.cities.2024.105535},
  url     = {https://doi.org/10.1016/j.cities.2024.105535}
}

@article{Zicari2021,
  author  = {Zicari, Roberto V. and Ahmed, Sheraz and Amann, Julia and Braun, Stephan Alexander and Brodersen, John and Bruneault, Fr{\'e}d{\'e}rick and Brusseau, James and Campano, Erik and Coffee, Megan and Dengel, Andreas and D{\"u}dder, Boris and Gallucci, Alessio and Gilbert, Thomas Krendl and Gottfrois, Philippe and Goffi, Emmanuel and Haase, Christoffer Bjerre and Hagendorff, Thilo and Hickman, Eleanore and Hildt, Elisabeth and Holm, Sune and Kringen, Pedro and K{\"u}hne, Ulrich and Lucieri, Adriano and Madai, Vince I. and Moreno-S{\'a}nchez, Pedro A. and Medlicott, Oriana and Ozols, Matiss and Schnebel, Eberhard and Spezzatti, Andy and Tithi, Jesmin Jahan and Umbrello, Steven and Vetter, Dennis and Volland, Holger and Westerlund, Magnus and Wurth, Renee},
  title   = {Co-Design of a Trustworthy {AI} System in Healthcare: Deep Learning Based Skin Lesion Classifier},
  journal = {Frontiers in Human Dynamics},
  year    = {2021},
  volume  = {3},
  pages   = {688152},
  doi     = {10.3389/fhumd.2021.688152},
  url     = {https://doi.org/10.3389/fhumd.2021.688152}
}

@book{Strathern2000,
  editor    = {Strathern, Marilyn},
  title     = {Audit Cultures: Anthropological Studies in Accountability, Ethics and the Academy},
  publisher = {Routledge},
  address   = {London},
  year      = {2000}
}

@article{Ercan02015,
  author    = {Ercan, Selen A. and Dryzek, John S.},
  title     = {The Reach of Deliberative Democracy},
  journal   = {Policy Studies},
  year      = {2015},
  volume    = {36},
  number    = {3},
  pages     = {241--248},
  doi       = {10.1080/01442872.2015.1065969},
  url       = {https://doi.org/10.1080/01442872.2015.1065969}
}

@article{larsson_governance_2020,
  author    = {Larsson, S.},
  title     = {On the Governance of Artificial Intelligence through Ethics Guidelines},
  journal   = {Asian Journal of Law and Society},
  year      = {2020},
  volume    = {7},
  number    = {3},
  pages     = {437--451},
  doi       = {10.1017/als.2020.19},
  url       = {https://doi.org/10.1017/als.2020.19}
}

@misc{Leslie2019,
  author       = {Leslie, D.},
  title        = {Understanding Artificial Intelligence Ethics and Safety: A Guide for the Responsible Design and Implementation of {AI} Systems in the Public Sector},
  year         = {2019},
  howpublished = {Zenodo. \url{https://doi.org/10.5281/ZENODO.3240529}}
}

@article{marmor2015privacy,
  author  = {Marmor, A.},
  title   = {What Is the Right to Privacy?},
  journal = {Philosophy \& Public Affairs},
  year    = {2015},
  volume  = {43},
  number  = {1},
  pages   = {3--26}
}

@inproceedings{nayak2024midspace,
  author    = {Nayak, Shravan and Mushkani, Rashid and Berard, Hugo and Cohen, Allison and Koseki, Shin and Bertrand, Hadrien},
  title     = {{MID}-Space: Aligning Diverse Communities' Needs to Inclusive Public Spaces},
  booktitle = {Pluralistic Alignment Workshop at NeurIPS 2024},
  year      = {2024},
  url       = {https://openreview.net/forum?id=kyfkMRT4Ao}
}

@book{Ostrom1990,
  author    = {Ostrom, Elinor},
  title     = {Governing the Commons: The Evolution of Institutions for Collective Action},
  publisher = {Cambridge University Press},
  address   = {Cambridge},
  year      = {1990}
}

@inproceedings{radford2021clip,
  author    = {Radford, Alec and Kim, Jong Wook and Hallacy, Chris and Ramesh, Aditya and Goh, Gabriel and Agarwal, Sandhini and Sastry, Girish and Askell, Amanda and Mishkin, Pamela and Clark, Jack and Krueger, Gretchen and Sutskever, Ilya},
  editor    = {Meila, Marina and Zhang, Tong},
  title     = {Learning Transferable Visual Models from Natural Language Supervision},
  booktitle = {Proceedings of the 38th International Conference on Machine Learning},
  series    = {Proceedings of Machine Learning Research},
  volume    = {139},
  pages     = {8748--8763},
  publisher = {PMLR},
  year      = {2021},
  url       = {https://proceedings.mlr.press/v139/radford21a.html}
}

@book{lefebvre1996writings,
  author    = {Lefebvre, Henri},
  title     = {Writings on Cities},
  publisher = {Blackwell Publishers},
  address   = {Oxford and Cambridge, MA},
  year      = {1996},
  isbn      = {9780631191889},
  note      = {Selected, translated, and introduced by Eleonore Kofman and Elizabeth Lebas},
  url       = {https://search.worldcat.org/title/Writings-on-cities/oclc/32349316}
}

@misc{mushkani2025urbanaigovernanceembed,
  author = {Mushkani, Rashid},
  title  = {Urban {AI} Governance Must Embed Legal Reasonableness for Democratic and Sustainable Cities},
  year   = {2025},
  url    = {https://arxiv.org/abs/2508.12174}
}

@article{wachter2019reasonable,
  author  = {Wachter, S. and Mittelstadt, B.},
  title   = {A Right to Reasonable Inferences: Re-Thinking Data Protection Law in the Age of Big Data and {AI}},
  journal = {Columbia Business Law Review},
  year    = {2019},
  volume  = {2019},
  number  = {2},
  pages   = {1--130}
}

\clearpage
\ifincludeappendices
\appendix


\chapter{Research instruments and protocols}
\label{app:instruments}

This chapter compiles participant-facing materials used to elicit (i) scenario prompts for public-space visualization and (ii) pairwise preference judgments for image-based evaluation.

\section{Use cases for multi-criteria visual prototyping}
\label{app:use_cases}

The multi-criteria alignment approach described in the main chapters can be adapted for practical urban-planning workflows. The use cases below summarize application contexts in which intersectional alignment and text-to-image generation can support early-stage deliberation.

\medskip

\paragraph{\textbf{Community consultations.}} Participatory design sessions often require communicating design alternatives to stakeholders with heterogeneous technical backgrounds. Preference-aligned image generation can be used to produce multiple scenario visualizations conditioned on locally defined criteria (e.g., accessibility and perceived safety), enabling structured discussion using a shared visual referent.

\medskip

\paragraph{\textbf{Peer-to-peer discussion.}} In contested projects, stakeholders can use generated scenario variations to clarify trade-offs between priorities (e.g., comfort and affordability). This can support peer-led discussion by externalizing assumptions and allowing participants to compare alternatives that differ on explicit criteria.

\medskip

\paragraph{\textbf{Iterative concept exploration.}} Early-stage design exploration benefits from low-cost iteration prior to committing resources to detailed drawings or renderings. Multi-criteria conditioning can be used to generate variations that intentionally probe specific constraints (e.g., inclusivity and invitingness), helping identify missing amenities or incompatible design elements.

\medskip

\paragraph{\textbf{Teaching and education.}} In planning and design curricula, the workflow provides a concrete setting for discussing how prompts, annotation signals, and optimization choices influence generated imagery. Students can compare outputs under different alignment objectives and examine the limits of visualizing abstract criteria.

\medskip

\paragraph{\textbf{Communicating requirements from marginalized perspectives.}} Stakeholders who are routinely underrepresented in planning processes can use visual prototypes to communicate requirements that are otherwise difficult to express in conventional planning documents (e.g., wheelchair navigation constraints, multilingual wayfinding needs). Images can function as boundary objects that support negotiation across expertise levels.

\section{Prompting workshop materials}
\label{app:prompting_workshop}

\subsection{Questions for prompt creation}
The following prompts were provided to support scenario specification during the prompt-writing workshop:
\begin{itemize}
\item What are the surroundings?
\item What decorative features and objects does the place have?
\item Is there nature present, and if yes, what kind?
\item How are the weather and light conditions?
\item What is the composition of the image?
\item What materials and surfaces are present?
\item How would you describe the atmosphere?
\item What amenities should be present in the space?
\end{itemize}

\subsection{Questions for evaluation of the images}
The following questions were used to guide reflection during the image-review stage:
\begin{itemize}
\item Can you imagine using the public space yourself?
\item Does the public space match what you had imagined when creating the prompt?
\item Are you satisfied with the image?
\item Can you see the image being used as a design for a public space in real life?
\end{itemize}

\subsection{Scenario briefs}
Table~\ref{tab:app-scenarios} reproduces the scenario briefs used during the workshop.

\small
\begin{longtable}{@{}
  >{\raggedright\arraybackslash}p{0.085\linewidth}
  >{\raggedright\arraybackslash}p{0.09\linewidth}
  >{\raggedright\arraybackslash}p{0.185\linewidth}
  >{\raggedright\arraybackslash}p{0.31\linewidth}
  >{\raggedright\arraybackslash}p{0.20\linewidth}
@{}}
\caption{Scenario briefs used in the prompting workshop.}
\label{tab:app-scenarios}\\

\toprule
Scenario & Optional & Public-space typology & Required amenities and features & Location brief \\
\midrule
\endfirsthead

\toprule
Scenario & Optional & Public-space typology & Required amenities and features & Location brief \\
\midrule
\endhead

\midrule
\multicolumn{5}{r}{\emph{Continued on next page}}\\
\endfoot

\bottomrule
\endlastfoot

A & No & Park &
Sitting areas, green space &
Low-density suburban Montréal \\

B & No & Pedestrian promenade &
Safe streets, community engagement spaces, green areas &
Historic neighborhood in Montréal \\

C & No & Street space &
Inclusive environments for all ages, genders, abilities, and identities &
Residential neighborhood in Montréal \\

D & No & Downtown plaza &
Meeting spaces, seating areas, rest areas, flexible-use spaces &
Downtown Montréal \\

E & No & Park &
Rest areas, community engagement spaces, waterfront access &
Dense urban area in Montréal \\

F & Yes & Urban garden &
Educational programs, community gardening spaces &
Near universities and colleges in Montréal \\

G & Yes & Waterfront sidewalk &
Outdoor cafés, art installations, pedestrian paths, bike lanes &
Along a river or lake in a mixed-use area of Montréal \\

H & Yes & Neighborhood square &
Playgrounds, outdoor fitness equipment, community noticeboards, seasonal markets &
Residential area in Montréal, near schools and local businesses \\

I & Yes & Transit plaza &
Sheltered seating, transit information displays, public art, bike-share stations &
Key transit hub in Montréal, adjacent to a metro station or major bus interchange \\

J & Yes & Alleyway &
Street murals, pedestrian lighting, small business kiosks &
Back alleys in a commercial district of Montréal \\

\end{longtable}

\subsection{Group assignment schedule}
Table~\ref{tab:app-workshop-groups} summarizes the group assignment schedule used during the workshop sessions.

\begin{table}[!htbp]
\centering
\begin{tabular}{@{}llll@{}}
Group & First hands-on session & Second hands-on session & Optional scenario \\ \hline
1 & Scenario A & Scenario B & Scenario F \\
2 & Scenario B & Scenario C & Scenario G \\
3 & Scenario C & Scenario D & Scenario H \\
4 & Scenario D & Scenario E & Scenario I \\
5 & Scenario E & Scenario A & Scenario J \\ \hline
\end{tabular}
\caption{Workshop group assignments to scenario briefs.}
\label{tab:app-workshop-groups}
\end{table}

\section{LIVS annotation protocol}
\label{app:livs_protocol}

This section reproduces the protocol provided to annotators contributing pairwise preference judgments for the LIVS dataset.

\subsection{Preparation}
Annotators were asked to review the criterion definitions in advance and to complete the task on a desktop or laptop computer. They were encouraged to take breaks to reduce fatigue and to prioritize careful judgments over speed. The expected time per comparison was approximately 15--30 seconds.

\subsection{Rating procedure}
Each task presented a pair of images and three criteria. For each criterion, annotators indicated whether the left image or the right image better satisfied the criterion, and the strength of that preference, using a continuous slider. Neutral placements were allowed when images appeared equally aligned with the criterion or when neither image matched the criterion well. Annotators were instructed to base judgments on properties of the depicted urban space rather than on the demographic characteristics of depicted people or animals. Highly distorted images could be skipped without attempting interpretation. A comment field was available for optional contextual notes.

\subsection{Session structure and batching}
The platform constrained sessions to a maximum of 90 comparisons per session, with typical session duration of approximately 25 minutes. The collection schedule was organized into four batches, with a target of 750 comparisons per batch per participant. Table~\ref{tab:app-livs-timeline} reproduces the batch periods used during data collection.

\begin{table}[!htbp]
\centering
\begin{tabular}{@{}lll@{}}
Batch & Period (2024) & Target number of comparisons \\ \hline
1 & 01 May--14 May & 750 \\
2 & 15 May--28 May & 750 \\
3 & 29 May--11 June & 750 \\
4 & 12 June--20 June & 750 \\ \hline
\end{tabular}
\caption{Annotation batches and targets for LIVS data collection.}
\label{tab:app-livs-timeline}
\end{table}

\subsection{Definitions used in the interface}
Table~\ref{tab:app-livs-definitions} lists the criterion definitions provided in the interface.

\begin{longtable}{@{}p{0.22\linewidth}p{0.74\linewidth}@{}}
\caption{Definitions displayed to annotators in the LIVS interface.}
\label{tab:app-livs-definitions}\\
Term & Definition \\ \hline
\endfirsthead
Term & Definition \\ \hline
\endhead
\hline
\multicolumn{2}{r}{\emph{Continued on next page}}\\
\endfoot
\hline
\endlastfoot

Public space & An area where people can go, such as parks and streets, intended for meeting, play, and everyday use in cities and towns. \\
Inclusion (inclusive) & Spaces in which people are welcome and treated with respect; spaces that do not discriminate across social groups. \\
Safe / secure & Spaces in which people feel calm and protected from hazards related to physical elements, pollution, or other factors that can diminish a sense of security. \\
Comfortable & Spaces with facilities and material conditions that support ease of use and protection from weather conditions (e.g., seating, shelter, surfaces). \\
Inviting & Spaces that attract and engage people through elements and activities that encourage participation and interaction. \\
Diverse & Spaces that support a range of social groups and offer multiple services, activities, and functions across cultures, ages, and abilities. \\
Accessibility & Spaces that are navigable for people with different physical abilities, including features such as ramps, wide walkways, clear signage, and tactile indicators. \\
\end{longtable}

\clearpage

\chapter{Technical documentation}
\label{app:technical}

This chapter documents supplementary technical details for dataset construction, optimization, and analysis. It also summarizes the public release of the Street Review dataset.

\section{LIVS dataset viewer}
\label{app:livs_viewer}
Figure~\ref{fig:app-livs-viewer} shows the internal dataset viewer used to inspect prompt-conditioned image pairs and their associated preference labels during curation and analysis.

\begin{figure}[!htbp]
\centering
\includegraphics[width=0.9\textwidth]{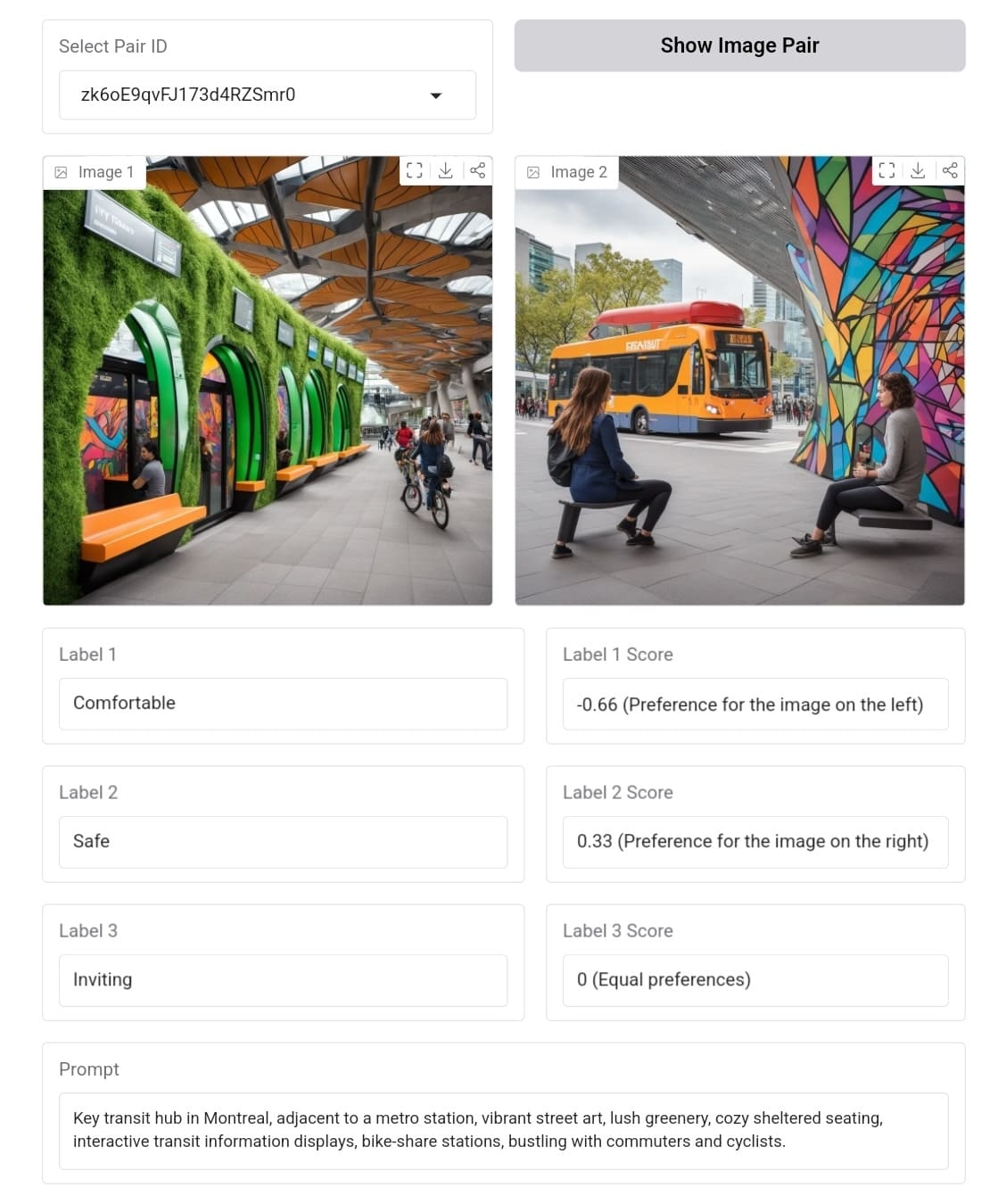}
\caption[LIVS dataset viewer]{Dataset viewer for LIVS, illustrating a prompt with a corresponding image pair and recorded preference information. \added{Source: LIVS dataset viewer; see \citet{pmlr-v267-mushkani25a}.}}
\label{fig:app-livs-viewer}
\label{fig:livs-viewer}
\end{figure}

\FloatBarrier

\clearpage

\section{Prompt augmentation procedures}
\label{app:prompt_augmentation}

We collected 440 prompts from annotators, each representing distinct scenarios related to public space typologies, amenities, and ambiances in Montreal. To expand the dataset, we supplemented these with synthetic captions generated by GPT-4 \citep{openai2024gpt4}. To ensure diversity, we incorporated a wide range of concepts drawn from public space literature, such as typologies like parks and wide walkways, amenities such as seating areas and streetlights, natural elements like forests and vegetation, locations such as suburban Montreal and old industrial ports, people such as First Nations and children, architectural elements like duplexes and houses, transportation modes including bike lanes and trams, artistic features such as street murals and art sculptures, different times like winter, nighttime, and Christmas, and animals such as dogs and raccoons. These concepts were used to increase the diversity of the concepts in the generated prompts. We employed three distinct prompting strategies to ensure that the synthetic prompts retained similarity to those generated by humans while covering a wide range of topics relevant to public space design. The specific strategies and prompts used for the LLM are outlined below.

\medskip

\paragraph{Method 1}
We randomly sampled 8-16 prompts from the human-generated set and used them as in-context examples for the LLM to generate new prompts.

\begin{tcolorbox}[
  colback=black!2!white,
  colframe=black!35,
  title=System prompt used Method 1,
  fonttitle=\bfseries,
  boxrule=0.5pt,
  sharp corners=all
]
Your task is to craft detailed and imaginative prompts suitable for diffusion models like Stable Diffusion. These prompts should generate images illustrating the variety of Montreal's public spaces, capturing the community's diverse aspirations and values.

Each prompt must be rooted in a specific scenario related to Montreal's public spaces. You will be provided with the scenario, keywords, and examples of prompts related to these scenarios. Using this information, your task is to create a series of diverse, contextually rich, and relevant prompts following a style similar to the ones given as examples. These should aim to generate images showcasing Montreal's public spaces from varied perspectives.
\end{tcolorbox}

\paragraph{Method 2}
We provided the LLM with a detailed scenario which was also provided to the annotators during the initial prompt collection phase. Along with this we also provided several keywords related to the public space concepts mentioned earlier. Additionally, we included 8 randomly selected in-context samples relevant to the scenario guiding the model to generate new prompts based on these concepts.

\begin{tcolorbox}[
  colback=black!2!white,
  colframe=black!35,
  title=System prompt used Method 1,
  fonttitle=\bfseries,
  boxrule=0.5pt,
  sharp corners=all
]
Your task is to craft detailed and imaginative prompts suitable for diffusion models like Stable Diffusion. These prompts should generate images illustrating the variety of Montreal's public spaces, capturing the community's diverse aspirations and values. To achieve this, you will construct prompts using specific keywords provided for the following categories:\\
Typology: The type of spaces you want to depict\\
Elements: Distinct elements to include in your scene\\
Context: The scenarios in which your elements are placed\\
Style: The artistic style or technique that the image should emulate, defining its visual appearance\\
Mood: The overall mood or atmosphere of the image\\

You will also be given a few examples that have been generated using these keywords. Using all this information create a complete, coherent prompt similar in style to the examples. Aim for creativity and diversity in your prompts, ensuring they cover several aspects of the keywords given. These should aim to generate images showcasing Montreal's public spaces from varied perspectives.\\

Note: \\
1. Ensure your prompts integrate some of the provided keywords to encapsulate the community's desired visions of Montreal's public spaces but ensure that style and length is same as the examples.\\
2. Do not mention the style and mood explicitly. Use keywords that bring out these attributes naturally.\\
3. The style and the length of the prompts should be similar to the examples given. The prompt should be less than 77 tokens.
\end{tcolorbox}

\paragraph{Method 3}
This method used a template-based approach where specific keywords related to public space concepts in the original prompts were masked. We instructed the LLM to replace these masked keywords with concepts from a wide variety of public space themes. This ensured the prompts closely followed the structure of the human-generated ones while incorporating a diverse range of concepts.

\medskip

\begin{tcolorbox}[
  colback=black!2!white,
  colframe=black!35,
  title=System prompt used Method 1,
  fonttitle=\bfseries,
  boxrule=0.5pt,
  sharp corners=all
]
Your task is to craft detailed and imaginative prompts suitable for diffusion models like Stable Diffusion. These prompts should generate images illustrating the variety of Montreal's public spaces, capturing the community's diverse aspirations and values.\\

For this task, you will be provided with a templated sentence containing several placeholders. Each placeholder represents a specific category (e.g., [Typology], [Location], [Activity], [Amenity]). Alongside the templated sentence, you will receive a list of words or phrases corresponding to each category. Your objective is to select the most appropriate word or phrase from each list to fill in the placeholders, creating a meaningful and grammatically correct sentence.\\

The structure of the templated sentence might require minimal modifications to ensure grammatical correctness and cohesiveness once the placeholders are filled.
\\

Example 1:\\

Template: a [Typology] for [People] in [Location]\\

Keywords:\\
Typology: artistic eco friendly park, pedestrian street, all identities, two-story residential street, park, neighbourhood public space, urban square, wide walkway\\
People: elderly person, adults, first nations, children, teenagers, adults and elderly people, black and white families, a mother and her child, people, various ethnicities\\
Location: plateau, wellington neighbourhood, old port, montreal `s chinatown, old montreal, Montreal, downtown montreal, mont royal street\\

Output: A neighbourhood public space for children, teenagers, adults and elderly people in Montreal\\

<more examples>
\end{tcolorbox}

\medskip

\subsection{Lexical deviation from human prompts}
To estimate lexical deviation between human-written prompts and synthetic prompts, Jensen--Shannon divergence (JSD) was computed between prompt token distributions \citep{plank2011jsd}. Methods 1 and 2 yielded higher divergence than Method 3 (0.53 and 0.58 vs.\ 0.40), consistent with the more constrained, template-based structure in Method 3.

\FloatBarrier

\section{Image generation and diversity filtering}
\label{app:image_generation}

For each prompt, a set of candidate images was generated using Stable Diffusion XL \citep{podell2023sdxlimprovinglatentdiffusion}. Twenty images per prompt were produced by varying the random seed, classifier-free guidance scale, and inference steps. To reduce redundancy in the final preference set, four images per prompt were selected to maximize visual diversity using CLIP-based similarity scores \citep{radford2021clip}.

\medskip

The selection procedure operated on an $n \times n$ similarity matrix $S$ computed from CLIP embeddings. A greedy max-min heuristic was used: the first image was selected as the one with the lowest mean similarity to the remaining candidates; subsequent images were selected to minimize the maximum similarity to the images already selected. The procedure used in curation is summarized below.

\medskip

\begin{center}
\fbox{%
\begin{minipage}{0.95\linewidth}
\small
\vspace{0.3em}

\textbf{Greedy max-min selection for diverse image subsets (CLIP similarity).}

\medskip

\setlength{\tabcolsep}{0pt}
\renewcommand{\arraystretch}{1.25}

\begin{tabularx}{\linewidth}{@{}X@{}}
\textbf{Input:} similarity matrix $S \in \mathbb{R}^{n \times n}$; target subset size $k = 4$.\\[0.6em]

\textbf{Output:} indices of the selected images.\\[0.8em]

1.\ Select $i_1 = \arg\min_i \;\frac{1}{n}\sum_{j=1}^{n} S_{ij}$.\\[0.6em]

2.\ Initialize $\mathcal{I} \leftarrow \{i_1\}$.\\[0.6em]

3.\ For $t = 2,\dots,k$: choose 
$i_t = \arg\min_{i \notin \mathcal{I}} \;\max_{j \in \mathcal{I}} S_{ji}$,  
and update $\mathcal{I} \leftarrow \mathcal{I} \cup \{i_t\}$.\\
\end{tabularx}

\vspace{0.3em}
\end{minipage}}
\end{center}

\FloatBarrier

\medskip

\section{Preference annotation interface}
\label{app:annotation_interface}

The web-based interface used for LIVS pairwise annotation is shown in Figure~\ref{fig:livs-interface} in the main chapters. Annotators compared two images per task using a slider, optionally consulting criterion definitions embedded in the interface.

\FloatBarrier

\section{Direct Preference Optimization configuration}
\label{app:dpo_config}

Stable Diffusion XL was fine-tuned using DPO \citep{rafailov2024}. Table~\ref{tab:app-dpo-hparams} reports the hyperparameter configuration used for the experiments summarized in this thesis.

\begin{table}[!htbp]
\centering
\begin{tabular}{@{}ll@{}}
Hyperparameter & Value \\ \hline
Batch size & 64 \\
Learning rate & $1\times10^{-8}$ with 20\% linear warmup \\
$\beta$ & 5{,}000 \\
Training steps & 500 (subset experiments); 1{,}500 (full preference dataset) \\
Hardware & Single NVIDIA A100 80GB GPU \\ \hline
\end{tabular}
\caption{Hyperparameters used for DPO fine-tuning of SDXL.}
\label{tab:app-dpo-hparams}
\end{table}

Continuous preference values recorded on sliders were discretized to binary labels (preferred vs.\ not preferred) for DPO compatibility. Multi-criterion conflicts within a comparison were resolved using a majority procedure.

\FloatBarrier

\section{Supplementary qualitative comparisons}
\label{app:qualitative}

Figures~\ref{fig:app-exp-bike-path}--\ref{fig:app-exp-diversity} provide qualitative comparisons between baseline SDXL generations and DPO-aligned variants trained on single criteria or multiple criteria. These examples are included to document typical visual changes observed during alignment and to illustrate recurring failure modes (e.g., inconsistent rendering of ramps or multilingual signage).

\begin{figure}[!htbp]
\centering
\includegraphics[width=\textwidth]{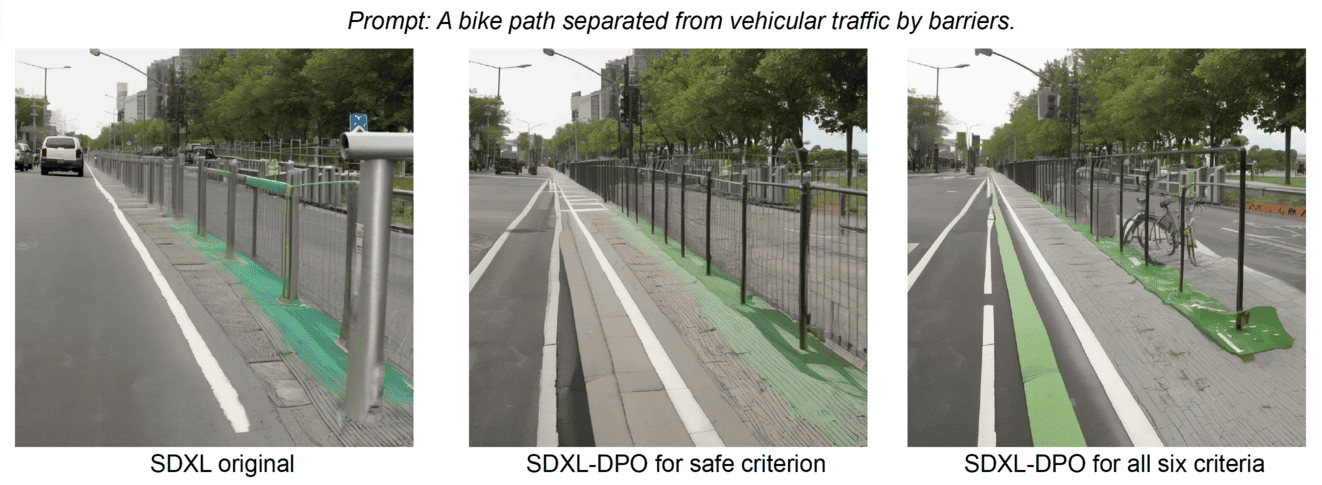}
\caption[Bike-path output comparison]{Bike-path scenario emphasizing safety. The example contrasts baseline SDXL output with single-criterion and multi-criterion DPO-aligned outputs. \added{Source: LIVS model outputs generated with SDXL; see \citet{pmlr-v267-mushkani25a}.}}
\label{fig:app-exp-bike-path}
\label{fig:bike-path-scenario}
\end{figure}

\begin{figure}[!htbp]
\centering
\includegraphics[width=\textwidth]{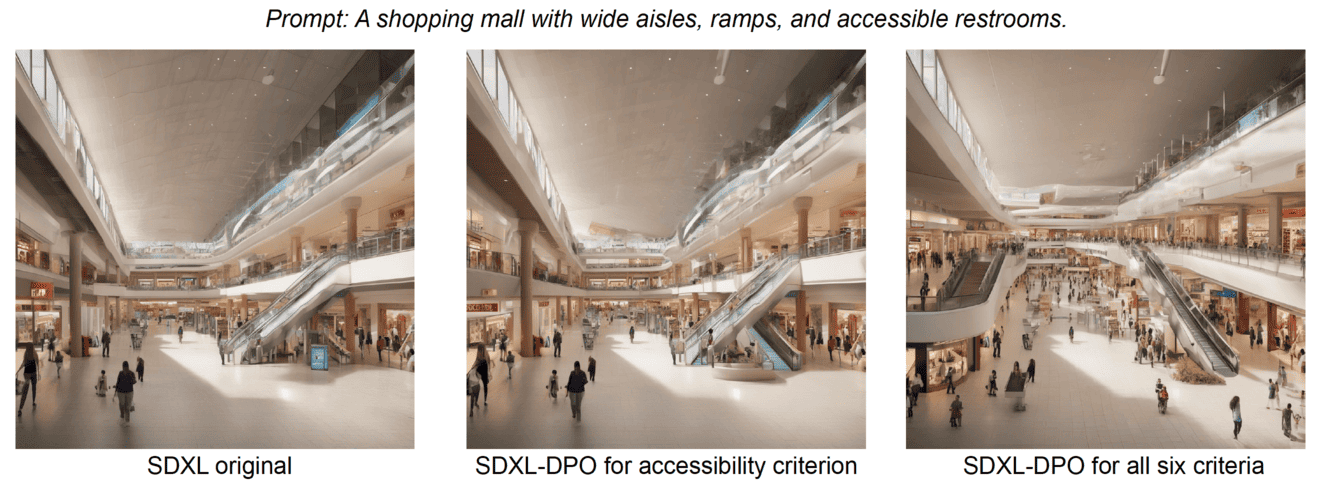}
\caption[Shopping-mall output comparison]{Shopping-mall scenario emphasizing accessibility. The example contrasts baseline SDXL output with single-criterion and multi-criterion DPO-aligned outputs. \added{Source: LIVS model outputs generated with SDXL; see \citet{pmlr-v267-mushkani25a}.}}
\label{fig:app-exp-mall}
\label{fig:shopping-mall-scenario}
\end{figure}

\begin{figure}[!htbp]
\centering
\includegraphics[width=\textwidth]{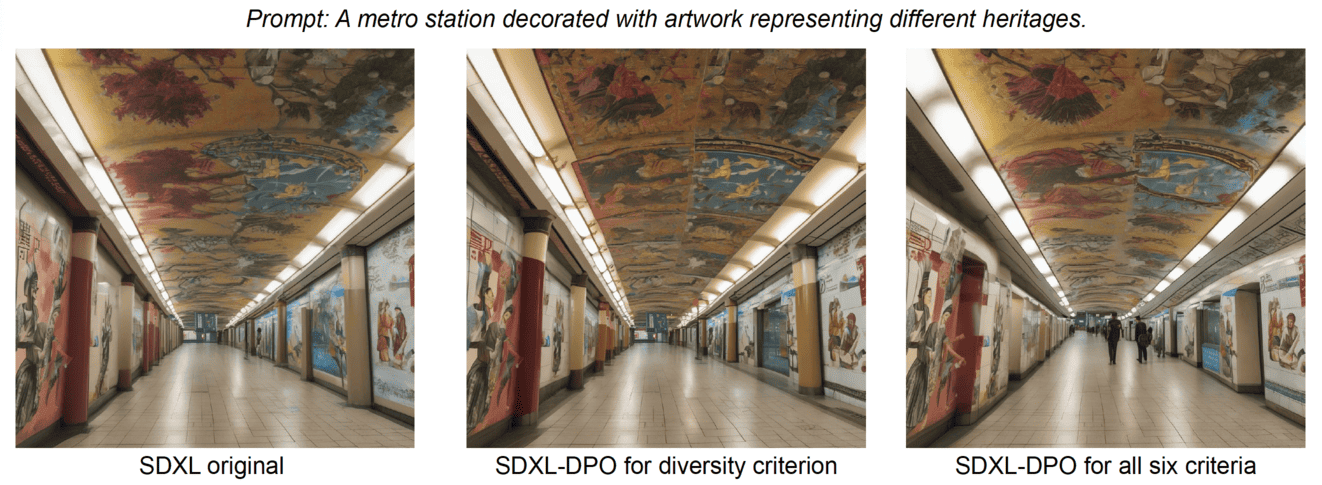}
\caption[Metro-station output comparison]{Metro-station scenario emphasizing diversity. The example contrasts baseline SDXL output with single-criterion and multi-criterion DPO-aligned outputs. \added{Source: LIVS model outputs generated with SDXL; see \citet{pmlr-v267-mushkani25a}.}}
\label{fig:app-exp-diversity}
\label{fig:metro-station-scenario}
\end{figure}

\FloatBarrier

\section{Additional analysis: intersectional scoring patterns}
\label{app:intersectional_scoring}

Figures~\ref{fig:app-dataset-scoring-pattern} and \ref{fig:app-eval-dataset-scoring-pattern} report criterion score summaries by intersectional identity group. Scores are shown in a centered form to facilitate comparison across criteria with different baselines. The plots are included as supplementary documentation of the groupwise scoring patterns discussed in the main chapters.

\begin{figure}[p]
\centering
\includegraphics[height=0.92\textheight]{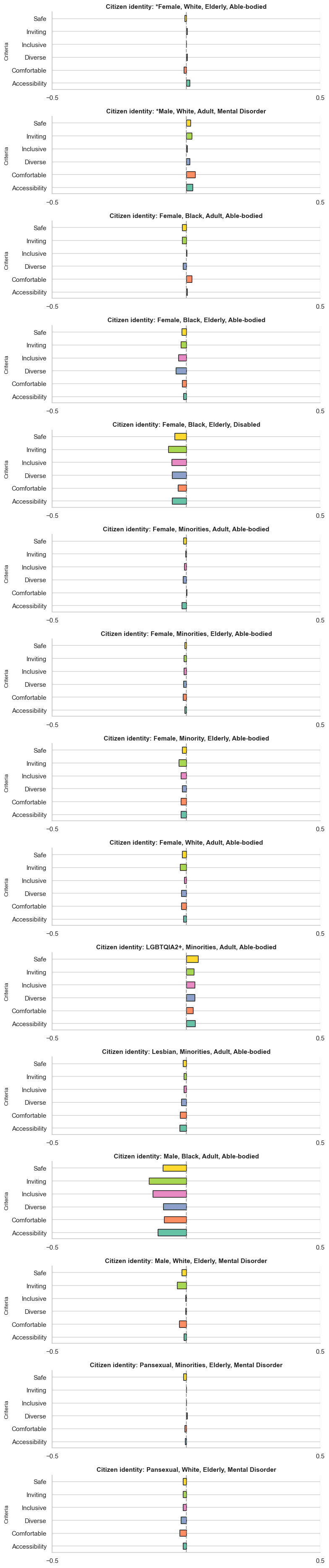}
\caption[Main-dataset centered scores]{Main dataset: centered mean criterion scores by intersectional identity group. \added{Source: analysis reported in \citet{pmlr-v267-mushkani25a}.}}
\label{fig:app-dataset-scoring-pattern}
\label{fig:main-centered-scores}
\end{figure}

\begin{figure}[p]
\centering
\includegraphics[height=0.92\textheight]{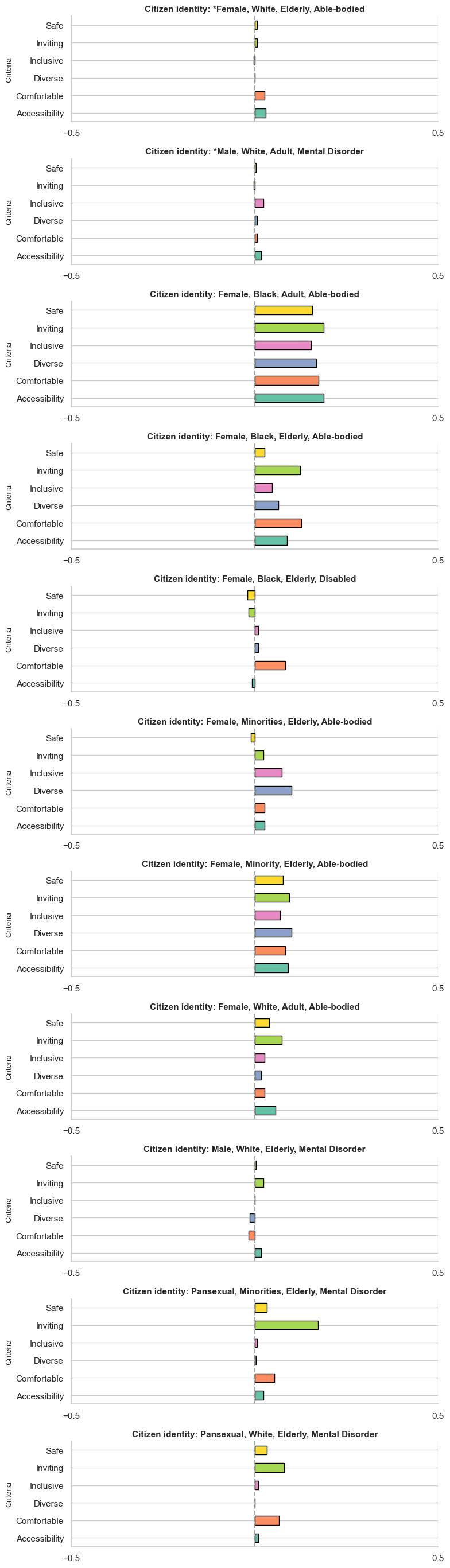}
\caption[Evaluation-dataset centered scores]{Evaluation dataset: centered mean criterion scores by intersectional identity group. \added{Source: analysis reported in \citet{pmlr-v267-mushkani25a}.}}
\label{fig:app-eval-dataset-scoring-pattern}
\label{fig:evaluation-centered-scores}
\end{figure}

\FloatBarrier

\section{Evaluation neutrality and score variance}
\label{app:variance}

Figure~\ref{fig:app-eval-neutral-variance} summarizes neutrality rates and score variance by criterion in the evaluation dataset. Neutral selections correspond to comparisons where annotators indicated no preference between the two images for the criterion presented.

\begin{figure}[!htbp]
\centering
\includegraphics[width=\textwidth]{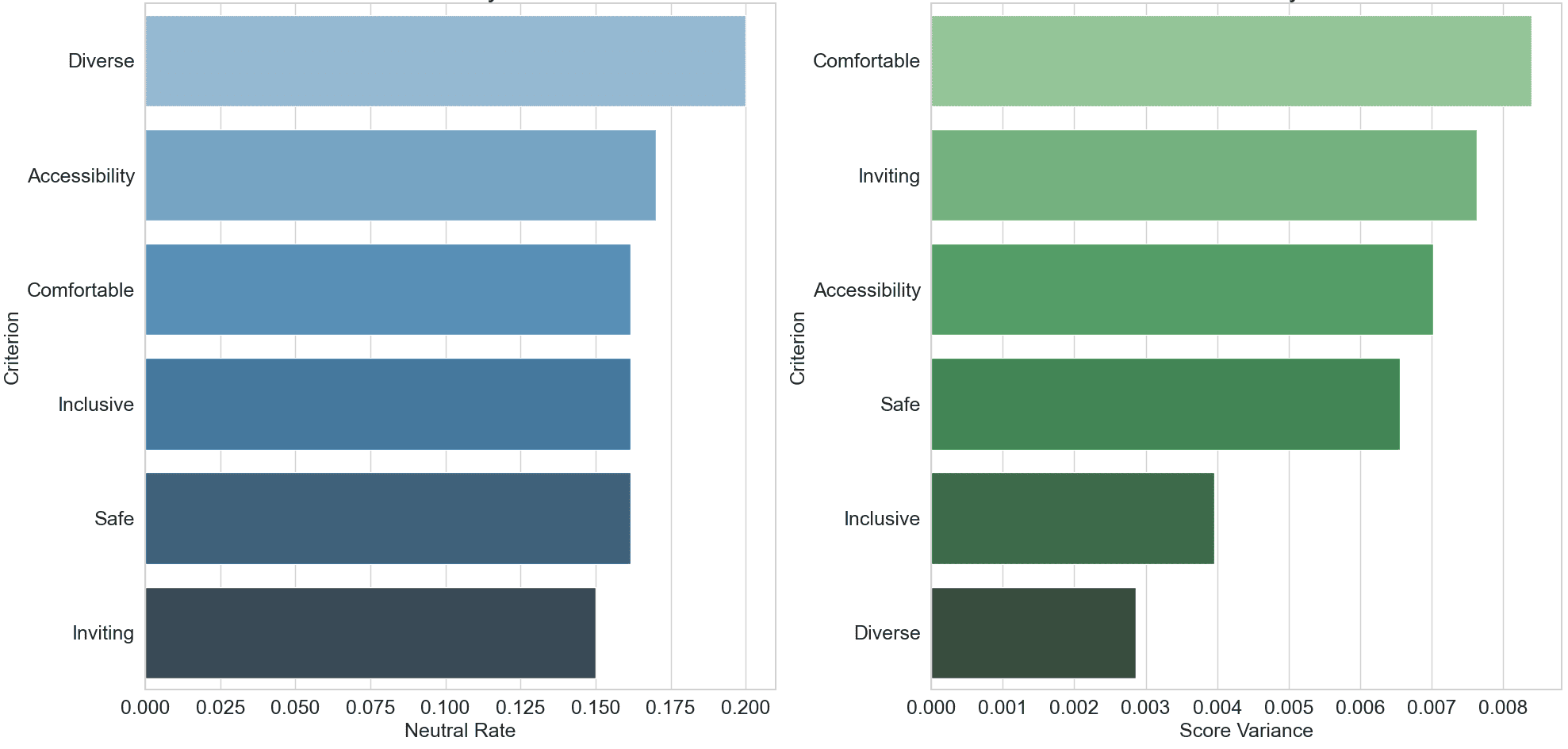}
\caption[Evaluation neutrality and variance]{Evaluation dataset: neutrality rate and score variance by criterion. \added{Source: analysis reported in \citet{pmlr-v267-mushkani25a}.}}
\label{fig:app-eval-neutral-variance}
\label{fig:evaluation-neutrality-variance}
\end{figure}

\FloatBarrier

\section{Street Review dataset}
\label{app:streetreview}

Street Review is publicly released on Hugging Face under the dataset repository name \texttt{rsdmu/streetreview}.\footnote{\url{https://huggingface.co/datasets/rsdmu/streetreview}} The dataset combines street-view imagery and structured evaluations of streetscapes in Montréal across four criteria: inclusivity, accessibility, aesthetics, and practicality. A companion repository with additional documentation and scripts is available on GitHub.\footnote{\url{https://github.com/rsdmu/streetreview}}

\subsection{Repository structure}
The dataset card describes the following top-level structure:
\begin{table}[!htbp]
\centering
\begin{tabularx}{\linewidth}{@{}lX@{}}
Path & Contents \\ \hline
\texttt{metadata.csv} & Metadata attributes for each evaluation point (e.g., identifiers and streetscape descriptors). \\
\texttt{street\_eval/} & Per-segment evaluation tables (CSV) containing group ratings by criterion. \\
\texttt{street\_img/} & Street-view images organized by segment and viewpoint. \\ \hline
\end{tabularx}
\caption{Street Review repository structure as documented in the dataset card.}
\label{tab:app-streetreview-structure}
\end{table}

Street-view images are stored under \texttt{street\_img/} and grouped by street segment and viewpoint (e.g., suffixes \texttt{\_main}, \texttt{\_head}, \texttt{\_tail}). Evaluation tables are stored under \texttt{street\_eval/} as per-segment CSV files (e.g., \texttt{i01\_evaluations.csv}). The dataset card reports 60 evaluation points across 20 streets, with two images per point, and an expanded image pool for diversity during presentation.\footnote{\url{https://huggingface.co/datasets/rsdmu/streetreview}}

\subsection{Rating scale and fields}
Ratings are reported on a 1--4 ordinal scale. For accessibility, the dataset card specifies that a score of 1 corresponds to ``not accessible,'' values of 2 or 3 correspond to average accessibility, and a score of 4 corresponds to the highest accessibility level.\footnote{\url{https://huggingface.co/datasets/rsdmu/streetreview}} Evaluation fields follow a \texttt{group\_criterion} naming scheme (e.g., \texttt{lgbtqia2+\_accessibility}, \texttt{elderly\_male\_practicality}, \texttt{group\_inclusivity}).

\FloatBarrier

\section{Supplementary Street Review Material}
\label{app:streetreview_supp_figs}

\subsection{Interview theme co-occurrence network}
A co-occurrence network was derived from coded interview transcripts to visualize how participants linked frequently discussed concepts (e.g., \enquote{safety}, \enquote{inclusivity}, \enquote{usage}). Nodes represent frequently mentioned concepts, while edges indicate concepts that were commonly discussed together in the same responses.

\begin{figure}[!htbp]
\centering
\includegraphics[width=\textwidth]{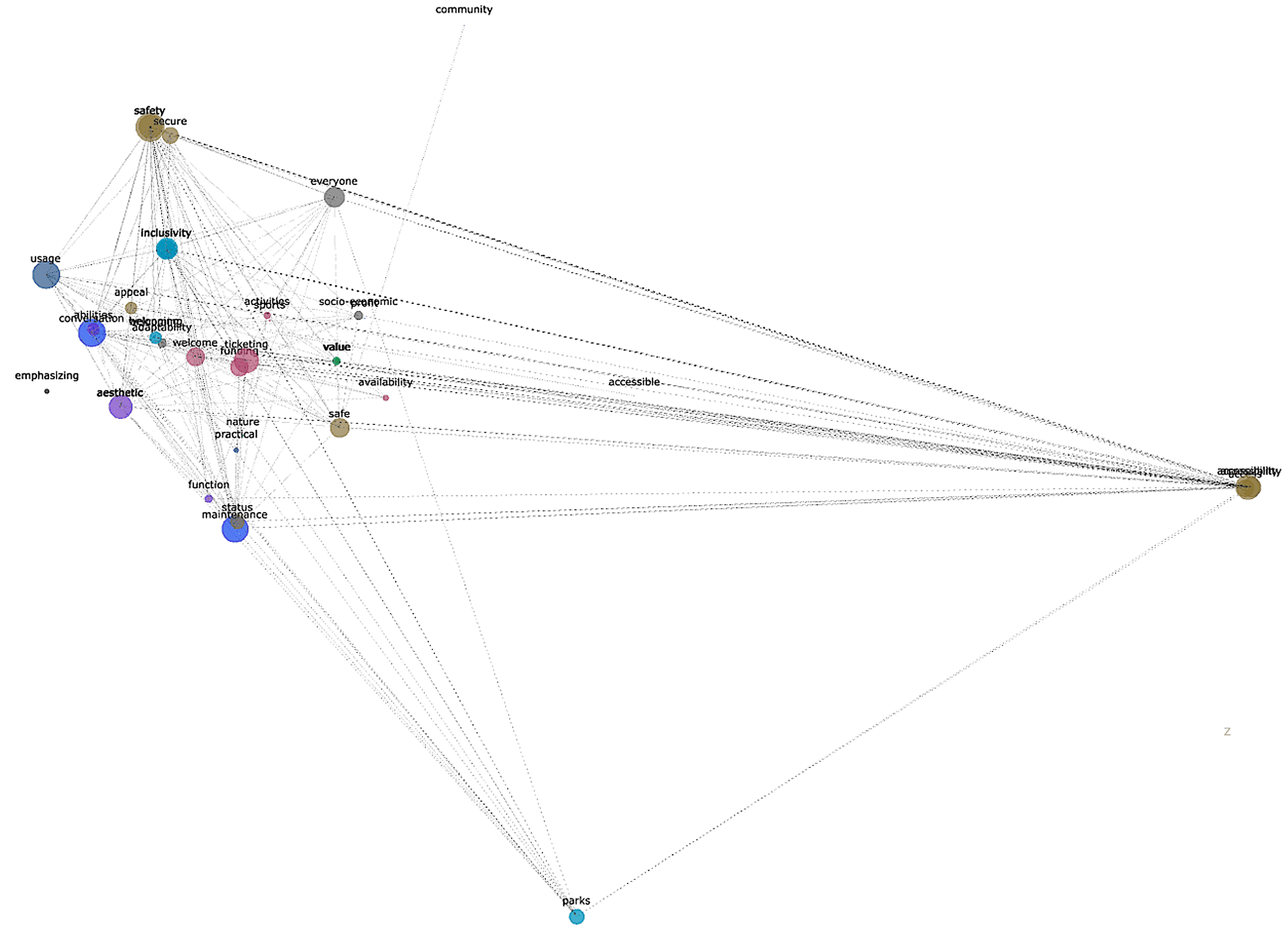}
\caption[Supplementary interview-theme network]{Network diagram of thematic connections derived from Street Review interview data. Nodes represent frequently discussed concepts, and edges indicate co-occurrence relationships in participant responses. \added{Source: author's coding of Street Review interview transcripts; see \citet{MushkaniKoseki2026}.}}
\label{fig:network_interview_themes2}
\label{fig:supplementary-theme-network}
\end{figure}

\subsection{Demographic rating profiles across criteria}
The radar chart below summarizes mean ratings across the four Street Review criteria (accessibility, inclusivity, practicality, aesthetics) for major demographic groups. It provides a compact view of how group profiles differ across dimensions when evaluating the same set of street stimuli.

\begin{figure}[!htbp]
\centering
\includegraphics[width=0.9\textwidth]{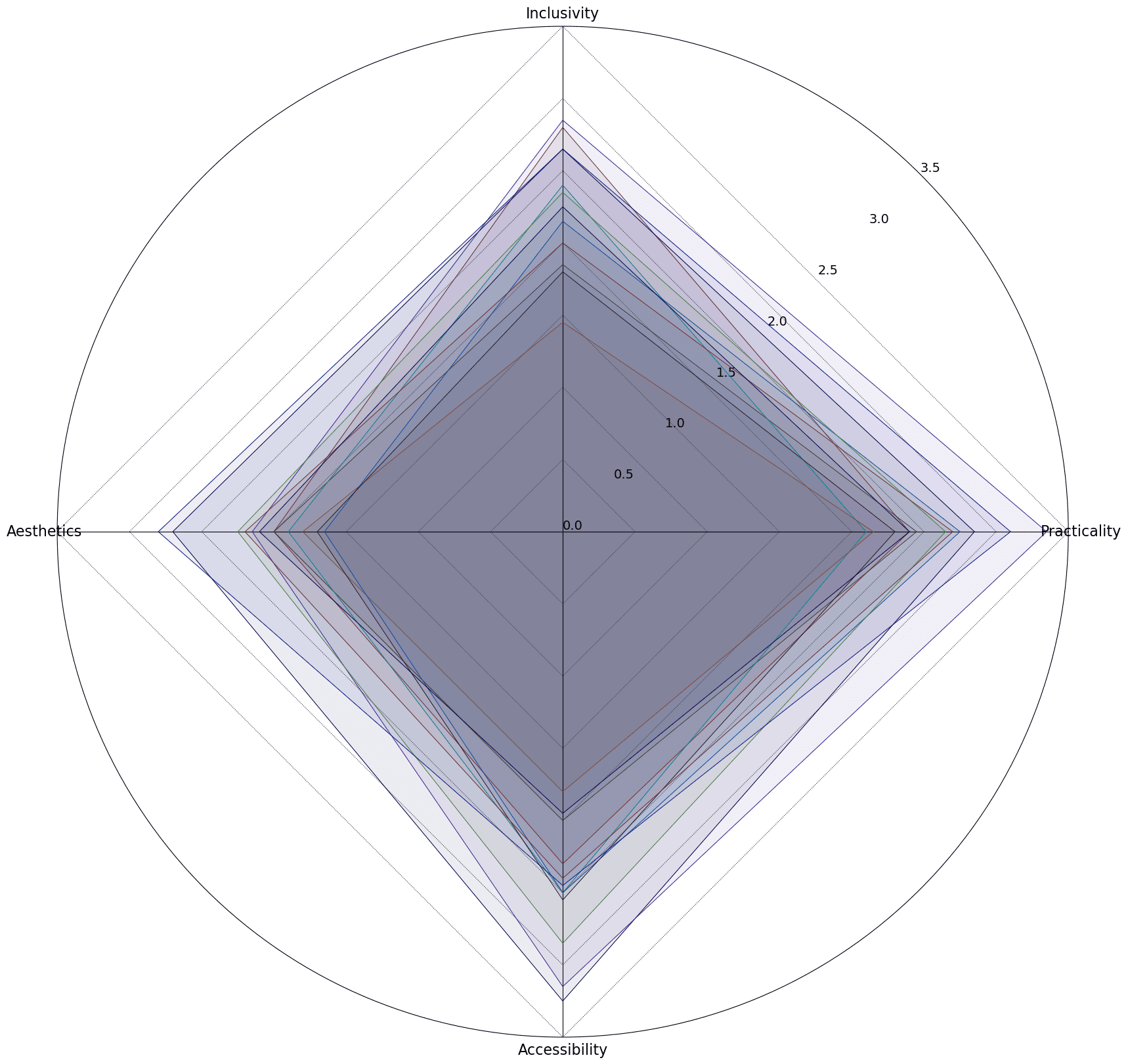}
\caption[Supplementary Street Review ratings]{Radar chart of mean Street Review ratings by demographic group across four criteria. Each line corresponds to a demographic group; farther outward values indicate higher mean ratings on that criterion. \added{Source: author analysis of Street Review ratings; see \citet{MushkaniKoseki2026}.}}
\label{fig:app-streetreview-radar}
\label{fig:supplementary-radar}
\end{figure}

\FloatBarrier

\subsection{Street Review agreement and correlation plots}

This section includes supplementary agreement analyses for the Street Review evaluation process. Pearson correlations and Kendall rank correlations are reported as descriptive measures of consistency across evaluation strata \citep{PuthEtAl2015}.

\begin{figure}[!htbp]
\centering
\includegraphics[width=\textwidth]{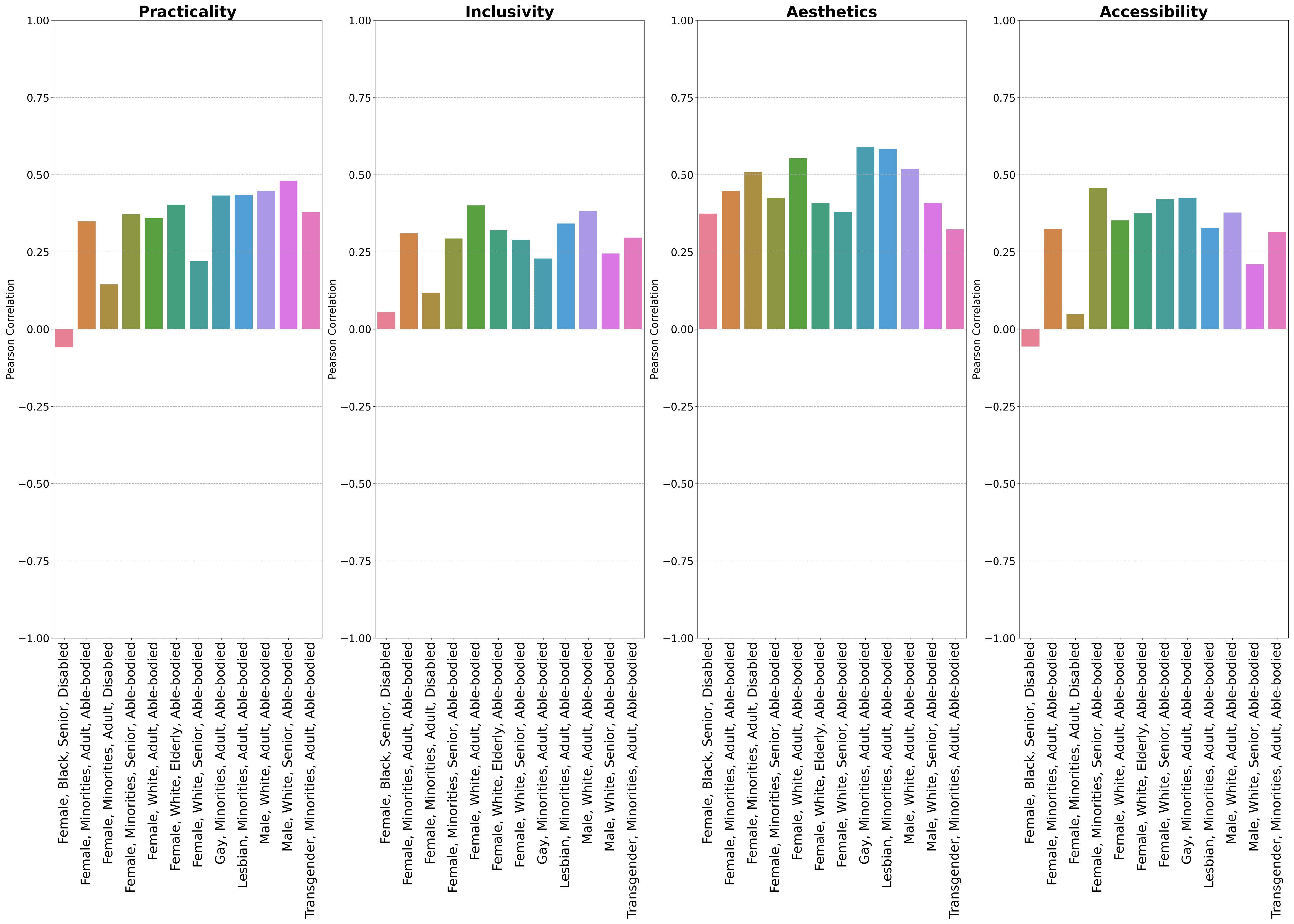}
\caption[Supplementary Street Review correlations]{Pearson correlation summaries by intersectional identity group for Street Review criteria (practicality, inclusivity, aesthetics, accessibility). \added{Source: author analysis; see \citet{MushkaniJUM2025}.}}
\label{fig:app-streetreview-pearson}
\label{fig:supplementary-pearson}
\end{figure}

\begin{figure}[!htbp]
\centering
\includegraphics[width=\textwidth]{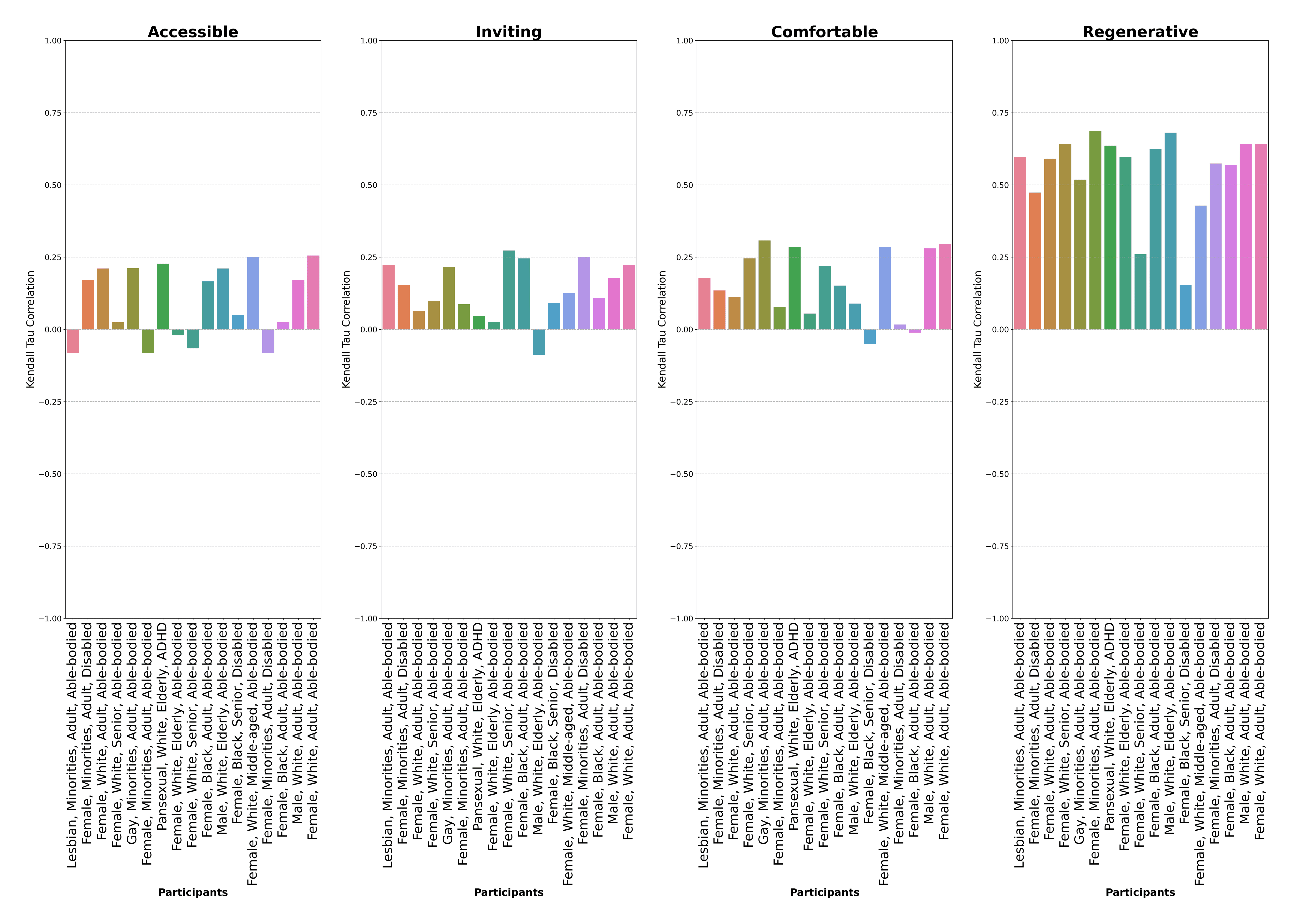}
\caption[Kendall correlations, criteria set 1]{Kendall's $\tau$ rank-correlation values for Group 1 criteria set (accessible, inviting, comfortable, regenerative). \added{Source: author analysis; see \citet{MushkaniJUM2025}.}}
\label{fig:app-streetreview-kendall-1}
\label{fig:kendall-group-1}
\end{figure}

\begin{figure}[!htbp]
\centering
\includegraphics[width=\textwidth]{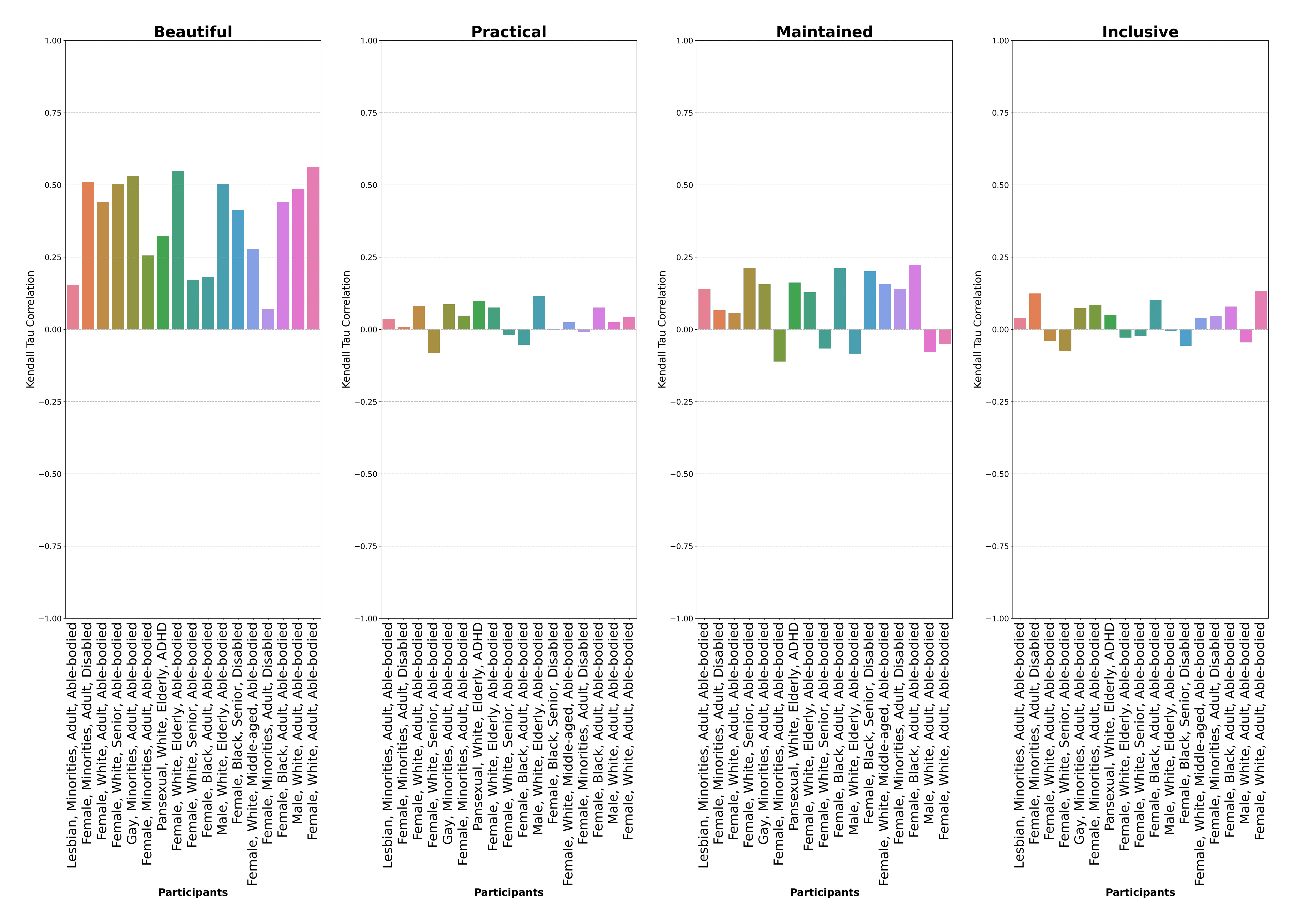}
\caption[Kendall correlations, criteria set 2]{Kendall's $\tau$ rank-correlation values for Group 2 criteria set (beautiful, practical, maintained, inclusive). \added{Source: author analysis; see \citet{MushkaniJUM2025}.}}
\label{fig:app-streetreview-kendall-2}
\label{fig:kendall-group-2}
\end{figure}

\begin{figure}[!htbp]
\centering
\includegraphics[width=\textwidth]{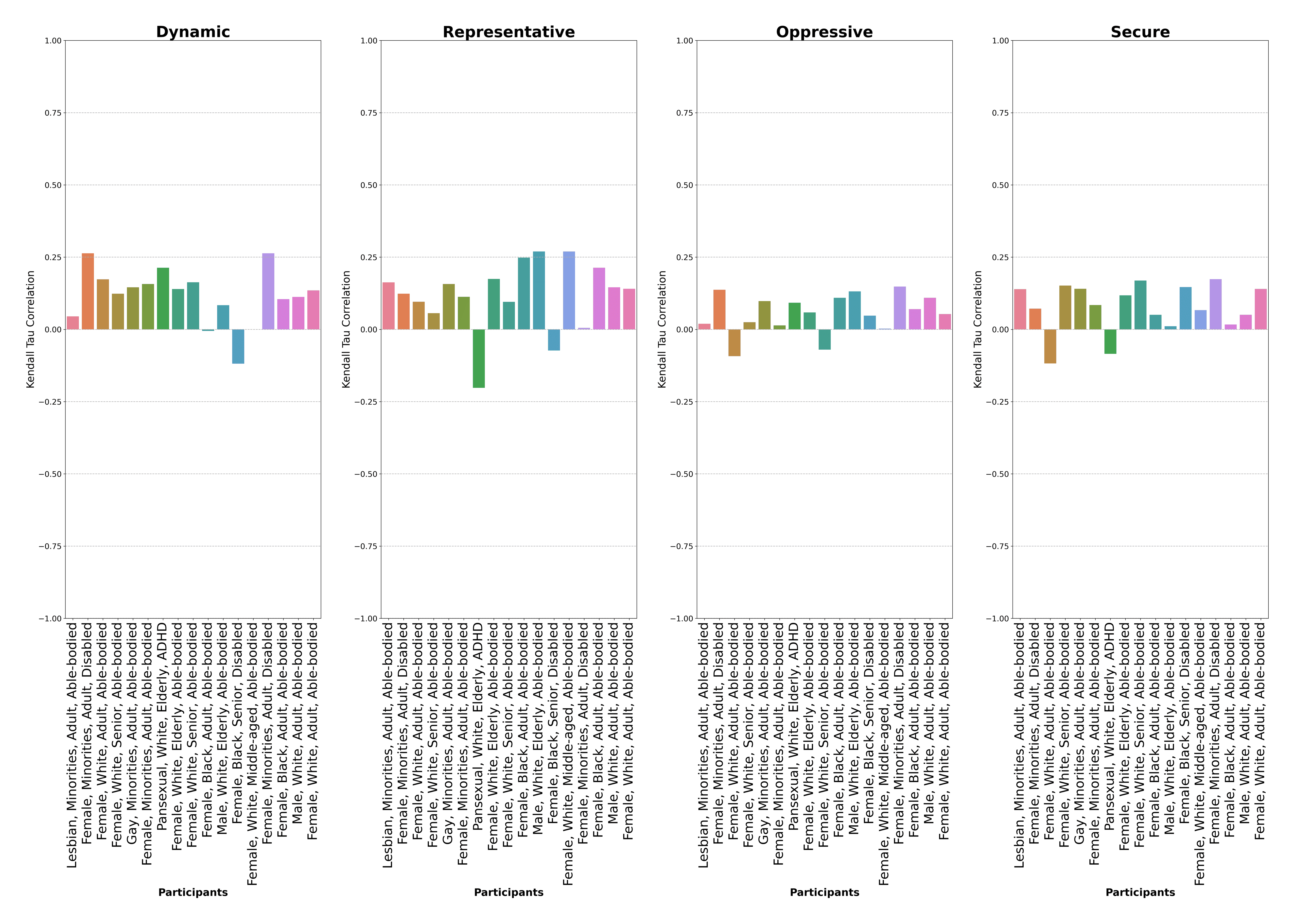}
\caption[Kendall correlations, criteria set 3]{Kendall's $\tau$ rank-correlation values for Group 3 criteria set (dynamic, representative, oppressive, secure). \added{Source: author analysis; see \citet{MushkaniJUM2025}.}}
\label{fig:app-streetreview-kendall-3}
\label{fig:kendall-group-3}
\end{figure}

\FloatBarrier

\fi

\end{document}